\documentclass[12pt]{report}

\usepackage{makeidx,amssymb,xspace}
\usepackage{graphicx}
\usepackage{epsfig,amsfonts,amssymb,amsmath}

\makeindex

\newcommand{\be}{\begin{equation}}\newcommand{\ee}{\end{equation}}
\newcommand{\bea}{\begin{eqnarray}}\newcommand{\eea}{\end{eqnarray}}
\newcommand{\brr}{\begin{array}}\newcommand{\err}{\end{array}}
\newcommand{\bit}{\begin{itemize}}\newcommand{\eit}{\end{itemize}}
\newcommand{\ben}{\begin{enumerate}}\newcommand{\een}{\end{enumerate}}

\def\lab{\label}\def\lan{\langle}\def\lar{\leftarrow}
\def\lf{\left}\def\lrar{\leftrightarrow}
\def\noi{\noindent}\def\non{\nonumber}
\def\pa{\partial}\def\ran{\rangle}\def\rar{\rightarrow}
\def\Rar{\Rightarrow}\def\ri{\right}\def\ti{\tilde}
\def\wti{\widetilde}
\def\al{\alpha}\def\bt{\beta}\def\ga{\gamma}
\def\de{\delta}\def\De{\Delta}\def\ep{\epsilon}
\def\te{\theta}\def\Te{\Theta}\def\la{\lambda}
\def\si{\sigma}\def\om{\omega}
\def\Om{\Omega}
\def\AB{{_{A,B}}}\newcommand{\mlab}[1]{\label{#1}}
\def\CP{{_{C\!P}}}\def\T{{_{T}}}
\def\AB{{_{A,B}}}\def\mass{{_{1,2}}}
\def\flav{{e,\mu}}\def\1{{_{1}}}\def\2{{_{2}}}
\begin{document}
$
\vspace{2cm}
$

 \centerline{\bf \huge Aspects of particle mixing} \vspace{2mm}
\centerline{\bf \huge in Quantum Field Theory}

\vspace{2cm}

\centerline{\bf \large Antonio Capolupo}

\vspace{8mm}

\centerline{Dipartimento di Fisica dell'Universita' di Salerno}
\centerline{and Istituto Nazionale di Fisica Nucleare}
\centerline{I-84100 Salerno, Italy}

% start pagenumbering from 1 in roman numerals

\pagenumbering{roman}

\vspace{2.5cm}
\newpage
\mbox{ }
\newpage
\centerline{\bf \large Abstract} \vspace{6mm}

The results obtained on the particle mixing in Quantum Field
Theory are reviewed.

 The Quantum Field Theoretical
formulation of fermion and boson mixed fields is analyzed in
detail and new oscillation formulas exhibiting corrections with
respect to the usual quantum mechanical ones are presented. It is
proved that the space for the mixed field states is unitary
inequivalent to the state space where the unmixed field operators
are defined. The structure of the currents and charges for the
charged mixed fields is studied. Oscillation formulas for neutral
fields are also derived. Moreover the study some aspects of three
flavor neutrino mixing is presented, particular emphasis is given
to the related algebraic structures and their deformation in the
presence of CP violation. The non-perturbative vacuum structure
associated with neutrino mixing is shown to lead to a non-zero
contribution to the value of the cosmological constant. Finally,
phenomenological aspects of the non-perturbative effects are
analyzed. The systems where this phenomena could be detected are
the $\eta-\eta'$ and $\phi-\omega$ mesons.

\vspace{8mm}

\newpage

\section*{Acknowledgments}
I would like to thank Prof.Giuseppe Vitiello, my advisor, for
introducing me to the Quantum Field Theory of particle mixing. He
motivated me to work at my best while allowing me the freedom to
pursue those areas that interested me the most. I further owe
thanks to Dr.Massimo Blasone for his advice, help and numerous
discussions. I am grateful to have had the opportunity to work
with the nuclear and particle theory group at North Carolina State
University under the leadership of Prof.Chueng-Ryong Ji and I
thank Yuriy Mishchenko of for his collaboration during my
permanence in USA. This work is dedicated to my family and to Ada.

% makes the table of contents automatically

\tableofcontents

% restart page numbering from 1 in arabic

\chapter*{Introduction}
\addcontentsline{toc}{chapter}{Introduction}
\noindent
\setcounter{page}{1} \pagenumbering{arabic}

 The particle mixing and oscillation \cite{Pontecorvo,christenson,Gell,mesons}
 are one of the most important topics in modern Particle Physics:
 these phenomena are
observed when a source create a particle that is a mixture of two
or more mass eigenstate and different mixing are observed in a
detector.

The oscillations are observed experimentally in the meson sector,
for kaons, $B^{0}$, $D^{0}$, and the system $\eta-\eta'$, and the
evidence of the neutrino oscillations seems now certain
\cite{kamiokande,experiments,Kam,SNO,kamland,K2K,solar,solar2,pascoli}.
The Standard Model incorporates the mixing of fermion fields
through the Kobayashi - Maskawa {\cite{KM}} mixing of 3 quark
flavors, i.e. a generalization of the original Cabibbo
{\cite{cabibbo}} mixing matrix between the $d$ and $s$ quarks.
Recently, the mixing phenomenon helped to provide a vital insight
into the puzzle of solar neutrinos \cite{solar} and neutrino
masses \cite{pascoli}. In the boson sector, the mixing of $K^0$
and $\bar{K^0}$ mesons via weak currents provided the first
evidence of $CP$ violation \cite{christenson} and $B^0-\bar B^0$
mixing plays an important role in determining the precise profile
of a CKM \cite{KM,cabibbo} unitarity triangle \cite{jichoi} in
Wolfenstein parameter space \cite{wolfenstein}. In this light,
beyond any doubt, the mixing of flavors at this date is the most
promising phenomenon of the physics beyond the Standard Model.

Regarding the vanishing magnitudes of the expected violation
effects (such as neutrino mass differences or unitarity violation
in CKM matrix), it is imperative that the theoretical aspects of
the quantum mixing were very well understood.

There are still many unanswered questions about the physics of the
mixing and the oscillations in particular from a theoretical point
of view.

Let us remind some basic facts about particle mixing and
oscillations.

In the Standard Model, the neutrinos appear among the fundamental
constituents, together with the corresponding charged leptons and
the quarks. The idea of the neutrino was first introduced by
W.Pauli in $1930$ in order to save the energy and momentum
conservation laws in beta-decay process of atomic nuclei and was
first observed in $1956$ by Reines and Cowen in the process
$\bar{\nu_{e}}+ p \rightarrow n + e^{+}$.

Traditionally, it has been supposed that neutrinos are massless
fermions, and therefore are described by two component Weyl
spinors, but according to recent experimental data
\cite{kamiokande,experiments,Kam,SNO,kamland,K2K,solar,solar2,pascoli},
neutrinos may have a mass. The fact that they are electrically
neutral, makes possible the existence of two different types of
massive neutrinos, namely Dirac or Majorana neutrinos. In the
first case, the massive neutrino would be described by a four
component Dirac spinor, similar to the one describing the
electron. In the case of Majorana neutrino, the spinor has only
two components, since neutrino and antineutrino are identified.

To have oscillations it is necessary that neutrino masses are not
zero and mixing is present, i.e. that the neutrinos belonging to
different generations do have a mixed mass term. Then the time
evolution of a neutrino mixed state would lead to flavor
oscillations, i.e. to a conversion of a neutrino of one flavor
into one of another flavor.

Several questions are still open. For example, the nature of the
neutrino mass (Dirac or Majorana) is not understood; it is not
justified the smallness of the neutrino masses with respect to
those of the other leptons; it is not clear how the mixing arises
and also it is difficult to understand the large mixing angles
necessary to fit the latest experimental data
\cite{kamiokande,experiments,Kam,SNO,kamland,K2K,solar,solar2,pascoli}.

On the other hand, from a theoretical point of view, mixing is
extremely interesting, since it appear to be a fundamental problem
in particle physics.

From a mathematical point of view, the problem in defining
properly the Hilbert space for mixed particles was resolved only
recently \cite{lathuile,currents,BHV98,Blasone:1999jb}.

It is indeed, a problem of unitarily inequivalent representations:
the choice of a proper Hilbert space, is involved when mixed
fields are considered. This is due to the peculiar mathematical
structure of Quantum Field Theory (QFT), where many inequivalent
representations (many different Hilbert space) are allowed for a
given dynamics \cite{Itz,Um1}.

This situation contrast the one of Quantum Mechanics, where, due
to the von Neumann theorem, only one Hilbert space is admitted due
to the finiteness of the number of the degrees of freedom of the
system under consideration.

It was noted, that in QFT the use of the usual perturbation
approach to mixing, where the flavor quantum states are defined in
the Fock space of the mass-eigenstate fields, leads to certain
difficulty \cite{BHV98}. That difficulty originates from the
unitary inequivalence between the quantum states of flavor and the
quantum states of energy \cite{BHV98}. Thus, it was suggested that
the problem may be fixed \cite{BHV98,ABIV96} by defining the
flavor states in a ("flavor") space "different", i.e. unitarily
inequivalent from the mass state space.

The non trivial nature of the mixing transformations manifests
itself also in the case of the mixing of boson fields. The mesons
that oscillates are: $K^{0}-\bar{K}^{0}$, $B^{0}-\bar{B}^{0}$,
$D^{0}-\bar{D}^{0}$, and the system $\eta-\eta'$.

 Extensive study of the mixing phenomenon with the new
definition of the flavor states both for neutrinos and mesons has
been carried out and the general formulation of such theory for
bosons and fermions has been also suggested
\cite{BHV98,Blasone:1999jb,ABIV96,BV95,BCRV01,JM01,fujii1,fujii2,yBCV02,comment,hannabus,JM011}.
Explicit form of the flavor vacuum has been found and it has been
shown that indeed the quantum states of flavor are always
unitarily inequivalent to those of energy.

In the framework of QFT, exact oscillation formulae has been
calculated for neutrinos \cite{BHV98,yBCV02} and bosons
\cite{BCRV01} case and it turned out to have an additional
oscillating term and energy dependent amplitudes, in contrast with
the usual quantum mechanical formulas, (Pontecorvo formula
\cite{Pontecorvo} in the neutrino case, Gell-Mann Pais formula
\cite{Gell} in the boson case), which are however recovered in the
relativistic limit.

Moreover, the non-perturbative vacuum structure associated with
neutrino mixing leads to a non zero contribution to the value of
the cosmological constant \cite{cosmolog2003}.

The thesis is organized in the following way:

Part 1 is devoted to the fermion mixing; in particular, in Chapter
1 we present the neutrino oscillations in Quantum Mechanics and
the Bilenky and Pontecorvo oscillation formulae \cite{Pontecorvo}.

The Quantum Field Theoretical formulation of fermion mixing is
analyzed in Chapter 2, where the unitary inequivalence of the
flavor and mass representations is proved, and the currents and
charges for mixed fields are introduced and used to derive the new
oscillation formulae, exhibiting corrections with respect to the
usual quantum mechanical ones
\cite{lathuile,currents,BHV98,Blasone:1999jb,ABIV96,BV95,fujii1,fujii2}.

The extension of the results to the case of three flavor neutrino
mixing and a study of the algebraic structures and their
deformations in the presence of CP violation
\cite{yBCV02,comment,BCV2003} are given in Chapters 3 and 4, where
by use of the generators of the mixing transformations, are
recovered all the known parametrizations of the three flavor
mixing matrix
\cite{KM,cabibbo,wolfenstein,Chaturvedi,Fritzsch,Maiani,Chau-Keung,Anselm}
and a number of new ones are found \cite{yBCV02}. In Chapter 5 it
is discussed the mixing and the oscillations of Majorana fermions
in Quantum Field Theory \cite{Blaspalm}. In Chapter 6 we show that
the neutrino mixing leads to a non-zero contribution to the value
of the cosmological constant \cite{cosmolog2003}.

In Part 2 the boson mixing is analyzed.

The usual representation of boson oscillations in Quantum
Mechanics, the Gell-Mann Pais model \cite{Gell}, is presented in
Chapter 7.

In Chapter 8 we treat the quantum field theory of boson mixing
\cite{BCRV01,binger,CJMV2003} and we derive the exact oscillation
formulae. In Chapter 9 we analyze phenomenological aspects of the
non-perturbative effect. We argue that the system where the new
oscillation formulae could be best checked are $\eta-\eta'$ and
$\phi-\omega$ mesons \cite{CJMV2003}.

Finally, details of the mathematical formalism are confined to the
Appendices A-K.

The corrections introduced by the present formalism to the usual
Quantum Mechanics formulae are in principle experimentally
testable. The fact that these corrections may be quantitatively
below the experimental accuracy reachable at the present state of
the art in the detection of the neutrino and boson oscillations,
does not justify neglecting them in the analysis of the particle
mixing and oscillation mechanism. The exact oscillation formulae
derived in QFT are the result of a mathematically consistent
analysis which cannot be ignored in a correct treatment of the
field mixing phenomenon. The QFT formalism accounts for all the
known parameterizations of the mixing matrix and explains their
origin and their reciprocal relations, thus unifying the
phenomenological proposals scattered in the literature where such
parameterizations have been presented. Moreover, the QFT formalism
clearly points to the truly non-perturbative character of the
particle mixing phenomenon. A lot of Physics must be there waiting
to be discovered.
\begin{center}

\newpage
\mbox{ }
\newpage

\chapter*{PART 1. FERMION MIXING}
\addcontentsline{toc}{chapter}{PART 1. FERMION MIXING} \noindent

\end{center}
\newpage
\mbox{ }
\newpage
\chapter{ Neutrino oscillations in Quantum Mechanics} \vspace{.4in}

In this chapter we present the theoretical model adopted to
describe the neutrino mixing and oscillations in Quantum
Mechanics. In particular, the neutrino oscillation will be treated
by using the formalism of Bilenky and Pontecorvo
\cite{Pontecorvo}.

\bigskip
\bigskip
\section{Introduction}

In the weak interaction it is assumed that the leptonic numbers
are strictly conserved. Then, the neutrino oscillations, that is
the mutation of the neutrino flavor, may arise if, in addition to
the usual weak interaction, a super-weak interaction which does
not conserve leptonic numbers is also taking place.

In such a case, the neutrino masses are assumed to be different
from zero, and the state vectors of the ordinary electron, muon
and tau neutrinos
 $\nu_{e}$, $\nu_{\mu}$, $\nu_{\tau}$ (flavor states), are
superpositions of the state vectors of neutrinos $\nu _{1}$, $\nu
_{2}$, $\nu _{3}$ (mass states), which propagate with different
frequencies due to different masses $m_{1}$, $m_{2}$, $m_{3}$.
Then, if a beam, of muon neutrinos, for example, is produced in a
weak process, at a certain distance from the production place, the
beam will be a coherent superposition of $\nu_{e}$, $\nu_{\mu}$
and $\nu _{\tau}$, that is, there arise oscillations
$\nu_{e}\leftrightarrows\nu_{\mu}\leftrightarrows\nu_{\tau}$.

The first theories of neutrino oscillations are formulated with
Majorana neutrinos and Dirac neutrinos and consider only two type
of neutrinos, $\nu_{e}$ and $\nu_{\mu}$. In both theories,
possible oscillations $\nu_{e}\leftrightarrows\nu_{\mu}$,
$\bar{\nu}_{e}\leftrightarrows\bar{\nu}_{\mu}$ are described by
identical expressions, in which two parameters are present, the
mixing angle $\theta$ and the difference $\Delta m^{2}\equiv\left|
m_{1}^{2}-m_{2}^{2}\right|$ of neutrino masses squared.

We will consider the theories in which the neutrino oscillations
arise \cite{Pontecorvo}. They are all based on the assumption
that, in addition to the usual weak interaction, there exist also
an interaction which does not conserve the leptonic numbers. In
agreement with experimental data, it is assumed that this
additional interaction is weaker (super-weak) than the usual weak
interaction.

Section 1.2 is devoted to describe the usual weak interaction, in
Sections 1.3 and 1.4, the theories with mixing of Majorana
neutrinos and Dirac neutrinos, respectively, are exposed and in
Section 1.5 the Pontecorvo oscillation formula is presented.

\section{Leptonic numbers}

 The usual weak interaction hamiltonian is
\bea\label{weak} H_{w}=H_{w}^{c}+H_{w}^{0} \eea with \bea
H_{w}^{c}=\frac{G}{\sqrt{2}}J^{\alpha }\bar{J}_{\alpha } \eea
where \bea\label{lepcurr} J_{\alpha }=\left( \bar{\nu }_{eL}\gamma
_{\alpha }e_{L}\right) +\left( \bar{\nu }_{\mu L}\gamma _{\alpha
}\mu _{L}\right) +J_{\alpha }^{h} \eea is the weak charged
current, $J_{\alpha }^{h}$ is the hadronic current and \bea \Psi
_{L}=\frac{1}{2}\left( 1+\gamma ^{5}\right) \Psi \eea is the
left-handed component of the field operator $\Psi$. $ G\simeq
10^{-5}M_{p}^{-2}$ is the weak interaction constant, with $M_{p}$
proton mass.

The second term of Eq.(\ref{weak}), $H_{w}^{0}$ is the neutral
current contribution, the structure of which is of no relevance in
this contest.

Let us remember that the neutral current for neutrinos is \bea
J_{(0)}^{\mu }\left( \nu \right) =\frac{1}{2}\left[ \bar{\nu }\text{ }%
\gamma ^{\mu }\left( 1-\gamma ^{5}\right) \nu \right] \eea and for
electrons \bea J_{(0)}^{\mu }\left( e\right) =\frac{1}{2}\left[
\bar{e}\text{ }\gamma ^{\mu }\left( g_{V}-g_{A\text{ }}\gamma
^{5}\right) e\right] \eea where \bea g_{V}=-1+4\sin ^{2}\theta
_{W} \eea with $\sin ^{2}\theta _{W}=(0.23\pm 0.02)$,  $\theta
_{W}$ is Weinberg angle and $g_{A\text{ }}=-1.$

The $\gamma _{j}$ are the Dirac matrices defined as \bea \gamma
_{j}=\left(
\begin{array}{ll}
0 & -i\sigma _{j} \\
i\sigma _{j} & 0
\end{array}
\right) \eea with $j=1,2,3$, $\sigma_{j}$ Pauli matrices: \bea
\sigma _{1}=\left(
\begin{array}{ll}
0 & 1 \\
1 & 0
\end{array}
\right) ,\text{ \qquad }\sigma _{2}=\left(
\begin{array}{ll}
0 & -i \\
i & 0
\end{array}
\right) ,\text{ \qquad }\sigma _{3}=\left(
\begin{array}{ll}
1 & 0 \\
0 & -1
\end{array}
\right) \eea moreover \bea \gamma _{4}=\left(
\begin{array}{ll}
I & 0 \\
0 & I
\end{array}
\right) \qquad \text{with} \qquad I=\left(
\begin{array}{ll}
1 & 0 \\
0 & 1
\end{array}
\right) , \eea and
\smallskip
\bea \gamma _{5}=\frac{i}{4!}\varepsilon _{\mu \nu \lambda \eta
}\gamma ^{\mu }\gamma ^{\nu }\gamma ^{\lambda }\gamma ^{\eta },
\eea where $\varepsilon _{\mu \nu \lambda \eta }$ is the totally
antisymmetric Levi-Civita tensor defined as

$\varepsilon _{\alpha \beta \gamma \delta }=0$ for any two equal
indices

$\varepsilon _{0123}=1$

$\varepsilon _{\alpha \beta \gamma \delta }$ changes sign under
interchange of two consecutive indices.

The interaction given by the Hamiltonian $H_{w}^{c}$, does
conserve separately the sum of the electron $L_{e},$ and muon
$L_{\mu }$ lepton numbers \bea {\text{ }\sum _{i}}L_{e}^{\left(
i\right) }=constant,\\ {\text{ }\sum_{i} }L_{\mu }^{\left(
i\right) }=constant. \eea

The experiments in which the neutrino oscillations are searched
for will allow one to test the hypothesis on the existence of an
interaction non-conserving the lepton numbers.

\section{Majorana neutrinos}

The first theory of neutrino oscillations was based on two
component neutrino theory \cite{Pontecorvo}.

According to the theory of the Majorana neutrinos oscillations,
only left-handed components of the neutrinos fields \bea \nu
_{eL}=\frac{1+\gamma _{5}}{2}\nu _{e},\qquad \nu _{\mu
L}=\frac{1+\gamma _{5}}{2}\nu _{\mu } \eea and right-handed
components of the antineutrinos fields \bea \nu
_{eR}^{C}=\frac{1-\gamma _{5}}{2}\nu _{e}^{C}=\left( \nu
_{eL}\right) ^{C},\qquad \nu _{\mu R}^{C}=\frac{1-\gamma
_{5}}{2}\nu _{\mu }^{C}=\left( \nu _{\mu L}\right) ^{C} \eea can
appear in the hamiltonian.

Here \bea \nu _{e,\mu }^{C}=C\bar{\nu }_{e,\mu } \eea is the
charge conjugated spinor. The  matrix $C$ satisfies the following
relations: \bea \non C^{\dagger }C &=&1
\\  C\gamma _{\alpha }^{T}C^{-1} &=&-\gamma _{\alpha }
\\ \non  C^{T} &=& -C \eea

The hamiltonian of the interaction which does not conserve the
lepton numbers, quadratic in the neutrino fields, has the
following general form \bea \label{Hsuperweak}
H=m_{\bar{e}e}\bar{\nu }_{eR}^{C}\nu _{eL}+m_{\bar{\mu }\mu }%
\bar{\nu }_{\mu R}^{C}\nu _{\mu L}+m_{\bar{\mu }e}\left( \bar{%
\nu }_{\mu R}^{C}\nu _{eL}+\bar{\nu }_{eR}^{C}\nu _{\mu L}\right)
+h.c. \eea where the parameters $m_{\bar{e}e}$, $m_{\bar{\mu}\mu
}$, $m_{\bar{\mu}e}$ have the dimensions of a mass.

The hamiltonian Eq.(\ref{Hsuperweak}) can be written more
compactly \bea H=\bar{\nu }_{R}^{C} M \nu _{L}+\bar{\nu }_{L}
M^{\dag}\nu _{R}^{C}, \eea where \bea \nu _{L}=\left(
\begin{array}{l}
\nu _{eL} \\
\nu _{\mu L}
\end{array}
\right) ,\text{ \qquad }\nu _{R}^{C}=\left(
\begin{array}{l}
\nu _{eR}^{C} \\
\nu _{\mu R}^{C}
\end{array}
\right) ,\text{ \qquad }M=\left(
\begin{array}{ll}
m_{\bar{e}e} & m_{\bar{\mu }e} \\
m_{\bar{\mu }e} & m_{\bar{\mu }\mu }
\end{array}
\right). \eea

If the interaction Eq.(\ref{Hsuperweak}) is invariant under the
$CP$-transformation, the parameters $m_{\bar{e}e}$, $m_{\bar{\mu
}\mu}$, $m_{\bar{\mu}e}$ are real and \bea M^{\dag}=M,\text{
\qquad }M^{T}=M\Longrightarrow M^{*}=M. \eea In such a case the
interaction hamiltonian can be expressed as follows:
\bea\label{ham} H=\bar{\nu }_{R}^{C} M \left( \nu _{L}+\nu
_{R}^{C}\right) +\bar{\nu }_{L} M \left( \nu _{R}^{C}+\nu
_{L}\right) \eea

\smallskip
To demonstrate Eq.(\ref{ham}) we note that the following equality
holds \bea
H=\bar{\nu }_{R}^{C}M\nu _{L}+\bar{\nu }_{L}M\nu _{R}^{C}=%
\bar{\nu }_{R}^{C}M\left( \nu _{L}+\nu _{R}^{C}\right) +\bar{\nu }%
_{L}M\left( \nu _{R}^{C}+\nu _{L}\right) . \eea Indeed, being \bea
\nu _{L}=\frac{1}{2}\left( 1+\gamma _{5}\right) \nu ,\text{ \qquad \qquad }%
\bar{\nu }_{L}=\frac{1}{2}\bar{\nu }\left( 1-\gamma _{5}\right) ,
\eea we have \bea \bar{\nu }_{L}\nu _{L}=\frac{\bar{\nu
}}{4}\left( 1-\gamma _{5}\right) \left( 1+\gamma _{5}\right) \nu
=0. \eea Likewise
\smallskip
\bea \bar{\nu }_{R}^{C}\nu _{R}^{C}=0. \eea In fact
\smallskip
\bea \left( 1-\gamma _{5}\right) \left( 1+\gamma _{5}\right) =0,
\eea since \bea \gamma _{5}=\left(
\begin{array}{llll}
 0 &  0 & -1 &  0 \\
 0 &  0 &  0 & -1 \\
-1 &  0 &  0 &  0 \\
 0 & -1 &  0 &  0
\end{array}
\right) \Longrightarrow \gamma _{5}\text{ }\gamma _{5}=\left(
\begin{array}{llll}
1 & 0 & 0 & 0 \\
0 & 1 & 0 & 0 \\
0 & 0 & 1 & 0 \\
0 & 0 & 0 & 1
\end{array}
\right) . \eea

Let we put \bea \chi =\nu _{L}+\nu _{R}^{C}=\left(
\begin{array}{l}
\nu _{eL}+\nu _{eR}^{C} \\
\nu _{\mu L}+\nu _{\mu R}^{C}
\end{array}
\right) =\left(
\begin{array}{l}
\chi _{e} \\
\chi _{\mu }
\end{array}
\right) , \eea with $\chi _{e},$ $\chi _{\mu }$ fields of Majorana
neutrinos, the Hamiltonian Eq.(\ref{ham}), then can be written as
\bea\label{ham2} H=\bar{\chi }M\chi , \eea moreover we have \bea
\chi ^{C}=C\bar{\chi }=\chi . \eea

The matrix $M$ in the Eq.(\ref{ham2}) can be diagonalized by using
the orthogonal matrix $U$ ($U^{T}U=1$): \bea\label{U} U=\left(
\begin{array}{ll}
\cos \theta & \sin \theta \\
-\sin \theta & \cos \theta
\end{array}
\right) . \eea We have \bea\label{mas} M=UM_{0}U^{-1} \eea with
\smallskip
\bea M_{0}=\left(
\begin{array}{ll}
m_{1} & 0 \\
0 & m_{2}
\end{array}
\right) . \eea
\smallskip

The expressions relating the masses $m_{1}$, $m_{2}$ and the
mixing angle $\theta $ to the values $m_{\bar{e}e}$, $m_{\bar{\mu
}\mu}$, $m_{\bar{\mu }e}$, are obtained by Eq.(\ref{mas}): \bea
\left(
\begin{array}{ll}
m_{\bar{e}e} & m_{\bar{\mu }e} \\
m_{\bar{\mu }e} & m_{\bar{\mu }\mu }
\end{array}
\right) =\left(
\begin{array}{ll}
\cos \theta & \sin \theta \\
-\sin \theta & \cos \theta
\end{array}
\right) \left(
\begin{array}{ll}
m_{1} & 0 \\
0 & m_{2}
\end{array}
\right) \left(
\begin{array}{ll}
\cos \theta & -\sin \theta \\
\sin \theta & \cos \theta
\end{array}
\right) \eea and \bea m_{\bar{e}e} &=&m_{1}\cos ^{2}\theta
+m_{2}\text{ }\sin ^{2}\theta
\\
m_{\bar{\mu }\mu } &=&m_{1}\sin ^{2}\theta +m_{2}\cos ^{2}\theta
\\
m_{\bar{\mu }e} &=&\left( -m_{1}+m_{2}\right) \sin \theta \cos
\theta. \eea Then

\begin{eqnarray}
tg2\theta &=&\frac{2m_{\bar{\mu }e}}{m_{\bar{\mu }\mu }-m_{%
\bar{e}e}}  \\
m_{1,2} &=&\frac{1}{2}\left[ m_{\bar{e}e}+m_{\bar{\mu }\mu }\pm
\sqrt{\left( m_{\bar{e}e}-m_{\bar{\mu }\mu }\right) ^{2}+4m_{%
\bar{\mu }e}^{2}}\right].
\end{eqnarray}

The oscillations take place if both $\theta \neq 0$ and $m_{1}\neq
m_{2}$.

Moreover for $\theta =\frac{\pi }{4}$ we have the maximum mixing and then $m_{%
\bar{e}e}=m_{\bar{\mu }\mu }$ with $m_{\bar{\mu }e}\neq 0$.

Letting \bea \Phi =U^{T}\chi, \eea the Hamiltonian \bea
H=\bar{\chi }M\chi \eea becomes \bea
H=\bar{\Phi }M_{0}\Phi ={\sum_{\sigma =1,2} }m_{\sigma }%
\bar{\Phi }_{\sigma }\Phi _{\sigma }. \eea

Then $\Phi _{1}$ and $\Phi _{2}$ are the fields of Majorana
neutrinos with masses $m_{1},m_{2}$ respectively.

Being \bea \Phi =U^{T}\chi \Longrightarrow \chi =U\Phi , \eea then
the fields $\nu _{eL},\nu _{\mu L}$ which are present in the
ordinary weak interaction hamiltonian, are connected with the
fields of Majorana neutrinos by the relations \bea \nu
_{eL}={\sum_{\sigma =1,2} }U_{1\sigma }\Phi _{\sigma L,\text{
\qquad \qquad }}\nu _{\mu L}={\sum_{\sigma =1,2} }U_{2\sigma }\Phi
_{\sigma L}. \eea

Thus, in the usual weak interaction hamiltonian there appear
orthogonal superpositions of the fields of Majorana neutrinos, the
mass of which $m_{1}$, $m_{2}$ are not equal to zero and in such a
case there arise the neutrino oscillations. Making use of the
expressions \bea \Phi =U^{T}\chi \text{ \qquad and\qquad }U=\left(
\begin{array}{ll}
\cos \theta & \sin \theta \\
-\sin \theta & \cos \theta
\end{array}
\right) , \eea we have \bea \non \Phi _{1}=\chi _{1}\cos \theta
-\chi _{2}\sin \theta , \\  \Phi _{2}=\chi _{1}\sin \theta +\chi
_{2}\cos \theta , \eea and \bea \non \nu _{eL}&=&\Phi _{1L}\cos
\theta +\Phi _{2L}\sin \theta, \\ \nu _{\mu L}&=& -\Phi _{1L}\sin
\theta +\Phi _{2L}\cos \theta . \eea

From these expressions it is clear that the angle $\theta $
characterizes the degree of mixing of the Majorana fields $\Phi
_{1}$ and $\Phi _{2}.$\smallskip
\smallskip

\section{Dirac neutrinos}

This theory is based on the analogy between leptons and quarks
\cite{Pontecorvo}.

The total charged hadronic current is  \bea \left( J_{\alpha
}^{h}\right) =\left( J_{\alpha }^{h}\right) _{C}+\left( J_{\alpha
}^{h}\right) _{GIM}, \eea where $\left( J_{\alpha }^{h}\right)
_{C}$ is the hadron Cabibbo current and $\left(
J_{\alpha}^{h}\right) _{GIM}$ is the hadron current of the G.I.M.
model.

In particular:
 $\left( J_{\alpha
}^{h}\right) _{C}$ is given by the expression \bea \left(
J_{\alpha }^{h}\right) _{C}=\bar{u}_{L}\gamma _{\alpha
}d_{L}^{^{\prime }} \eea
\smallskip where
\bea d^{\prime }=d\cos \theta _{c}+s\sin \theta _{c},\eea with
$\theta_{c}$ Cabibbo angle, $u$, $d$, $s$ field operators of the
$u$-quarks $\left( Q=\frac{2}{3},\text{ }T_{3}=\frac{1}{2},\text{
}S=0\right)$, $d$-quarks $\left( Q=-\frac{1}{3},\text{
}T_{3}=-\frac{1}{2}, \text{ }S=0\right)$, $s$-quarks $\left(
Q=-\frac{1}{3},\text{ }T=0,\text{ } S=-1\right)$ respectively.
Moreover, the hadron current of the G.I.M. model $\left( J_{\alpha
}^{h}\right) _{GIM}$ is \bea \left( J_{\alpha }^{h}\right)
_{GIM}=\bar{c}_{L}\gamma _{\alpha }s_{L}^{\prime } \eea where \bea
s^{\prime }=-d\sin \theta _{c}+s\cos \theta _{c}, \eea $c$ field
operator of the $c$ quark $ \left( Q=\frac{2}{3},\text{
}T=0,\text{ }S=0,\text{ }C=1\right)$.

The leptonic current has the same structure as the charged hadron
current: \bea\label{lepcurr} J_{\alpha }=\left( \bar{\nu
}_{eL}\gamma _{\alpha }e_{L}\right) +\left( \bar{\nu }_{\mu
L}\gamma _{\alpha }\mu_{L}\right). \eea
\smallskip Only the left-handed field components are present in both
currents.

Comparing the hadron current with the lepton current
Eq.(\ref{lepcurr}), we see an important difference between them:
whereas orthogonal superpositions of the $d$ and $s$ quark fields
are present in the hadronic current, in the lepton current the
electron neutrino and muon neutrino fields appear unmixed, (the
charged hadron current does not conserve strangeness, while the
lepton current is conserving the lepton numbers).

In order to remove this difference, let us assume that there exist
two neutrinos $(\nu _{1}$, $\nu _{2})$ with finite masses
$(m_{1}$, $m_{2})$ and that the operators $\nu _{e}$, $\nu _{\mu}$
are orthogonal combinations of $\nu _{1}$, $\nu _{2}$
\begin{eqnarray}
\non \nu _{e} &=&\nu _{1}\cos \theta +\nu _{2}\sin \theta   \\
\nu _{\mu } &=&-\nu _{1}\sin \theta +\nu _{2}\cos \theta
\end{eqnarray}
where  $\nu _{1}$, $\nu _{2}$ are the fields of Dirac neutrinos
with masses $m_{1}$ and $m_{2}$, $\theta $ is the mixing angle.
Let us note that $\theta$ is different to the Cabibbo angle
$\theta _{c}$.

Two values of the mixing angle are of special significance,
$\theta =0$ and $\theta =\frac{\pi }{4}$. The case $\theta =0$ (no
mixing) corresponds to the theory with strict conservation of the
electron and muon lepton number. The case $\theta =\frac{\pi }{4}$
corresponds to maximum amplitude of oscillations. Thus in the
Dirac theory there is a full analogy between lepton and quarks
weak currents.

In this scheme there is no lepton number distinguishing the two
types of neutrinos
 $\nu _{1}$ and $\nu _{2}$;
the neutrinos $\nu _{1}$ and $\nu _{2}$  differ only in their mass
values (just as the $d$ and $s$ quarks do).

The mass term of the hamiltonian in the Dirac theory has the form
\bea H_{1}=m_{1}\bar{\nu }_{1}\nu _{1}+m_{2}\bar{\nu }_{2}\nu
_{2}. \eea

Let us express $\nu _{1}$ and $\nu _{2}$ through $\nu _{e}$ and
$\nu _{\mu }$. We get \bea H_{1}=m_{ee}\bar{\nu}_{e}\nu
_{e}+m_{\mu \mu }\bar{\nu}_{\mu }\nu _{\mu }+m_{\mu e}(\bar{\nu
}_{\mu }\nu _{e}+\bar{\nu }_{e}\nu _{\mu }) \eea with \bea
m_{ee} &=&m_{1}\cos ^{2}\theta +m_{2}\sin ^{2}\theta   \\
m_{\mu \mu } &=&m_{1}\sin ^{2}\theta +m_{2}\cos ^{2}\theta   \\
m_{\mu e} &=&\left( -m_{1}+m_{2}\right) \sin \theta \cos \theta.
\eea

Where $m_{ee}$ and $m_{\mu \mu }$ are the bare masses of electron
and muon neutrinos.

The term \bea H=m_{\mu e}(\bar{\nu }_{\mu }\nu _{e}+\bar{\nu
}_{e}\nu _{\mu }), \eea in $H_{1}$, does not conserve separately
$L_{e}$ and $L_{\mu}$, but conserve $L_{e}+L_{\mu}$; then $H$ is a
super-weak interaction that does not conserve the leptonic number.

The masses $m_{1}$, $m_{2}$ and the mixing angle $\theta$ are
related to the values $m_{ee}$, $m_{\mu\mu}$, $m_{\mu e}$, by
\begin{eqnarray}
tg2\theta &=&\frac{2m_{\mu e}}{m_{\mu \mu }-m_{ee}}   \\
m_{1,2} &=&\frac{1}{2}\left[ m_{ee}+m_{\mu \mu }\pm \sqrt{\left(
m_{ee}-m_{\mu \mu }\right) ^{2}+4m_{\mu e}^{2}}\right]
\end{eqnarray}

These results are to be compared with the corresponding ones for
the Majorana case in Section 1.3.

We have been considering the simplest theories with mixing of two
neutrinos with finite masses.

In both the theories of Majorana and of Dirac, the neutrino masses
are not equal to zero and the operators of neutrino fields are
present in the hamiltonian in the form of orthogonal combinations.

The hamiltonian of the Majorana theory does not conserve the muon
$L_{\mu}$ and the electron $L_{e}$ lepton numbers. The hamiltonian
of the Dirac theory is conserving the sum $L_{e}+L_{\mu}$. The
neutrino oscillations are described by identical expressions in
both theories. Contrarily to the Dirac theory, the Majorana theory
allows in principle the existence of neutrino-less double beta
decay and of other processes in which $L_{e}+L_{\mu}$ is not
conserved (see Appendix A).

In the Majorana theory every type of neutrino is associated with
two states; in the Dirac theory there are four states for every
neutrino type. In this sense, in the Majorana theory there is no
analogy between leptons and quarks.

\section{Neutrino oscillations}

The theories we have considered above lead to neutrino
oscillations \cite{Pontecorvo}. Let us denote by
 $|\nu _{e}\rangle$, and $|\nu _{\mu}\rangle$
the state vectors of the electron and muon neutrinos with momentum
$\mathbf{p}$ and helicity equal to $-1$. From
\begin{eqnarray*}
\nu _{eL} &=&{\sum_{\sigma =1,2} }U_{1\sigma }\Phi _{\sigma L},%
\text{ \qquad \qquad }\nu _{\mu L}={\sum_{\sigma =1,2} }U_{2\sigma
}\Phi _{\sigma L}\text{ \qquad } \\
\nu _{e} &=&\nu _{1}\cos \theta +\nu _{2}\sin \theta ,\text{
\qquad }\nu _{\mu }=-\nu _{1}\sin \theta +\nu _{2}\cos \theta ,
\end{eqnarray*}
it follows that \bea |\nu _{l}\rangle ={\sum_{\sigma =1,2}
}U_{l\sigma }|\nu _{\sigma }\rangle \eea with $l=e$, $\mu$.

Here $|\nu _{\sigma }\rangle$, with $\sigma =1,2$, is the state
vector of the neutrino with mass $m_{\sigma }$, momentum
$\mathbf{p}$, helicity $-1$. The orthogonal matrix $U$ has the
form of the Eq.(\ref{U}).

The vectors $|\nu_{\sigma }\rangle$ describe both Majorana
neutrinos and Dirac neutrinos.

We have \bea H|\nu _{\sigma }\rangle =E|\nu _{\sigma }\rangle \eea
where $H$ is the total hamiltonian and $E_{\sigma
}=\sqrt{m_{\sigma }^{2}+p^{2}}$.

We also have \bea |\nu _{\sigma }\rangle = {\sum_{l=e,\mu}
}U_{l\sigma }|\nu _{l}\rangle. \eea

Let us consider the behavior of the beam of neutrinos, at initial
time $(t=0)$ such a beam is described by the vector $|\nu
_{l}\rangle$. At time $t$ the state vector of the beam is given by
the expression \bea\label{vector} |\nu _{l}(t)\rangle
=e^{-iHt}|\nu _{l}\rangle ={\sum_{\sigma =1,2} }U_{l\sigma
}e^{-iE_{\sigma }t}|\nu _{\sigma }\rangle . \eea

Then the neutrino beam is not described by a stationary state, but
by a superpositions of stationary states. This happens because
$|\nu _{l}\rangle$ is not an eigenstate of the hamiltonian $H$.

If we denote with \bea a_{\nu _{l^{\prime }};\nu
_{l}}(t)={\sum_{\sigma =1,2} }U_{l\sigma }e^{-iE_{\sigma
}t}U_{l^{\prime }\sigma } \eea the probability amplitude of
finding $\nu _{l^{\prime}}$ at a time $t$ after the generation of
$\nu _{l}$, then we can expand the state vector Eq.(\ref{vector})
in terms of vectors $|\nu _{l^{\prime}}\rangle$ \bea |\nu
_{l}(t)\rangle ={\sum_{l^{\prime }=e,\mu} }a_{\nu _{l^{\prime
}};\nu _{l}}(t)|\nu _{l^{\prime }}\rangle. \eea

We have also \bea a_{\nu _{l^{\prime }};\nu _{l}}(0)={\sum_{\sigma
=1,2} }U_{l\sigma }U_{l^{\prime }\sigma }=\delta _{l^{\prime }l}.
\eea

Clearly in the case $m_{1}\neq m_{2}$ and $U_{l\sigma}\neq
\delta_{l\sigma}$, we have $a_{\nu _{e};\nu _{\mu }}(t)=a_{\nu
_{\mu };\nu e}(t)\neq 0$ that is there arise oscillations $\nu
_{e}\rightleftarrows \nu _{\mu }$.

The probability of transitions $\nu _{l}\rightleftarrows \nu
_{l^{\prime }}$ is given by the expression \bea P_{\nu _{l^{\prime
}}\rightarrow \nu _{l}}(t)=P_{\nu _{l}\rightarrow \nu _{l^{\prime
}}}(t)={\sum_{\sigma ,\sigma ^{\prime }} }U_{l\sigma }U_{l^{\prime
}\sigma }U_{l\sigma ^{\prime }}U_{l^{\prime }\sigma ^{\prime
}}\cos \left( E_{\sigma }-E_{\sigma ^{\prime }}\right) t. \eea

It easy to see that the probabilities $P_{\nu
_{l^{\prime}}\rightarrow \nu _{l}}(t)$ satisfy the relation \bea
{\sum_{l^{\prime }=e,\mu} }P_{\nu _{l^{\prime }}\rightarrow \nu
_{l}}(t)=1. \eea

In the case $p\gg m_{1},$ $m_{2\text{ }}$, which is of interest,
we have \bea
E_{i} &=&\sqrt{p^{2}c^{2}+m_{i}^{2}c^{4}}\approxeq pc\left( 1+\frac{%
m_{i}^{2}c^{2}}{2p^{2}}\right) =pc+\frac{m_{i}^{2}c^{3}}{2p}\text{
,\qquad
and setting \quad}c=1:  \nonumber \\
E_{i} &=&p+\frac{m_{i}^{2}}{2p}\Longrightarrow E_{1}-E_{2}=\frac{%
m_{1}^{2}-m_{2}^{2}}{2p}. \eea

Making use of expressions of $U$ and $P_{\nu _{l^{\prime
}}\rightarrow \nu _{l}}(t)$, and indicating with \bea L=\frac{4\pi
p}{\left| m_{1}^{2}-m_{2}^{2}\right| } \eea the oscillation
length, the probability of finding $\nu _{l^{\prime}}$ at a
distance $R$ from a source of $\nu _{l}$ is \bea P_{\nu
_{e}\rightarrow \nu e}(R) &=&P_{\nu _{\mu }\rightarrow \nu _{\mu
}}(R)=1-\frac{1}{2}\sin ^{2}2\theta \text{ }\left( 1-\cos \frac{2\pi R}{L}%
\right) ,  \\
P_{\nu _{e}\rightarrow \nu _{\mu }}(R) &=&P_{\nu _{\mu
}\rightarrow \nu
_{e}}(R)=\frac{1}{2}\sin ^{2}2\theta \text{ }\left( 1-\cos \frac{2\pi R}{L}%
\right) . \eea \qquad \qquad \qquad \qquad \qquad

\smallskip Which can be expressed also as \cite{Pontecorvo}
\bea\label{Pontecorvoee} P_{\nu _{e}\rightarrow \nu e}(t)
&=&P_{\nu _{\mu }\rightarrow \nu _{\mu
}}(t)=1-\sin ^{2}2\theta \text{ }\sin ^{2}\left( \frac{\Delta \omega }{2}%
t\right) ,   \\\label{Pontecorvoem} P_{\nu _{e}\rightarrow \nu
_{\mu }}(t) &=&P_{\nu _{\mu }\rightarrow \nu
_{e}}(t)=\sin ^{2}2\theta \text{ }\sin ^{2}\left( \frac{\Delta \omega }{2}%
t\right) , \eea where $\Delta \omega =\omega_{1}-\omega_{2}$, is
the energy difference of the mass components. The probability
conservation is satisfied \bea P_{\nu _{e}\rightarrow \nu
_{e}}(t)+P_{\nu _{e}\rightarrow \nu _{\mu }}(t)=1. \eea The
Eqs.(\ref{Pontecorvoee}), (\ref{Pontecorvoem}) are the
Bilenky-Pontecorvo formulas for neutrino oscillations.
\bigskip
\bigskip
\chapter{Quantum field theory of fermion mixing}

\vspace{.4in}

The fermion mixing transformations are studied in the quantum
field theory framework. In particular neutrino mixing is
considered and the Fock space of definite flavor states is shown
to be unitarily inequivalent to the Fock space of definite mass
states. We study the structure of the currents and charges for the
mixed fermion fields and the flavor oscillation formula is
computed for two flavors mixing, the oscillation amplitude is
found to be momentum dependent. The flavor vacuum state exhibits
the structure of $SU(2)$ generalized coherent state.

\section{Introduction}

The purpose of the present chapter is the study of the quantum
field theory (QFT) framework of the fermion mixing
transformations, thus focusing the attention on the theoretical
structure of fermion mixing.

We will consider, in particular, neutrino mixing transformations
and this analysis will lead to some modifications of the quantum
mechanics neutrino oscillation formulas.

The chapter is organized as follows. In Section 2.2 we study the
generator of the Pontecorvo neutrino mixing transformations (two
flavors mixing for Dirac fields). We show that in the
Lehmann-Symanzik-Zimmermann (LSZ) formalism of quantum field
theory \cite{Itz,Um1} the Fock space of the flavor states is
unitarily inequivalent to the Fock space of the mass eigenstates
in the infinite volume limit \cite{BV95}. The flavor states are
obtained as condensate of massive neutrino pairs and exhibit the
structure of $SU(2)$ coherent states \cite{Per}. In Section 2.3 we
generalize the mixing transformations, indeed we expand the flavor
fields in a basis with arbitrary masses \cite{fujii1,fujii2}. In
Section 2.4 we analyze the structure of currents for mixed fermion
fields \cite{currents} and in Section 2.5 we derive the neutrino
flavor oscillation formulae \cite{BHV98,Blasone:1999jb}. Finally,
the Section 2.6 is devoted to the conclusions. We confine
mathematical details to the Appendices B, C, D.

\section{The vacuum structure for fermion mixing}

We consider the Pontecorvo mixing relations \cite{Pontecorvo},
although the following discussion applies to any Dirac fields.

The mixing relations are \bea\label{mix} \nu _{e}(x) &=&\nu
_{1}(x)\cos \theta +\nu _{2}(x)\sin \theta \\ \non \nu _{\mu }(x)
&=&-\nu _{1}(x)\sin \theta +\nu _{2}(x)\cos \theta \eea where
$\nu_{e}(x)$ and $\nu_{\mu}(x)$ are the Dirac neutrino fields with
definite flavors. $\nu_{1}(x)$ and $\nu_{2}(x)$ are the free
 neutrino
fields with definite masses $m_{1}$ and $m_{2}$, respectively. The
fields $\nu_{1}(x)$ and $\nu_{2}(x)$ are written as
\bea\label{freefi} \nu _{i}(x)=\frac{1}{\sqrt{V}}{\sum_{{\bf k} ,
r}} \left[ u^{r}_{{\bf k},i}(t) \al^{r}_{{\bf k},i}(t) +
v^{r}_{-{\bf k},i}(t) \bt^{r\dag}_{-{\bf k},i}(t) \ri] e^{i {\bf
k}\cdot{\bf x}},\text{ \qquad \qquad }i=1,2 \eea with

\bea u_{{\bf k},i}^{r}(t)=e^{-i\omega _{k,i}t}u_{{\bf
k},i}^{r}\qquad \text{and}\qquad v_{{\bf k},i}^{r}(t)=e^{i\omega
_{k,i}t}v_{{\bf k},i}^{r}, \eea
\[
u_{{\bf k},i}^{1}=\left( \frac{\omega _{k,i}+m_{i}}{2\omega
_{k,i}} \right) ^{\frac{1}{2}}\left(
\begin{array}{l}
1 \\
0 \\
\frac{k_{3}}{\omega _{k,i}+m_{i}} \\
\frac{k_{1}+ik_{2}}{\omega _{k,i}+m_{i}}
\end{array}
\right) ;\text{ \qquad }u_{{\bf k},i}^{2}=\left( \frac{\omega _{k,i}+m_{i}%
}{2\omega _{k,i}}\right) ^{\frac{1}{2}}\left(
\begin{array}{l}
0 \\
1 \\
\frac{k_{1}-ik_{2}}{\omega _{k,i}+m_{i}} \\
\frac{-k_{3}}{\omega _{k,i}+m_{i}}
\end{array}
\right)
\]
\[
v_{-{\bf k},i}^{1}=\left( \frac{\omega _{k,i}+m_{i}}{2\omega _{k,i}}%
\right) ^{\frac{1}{2}}\left(
\begin{array}{l}
\frac{-k_{3}}{\omega _{k,i}+m_{i}} \\
\frac{-k_{1}-ik_{2}}{\omega _{k,i}+m_{i}} \\
1 \\
0
\end{array}
\right) ;\text{ \qquad }v_{-{\bf k} ,i}^{2}=\left( \frac{\omega
_{k,i}+m_{i}}{2\omega _{k,i}}\right) ^{\frac{1}{2}}\left(
\begin{array}{l}
\frac{-k_{1}+ik_{2}}{\omega _{k,i}+m_{i}} \\
\frac{k_{3}}{\omega _{k,i}+m_{i}} \\
0 \\
1
\end{array}
\right)
\]
and

\bea \omega _{k,i}=\sqrt{{\bf k}^{2} + m_{i}^{2}}. \eea

The operator
 $\alpha ^{r}_{{\bf k},i}$ and $ \beta ^{r }_{{\bf k},i}$, $
i=1,2 \;, \;r=1,2$ are the annihilator operators for the vacuum
state $|0\rangle_{1,2}\equiv|0\rangle_{1}\otimes |0\rangle_{2}$:
$\alpha ^{r}_{{\bf k},i}|0\rangle_{12}= \beta ^{r }_{{\bf
k},i}|0\rangle_{12}=0$.

 The anticommutation relations are:
\bea\label{neutcomm} \left\{ \nu _{i}^{\alpha }(x),\nu _{j}^{\beta
\dagger }(y)\right\} _{t=t^{\prime }}=\delta ^{3}({\bf x-y})\delta
_{\alpha \beta } \delta _{ij}, \text{ \qquad }\alpha ,\beta =
1,...4, \eea \bea \left\{ \alpha _{{\bf k},i}^{r},\alpha _{{\bf
q},j}^{s\dagger }\right\} =\delta _{{\bf kq}}\delta _{rs}\delta
_{ij};\text{ \qquad }\left\{ \beta _{{\bf k},i}^{r},\beta _{{\bf
q,}j}^{s\dagger }\right\} =\delta _{{\bf kq}}\delta _{rs}\delta
_{ij}, \text{ \qquad } i,j=1,2. \eea

All other anticommutators are zero. The orthonormality and
completeness relations are
\begin{eqnarray}\label{orthocompl}
u_{{\bf k},i}^{r\dagger }u_{{\bf k},i}^{s} = v_{{\bf
k},i}^{r\dagger }v_{{\bf k},i}^{s}&=& \delta _{rs},\text{\qquad }
\\
u_{{\bf k},i}^{r\dagger }v_{-{\bf k},i}^{s} = v_{-{\bf k}
,i}^{r\dagger }u_{{\bf k},i}^{s}&=&0,   \\
\text{ }\sum_{r}(u_{{\bf k},i}^{r}u_{{\bf k},i}^{r\dagger
}+v_{-{\bf k},i}^{r}v_{-{\bf k},i}^{r\dagger }) &=& 1.
\end{eqnarray}
The Eqs.(\ref{mix}) relate the respective hamiltonians $H_{1,2}$
(we consider only the mass terms) and $H_{e,\mu}$
\cite{Pontecorvo}:
\bea H_{1,2}=m_{1}\;\nu^{\dag}_{1} \nu_{1} + m_{2}\;\nu^{\dag}_{2}
\nu_{2} \eea \bea
 H_{e,\mu}=m_{ee}\;\nu^{\dag}_{e} \nu_{e} +
m_{\mu\mu}\;\nu^{\dag}_{\mu} \nu_{\mu}+
m_{e\mu}\left(\nu^{\dag}_{e} \nu_{\mu} + \nu^{\dag}_{\mu}
\nu_{e}\right) \eea
where \bea m_{ee}&=& m_{1}\cos^{2}\theta + m_{2} \sin^{2}
\theta,\\
m_{\mu\mu}&=& m_{1}\sin^{2}\theta + m_{2} \cos^{2} \theta,\\
m_{e\mu}&=&(m_{2}-m_{1})\sin\theta \cos \theta. \eea

In QFT the basic dynamics, i.e. the Lagrangian and the resulting
field equations, is given in terms of Heisenberg (or interacting)
fields. The physical observables are expressed in terms of
asymptotic in- (or out-) fields, also called physical or free
fields. In the LSZ formalism of QFT \cite{Itz,Um1}, the free
fields, say for definitiveness the in-fields, are obtained by the
weak limit of the Heisenberg fields for time $t \rightarrow -
\infty$. The meaning of the weak limit is that the realization of
the basic dynamics in terms of the in-fields is not unique so that
the limit for $t \rightarrow -
 \infty$ (or $t \rightarrow + \infty$ for the out-fields) is representation
dependent. The representation dependence of the asymptotic limit
arises from the existence in QFT of infinitely many unitarily
non-equivalent representations of the canonical (anti-)commutation
relations \cite{Itz,Um1}. Of course, since observables are
described in terms of asymptotic fields, unitarily inequivalent
representations describe different, i.e. physically inequivalent,
phases. It is therefore of crucial importance, in order to get
physically meaningful results, to investigate with much care the
mapping among Heisenberg or interacting fields and free fields.
Such a mapping is usually called the Haag expansion or the
dynamical map \cite{Itz,Um1}. Only in a very rude and naive
approximation we may assume that interacting fields and free
fields share the same vacuum state and the same Fock space
representation.

We stress that the above remarks apply to QFT, namely to systems
with infinite number of degrees of freedom. In quantum mechanics,
where finite volume systems are considered, the von Neumann
theorem ensures that the representations of the canonical
commutation relations are each other unitary equivalent and no
problem arises with uniqueness of the asymptotic limit. In QFT,
however, the von Neumann theorem does not hold and much more
careful attention is required when considering any mapping among
interacting and free fields \cite{Itz,Um1}.

With this warnings, mixing relations such as the relations
Eqs.(\ref{mix}) deserve a careful analysis. In fact we will
investigate the structure of the Fock spaces ${\cal H}_{1,2}$ and
${\cal H}_{e,\mu}$ relative to $\nu_{1}(x)$, $\nu_{2}(x)$ and
$\nu_{e}(x)$,  $\nu_{\mu}(x)$, respectively. In particular we will
study the relation among these spaces in the infinite volume
limit. We expect that ${\cal H}_{1,2}$ and ${\cal H}_{e,\mu}$
become orthogonal in such a limit, since they represent the
Hilbert spaces for free and interacting fields, respectively
\cite{Itz,Um1}. In the following, we will perform all computations
at finite volume $V$ and only at the end we will put $V
\rightarrow \infty$.

We construct the generator for the mixing transformation
Eqs.(\ref{mix}) and define \cite{BV95}:
\bea \label{mixG} \nu_{e}^{\alpha}(x) = G^{-1}_{\bf \te}(t)\;
\nu_{1}^{\alpha}(x)\; G_{\bf \te}(t) \\ \non \nu_{\mu}^{\alpha}(x)
= G^{-1}_{\bf \te}(t)\; \nu_{2}^{\alpha}(x)\; G_{\bf \te}(t) \eea
where $G_{\bf \te}(t)$ is given by
\bea\label{generator12} G_{\bf \te}(t) = exp\left[\theta \int
d^{3}{\bf x} \left(\nu_{1}^{\dag}(x) \nu_{2}(x) -
\nu_{2}^{\dag}(x) \nu_{1}(x) \right)\right]\;, \eea
 and is, at finite volume, an unitary
operator: $G^{-1}_{\bf \te}(t)=G_{\bf -\te}(t)=G^{\dag}_{\bf
\te}(t)$, preserving the canonical anticommutation relations
Eqs.(\ref{neutcomm}). To obtain the Eq.(\ref{generator12}), we
observe that, from Eqs.(\ref{mixG}), \bea
d^{2}\nu^{\alpha}_{e}/d\theta^{2}=-\nu^{\alpha}_{e}\;,\;\;\;
d^{2}\nu^{\alpha}_{\mu}/d\theta^{2}=-\nu^{\alpha}_{\mu},\eea
by using the initial conditions \bea
\nu^{\alpha}_{e}|_{\theta=0}=\nu^{\alpha}_{1},\quad
d\nu^{\alpha}_{e}/d\theta|_{\theta=0}=\nu^{\alpha}_{2}\quad
\text{and}
\quad\nu^{\alpha}_{\mu}|_{\theta=0}=\nu^{\alpha}_{2},\quad
d\nu^{\alpha}_{\mu}/d\theta|_{\theta=0}=-\nu^{\alpha}_{1},\eea the
operator $G_{\bf \te}(t)$ generates Eqs.(\ref{mix}).

By introducing the operators
\bea S_{+}(t) \equiv  \int d^{3}{\bf x} \; \nu_{1}^{\dag}(x)
\nu_{2}(x) \;\;,\;\;\; S_{-}(t) \equiv  \int d^{3}{\bf x} \;
\nu_{2}^{\dag}(x) \nu_{1}(x)\;= \left( S_{+}\right)^{\dag}\;, \eea
$G_{\bf \te}(t)$ can be written as \bea G_{\bf \te}(t) =
exp[\theta(S_{+} - S_{-})]\;. \eea
Introducing $S_{3}$ and the Casimir operator $S_{0}$ (proportional
to the total charge) as follows
\bea S_{3} \equiv \frac{1}{2} \int d^{3}{\bf x}
\left(\nu_{1}^{\dag}(x)\nu_{1}(x) -
\nu_{2}^{\dag}(x)\nu_{2}(x)\right)\;, \eea
\smallskip
\bea S_{0} \equiv \frac{1}{2} \int d^{3}{\bf x}
\left(\nu_{1}^{\dag}(x)\nu_{1}(x) +
\nu_{2}^{\dag}(x)\nu_{2}(x)\right)\;, \eea
the $su(2)$ algebra is closed:
\bea\label{su(2)alg} [S_{+}(t) , S_{-}(t)]=2S_{3} \;\;\;,\;\;\;
[S_{3} , S_{\pm}(t) ] = \pm S_{\pm}(t) \;\;\;,\;\;\;[S_{0} ,
S_{3}]= [S_{0} , S_{\pm}(t) ] = 0\;. \eea

The momentum expansion of $S_{+}(t)$, $S_{-}(t)$, $S_{3}$, and
$S_{0}$, is given by using the Eq.(\ref{freefi}):

\begin{eqnarray}
\label{s+} S_{+}(t) &\equiv &{\sum_{{\bf k}} }S_{+}^{{\bf k}}(t)=%
{\sum_{{\bf k}} }{\sum_{r,s}}\left( u_{{\bf k},1}^{r\dagger
}(t)u_{{\bf k},2}^{s}(t)\alpha _{{\bf k},1}^{r\dagger }\alpha
_{{\bf k},2}^{s}+v_{-{\bf k},1}^{r\dagger}(t)u_{{\bf k},
2}^{s}(t)\beta _{-{\bf k},1}^{r}\alpha _{{\bf k},2}^{s}+\right.
\nonumber \\
&&\left. +u_{{\bf k},1}^{r\dagger}(t)v_{-{\bf k},2}^{s}(t)\alpha
_{{\bf k},1}^{r\dagger }\beta_{-{\bf k},2}^{s\dagger}+v_{-{\bf
k},1}^{r\dagger } (t)v_{-{\bf k},2}^{s}(t)\beta _{-{\bf
k},1}^{r}\beta _{-{\bf k},2}^{s\dagger }\right) ,
\end{eqnarray}
\begin{eqnarray}
\label{s-} S_{-}(t) &\equiv &{\sum_{{\bf k}} }S_{-}^{{\bf k}}(t)=%
{\sum_{{\bf k}}}{\sum_{r,s} }\left( u_{{\bf k}, 2}^{r\dagger
}(t)u_{{\bf k},1}^{s}(t)\alpha _{{\bf k},2}^{r\dagger }\alpha
_{{\bf k},1}^{s}+v_{-{\bf k},2}^{r\dagger }(t)u_{{\bf k},
1}^{s}(t)\beta _{-{\bf k},2}^{r}\alpha _{{\bf k},1}^{s}+\right.
\nonumber \\
&&\left. +u_{{\bf k},2}^{r\dagger }(t)v_{-{\bf k},1}^{s}(t)\alpha
_{{\bf k}, 2}^{r\dagger }\beta_{-{\bf k},1}^{s\dagger }+v_{-{\bf
k},2}^{r\dagger }(t)v_{-{\bf k},1}^{s}(t)\beta _{-{\bf
k},2}^{r}\beta _{-{\bf k},1}^{s\dagger }\right) ,
\end{eqnarray}
\bea S_{3}\equiv {\sum_{{\bf k}} }S_{3}^{{\bf k}}=\frac{1}{2}
{\sum_{{\bf k},r} }\left( \alpha _{{\bf k},1}^{r\dagger }\alpha
_{{\bf k},1}^{r}-\beta _{-{\bf k},1}^{r\dagger }\beta _{-{\bf k},
1}^{r}-\alpha _{{\bf k},2}^{r\dagger }\alpha _{{\bf k},
2}^{r}+\beta _{-{\bf k},2}^{r\dagger }\beta _{-{\bf
k},2}^{r}\right) , \eea \bea
S_{0}\equiv {\sum_{{\bf k}} }S_{0}^{{\bf k}}=\frac{1}{2}%
{\sum_{{\bf k},r} }\left( \alpha _{{\bf k},1}^{r\dagger }\alpha
_{{\bf k},1}^{r}-\beta _{-{\bf k},1}^{r\dagger }\beta _{-{\bf
k},1}^{r}+\alpha _{{\bf k},2}^{r\dagger }\alpha _{{\bf k},
2}^{r}-\beta _{-{\bf k},2}^{r\dagger }\beta _{-{\bf
k},2}^{r}\right) . \eea

We observe that the operatorial structure of Eqs.(\ref{s+}) and
(\ref{s-}) is the one of the rotation generator and of the
Bogoliubov generator. Using these expansions it is easy to show
that the $su(2)$ algebra does hold for each momentum component:
\bea\non
\left[ S_{+}^{\mathbf{k}}(t),S_{-}^{\mathbf{k}}(t)\right] =2S_{3}^{\mathbf{k}%
},\text{ \quad }\left[ S_{3}^{\mathbf{k}}(t),S_{\pm }^{\mathbf{k}%
}(t)\right] =\pm S_{\pm }^{\mathbf{k}}(t),\text{ \quad }\left[ S_{0}^{%
\mathbf{k}},S_{3}^{\mathbf{k}}\right] =\left[ S_{0}^{\mathbf{k}},S_{\pm }^{%
\mathbf{k}}\right] =0, \\ \label{su(2)k1}\eea \bea\label{su(2)k}
\left[ S_{\pm }^{\mathbf{k}}(t),S_{\pm }^{\mathbf{p}}(t)\right]
=\left[
S_{3}^{\mathbf{k}}(t),S_{\pm }^{\mathbf{p}}(t)\right] =\left[ S_{3}^{\mathbf{%
k}},S_{3}^{\mathbf{p}}\right] =0,\text{ \qquad }\mathbf{k\neq p.}
\eea

This means that the original $su(2)$ algebra given in
Eqs.(\ref{su(2)alg}) splits into ${\bf k}$ disjoint $su_{{\bf
k}}(2)$ algebras, given by Eqs.(\ref{su(2)k1}), i.e. we have the
group structure $ \bigotimes_{{\bf k}} SU_{{\bf k}}(2)$.

To establish the relation between the Hilbert spaces for free
fields ${\cal H}_{1,2}$ and interacting fields ${\cal H}_{e,\mu}$
we consider the generic matrix element $_{1,2}\langle
a|\nu^{\alpha}_{1}(x)|b \rangle_{1,2}$ (a similar argument holds
for $\nu^{\alpha}_{2}(x)$), where $|a \rangle_{1,2}$ is the
generic element of ${\cal H}_{1,2}$. Using the inverse of the
first of the Eqs.(\ref{mixG}), we obtain:
\bea \label{gena} \;_{1,2}\langle a|G_{\bf \te}(t)\;
\nu^{\alpha}_{e}(x)\; G^{-1}_{\bf \te}(t) |b \rangle_{1,2}\;
=\;_{1,2}\langle a|\nu^{\alpha}_{1}(x)|b \rangle_{1,2}\;. \eea

Since the operator field $\nu_{e}$ is defined on the Hilbert space
${\cal H}_{e,\mu}$, Eq.(\ref{gena}) shows that $G^{-1}_{\bf
\te}(t) |a \rangle_{1,2}$ is a vector of ${\cal H}_{e,\mu}$, so
$G^{-1}_{\bf \te}(t)$  maps ${\cal H}_{1,2}$ to  ${\cal
H}_{e,\mu}$: \bea G^{-1}_{\bf \te}(t): {\cal H}_{1,2} \mapsto
{\cal H}_{e,\mu}.\eea In particular for the vacuum $|0
\rangle_{1,2}$ we have, at finite volume $V$:
\bea\label{flavvac} |0(t) \rangle_{e,\mu} = G^{-1}_{\bf \te}(t)\;
|0 \rangle_{1,2}\;. \eea
$|0 \rangle_{e,\mu}$ is the vacuum for ${\cal H}_{e,\mu}$, which
we will refer to as the flavor vacuum. Due to the linearity of
$G_{\bf \te}(t)$, we can define the flavor annihilators, relative
to the fields $\nu_{e}(x)$ and $\nu_{\mu}(x)$ at each time
expressed as \bea \non \alpha _{{\bf k},e}^{r}(t)\;|0(t)\rangle
_{e,\mu} &=& G^{-1}_{\bf \te}(t)\;\alpha _{{\bf
k},1}^{r}\;|0\rangle _{1,2}=0,
\\ \non
\alpha _{{\bf k},\mu }^{r}(t)\;|0(t)\rangle _{e,\mu }&=&
G^{-1}_{\bf \te}(t)\;\alpha _{{\bf k},2}^{r}\;|0\rangle _{1,2}=0,
\\
\beta _{{\bf k},e}^{r}(t)\;|0(t)\rangle _{e,\mu }&=& G^{-1}_{\bf
\te}(t)\;\beta _{{\bf k},1}^{r}\;|0\rangle _{1,2}=0,
\\ \non
\beta _{{\bf k},\mu }^{r}(t)\;|0(t)\rangle _{e,\mu }&=&
G^{-1}_{\bf \te}(t)\;\beta _{{\bf k},2}^{r}\;|0\rangle _{1,2}=0,
\eea in the following way \begin{eqnarray}\label{flavannich}
\alpha _{{\bf k},e}^{r}(t) &\equiv &G^{-1}_{\bf \te}(t)\;\alpha
_{{\bf k},1}^{r}\;G_{\bf \te}(t),  \nonumber \\
\alpha _{{\bf k},\mu }^{r}(t) &\equiv &G^{-1}_{\bf \te}(t)\;
\alpha _{{\bf k},2}^{r}\;G_{\bf \te}(t),
\nonumber \\
\beta _{{\bf k},e}^{r}(t) &\equiv &G^{-1}_{\bf \te}(t)\;\beta
_{{\bf
k},1}^{r}\;G_{\bf \te}(t),   \\
\beta _{{\bf k},\mu }^{r}(t) &\equiv &G^{-1}_{\bf \te}(t)\;\beta
_{{\bf k},2}^{r}\;G_{\bf \te}(t).  \nonumber
\end{eqnarray}

The flavor fields are then rewritten into the form:
\begin{eqnarray}
\nu _{e}({\bf x},t) &=&\frac{1}{\sqrt{V}}{\sum_{{\bf k},r} }
e^{i{\bf k.x}}\left[ u_{{\bf k},1}^{r}(t)\alpha _{{\bf
k},e}^{r}(t)+v_{-{\bf k},1}^{r}(t)\beta _{-{\bf k},e}^{r\dagger
}(t)\right] ,   \\
\nu _{\mu }({\bf x},t) &=&\frac{1}{\sqrt{V}}{\sum_{{\bf
k},r}}e^{i{\bf k.x}}\left[ u_{{\bf k},2}^{r}(t)\alpha _{{\bf
k},\mu }^{r}(t)+v_{-{\bf k},2}^{r}(t)\beta _{-{\bf
k},\mu}^{r\dagger }(t)\right] ,
\end{eqnarray}
i.e. they can be expanded in the same bases as $\nu_{1}$ and
$\nu_{2}$, respectively.

The flavor annihilation operators can be calculated explicitly, we
have

\bea \non \alpha^{r}_{{\bf k},e}(t)&=&\cos\theta\;\alpha^{r}_{{\bf
k},1}\;+\;\sin\theta\;\sum_{s}\left[u^{r\dag}_{{\bf k},1}(t)
u^{s}_{{\bf k},2}(t)\; \alpha^{s}_{{\bf k},2}\;+\; u^{r\dag}_{{\bf
k},1}(t) v^{s}_{-{\bf k},2}(t)\; \beta^{s\dag}_{-{\bf k},2}\right]
\\ \non
\alpha^{r}_{{\bf k},\mu}(t)&=&\cos\theta\;\alpha^{r}_{{\bf
k},2}\;- \;\sin\theta\;\sum_{s}\left[u^{r\dag}_{{\bf k},2}(t)
u^{s}_{{\bf k},1}(t)\; \alpha^{s}_{{\bf k},1}\;+\; u^{r\dag}_{{\bf
k},2}(t) v^{s}_{-{\bf k},1}(t)\; \beta^{s\dag}_{-{\bf k},1}\right]
\\ \non
\beta^{r}_{-{\bf k},e}(t)&=&\cos\theta\;\beta^{r}_{-{\bf k},1}\;+
\;\sin\theta\;\sum_{s}\left[v^{s\dag}_{-{\bf k},2}(t) v^{r}_{-{\bf
k},1}(t)\; \beta^{s}_{-{\bf k},2}\;+\; u^{s\dag}_{{\bf k},2}(t)
v^{r}_{-{\bf k},1}(t)\; \alpha^{s\dag}_{{\bf k},2}\right]
\\ \non
\beta^{r}_{-{\bf k},\mu}(t)&=&\cos\theta\;\beta^{r}_{-{\bf
k},2}\;- \;\sin\theta\;\sum_{s}\left[v^{s\dag}_{-{\bf k},1}(t)
v^{r}_{-{\bf k},2}(t)\; \beta^{s}_{-{\bf k},1}\;+\;
u^{s\dag}_{{\bf k},1}(t) v^{r}_{-{\bf k},2}(t)\;
\alpha^{s\dag}_{{\bf k},1}\right].
\\\label{annih1} \eea

Without loss of generality, we can choose  the reference frame
such that ${\bf k}=(0,0,|{\bf k}|)$. This implies that only the
products of wave functions with $r=s$ will survive, i.e. the spins
decouple and the Eqs.(\ref{annih1}) assume the simpler form:
\bea\label{annihilator} \non \alpha^{r}_{{\bf
k},e}(t)&=&\cos\theta\;\alpha^{r}_{{\bf
k},1}\;+\;\sin\theta\;\left( U_{{\bf k}}^{*}(t)\; \alpha^{r}_{{\bf
k},2}\;+\;\epsilon^{r}\; V_{{\bf k}}(t)\; \beta^{r\dag}_{-{\bf
k},2}\right)
\\ \non
\alpha^{r}_{{\bf k},\mu}(t)&=&\cos\theta\;\alpha^{r}_{{\bf
k},2}\;-\;\sin\theta\;\left( U_{{\bf k}}(t)\; \alpha^{r}_{{\bf
k},1}\;-\;\epsilon^{r}\; V_{{\bf k}}(t)\; \beta^{r\dag}_{-{\bf
k},1}\right)
\\
\beta^{r}_{-{\bf k},e}(t)&=&\cos\theta\;\beta^{r}_{-{\bf
k},1}\;+\;\sin\theta\;\left( U_{{\bf k}}^{*}(t)\; \beta^{r}_{-{\bf
k},2}\;-\;\epsilon^{r}\; V_{{\bf k}}(t)\; \alpha^{r\dag}_{{\bf
k},2}\right)
\\ \non
\beta^{r}_{-{\bf k},\mu}(t)&=&\cos\theta\;\beta^{r}_{-{\bf
k},2}\;-\;\sin\theta\;\left( U_{{\bf k}}(t)\; \beta^{r}_{-{\bf
k},1}\;+\;\epsilon^{r}\; V_{{\bf k}}(t)\; \alpha^{r\dag}_{{\bf
k},1}\right), \eea with $\epsilon^{r}=(-1)^{r}$ and
\bea\label{Vk2} \non U_{{\bf k}}(t)&\equiv &u^{r\dag}_{{\bf
k},2}(t)u^{r}_{{\bf k},1}(t)= v^{r\dag}_{-{\bf
k},1}(t)v^{r}_{-{\bf k},2}(t)
\\
V_{{\bf k}}(t)&\equiv & \epsilon^{r}\; u^{r\dag}_{{\bf
k},1}(t)v^{r}_{-{\bf k},2}(t)= -\epsilon^{r}\; u^{r\dag}_{{\bf
k},2}(t)v^{r}_{-{\bf k},1}(t) \eea

We have: \bea\label{Vk1} V_{{\bf k}}(t)=|V_{{\bf
k}}|\;e^{i(\omega_{k,2}+\omega_{k,1})t}\;\;\;\;,\;\;\;\; U_{{\bf
k}}(t)=|U_{{\bf k}}|\;e^{i(\omega_{k,2}-\omega_{k,1})t}; \eea \bea
\non |U_{{\bf
k}}|=\left(\frac{\omega_{k,1}+m_{1}}{2\omega_{k,1}}\right)^{\frac{1}{2}}
\left(\frac{\omega_{k,2}+m_{2}}{2\omega_{k,2}}\right)^{\frac{1}{2}}
\left(1+\frac{{\bf
k}^{2}}{(\omega_{k,1}+m_{1})(\omega_{k,2}+m_{2})}\right)
\\
\label{Vk}|V_{{\bf
k}}|=\left(\frac{\omega_{k,1}+m_{1}}{2\omega_{k,1}}\right)^{\frac{1}{2}}
\left(\frac{\omega_{k,2}+m_{2}}{2\omega_{k,2}}\right)^{\frac{1}{2}}
\left(\frac{k}{(\omega_{k,2}+m_{2})}-\frac{k}{(\omega_{k,1}+m_{1})}\right)
\eea \bea|U_{{\bf k}}|^{2}+|V_{{\bf k}}|^{2}=1.\eea

We thus see that, at level of annihilation operators, the
structure of the mixing transformation is that of a Bogoliubov
transformation nested into a rotation. The two transformations
however cannot be disentangled, thus the mixing transformations
Eq.(\ref{annihilator}) are essentially different from the usual
Bogoliubov transformations.

We observe that $G^{-1}_{\bf \te}(t) = exp[\theta(S_{-} - S_{+})]$
is just the generator for generalized coherent states of $SU(2)$:
the flavor vacuum state is therefore an $SU(2)$ (time dependent)
coherent state. Let us now obtain the explicit expression for
$|0\rangle_{e,\mu}$ and investigate the infinite volume limit of
Eq.(\ref{flavvac}).

Using the Gaussian decomposition, $G^{-1}_{\bf \te}(t)$ can be
written as  \bea\non G^{-1}_{\bf \te}(t) = exp[\theta(S_{-} -
S_{+})]= exp(-tan\theta \; S_{+}) \;exp(-2 ln \: cos\theta \;
S_{3}) \;exp(tan\theta \; S_{-})\\ \eea
where $0\leq \theta < \frac{\pi}{2}$. Eq.(\ref{flavvac}) then
becomes
\bea\non |0\rangle_{e,\mu} =\prod_{{\bf k}}|0\rangle_{e,\mu}^{{\bf
k}}= \prod_{{\bf k}} exp(-tan\theta \; S_{+}^{{\bf k}}) exp(-2 ln
\: cos\theta \; S_{3}^{{\bf k}}) \;exp(tan\theta \; S_{-}^{{\bf
k}})|0\rangle_{1,2}\;.\\ \label{flavvac2} \eea
The right hand side of Eq.(\ref{flavvac2}) may be computed by
using the relations
\bea\label{S3Spm} S_{3}^{{\bf
k}}|0\rangle_{1,2}=0\;\;,\;\;S_{\pm}^{{\bf k}}|0\rangle_{1,2}\neq
0 \;\;,\;\; (S_{\pm}^{{\bf k}})^{2}|0\rangle_{1,2}\neq 0 \;\;,\;\;
(S_{\pm}^{{\bf k}})^{3}|0\rangle_{1,2}=0\;, \eea and other useful
relations which are given in the Appendix B. The expression for
$|0\rangle_{e,\mu}$ in terms of $S^{{\bf k}}_{\pm}$ and $S^{{\bf
k}}_{3}$ is:
\bea\label{flavvac3} |0\rangle_{e,\mu}=\prod_{{\bf
k}}|0\rangle_{e,\mu}^{{\bf k}}= \prod_{{\bf k}}\left[ 1 +
\sin\theta \cos\theta \left(S_{-}^{{\bf k}} - S_{+}^{{\bf
k}}\right)+\frac{1}{2}\sin^{2}\theta \cos^{2}\theta
\left((S_{-}^{{\bf k}})^{2} + (S_{+}^{{\bf k}})^{2}\right)+\right.
\nonumber \\\non \left.
 -\sin^{2}\theta S_{+}^{{\bf k}}S_{-}^{{\bf k}} +
\frac{1}{2}\sin^{3}\theta \cos\theta \left(S_{-}^{{\bf
k}}(S_{+}^{{\bf k}})^{2} - S_{+}^{{\bf k}}(S_{-}^{{\bf
k}})^{2}\right)+ \frac{1}{4} \sin^{4}\theta (S_{+}^{{\bf
k}})^{2}(S_{-}^{{\bf k}})^{2}\right]|0\rangle_{1,2}\;.\\ \eea

The state $|0\rangle_{e,\mu}$ is normalized to 1 (see
Eq.(\ref{flavvac})). Eq.(\ref{flavvac3}) and Eqs.(\ref{s+}) and
(\ref{s-}) exhibit the rich coherent state structure of
$|0\rangle_{e,\mu}$.

Let us now compute $_{1,2}\langle0|0\rangle_{e,\mu}$. We obtain
\bea\non _{1,2}\langle0|0\rangle_{e,\mu}  = \prod_{{\bf
k}}\left(1-\sin^{2}\theta\; _{1,2}\langle0|S_{+}^{{\bf k}}
S_{-}^{{\bf k}}|0\rangle_{1,2} + \frac{1}{4} \sin^{4}\theta\;
_{1,2}\langle0|(S_{+}^{{\bf k}})^{2} (S_{-}^{{\bf
k}})^{2}|0\rangle_{1,2}\right)\\\eea
where (see Appendix B)
\bea\non && _{1,2}\langle0|S_{+}^{{\bf k}}S_{-}^{{\bf
k}}|0\rangle_{1,2}=
\\ \non
&=&_{1,2}\langle0|\left(\sum_{\sigma,\tau}\sum_{r,s}
\left[v^{\sigma\dag}_{-{\bf k},1}(t) u^{\tau}_{{\bf
k},2}(t)\right] \left[u^{s\dag}_{{\bf k},2}(t) v^{r}_{-{\bf
k},1}(t)\right] \beta^{\sigma}_{-{\bf k},1}\alpha^{\tau}_{{\bf
k},2} \alpha^{s\dag}_{{\bf k},2}\beta^{r\dag}_{-{\bf
k},1}\right)|0\rangle_{1,2} =
\\  &=& \sum_{r,s} |\;v^{r\dag}_{-{\bf k},1}(t) u^{s}_{{\bf k},2}(t)\; |^{2}
 \equiv 2|V_{{\bf k}}|^{2}\;.
\eea
\smallskip
In a similar way we find
\bea\label{s+s-} _{1,2}\langle0|(S_{+}^{{\bf k}})^{2}(S_{-}^{{\bf
k}})^{2}|0\rangle_{1,2}= 2|V_{{\bf k}}|^{4}\;. \eea
\medskip
 The function $|V_{{\bf k}}|^{2}$ depends on $|{\bf k}|$ only
through its modulus and it is always in the interval
$[0,\frac{1}{2}[$. It has a maximum at $|{\bf k}|=\sqrt{m_1 m_2}$
($\sqrt{m_{1} m_{2}}$ is the scale of the condensation density)
and $|V_{{\bf k}}|^{2}=0$ when $m_{1} = m_{2}$. Also, $|V_{{\bf
k}}|^{2} \rightarrow 0$ when $k \rightarrow \infty$.

In conclusion we have
\bea_{1,2}\langle0|0\rangle_{e,\mu}  &=&\prod_{{\bf k}}\left(1-
 \sin^{2}\theta\;|V_{{\bf k}}|^{2}\right)^{2}\equiv
\prod_{k}\Gamma({\bf k})= \\ \non &=& \prod_{{\bf k}}e^{ln\;
\Gamma({\bf k})}=e^{\sum_{{\bf k}}ln\; \Gamma({\bf k})}.\eea
{}From the properties of $|V_{{\bf k}}|^{2}$ we have that
$\Gamma({\bf k}) < 1$ for any value of ${\bf k}$ and of the
parameters $m_{1}$ and $m_{2}$. By using the continuous limit
relation $\sum_{{\bf k}}\;\rightarrow \; \frac{V}{(2\pi)^{3}}\int
d^{3}{\bf k}$, in the infinite volume limit we obtain
\bea \label{orto} \lim_{V \rightarrow \infty}\;
_{1,2}\langle0|0\rangle_{e,\mu} = \lim_{V \rightarrow \infty}\;
e^{\frac{V}{(2\pi)^{3}}\int d^{3}{\bf k} \;ln\; \Gamma({\bf k})}=
0 \eea
Notice that Eq.(\ref{orto}) shows that the orthogonality between
$|0\rangle_{e,\mu}$ and $|0\rangle_{1,2}$ is due to the infrared
contributions which are taken in care by the infinite volume limit
and therefore high momentum contributions do not influence the
result (for this reason here we do not need to consider the
regularization problem of the UV divergence of the integral of
$ln\;\Gamma({\bf k})$).

Of course, this orthogonality disappears when $\theta =0$ and/or
when $m_{1} = m_{2}$ (because in this case $|V_{{\bf k}}|^{2}=0$
and no mixing occurs in Pontecorvo theory).

Eq.(\ref{orto}) expresses the unitary inequivalence in the
infinite volume limit of the flavor and the mass representations
and shows the absolutely non-trivial nature of the mixing
transformations Eq.(\ref{mix}). In other words, the mixing
transformations induce a physically non-trivial structure in the
flavor vacuum which indeed turns out to be an $SU(2)$ generalized
coherent state. In Section 2.5 we will see how such a vacuum
structure may lead to phenomenological consequences in the
neutrino oscillations. From Eq.(\ref{orto}) we also see that
Eq.(\ref{flavvac}) is a purely formal expression which only holds
at finite volume.

We thus realize the limit of validity of the approximation usually
adopted when the mass vacuum state (representation for definite
mass operators) is identified with the vacuum for the flavor
operators. We point out that even at finite volume the vacua
identification is actually an approximation since the flavor
vacuum is an $SU(2)$ generalized coherent state. In such an
approximation, the coherent state structure and many physical
features are missed.

It is also interesting to exhibit the explicit expression of
$|0\rangle_{e,\mu}^{{\bf k}}$, at time $t=0$, in the reference
frame for which ${\bf k}=(0,0,|{\bf k}|)$ (see Appendix D):
\bea\non |0\rangle_{e,\mu}^{{\bf k}}&=& \prod_{r}
\Big[(1-\sin^{2}\theta\;|V_{{\bf k}}|^{2})
-\epsilon^{r}\sin\theta\;\cos\theta\; |V_{{\bf k}}|
(\alpha^{r\dag}_{{\bf k},1}\beta^{r\dag}_{-{\bf k},2}+
\alpha^{r\dag}_{{\bf k},2}\beta^{r\dag}_{-{\bf k},1})+  \\
\non &+&\epsilon^{r}\sin^{2}\theta \;|V_{{\bf k}}||U_{{\bf
k}}|(\alpha^{r\dag}_{{\bf k},1}\beta^{r\dag}_{-{\bf k},1}-
\alpha^{r\dag}_{{\bf k},2}\beta^{r\dag}_{-{\bf k},2})
+\sin^{2}\theta \; |V_{{\bf k}}|^{2}\alpha^{r\dag}_{{\bf
k},1}\beta^{r\dag}_{-{\bf k},2} \alpha^{r\dag}_{{\bf
k},2}\beta^{r\dag}_{-{\bf k},1} \Big]|0\rangle_{1,2}\\\label{0emu}
\eea

We see that the expression of the flavor vacuum
$|0\rangle_{e,\mu}$ involves four different particle-antiparticle
"couples".

The condensation density is given by
\bea _{e,\mu}\langle 0| \al_{{\bf k},i}^{r \dag} \al^r_{{\bf k},i}
|0\rangle_{e,\mu}\,= \;_{e,\mu}\langle 0| \bt_{{\bf k},i}^{r \dag}
\bt^r_{{\bf k},i} |0\rangle_{e,\mu}\,=\, \sin^{2}\te\; |V_{{\bf
k}}|^{2} \;, \qquad i=1,2\,. \eea

\section{Generalization of mixing transformations}

In Section 2.2 we have expressed the flavor fields $\nu_e$ and
$\nu_\mu$ in the same bases as the (free) fields with definite
masses $\nu_1$ and $\nu_2$, respectively. However, it has been
noticed \cite{fujii1,fujii2}, that this is actually a special
choice, and that a more general possibility exists.

Let us introduce the notation $(\sigma,j)=(e,1) , (\mu,2)$, the
fields $\nu_e$ and $\nu_\mu$ can be rewritten in the following
form:
\begin{eqnarray}\non
\nu_{\sigma}(x) &=& G^{-1}_{\theta}(t)\, \nu_{j}(x)\,
G_{\theta}(t) = \frac{1}{\sqrt{V}} \sum_{{\bf k},r}\left[
u^{r}_{{\bf k},j} \alpha^{r}_{{\bf k},\sigma}(t)+ v^{r}_{-{\bf
k},j} \beta^{r\dag}_{-{\bf k},\sigma}(t) \right] e^{i {\bf
k}\cdot{\bf x}},\\\label{exnue12}
\end{eqnarray}

The flavor annihilation operators are rewritten as
\begin{equation}\label{BVoper}
\left(\begin{array}{c} \alpha^{r}_{{\bf k},\sigma}(t)\\
\beta^{r\dag}_{{-\bf k},\sigma}(t)
\end{array}\right)
= G^{-1}_{\theta}(t)  \left(\begin{array}{c} \alpha^{r}_{{\bf k},j}(t)\\
\beta^{r\dag}_{{-\bf k},j}(t)
\end{array}\right)
G_{\theta}(t)
\end{equation}
and the explicit expression of the flavor annihilation operators
is (we choose ${\bf k}=(0,0,|{\bf k}|)$):
\begin{equation}\label{BVmatrix}
\left(\begin{array}{c} {\alpha}^{r}_{{\bf k},e}(t)\\
{\alpha}^{r}_{{\bf k},\mu}(t)\\ {\beta}^{r\dag}_{{-\bf k},e}(t)\\
{\beta}^{r\dag}_{{-\bf k},\mu}(t)
\end{array}\right)
\, = \, \left(\begin{array}{cccc} c_\theta &  s_\theta\, |U_{{\bf
k}}| &0 & s_\theta \, \epsilon^{r} \,|V_{{\bf k}}| \\ - s_\theta\,
|U_{{\bf k}}|& c_\theta &  s_\theta \,\epsilon^{r} \,|V_{{\bf k}}|
& 0  \\ 0& - s_\theta \,\epsilon^{r} \,|V_{{\bf k}}| &c_\theta &
s_\theta \,|U_{{\bf k}}| \\ - s_\theta \,\epsilon^{r} \,|V_{{\bf
k}}| & 0 &
- s_\theta\, |U_{{\bf k}}| & c_\theta \\
\end{array}\right)
\left(\begin{array}{c} \alpha^{r}_{{\bf k},1}(t)\\
\alpha^{r}_{{\bf k},2}(t)\\ \beta^{r\dag}_{{-\bf k},1}(t)\\
\beta^{r\dag}_{{-\bf k},2}(t)
\end{array}\right)
\end{equation}
where $c_\theta\equiv \cos\theta$, $s_\theta\equiv \sin\theta$,
$\epsilon^r\equiv (-1)^r$

In the expansion Eq.(\ref{exnue12}) one could use eigenfunctions
with arbitrary masses $\mu_\sigma$, and therefore not necessarily
the same as the masses which appear in the Lagrangian.  Indeed,
the transformation Eq.(\ref{BVoper}) can be generalized
\cite{fujii1,fujii2} by writing the flavor fields as
\begin{eqnarray}\label{exnuf22}
\nu_{\sigma}(x)     &=& \frac{1}{\sqrt{V}} \sum_{{\bf k},r}
\left[ u^{r}_{{\bf k},\sigma} {\widetilde \alpha}^{r}_{{\bf
k},\sigma}(t) + v^{r}_{-{\bf k},\sigma} {\widetilde
\beta}^{r\dag}_{-{\bf k},\sigma}(t) \right]  e^{i {\bf k}\cdot{\bf
x}} ,
\end{eqnarray}
where $u_{\sigma}$ and $v_{\sigma}$ are the helicity
eigenfunctions with mass $\mu_\sigma$\footnote{The use of such a
basis simplifies considerably calculations with respect to the
original choice of ref.\cite{BV95}.}. We denote by a tilde the
generalized flavor operators \cite{fujii1,fujii2} in order to
distinguish them from the ones defined in Eq.(\ref{BVoper}).  The
expansion Eq.(\ref{exnuf22}) is more general than the one in
Eq.(\ref{exnue12}) since the latter corresponds to the particular
choice $\mu_e\equiv m_1$, $\mu_\mu \equiv m_2$.

The relation between the general flavor and the mass operators is
now:
\begin{equation}\label{FHYoper}
\left(\begin{array}{c} {\widetilde \alpha}^{r}_{{\bf k},\sigma}(t)\\
{\widetilde \beta}^{r\dag}_{{-\bf k},\sigma}(t)
\end{array}\right)
= K^{-1}_{\theta,\mu}(t)  \left(\begin{array}{c} \alpha^{r}_{{\bf
k},j}(t)\\ \beta^{r\dag}_{{-\bf k},j}(t)
\end{array}\right)
K_{\theta,\mu}(t) ~~,
\end{equation}
with $(\sigma,j)=(e,1) , (\mu,2)$, where $K_{\theta,\mu}(t)$ is
the generator of the transformation Eq.(\ref{mix}) and can be
expressed as
\begin{eqnarray}\label{FHYgen}
K_{\theta,\mu}(t)&=& I_{\mu}(t)\, G_{\theta}(t) \\ \label{Igen}
I_{\mu}(t)&=& \prod_{{\bf k}, r}\, \exp\left\{ i
\mathop{\sum_{(\sigma,j)}} \xi_{\sigma,j}^{\bf k}\left[
\alpha^{r\dag}_{{\bf k},j}(t)\beta^{r\dag}_{{-\bf k},j}(t) +
\beta^{r}_{{-\bf k},j}(t)\alpha^{r}_{{\bf k},j}(t) \right]\right\}
\end{eqnarray}
with \bea \xi_{\sigma,j}^{\bf k}\equiv (\chi_\sigma -
\chi_j)/2,\quad \,\quad \cot\chi_\sigma = |{\bf
k}|/\mu_\sigma,\quad \,\quad \cot\chi_j = |{\bf k}|/m_j.\eea For
$\mu_e\equiv m_1$, $\mu_\mu \equiv m_2$, we have $I_{\mu}(t)=1$.

The explicit  matrix form of the flavor operators is
\cite{fujii1,fujii2}:
\begin{equation}\label{FHYmatrix}
\left(\begin{array}{c} {\widetilde \alpha}^{r}_{{\bf k},e}(t)\\
{\widetilde \alpha}^{r}_{{\bf k},\mu}(t)\\ {\widetilde
\beta}^{r\dag}_{{-\bf k},e}(t)\\ {\widetilde \beta}^{r\dag}_{{-\bf
k},\mu}(t)
\end{array}\right)
\, = \, \left(\begin{array}{cccc} c_\theta\, \rho^{\bf k}_{e1} &
s_\theta \,\rho^{\bf k}_{e2}  &i c_\theta \,\lambda^{\bf k}_{e1} &
i s_\theta \,\lambda^{\bf k}_{e2}  \\ - s_\theta \,\rho^{\bf
k}_{\mu 1} & c_\theta \,\rho^{\bf k}_{\mu 2} &- i s_\theta
\,\lambda^{\bf k}_{\mu 1} & i c_\theta \,\lambda^{\bf k}_{\mu 2}
\\ i c_\theta \,\lambda^{\bf k}_{e1} & i s_\theta \,\lambda^{\bf
k}_{e2} &c_\theta\, \rho^{\bf k}_{e1} & s_\theta \,\rho^{\bf
k}_{e2}  \\ - i s_\theta \,\lambda^{\bf k}_{\mu 1} & i c_\theta\,
\lambda^{\bf k}_{\mu 2} &- s_\theta\,
\rho^{\bf k}_{\mu 1} & c_\theta\, \rho^{\bf k}_{\mu 2} \\
\end{array}\right)
\left(\begin{array}{c} \alpha^{r}_{{\bf k},1}(t)\\
\alpha^{r}_{{\bf k},2}(t)\\ \beta^{r\dag}_{{-\bf k},1}(t)\\
\beta^{r\dag}_{{-\bf k},2}(t)
\end{array}\right)
\end{equation}
where $c_\theta\equiv \cos\theta$, $s_\theta\equiv \sin\theta$ and
\begin{eqnarray}\label{rho}
 \rho^{\bf k}_{a b} \delta_{rs}&\equiv& \cos\lf(\frac{\chi_a -
 \chi_b}{2}\ri)
\delta_{rs}= u^{r\dag}_{{\bf k},a} u^{s}_{{\bf k},b} =
v^{r\dag}_{-{\bf k},a} v^{s}_{-{\bf k},b} \\ \label{lambda} i
\lambda^{\bf k}_{a b}\delta_{rs} &\equiv& i \sin\lf(\frac{\chi_a -
\chi_b}{2}\ri) \delta_{rs} = u^{r\dag}_{{\bf k},a} v^{s}_{-{\bf
k},b} = v^{r\dag}_{-{\bf k},a} u^{s}_{{\bf k},b}
\end{eqnarray}
with $a,b = 1,2,e,\mu$.

Since $\rho^{\bf k}_{1 2}=|U_{{\bf k}}|$ and $i \lambda^{\bf k}_{1
2}=\epsilon^r |V_{{\bf k}}|$, etc., the operators
Eq.(\ref{FHYmatrix}) reduce to the ones in Eqs.(\ref{BVmatrix})
when $\mu_e\equiv m_1$ and $\mu_\mu \equiv m_2$\footnote{In
performing such an identification, one should take into account
that the operators for antiparticles differ for a minus sign,
related to the different spinor bases used in the expansions
Eqs.(\ref{exnue12}) and (\ref{exnuf22}). Such a sign difference is
however irrelevant in what follows.}.

%%%%% Finally,
The generalization of the flavor vacuum, which is annihilated by
the general flavor operators given by Eq.(\ref{FHYoper}), is
\cite{fujii1,fujii2}:
\begin{equation}\label{FHYvac}
|{\widetilde 0(t)}\rangle_{e,\mu}\equiv
K^{-1}_{\theta,\mu}(t)|0\rangle_{1,2} ~~.
\end{equation}
For $\mu_e\equiv m_1$ and $\mu_\mu \equiv m_2$, this state reduces
to the flavor vacuum $| 0(t)\rangle_{e,\mu}$ above defined.

%%%%%

The relation between the general flavor operators of
Eq.(\ref{FHYoper}) and the flavor operators of Eq.(\ref{BVoper})
is \cite{fujii1,fujii2}:
\begin{eqnarray}\label{FHYBVa}
\left(\begin{array}{c} {\widetilde \alpha}^{r}_{{\bf k},\sigma}(t)\\
{\widetilde \beta}^{r\dag}_{{-\bf k},\sigma}(t)
\end{array}\right)
&=& J^{-1}_{\mu}(t)  \left(\begin{array}{c} \alpha^{r}_{{\bf
k},\sigma}(t)\\ \beta^{r\dag}_{{-\bf k},\sigma}(t)
\end{array}\right)J_{\mu}(t) ~~,
\\ \label{FHYBVb12}
J_{\mu}(t)&=& \prod_{{\bf k}, r}\, \exp\left\{ i
\mathop{\sum_{(\sigma,j)}} \xi_{\sigma,j}^{\bf k}\left[
\alpha^{r\dag}_{{\bf k},\sigma}(t)\beta^{r\dag}_{{-\bf
k},\sigma}(t) + \beta^{r}_{{-\bf k},\sigma}(t)\alpha^{r}_{{\bf
k},\sigma}(t) \right]\right\}\,.
\end{eqnarray}

We have shown that the Hilbert space for the flavor fields is not
unique: an infinite number of vacua (and consequently infinitely
many Hilbert spaces) can be generated by introducing the arbitrary
mass parameters $\mu_{e}$, $\mu_{\mu}$. It is obvious that
physical quantities must not depend on these parameters. In
Section 2.6 we show indeed that the observable charge operators
are invariant under the transformations generated by
Eq.(\ref{FHYBVb12}). Thus, physical observables are fully
independent of the arbitrary mass parameters $\mu_\sigma$. See
also Appendix I.

\section{The current structure for field mixing}

We now analize the transformations acting on a doublet of free
fields with different masses. The results of this Section clarify
the meaning of the $su(2)$ algebraic structure found before and
will be useful in the discussion of neutrino oscillations.

Let us consider the Lagrangian describing two free Dirac fields
with masses $m_{1}$ and $m_{2}$: \bea\label{lagrmas} {\cal
L}(x)\,=\,  {\bar \Psi_m}(x) \lf( i \not\!\partial -
\textsf{M}_{d} \ri) \Psi_m(x)\, , \eea where
$\Psi_m^T=(\nu_1,\nu_2)$ and $\textsf{M}_{d}=diag(m_{1},m_{2})$.
We introduce a subscript $m$, in order to distinguish the
quantities here introduced, which are in terms of fields with
definite masses, from the ones which are in terms of flavor
fields.

The Lagrangian ${\cal L}(x)$ is invariant under global $U(1)$
phase transformations of the type \bea \Psi _{m}^{^{\prime
}}(x)=e^{i\alpha}\Psi _{m}(x), \eea then, we have the conservation
of the Noether charge \bea Q_=\int I^{0}(x)d^{3}{\bf x} \eea (with
$I^{\mu}(x)=\bar{\Psi}_{m}(x)\gamma ^{\mu }\Psi _{m}(x)$) which is
indeed the total charge of the system, i.e. the total lepton
number.

Consider then the global $SU(2)$ transformation \cite{currents}:
\bea \Psi _{m}^{^{\prime }}(x)=e^{i\alpha_{j}\cdot \tau_{j}}\Psi
_{m}(x)\text{ \qquad \qquad } j=1,2,3. \eea with $\alpha_{j}$ real
constants, ${\tau_{j}} =\sigma_{j}/2$ and $\sigma_{j}$ being the
Pauli matrices.

Since the masses $m_{1}$ and $m_{2}$ are different, the Lagrangian
is not invariant under the above transformations. By use of
equations of motion, we obtain the variation of the Lagrangian:
\bea \delta {\cal L}=i\alpha_{j}\bar{\Psi}_{m}(x)\left[\tau_{j}
,\;M_{d} \right] \Psi _{m}(x)=-\alpha_{j}\partial _{\mu
}J_{m,j}^{\mu}(x), \eea where the currents for a complex field
$\Psi ^{\alpha }(x)$ are given by:
 \bea J^{\mu }(x)=i\left\{ \Psi ^{\dagger \alpha
}(x)\frac{\partial {\cal L}}{\partial \left( \partial _{\mu }\Psi
^{\dagger \alpha }(x)\right) }-\frac{\partial {\cal L}}{
\partial \left( \partial _{\mu }\Psi ^{\alpha }(x)\right) }\Psi ^{\alpha
}(x)\right\} , \eea and then, in our case \bea J_{m,j}^{\mu
}(x)=\bar{\Psi}_{m}(x)\;\gamma ^{\mu }\;\tau _{j}\;\Psi
_{m}(x),\quad\;\quad\;\quad j=1,2,3. \eea

We thus have the following currents: \bea J_{m,1}^{\mu
}(x)=\frac{1}{2}\left[ \bar{\nu }_{1}(x)\;\gamma ^{\mu }\;\nu
_{2}(x)+ \bar{\nu }_{2}(x)\;\gamma ^{\mu }\;\nu _{1}(x)\right] ,
\eea \bea J_{m,2}^{\mu }(x)=\frac{i}{2}\left[ \bar{\nu
}_{1}(x)\;\gamma ^{\mu }\;\nu _{2}(x)- \bar{\nu }_{2}(x)\;\gamma
^{\mu }\;\nu _{1}(x)\right] , \eea \bea J_{m,3}^{\mu
}(x)=\frac{1}{2}\left[ \bar{\nu }_{1}(x)\;\gamma ^{\mu }\;\nu
_{1}(x)- \bar{\nu }_{2}(x)\;\gamma ^{\mu }\;\nu _{2}(x)\right] .
\eea

The related charges, defined as \bea Q_{m,j}(t)=\int
J_{m,j}^{0}(x)d^{3}{\bf x},\quad\;\quad\;\quad j=1,2,3, \eea
satisfy the $su(2)$ algebra: \bea \left[
Q_{m,i}(t),Q_{m,j}(t)\right] =i\varepsilon _{ijk}Q_{m,k}(t). \eea

We note that the Casimir operator is proportional to the total
(conserved) charge: \bea Q_{m,0}=\frac{1}{2}Q. \eea

Also $Q_{m,3}$ is conserved, due to the fact that the mass matrix
 $\textsf{M}_{d}$ is diagonal. This implies the conservation of
 charge separately for $\nu _{1}$ and $\nu _{2}$, which is what we
 expect for a system of two non-interacting fields.
 We can thus define
the combinations: \bea Q_{1}=\frac{1}{2}Q +
Q_{m,3},\\Q_{2}=\frac{1}{2}Q - Q_{m,3}, \eea

\bea Q _{i}=\sum_{{\bf k},r}\left( \alpha _{{\bf k},i }^{r\dagger
}(t)\alpha _{{\bf k},i }^{r}(t)-\beta _{-{\bf k},i}^{r\dagger
}(t)\beta _{-{\bf k},i}^{r}(t)\right) ,\qquad i=1,2.\eea

These are the Noether charges associated with the non interacting
fields $\nu_{1}$ and $\nu_{2}$. Explicitly, the transformations
induced by the three above generators $\tau_{1},\tau_{2},\tau_{3}$
are \bea \Psi _{m}^{^{\prime }}=\left(
\begin{array}{ll}
\cos \theta _{1} & i\sin \theta _{1} \\
i\sin \theta _{1} & \cos \theta _{1}
\end{array}
\right) \Psi _{m}, \eea \bea \Psi _{m}^{^{\prime }}=\left(
\begin{array}{ll}
\cos \theta _{2} & \sin \theta _{2} \\
-\sin \theta _{2} & \cos \theta _{2}
\end{array}
\right) \Psi _{m}, \eea \bea \Psi _{m}^{^{\prime }}=\left(
\begin{array}{ll}
e^{i\theta _{3}} & 0 \\
0 & e^{-i\theta _{3}}
\end{array}
\right) \Psi _{m}. \eea

We observe that the transformation induced by  $Q_{m,2}(t)$, \bea
\Psi _{f}(x)=e^{-2i \theta Q_{m,2}(t)}\; \Psi _{m}(x)\;e^{2i
\theta Q_{m,2}(t)} \eea is just the mixing transformation
Eq.(\ref{mix}).

Let us consider the Lagrangian written in the flavor basis (the
subscript $f$ denotes here flavor) \bea\label{lagrflav} {\cal
L}(x)\,=\,  {\bar \Psi_f}(x) \lf( i \not\!\partial - \textsf{M}
\ri) \Psi_f(x)\, , \eea where $\Psi _{f}^{T}=(\nu _{e},$ $\nu
_{\mu }),$ and $\textsf{M} =\left(
\begin{array}{ll}
m_{e} & m_{e\mu } \\
m_{e\mu } & m_{\mu }
\end{array}\right). $

In analogy with was done above, consider now the variation of the
Lagrangian Eq.(\ref{lagrflav}) under the $SU(2)$ transformation:
\bea \Psi _{f}^{^{\prime }}(x)=e^{i\alpha_{j}\cdot \tau_{j}}\Psi
_{f}(x) \text{ \qquad \qquad } j=1,2,3. \eea

We have \bea \delta {\cal L}(x)=i\alpha_{j}\bar{\Psi}_{f}(x)\left[
\tau_{j}, \textsf{M}\right] \Psi _{f}(x)=-\alpha_{j}\partial _{\mu
}J_{f,j}^{\mu }(x), \eea \bea J_{f,j}^{\mu
}(x)=\bar{\Psi}_{f}(x)\;\gamma ^{\mu }\;\tau_{j}\; \Psi _{f}(x)
\eea and obtain the currents: \bea J_{f,1}^{\mu
}(x)=\frac{1}{2}\left[ \bar{\nu }_{e}(x)\;\gamma ^{\mu }\;\nu
_{\mu }(x)+\bar{\nu }_{\mu }(x)\;\gamma ^{\mu }\;\nu
_{e}(x)\right] , \eea \bea J_{f,2}^{\mu }(x)=\frac{i}{2}\left[
\bar{\nu }_{e}(x)\;\gamma ^{\mu }\;\nu _{\mu }(x)-\bar{\nu }_{\mu
}(x)\;\gamma ^{\mu }\;\nu _{e}(x)\right] , \eea \bea J_{f,3}^{\mu
}(x)=\frac{1}{2}\left[ \bar{\nu }_{e}(x)\;\gamma ^{\mu }\;\nu
_{e}(x)\;- \bar{\nu }_{\mu }(x)\gamma ^{\mu }\;\nu _{\mu
}(x)\right] , \eea and \bea J_{f,0}^{\mu }(x)=\frac{1}{2}\left[
\bar{\nu }_{e}(x)\;\gamma ^{\mu }\;\nu _{e}(x)+ \bar{\nu }_{\mu
}(x)\;\gamma ^{\mu }\;\nu _{\mu }(x)\right] . \eea

Again, the charges \bea Q_{f,j}(t)=\int J_{f,j}^{0}(x)d^{3}{\bf x}
\quad \;\quad j=1,2,3,\eea satisfy the $su(2)$ algebra: $\left[
Q_{f,i}(t),Q_{f,j}(t)\right] =i\varepsilon _{i,jk}Q_{f,k}(t)$.

The Casimir $Q_{f,0}$ is proportional to the total charge $Q_{f,0}
= Q_{0}= \frac{1}{2}Q$.

However, now, because of the off-diagonal (mixing) terms in the
mass matrix $\textsf{M}$, $Q_{f,3}$ is not conserved anymore. This
implies an exchange of charge between $\nu_{e}$ and $\nu_{\mu }$,
resulting in the phenomenon of neutrino oscillations.

Let us indeed define the $flavor$ $charges$ as \bea Q_{e}(t)\equiv
\frac{1}{2}Q + Q_{f,3}(t), \eea \bea Q_{\mu}(t)\equiv \frac{1}{2}Q
-Q_{f,3}(t), \eea where \bea Q_{e}(t)+Q_{\mu }(t)=Q, \eea the
oscillation formulas are obtained by taking expectation values of
the above charges on the neutrino state.

\section{The exact formula for neutrino mixing}

In terms of the flavor operators, the flavor charge operators are
\bea\label{charneut} Q_{\sigma }(t)=\sum_{{\bf k},r}\left( \alpha
_{{\bf k},\sigma }^{r\dagger }(t)\alpha _{{\bf k},\sigma
}^{r}(t)-\beta _{-{\bf k} ,\sigma }^{r\dagger }(t)\beta _{-{\bf
k},\sigma }^{r}(t)\right) ,\text{ \qquad }\sigma =e,\mu. \eea

At time $t=0$, the vacuum state is $|0\rangle_{e,\mu}$ and the one
electron neutrino state is (for ${\bf k}=(0,0,|{\bf k}|)$):
\bea\non |\nu_e \rangle \equiv \alpha_{{\bf k},e}^{r
\dag}|0\rangle_{e,\mu} = \left[ \cos\theta\,\alpha_{{\bf k},1}^{r
\dag} + |U_{\bf k}|\; \sin\theta\;\alpha_{{\bf k},2}^{r \dag} +
\epsilon^r \; |V_{\bf k}| \,\sin\theta \; \alpha_{{\bf k},2}^{r
\dag}\alpha_{{\bf k},1}^{r \dag} \beta_{-{\bf k},1}^{r \dag}
\right] |0\rangle_{1,2} \,. \\\label{h1}\eea
In this state a multiparticle component is present, disappearing
in the relativistic limit $|{\bf k}|\gg \sqrt{m_1m_2}\,$: in this
limit the (quantum-mechanical) Pontecorvo state is recovered.

If we now assume that the neutrino state at time $t$ is given by
$|\nu_e (t)\rangle = e^{-iH t} |\nu_e\rangle$, we see that it is
not possible to compare directly this state with the one at time
$t=0$ given in Eq.(\ref{h1}).

In fact, $|\nu_e (t)\rangle$ is given by:
\bea \non |\nu_{e}(t)\ran &=& e^{-i H t} \,\al_{{\bf k},e}^{r
\dag}\,|0\ran_\flav =e^{-i \om_{\1} t}\Big[ \cos\te\,\al_{{\bf
k},1}^{r \dag} + \,e^{-i (\om_{\2}-\om_{\1}) t}\, |U_{\bf k}|\,
\sin\te\,\al_{{\bf k},2}^{r \dag} + \\
 &+& \epsilon^{r} \, e^{-i
(\om_{\2}+\om_{\1}) t}\, |V_{\bf k}|
  \,\sin\te \,
\al_{{\bf k},2}^{r \dag}\al_{{\bf k},1}^{r \dag}\bt_{{\bf k},1}^{r
\dag}
 \Big] |0\ran_\mass
\eea
and then (see the Appendix E) \bea \lim_{V \to \infty }\,\lan
\nu_e(t)| \nu_e(0)\ran\,=\,0. \eea

 The reason is that the flavor vacuum
$|0\rangle_{e,\mu}$ is not eigenstate of the free Hamiltonian $H$
and it ``rotates'' under the action of the time evolution
generator: one indeed finds $\lim_{V \rightarrow
\infty}\;_{e,\mu}\langle 0\;|\;0(t)\rangle_{e,\mu} = 0$ (Appendix
E). Thus at different times we have unitarily inequivalent flavor
vacua (in the limit $V\rightarrow \infty$): this expresses the
different particle content of these (coherent) states and it is
direct consequence of the fact that flavor states are not mass
eigenstates.

As already observed, this implies that we cannot directly compare
flavor states at different times. However we can consider the
flavor charge operators, defined as in Eq.(\ref{charneut}). We
then have (in the Heisenberg representation) \bea
 _{e,\mu }\langle 0|Q_{e}(t)|0\rangle _{e,\mu }=_{e,\mu
}\langle 0|Q_{\mu }(t)|0\rangle _{e,\mu }=0, \eea \bea {\cal
Q}^e_{{\bf k},e}(t)=\langle \nu _{e}|Q_{e}(t)|\nu _{e}\rangle
=\lf|\lf \{\al^{r}_{{\bf k},e}(t), \al^{r \dag}_{{\bf k},e}(0)
\ri\}\ri|^{2} \;+ \;\lf|\lf\{\bt_{{-\bf k},e}^{r \dag}(t), \al^{r
\dag}_{{\bf k},e}(0) \ri\}\ri|^{2}, \eea \bea {\cal Q}^e_{{\bf
k},\mu}(t) =\langle \nu _{e}|Q_{\mu }(t)|\nu _{e}\rangle =\lf|\lf
\{\al^{r}_{{\bf k},\mu}(t), \al^{r \dag}_{{\bf k},e}(0)
\ri\}\ri|^{2} \;+ \;\lf|\lf\{\bt_{{-\bf k},\mu}^{r \dag}(t),
\al^{r \dag}_{{\bf k},e}(0) \ri\}\ri|^{2}. \eea

Charge conservation is obviously ensured at any time: \bea {\cal
Q}^e_{{\bf k},e}(t)+{\cal Q}^e_{{\bf k},\mu}(t)=1. \eea

The oscillation formula for the flavor charges are then
\cite{BHV98}:
\begin{eqnarray}\non\label{oscillfor1}
{\cal Q}^e_{{\bf k},e}(t)&=&\lf|\lf \{\al^{r}_{{\bf k},e}(t),
\al^{r \dag}_{{\bf k},e}(0) \ri\}\ri|^{2} \;+
\;\lf|\lf\{\bt_{{-\bf k},e}^{r \dag}(t), \al^{r \dag}_{{\bf
k},e}(0) \ri\}\ri|^{2}=  \\\nonumber &=&1-\sin ^{2}(2\theta
)\left[ \left| U_{\mathbf{k}}\right| ^{2}\sin
^{2}\left( \frac{\omega _{k,2}-\omega _{k,1}}{2}t\right) +\left| V_{\mathbf{k%
}}\right| ^{2}\sin ^{2}\left( \frac{\omega _{k,2}+\omega
_{k,1}}{2}t\right) \right] ,\\
\end{eqnarray}
\begin{eqnarray}\non\label{oscillfor2}
{\cal Q}^e_{{\bf k},\mu}(t) &=& \lf|\lf \{\al^{r}_{{\bf
k},\mu}(t), \al^{r \dag}_{{\bf k},e}(0) \ri\}\ri|^{2} \;+
\;\lf|\lf\{\bt_{{-\bf k},\mu}^{r \dag}(t), \al^{r \dag}_{{\bf
k},e}(0) \ri\}\ri|^{2}=   \\\nonumber &=&\sin ^{2}(2\theta )\left[
\left| U_{\mathbf{k}}\right| ^{2}\sin
^{2}\left( \frac{\omega _{k,2}-\omega _{k,1}}{2}t\right) +\left| V_{\mathbf{k%
}}\right| ^{2}\sin ^{2}\left( \frac{\omega _{k,2}+\omega
_{k,1}}{2}t\right) \right] .\\
\end{eqnarray}

This result is exact. There are two differences with respect to
the usual formula for neutrino oscillations: the amplitudes are
energy dependent, and there is an additional oscillating term.

For $\left| \mathbf{k}\right| \gg \sqrt{m_{1}m_{2}}$ we have
$\left| U_{\mathbf{k}}\right| ^{2}\longrightarrow 1$ and $\left|
V_{\mathbf{k}}\right| ^{2}\longrightarrow 0$ and the traditional
formula is recovered.

\section{Discussion}

We conclude the chapter with a number of considerations about the
oscillation formulas Eqs.(\ref{oscillfor1}), (\ref{oscillfor2}).

The Eqs.(\ref{oscillfor1}), (\ref{oscillfor2}) have a sense as
statistical averages, i.e. as mean values. This is because the
structure of the theory for mixed field is that of a many-body
theory, where does not make sense to talk about single particle
states.

This situation contrast with the quantum mechanical picture, which
however is recovered in the ultra-relativistic limit. There, the
approximate Pontecorvo result is recovered.

We now show \cite{Blasone:1999jb} that the exact oscillation
formulae are independent of the arbitrary mass parameters, indeed
we have \bea\non\label{miracle} \left| \left\{
\widetilde{\alpha}_{{\bf k},e}^{r}(t), \widetilde{\alpha}_{{\bf
k},e}^{r\dagger }(0)\right\}
 \right| ^{2}+\left| \left\{ \widetilde{\beta}_{{\bf k},e}^{r\dagger
}(t),\widetilde{\alpha}_{{\bf k},e}^{r\dagger }(0)\right\} \right|
^{2}=\\=\left| \left\{ \alpha _{{\bf k}, e}^{r}(t),\alpha _{{\bf
k},e}^{r\dagger }(0)\right\} \right| ^{2}+\left| \left\{ \beta
_{-{\bf k},e}^{r\dagger }(t),\alpha _{{\bf k}, e}^{r\dagger
}(0)\right\} \right| ^{2},\eea \bea\non\label{miracle2} \left|
\left\{ \widetilde{\alpha}_{{\bf k},\mu }^{r}(t),
\widetilde{\alpha}_{{\bf k},e}^{r\dagger }(0)\right\}
 \right| ^{2}+\left| \left\{ \widetilde{\beta}_{-{\bf k},\mu }^{r\dagger
}(t),\widetilde{\alpha}_{{\bf k},e}^{r\dagger }(0)\right\} \right|
^{2}=\\=\left| \left\{ \alpha _{{\bf k},\mu }^{r}(t),\alpha _{{\bf
k},e}^{r\dagger }(0)\right\} \right| ^{2}+\left| \left\{ \beta
_{-{\bf k},\mu }^{r\dagger }(t),\alpha _{{\bf k}, e}^{r\dagger
}(0)\right\} \right| ^{2}, \eea which ensure the cancellation of
the arbitrary mass parameters.

However, the important point for the full understanding of the
result Eqs.(\ref{miracle}), (\ref{miracle2}) is that the charge
operators $Q_\sigma$ are {\em invariant} under the action of the
Bogoliubov generator Eq.(\ref{FHYBVb12}), i.e. ${\widetilde
Q}_\sigma = Q_\sigma$, where ${\widetilde Q}_{\sigma}\equiv
{\widetilde \alpha}_{\sigma}^{\dag}{\widetilde \alpha}_{\sigma} -
{\widetilde \beta}^{\dag}_{\sigma}{\widetilde \beta}_{\sigma}$.
Besides the direct computations leading to Eqs.(\ref{miracle}),
(\ref{miracle2}), such an invariance provides a strong and
immediate proof of the independence of the oscillation formula
from the $\mu_\sigma$ parameters. Thus, the expectation values of
the flavor charge operators are the only physical relevant
quantities in the context of the above theory, all other operators
having expectation values depending on the arbitrary parameters
above introduced.

\bigskip
\bigskip
\chapter{Quantum Field Theory of three flavor neutrino mixing and
oscillations with  CP violation}

 \vspace{.4in}

We analyze the Quantum Field Theory of mixing among three
generations of Dirac fermions (neutrinos). We construct the
Hilbert space for the flavor fields and determine the generators
of the mixing transformations. By use of these generators, we
recover all the known parameterizations of the three-flavor mixing
matrix and we find a number of new ones. The algebra of the
currents associated with the mixing transformations is shown to be
a deformed $su(3)$ algebra, when CP violating phases are present.
We then derive the exact oscillation formulas, exhibiting new
features with respect to the usual ones. CP and T violation are
also discussed.

\section{Introduction}

In this chapter we study in detail the case of three flavor
fermion (neutrino) mixing. This is not a simple extension of the
previous results \cite{BHV98,Blasone:1999jb,BV95} since the
existence of a CP violating phase in the parameterization of the
three-flavor mixing matrix introduces novel features which are
absent in the two-flavor case. We determine the generators of the
mixing transformations and by use of them, we recover the known
parameterizations of the three-flavor mixing matrix and  find a
number of new ones. We construct the flavor Hilbert space, for
which the ground state (flavor vacuum) turns out to be a
generalized coherent state. We also study the algebraic structure
of currents and charges associated with the mixing transformations
and we find (as will show in the next chapter) that, in presence
of CP violation, it is that of a deformed $su(3)$. The
construction of the flavor Hilbert space is an essential step in
the derivation of exact oscillation formulas, which account for CP
violation and reduce to the corresponding quantum--mechanical ones
in the relativistic limit.

The chapter is organized as follows. In Section 3.2  we construct
the Hilbert space for three-flavor mixed fermions. In Section 3.3
we study the various parameterization of the unitary $3\times3$
mixing matrix obtained by use of the algebraic generators. In
Section 3.4, we study the currents and charges for three-flavor
mixing, which are then used in Section 3.5 to derive the exact
neutrino oscillation formulas. Finally in Section 3.6, CP an T
violation in QFT neutrino oscillations are discussed. The Section
3.7 is devoted to conclusions. In the Appendices F, G, H, I we put
some useful formulas and a discussion of the arbitrary mass
parameterization in the expansion of flavor fields as recently
reported in \cite{fujii2,yBCV02,comment}.

\section{Three flavor fermion mixing}

We start by considering the following Lagrangian density
describing three Dirac fields with a mixed mass term:
\bea\label{lagemu123} {\cal L}(x)\,=\,  {\bar \Psi_f}(x) \lf( i
\not\!\partial - \textsf{M} \ri) \Psi_f(x)\, , \eea
where $\Psi_f^T=(\nu_e,\nu_\mu,\nu_{\tau})$ and $\textsf{M} =
\textsf{M}^\dag$ is the mixed mass matrix.

Among the various possible parameterizations of the mixing matrix
for three fields, we choose to work with the following one since
it is the familiar parameterization of the CKM matrix
\cite{mesons}:
\bea\non\label{fermix123} \Psi_f(x) \, = {\cal U} \, \Psi_m
(x)=\begin{pmatrix}
c_{12}c_{13} & s_{12}c_{13} & s_{13}e^{-i\de} \\
-s_{12}c_{23}-c_{12}s_{23}s_{13}e^{i\de} &
c_{12}c_{23}-s_{12}s_{23}s_{13}e^{i\de} & s_{23}c_{13} \\
s_{12}s_{23}-c_{12}c_{23}s_{13}e^{i\de} &
-c_{12}s_{23}-s_{12}c_{23}s_{13}e^{i\de} & c_{23}c_{13}
\end{pmatrix}\,\Psi_m (x),\\ \eea
with $c_{ij}=\cos\te_{ij}$ and  $s_{ij}=\sin\te_{ij}$, being
$\te_{ij}$ the mixing angle between $\nu_{i},\nu_{j}$ and
$\Psi_m^T=(\nu_1,\nu_2,\nu_3)$.

Using Eq.(\ref{fermix123}), we diagonalize the quadratic form of
Eq.(\ref{lagemu123}), which then reduces to the Lagrangian for
three Dirac fields, with masses $m_1$, $m_2$ and $m_3$:
\bea\label{lage123} {\cal L}(x)\,=\,  {\bar \Psi_m}(x) \lf( i
\not\!\partial -  \textsf{M}_d\ri) \Psi_m(x)  \, , \eea
where $\textsf{M}_d = diag(m_1,m_2,m_3)$.

As in the case of two flavor fermion mixing \cite{BV95}, we
construct the generator for the mixing transformation
(\ref{fermix123}) and define\footnote{Let us consider for example
the generation of the first row of the mixing matrix ${\cal U}$.
We have (see also Appendix H) $ \pa \nu_e /\pa\te_{23}\, =\, 0; $
and \bea \non\pa\nu_e /\pa \te_{13} \, =
  \,G_{12}^{-1}G_{13}^{-1}[\nu_1,L_{13}]G_{13}G_{12}\,=\,
  G_{12}^{-1}G_{13}^{-1} e^{-i \de}\nu_3 G_{13}G_{12},\eea thus:
\bea
 \non \pa^2\nu_e/\pa \te_{13}^2  \, = \, -\nu_e \quad \Rar \quad
\nu_e \, = \, f(\te_{12}) \cos\te_{13} + g(\te_{12}) \sin\te_{13};
\eea with the initial conditions (from Eq.(\ref{incond})):
$f(\te_{12}) = \nu_e|_{\te_{13}=0}  $ and $ g(\te_{12}) =
\pa\nu_e/\pa \te_{13}|_{\te_{13}=0} = e^{-i \de}\nu_3 $. We also
have \bea \non \pa^2f(\te_{12})/\pa \te_{13}^2  \,  = \,
-f(\te_{12}) \quad \Rar \quad f(\te_{12}) \, = \, A \cos\te_{12}
\,+ B \sin\te_{12} \eea with the initial conditions
$A=\nu_e|_{\te=0}=\nu_1$ and $B=\pa f(\te_{12})
/\pa\te_{12}|_{\te=0}=\nu_2$, and $\te=(\te_{12}, \te_{13},
\te_{23}).$}
\bea\label{incond} &&\nu_{\si}^{\al}(x)\equiv G^{-1}_{\bf \te}(t)
\, \nu_{i}^{\al}(x)\, G_{\bf \te}(t), \eea where $(\si,i)=(e,1),
(\mu,2), (\tau,3)$, and
\bea\label{generator} &&G_{\bf
\te}(t)=G_{23}(t)\;G_{13}(t)\;G_{12}(t)\, , \eea
where
\bea\label{generators1} && G_{12}(t)\equiv
\exp\Big[\te_{12}L_{12}(t)\Big];\;\;\;\;\;\;\;\; L_{12}(t)=\int
d^{3}{\bf
x}\lf[\nu_{1}^{\dag}(x)\nu_{2}(x)-\nu_{2}^{\dag}(x)\nu_{1}(x)\ri],
\\ \label{generators2}
&&G_{23}(t)\equiv\exp\Big[\te_{23}L_{23}(t)\Big];\;\;\;\;\;\;\;\;
L_{23}(t)=\int d^{3}{\bf
x}\lf[\nu_{2}^{\dag}(x)\nu_{3}(x)-\nu_{3}^{\dag}(x)\nu_{2}(x)\ri],
\\ \non
&&G_{13}(t)\equiv\exp\Big[\te_{13}L_{13}(\de,t)\Big];
\;\;\;\;\;L_{13}(\de,t)=\int d^{3}{\bf
x}\lf[\nu_{1}^{\dag}(x)\nu_{3}(x)e^{-i\de}-\nu_{3}^{\dag}(x)
\nu_{1}(x)e^{i\de}\ri].\\\label{generators3} \eea
It is evident from the above form of the generators, that the
phase $\de$ is unavoidable for three field mixing, while it can be
incorporated in the definition of the fields in the two flavor
case.

The free fields  $\nu_i$ (i=1,2,3) can be quantized in the usual
way \cite{Itz} (we use $t\equiv x_0$):
\bea\label{2.2} \nu_{i}(x) = \sum_{r} \int \frac{d^3 {\bf
k}}{(2\pi)^{\frac{3}{2}}} \lf[u^{r}_{{\bf k},i}(t) \al^{r}_{{\bf
k},i}\:+ v^{r}_{-{\bf k},i}(t) \bt^{r\dag }_{-{\bf k},i}  \ri]
e^{i {\bf k}\cdot{\bf x}} ,\qquad i=1,2,3\,, \eea
with $u^{r}_{{\bf k},i}(t)=e^{-i\om_{k,i} t}u^{r}_{{\bf k},i}$,
$v^{r}_{{\bf k},i}(t)=e^{i\om_{k,i} t}v^{r}_{{\bf k},i}$ and
$\om_{k,i}=\sqrt{{\bf k}^2+m_i^2}$. The vacuum for the mass
eigenstates is denoted by $|0\ran_{m}$:  $\; \; \al^{r}_{{\bf
k},i}|0\ran_{m}= \bt^{r }_{{\bf k},i}|0\ran_{m}=0$.   The
anticommutation relations are the usual ones; the wave function
orthonormality and completeness relations are given in
Eqs.(\ref{orthocompl}).

An important result of previous chapter is the unitary
inequivalence \cite{BV95} (in the infinite volume limit) of the
vacua for the flavor fields and for the fields with definite
masses. There such an inequivalence was proved for the case of two
generations; subsequently, in Ref.\cite{hannabus}, a rigorous
general proof of such inequivalence for any number of generations
has been given (see also Ref.\cite{JM011}). Thus we define the
{\em flavor vacuum} as:
\bea\label{flavac123}
|0(t)\ran_{f}\,\equiv\,G_{\te}^{-1}(t)\;|0\ran_{m} \;. \eea

The form of this state is considerably more complicated of the one
for two generations. When $\de=0$, the generator $G_\te$ is an
element of the $SU(3)$ group (see Chapter 4) and  the flavor
vacuum is classified as an $SU(3)$ generalized coherent state \`a
la Perelomov \cite{Per}. A nonzero CP violating phase introduces
an interesting modification of the algebra associated with the
mixing transformations Eq.(\ref{fermix123}): we discuss this in
Chapter 4.

By use of $G_{\bf \te}(t)$, the flavor fields can be expanded as:
\bea\label{exnue123} &&{}\quad\qquad \nu_\si(x)= \sum_{r} \int
\frac{d^3 {\bf k}}{(2\pi)^{\frac{3}{2}}} \lf[ u^{r}_{{\bf k},i}(t)
\al^{r}_{{\bf k},\si}(t) + v^{r}_{-{\bf k},i}(t)
\bt^{r\dag}_{-{\bf k},\si}(t) \ri]  e^{i {\bf k}\cdot{\bf x}},\eea
with $(\si,i)=(e,1),(\mu,2),(\tau,3)$.

The flavor annihilation operators are defined as \bea
\al^{r}_{{\bf k},\si}(t) &\equiv& G^{-1}_{\bf
\te}(t)\;\al^{r}_{{\bf k},i}\; G_{\bf \te}(t)\\\bt^{r\dag}_{{-\bf
k},\si}(t)&\equiv&
 G^{-1}_{\bf \te}(t)\; \bt^{r\dag}_{{-\bf k},i}\;
G_{\bf \te}(t).\eea They clearly act as annihilators for the
flavor vacuum Eq.(\ref{flavac123}). For further reference, it is
useful to list explicitly the flavor annihilation/creation
operators \cite{yBCV02}. In the reference frame ${\bf
k}=(0,0,|{\bf k}|)$ the spins decouple and their form is
particularly simple:
\bea\non \al_{{\bf k},e}^{r}(t)&=&c_{12}c_{13}\;\al_{{\bf
k},1}^{r} + s_{12}c_{13}\lf(U^{{\bf k}*}_{12}(t)\;\al_{{\bf
k},2}^{r} +\epsilon^{r} V^{{\bf k}}_{12}(t)\;\bt_{-{\bf
k},2}^{r\dag}\ri)+
\\&&+e^{-i\de}\;s_{13}\lf(U^{{\bf k}*}_{13}(t)\;\al_{{\bf k},3}^{r}
+\epsilon^{r} V^{{\bf k}}_{13}(t)\;\bt_{-{\bf k},3}^{r\dag}\ri)\;,
\\[2mm]\non
\al_{{\bf k},\mu}^{r}(t)&=&\lf(c_{12}c_{23}- e^{i\de}
\;s_{12}s_{23}s_{13}\ri)\;\al_{{\bf k},2}^{r} -
\lf(s_{12}c_{23}+e^{i\de}\;c_{12}s_{23}s_{13}\ri)\times
\\\non&&\times
\lf(U^{{\bf k}}_{12}(t)\;\al_{{\bf k},1}^{r} -\epsilon^{r} V^{{\bf
k}}_{12}(t)\;\bt_{-{\bf k},1}^{r\dag}\ri)
+\;s_{23}c_{13}\lf(U^{{\bf k}*}_{23}(t)\;\al_{{\bf k},3}^{r} +
\epsilon^{r} V^{{\bf k}}_{23}(t)\;\bt_{-{\bf
k},3}^{r\dag}\ri)\;,\\
\\[2mm]\non
\al_{{\bf k},\tau}^{r}(t)&=&c_{23}c_{13}\;\al_{{\bf k},3}^{r} -
\lf(c_{12}s_{23}+e^{i\de}\;s_{12}c_{23}s_{13}\ri) \lf(U^{{\bf
k}}_{23}(t)\;\al_{{\bf k},2}^{r} -\epsilon^{r} V^{{\bf
k}}_{23}(t)\;\bt_{-{\bf k},2}^{r\dag}\ri)+
\\
&&+\;\lf(s_{12}s_{23}- e^{i\de}\;c_{12}c_{23}s_{13}\ri)
\lf(U^{{\bf k}}_{13}(t)\;\al_{{\bf k},1}^{r} -\epsilon^{r} V^{{\bf
k}}_{13}(t)\;\bt_{-{\bf k},1}^{r\dag}\ri)\;, \eea

\bea\non \bt^{r}_{-{\bf k},e}(t)&=&c_{12}c_{13}\;\bt_{-{\bf
k},1}^{r} + s_{12}c_{13}\lf(U^{{\bf k}*}_{12}(t)\;\bt_{-{\bf
k},2}^{r} -\epsilon^{r}V^{{\bf k}}_{12}(t)\;\al_{{\bf
k},2}^{r\dag}\ri)+
\\&&+e^{i\de}\; s_{13}\lf(U^{{\bf k}*}_{13}(t)\;\bt_{-{\bf k},3}^{r}
-\epsilon^{r} V_{13}^{{\bf k}}(t)\;\al_{{\bf k},3}^{r\dag}\ri)\;,
\\[2mm] \non
\bt^{r}_{-{\bf k},\mu}(t)&=&\lf(c_{12}c_{23}- e^{-i\de}\;
s_{12}s_{23}s_{13}\ri)\;\bt_{-{\bf k},2}^{r} -
\lf(s_{12}c_{23}+e^{-i\de}\;c_{12}s_{23}s_{13}\ri)\times
\\\non && \times \lf(U^{{\bf k}}_{12}(t)\;\bt_{-{\bf k},1}^{r}
+\epsilon^{r}\; V^{{\bf k}}_{12}(t)\;\al_{{\bf k},1}^{r\dag}\ri) +
\; s_{23}c_{13}\lf(U^{{\bf k}*}_{23}(t)\;\bt_{-{\bf k},3}^{r} -
\epsilon^{r} V^{{\bf k}}_{23}(t)\;\al_{{\bf
k},3}^{r\dag}\ri)\;,\\
\\[2mm] \non
\bt^{r}_{-{\bf k},\tau}(t)&=&c_{23}c_{13}\;\bt_{-{\bf k},3}^{r} -
\lf(c_{12}s_{23}+e^{-i\de}\;s_{12}c_{23}s_{13}\ri) \lf(U^{{\bf
k}}_{23}(t)\;\bt_{-{\bf k},2}^{r} + \epsilon^{r} V^{{\bf
k}}_{23}(t)\;\al_{{\bf k},2}^{r\dag}\ri)
\\
&&+\;\lf(s_{12}s_{23}- e^{-i\de}\;c_{12}c_{23}s_{13}\ri)
\lf(U^{{\bf k}}_{13}(t)\;\bt_{-{\bf k},1}^{r} + \epsilon^{r}
V^{{\bf k}}_{13}(t)\;\al_{{\bf k},1}^{r\dag}\ri)\;. \eea

These operators satisfy canonical (anti)commutation relations at
equal times. The main difference with respect to their ``naive''
quantum-mechanical counterparts is in the anomalous terms
proportional to the $V_{ij}$ factors. In fact, $U^{{\bf k}}_{ij}$
and $V^{{\bf k}}_{ij}$ are Bogoliubov coefficients defined as:
\bea V^{{\bf k}}_{ij}(t)=|V^{{\bf
k}}_{ij}|\;e^{i(\om_{k,j}+\om_{k,i})t}\;\;\;\;,\;\;\;\; U^{{\bf
k}}_{ij}(t)=|U^{{\bf k}}_{ij}|\;e^{i(\om_{k,j}-\om_{k,i})t} \eea

\bea\non &&|U^{{\bf
k}}_{ij}|=\lf(\frac{\om_{k,i}+m_{i}}{2\om_{k,i}}\ri)
^{\frac{1}{2}}
\lf(\frac{\om_{k,j}+m_{j}}{2\om_{k,j}}\ri)^{\frac{1}{2}}
\lf(1+\frac{|{\bf k}|^{2}}{(\om_{k,i}+m_{i})
(\om_{k,j}+m_{j})}\ri)=\cos(\xi_{ij}^{{\bf k}})
\\\eea\bea\non
&&|V^{{\bf k}}_{ij}|=\lf(\frac{\om_{k,i}+m_{i}}{2\om_{k,i}}\ri)
^{\frac{1}{2}}
\lf(\frac{\om_{k,j}+m_{j}}{2\om_{k,j}}\ri)^{\frac{1}{2}}
\lf(\frac{|{\bf k}|}{(\om_{k,j}+m_{j})}-\frac{|{\bf
k}|}{(\om_{k,i}+m_{i})}\ri)=\sin(\xi_{ij}^{{\bf k}})\\ \eea \bea
|U^{{\bf k}}_{ij}|^{2}+|V^{{\bf k}}_{ij}|^{2}=1 \eea where
$i,j=1,2,3$ and $j>i$.
The following identities hold:

\bea \label{ident1} &&V^{{\bf k}}_{23}(t)V^{{\bf
k}*}_{13}(t)+U^{{\bf k}*}_{23}(t)U^{{\bf k}}_{13}(t) = U^{{\bf
k}}_{12}(t),\\ &&V^{{\bf k}}_{23}(t)U^{{\bf k}*}_{13}(t)-U^{{\bf
k}*}_{23}(t)V^{{\bf k}}_{13}(t) =- V^{{\bf k}}_{12}(t),
\\ \label{ident2}
&&U^{{\bf k}}_{12}(t)U^{{\bf k}}_{23}(t)-V^{{\bf
k}*}_{12}(t)V^{{\bf k}}_{23}(t) = U^{{\bf k}}_{13}(t),\\
&&U^{{\bf k}}_{23}(t)V^{{\bf k}}_{12}(t)+U^{{\bf
k}*}_{12}(t)V^{{\bf k}}_{23}(t) = V^{{\bf k}}_{13}(t),
\\ \label{ident3}
&&V^{{\bf k}*}_{12}(t)V^{{\bf k}}_{13}(t)+U^{{\bf
k}*}_{12}(t)U^{{\bf k}}_{13}(t) = U^{{\bf k}}_{23}(t),
\\
&&V^{{\bf k}}_{12}(t)U^{{\bf k}}_{13}(t)-U^{{\bf
k}}_{12}(t)V^{{\bf k}}_{13}(t) =- V^{{\bf k}}_{23}(t)\;,
\\ \label{ident4}
&& \xi_{13}^{{\bf k}}=\xi_{12}^{{\bf k}} +\xi_{23}^{{\bf k}}
\qquad, \qquad
 \xi_{ij}^{{\bf k}}= \arctan\lf(|V^{{\bf k}}_{ij}|\,/\,|U^{{\bf k}}_{ij}|\ri)\,.
\eea

 We observe that, in contrast with the
case of two flavor mixing, the condensation densities are now
different for particles of different masses \cite{yBCV02}:
\bea\non {\cal N}^{\bf k}_1\, &=& \,_{f}\langle0(t)|N^{{\bf
k},r}_{\al_{1}} |0(t)\ran_{f}= \,_{f}\langle0(t)|N^{{\bf
k},r}_{\bt_{1}}|0(t)\ran_{f}= s^{2}_{12}c^{2}_{13}\,|V^{{\bf
k}}_{12}|^{2}+ s^{2}_{13}\,|V^{{\bf k}}_{13}|^{2}\,,
\\\eea
\bea\non{\cal N}^{\bf k}_2\, &=& \,_{f}\langle0(t)|N^{{\bf
k},r}_{\al_{2}}|0(t)\ran_{f}= \,_{f}\langle0(t)|N^{{\bf
k},r}_{\bt_{2}}|0(t)\ran_{f}=
\\&=&\lf|-s_{12}c_{23}+e^{i\de}\,c_{12}s_{23}s_{13}\ri|^{2}
\,|V^{{\bf k}}_{12}|^{2}+ s^{2}_{23}c^{2}_{13}\;|V^{{\bf
k}}_{23}|^{2}\,, \eea\bea\non {\cal N}^{\bf k}_3\, &=&
\,_{f}\langle0(t)|N^{{\bf k},r}_{\al_{3}}|0(t)\ran_{f}=
\,_{f}\langle0(t)|N^{{\bf k},r}_{\bt_{3}}|0(t)\ran_{f}=
\\&=&\lf|-c_{12}s_{23}+e^{i\de}\,s_{12}c_{23}s_{13}\ri|^{2} |V^{{\bf
k}}_{23}|^{2} + \lf|s_{12}s_{23}+
e^{i\de}\,c_{12}c_{23}s_{13}\ri|^{2} |V^{{\bf k}}_{13}|^{2}\,.
\eea

We plot in Figure 3.1 the condensation densities for sample values
of parameters as given in the table 3.1\footnote{Here and in the
following plots, we use the same (energy) units for the values of
masses and momentum.}: \vspace{0.5cm}

$$
\begin{tabular}{|c|c|c|c|c|c|c|}
\hline \;\;{\em $m_{1}^{}$} \; & \;\;{\em $m_{2}$}\; & \;\;{\em
$m_{3}$}\; & \;\;{\em $\te_{12}$}\; &\;\;{\em $\te_{13}$} \;&
\;\;{\em $\te_{23}$}\;&
\;\;{\em $\de$}\; \\[1.5mm] \hline
$ 1^{}$ & $ 200$   &  $ 3000 $    &  $  \pi/4  $  &  $ \pi/4  $ &
$ \pi/4  $&  $\; \pi/4  $
\\[1.5mm] \hline
\end{tabular}
$$
\centerline{Table 3.1: The values of masses and mixing angles used
for plots}

\vspace{0.5cm}

\begin{figure}
\centerline{\epsfysize=3.0truein\epsfbox{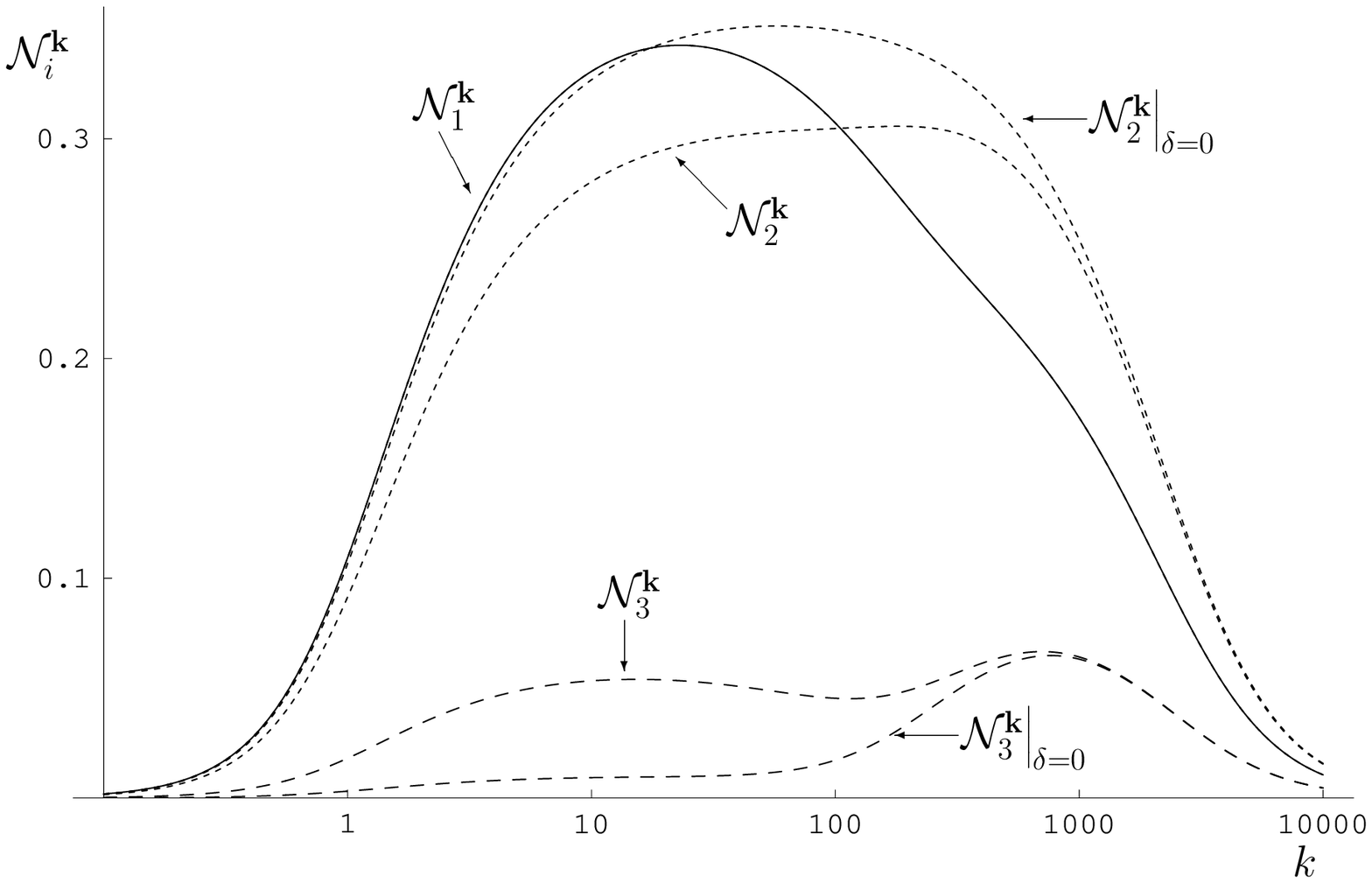}}
\vspace{.2cm}
\caption{Plot of the condensation densities ${\cal N}^{\bf k}_i$
in function of $|{\bf k}|$  for the values of parameters as in
Tab.(1).}

\vspace{0.5cm}

\hrule
\end{figure}
\normalsize

\newpage

%\vspace{8cm}
\section{The  parameterizations of the three flavor mixing
matrix}

In Section 3.2 we have studied the generator of the mixing matrix
${\cal U}$ of Eq.(\ref{fermix123}). However, this matrix is only
one of the various forms in which a $3\times 3$ unitary matrix can
be parameterized. Indeed, the generator Eq.(\ref{generator}) can
be used for generating such alternative parameterizations. To see
this, let us first  define in a more general way the generators
$G_{ij}$ including phases for all of them:

\bea\non G_{12}(t)\equiv\exp\Big[\te_{12}L_{12}(\de_{12},t)\Big]
\;;\;\;\; L_{12}(\de_{12},t)=\int d^{3}{\bf
x}\lf[\nu_{1}^{\dag}(x)\nu_{2}(x)e^{-i\de_{12}}-
\nu_{2}^{\dag}(x)\nu_{1}(x)e^{i\de_{12}}\ri],
\\\eea \bea\non G_{23}(t)\equiv \exp\Big[\te_{23}L_{23}(\de_{23},t)\Big]
\;;\;\;\; L_{23}(\de_{23},t)=\int d^{3}{\bf
x}\lf[\nu_{2}^{\dag}(x)\nu_{3}(x)e^{-i\de_{23}}-
\nu_{3}^{\dag}(x)\nu_{2}(x)e^{i\de_{23}}\ri],
\\\eea \bea\non
G_{13}(t)\equiv\exp\Big[\te_{13}L_{13}(\de_{13},t)\Big] \;;\;\;\;
L_{13}(\de_{13},t)=\int d^{3}{\bf
x}\lf[\nu_{1}^{\dag}(x)\nu_{3}(x)e^{-i\de_{13}}-
\nu_{3}^{\dag}(x)\nu_{1}(x)e^{i\de_{13}}\ri]\,.\\ \eea

Six different matrices can be obtained by permuting the order of
the $G_{ij}$ (useful relations are listed in Appendices F and H)
in Eq.(\ref{generator}). We obtain:

\bea G_{1} \,\equiv\, G_{23}G_{13}G_{12} \eea
\bea\non
 {\cal U}_1\,=\,\begin{pmatrix}
c_{12}c_{13} & s_{12}c_{13}e^{-i\de_{12}} & s_{13}e^{-i\de_{13}}
\\\ -s_{12}c_{23}e^{i\de_{12}}- s_{23}s_{13}c_{12}e^{i
(\de_{13}-\de_{23}) } & c_{12}c_{23}-
s_{23}s_{13}s_{12}e^{-i(\de_{23} - \de_{13} +\de_{12})}&
s_{23}c_{13}e^{-i\de_{23}} \\
-s_{13}c_{12}c_{23}e^{i\de_{13}}+ s_{12}s_{23}e^{i (\de_{12} +
\de_{23})}&
-c_{23}s_{13}s_{12}e^{i(\de_{13}-\de_{12})}-s_{23}c_{12}e^{i\de_{23}}&
c_{23}c_{13}
\end{pmatrix}\
\\\non
\eea

\bea G_{2}\,\equiv\,G_{23}G_{12}G_{13} \eea
\bea\non {\cal
U}_2\,=\,\begin{pmatrix} c_{12}c_{13} & s_{12}e^{-i\de_{12}}
& s_{13}c_{12}e^{-i\de_{13}} \\
-s_{12}c_{13}c_{23}e^{i\de_{12}}-
s_{23}s_{13}e^{i(\de_{13}-\de_{23}) } & c_{12}c_{23} & -
s_{13}c_{23}s_{12}e^{i(\de_{12}-\de_{13})} +
s_{23}c_{13}e^{-i\de_{23}} \\
-s_{13}c_{23}e^{i\de_{13}}+
s_{12}s_{23}c_{13}e^{i(\de_{12}+\de_{23})}&
-c_{12}s_{23}e^{i\de_{23}} & c_{23}c_{13}+
s_{12}s_{13}s_{23}e^{i(\de_{12}+\de_{23}-\de_{13})}
\end{pmatrix}\
\\\non\eea

\bea  G_{3}\,\equiv\,G_{13}G_{23}G_{12} \eea
\bea\non {\cal
U}_3\,=\,\begin{pmatrix} c_{12}c_{13}+
s_{13}s_{23}s_{12}e^{i(\de_{12}-\de_{13}+\de_{23})} &
s_{12}c_{13}e^{-i\de_{12}}-
s_{13}s_{23}c_{12}e^{i(\de_{23}-\de_{13})}&
s_{13}c_{23}e^{-i\de_{13}} \\
-s_{12}c_{23}e^{i\de_{12}} &
c_{12}c_{23} & s_{23}e^{-i\de_{23}} \\
c_{13}s_{23}s_{12}e^{i(\de_{23}+\de_{12})}
-s_{13}c_{12}e^{i\de_{13}} & -c_{13}s_{23}c_{12}e^{i\de_{23}} -
s_{12}s_{13}e^{i(\de_{13}-\de_{12})} & c_{23}c_{13}
\end{pmatrix}\
\\\non\eea

\bea G_{4}\,\equiv\,G_{13}G_{12}G_{23} \eea
\bea\non {\cal
U}_4\,=\,\begin{pmatrix} c_{12}c_{13} &
s_{12}c_{13}c_{23}e^{-i\de_{12}} -
s_{13}s_{23}e^{i(\de_{23}-\de_{13})} &
s_{12}s_{23}c_{13}e^{-i(\de_{12}+\de_{23})}
+ s_{13}c_{23}e^{-i\de_{13}} \\
-s_{12}e^{i\de_{12}} &   c_{12}c_{23} &
s_{23}c_{12}e^{-i\de_{23}} \\
-c_{12}s_{13}e^{i\de_{13}} & -c_{13}s_{23}e^{i\de_{23}} -
s_{12}c_{23}s_{13}e^{i(\de_{13}-\de_{12})} & c_{23}c_{13} -
s_{12}s_{23}s_{13}e^{-i(\de_{12}+\de_{23}-\de_{13})}
\end{pmatrix}\
\\\non\eea

\bea G_5 \,\equiv\, G_{12}G_{13}G_{23} \eea
\bea\non {\cal
U}_5\,=\,\begin{pmatrix} c_{12}c_{13} &
s_{12}c_{23}e^{-i\de_{12}}- s_{13}c_{12}s_{23}
e^{-i(\de_{13}-\de_{23})} & s_{13}c_{12}c_{23}e^{-i\de_{13}}+
s_{12}s_{23}e^{-i(\de_{12}+\de_{23})}\\
-s_{12}c_{13}e^{i\de_{12}} & c_{12}c_{23}+
s_{12}s_{23}s_{13}e^{i(\de_{12}-\de_{13}+\de_{23})}
&s_{23}c_{12}e^{-i\de_{23}}- s_{12}c_{23}s_{13}
e^{i(\de_{12}-\de_{13})} \\
-s_{13}e^{i\de_{13}} & -c_{13}s_{23}e^{i\de_{23}} & c_{23}c_{13}
\end{pmatrix}\
\\\non\eea

\bea\ \label{matrix} G_6\,\equiv\,G_{12}G_{23}G_{13} \eea\bea\non
{\cal U}_6\,=\,\begin{pmatrix} c_{12}c_{13}-
s_{12}s_{23}s_{13}e^{-i(\de_{12}+\de_{23} -\de_{13})} &
s_{12}c_{23} e^{-i\de_{12}} & c_{12}s_{13}e^{-i\de_{13}}+
s_{12}s_{23}c_{13}e^{-i(\de_{12}+\de_{23})} \\
-c_{12}s_{23}s_{13}e^{i(\de_{13} -\de_{23})}- s_{12}c_{13}
e^{i\de_{12}}&c_{12}c_{23} & c_{12}s_{23}c_{13}e^{-i\de_{23}}
- s_{12}s_{13}e^{i(\de_{12}-\de_{13})}\\
-c_{23}s_{13}e^{i\de_{13}} & -s_{23}e^{i\de_{23}} & c_{23}c_{13}
\end{pmatrix}\
\\\non\eea

The above matrices are generated for a particular set of initial
conditions, namely for those of Eq.(\ref{incond}). The freedom in
the choice of the initial conditions reflects into the possibility
of obtaining other unitary matrices from the above ones by
permuting rows and columns and by multiplying row or columns for a
phase factor.

We thus can easily recover all the existing parameterizations of
the CKM matrix
\cite{mesons,wolfenstein,Chaturvedi,Fritzsch,Maiani,Chau-Keung,Anselm}:

\noi - the Maiani parameterization \cite{Chaturvedi,Maiani} is
obtained from ${\cal U}_1$ by setting $\te_{12}\rar \te$,
$\te_{13}\rar\bt$, $\te_{23}\rar \ga$, $\de_{12}\rar 0$,
$\de_{13}\rar 0$, $\de_{23}\rar -\de$;

\noi - the Chau--Keung parameterization
\cite{Chaturvedi,Chau-Keung} is recovered from ${\cal U}_1$ by
setting $\de_{12}\rar 0$ and $\de_{23}\rar 0$;

\noi - the Kobayshi--Maskawa \cite{mesons,Chaturvedi} is recovered
from ${\cal U}_5$ by setting $\te_{12}\rar \te_{2}$,
$\te_{13}\rar\te_{1}$, $\te_{23}\rar \te_{3}$, $\de_{12}\rar
-\de$, $\de_{13}\rar 0$ and $\de_{23}\rar 0$, $\te_{i}\rar
\frac{3}{2}\pi-\te_{i}$, with $i=1,2,3 $, and multiply the last
column for $(-1)$;

\noi - the Anselm parameterization \cite{Chaturvedi,Anselm} is
obtained from ${\cal U}_1$ by setting $\te_{12} \leftrightarrow
\te_{13}$, then $\de_{12}\rar 0$, $\de_{13}\rar 0$, $ \te_{12}\rar
\pi + \te_{12}$, $ \te_{13}\rar
 \pi -\te_{13}$, $ \te_{23}\rar \frac{3}{2}\pi+\te_{23}$,
exchanging second and third column and multiplying the last row
for $(-1)$.

From the above analysis it is clear that a number of new
parameterizations of the mixing matrix can be generated and that a
clear physical meaning can be attached to each of them, by
considering the order in which the generators $G_{ij}$ act and the
initial conditions used for getting that particular matrix.

\section{Currents and charges for three flavor fermion mixing}

In this Section we study the currents associated to the
Lagrangians Eqs.(\ref{lagemu123}) and (\ref{lage123}). To this
end, let us consider the transformations acting on the triplet of
free fields with different masses $\Psi_m$ \cite{currents,yBCV02}.

${\cal L}$ is invariant under global $U(1)$ phase transformations
of the type $\Psi_m' \, =\, e^{i \al }\, \Psi_m$: as a result, we
have the conservation of the Noether charge $Q=\int d^3{\bf x} \,
I^0(x) $ (with $I^\mu(x)={\bar \Psi}_m(x) \, \ga^\mu \,
\Psi_m(x)$) which is indeed the total charge of the system (i.e.
the total lepton number).

Consider then the $SU(3)$ global transformations acting on
$\Psi_m$:
\bea \label{massu3} \Psi_m'(x) \, =\, e^{i \al_j  F_j}\, \Psi_m
(x) \, \qquad, \qquad
 j=1, 2,..., 8.
\eea
with $\al_j$ real constants, $F_{j}=\frac{1}{2}\la_{j}$ being the
generators  of $SU(3)$ and $\lambda_j$ the Gell-Mann matrices
\cite{mesons}.

The Lagrangian is not generally invariant under (\ref{massu3}) and
we obtain, by use of the equations of motion,
\bea \non &&\de {\cal L}(x)\,= \,  i \al_j \,{\bar \Psi_m}(x)\,
[F_j,\textsf{M}_d ]\, \Psi_m(x) \, =\,  - \al_j \,\pa_\mu
J_{m,j}^\mu (x)
\\ [3mm]\label{fermacu1}
&&J^\mu_{m,j}(x)\, =\, {\bar \Psi_m}(x)\, \ga^\mu\, F_j\,
\Psi_m(x) \qquad, \qquad j=1, 2,..., 8. \eea

It is useful to list explicitly the eight currents:
\bea &&J_{m,1}^\mu(x)\; =\; \frac{1}{2} \lf[ {\bar \nu}_1
(x)\;\ga^\mu\; \nu_2 (x)\; + \,{\bar \nu}_2(x)\; \ga^\mu\; \nu_1 (x) \ri] \\
&&J_{m,2}^\mu (x)\; =\; -\frac{i}{2} \lf[  {\bar \nu}_1(x)\;
\ga^\mu\; \nu_2 (x)\; - \,{\bar \nu}_2(x)\; \ga^\mu\; \nu_1 (x) \ri] \\
&&J_{m,3}^\mu (x)\; =\; \frac{1}{2} \lf[ {\bar \nu}_1(x)\;
\ga^\mu\; \nu_1 (x)\; - \,{\bar \nu}_2(x)\; \ga^\mu\; \nu_2 (x)\ri] \\
&&J_{m,4}^\mu (x)\; =\; \frac{1}{2} \lf[ {\bar \nu}_1(x)\;
\ga^\mu\; \nu_3(x)\; + \,{\bar \nu}_3(x)\; \ga^\mu\; \nu_1 (x) \ri] \\
&&J_{m,5}^\mu (x)\; =\; -\frac{i}{2} \lf[ {\bar \nu}_1(x)\;
\ga^\mu\; \nu_3(x)\; - \,{\bar \nu}_3(x)\; \ga^\mu\; \nu_1(x) \ri] \\
&&J_{m,6}^\mu (x)\; =\; \frac{1}{2} \lf[ {\bar \nu}_2(x)\;
\ga^\mu\; \nu_3 (x)\; + \,{\bar \nu}_3 (x)\;\ga^\mu\; \nu_2 (x)\;
\ri]
\\ &&J_{m,7}^\mu (x)\;=\; -\frac{i}{2} \lf[  {\bar \nu}_2(x)\;
\ga^\mu\; \nu_3 (x)\; - \,{\bar \nu}_3(x)\; \ga^\mu\; \nu_2(x) \ri]
\\\non
&&J_{m,8}^\mu (x)\; =\; \frac{1}{2\sqrt{3}} \lf[ {\bar \nu}_1(x)\;
\ga^\mu\; \nu_1 (x)\; + \,{\bar \nu}_2(x)\; \ga^\mu\; \nu_2(x)\;-
2{\bar \nu}_3(x)\; \ga^\mu\; \nu_3(x) \ri].\\ \eea

The related charges  $Q_{m,j}(t)\equiv \int d^3 {\bf x}
\,J^0_{m,j}(x) $, satisfy the  $su(3)$  algebra \bea[Q_{m,j}(t),
Q_{m,k}(t)]\, =\, i \,f_{jkl}\,Q_{m,l}(t).\eea

Note that only two of the above charges are time-independent,
namely $Q_{m,3}$ and $Q_{m,8}$. We can thus define the
combinations:
\bea\non Q_{1}& \equiv &\frac{1}{3}Q \,+ \,Q_{m,3}+
\,\frac{1}{\sqrt{3}}Q_{m,8},
\\ \lab{noether2}
Q_{2}& \equiv & \frac{1}{3}Q \,-
\,Q_{m,3}+\,\frac{1}{\sqrt{3}}Q_{m,8},
\\\non
Q_{3}& \equiv &\frac{1}{3}Q \,- \,\frac{2}{\sqrt{3}}Q_{m,8}, \eea
\bea
 &&\lab{charge}Q_i \, = \,\sum_{r} \int d^3 {\bf k}\lf(
\al^{r\dag}_{{\bf k},i} \al^{r}_{{\bf k},i}\, -\,
\bt^{r\dag}_{-{\bf k},i}\bt^{r}_{-{\bf k},i}\ri),\,\,\,\qquad i=1,
2, 3 . \eea

These are nothing but  the Noether charges associated with the
non-interacting fields $\nu_1$, $\nu_2$ and $\nu_3$: in the
absence of mixing, they are the flavor charges,  separately
conserved for each generation.

As already observed in Section 3.2, in the case when CP is
conserved ($\de=0$), the mixing generator  Eq.(\ref{generator}) is
an element of the $SU(3)$ group and can be expressed in terms of
the above charges as:
\bea \lf.G_{\bf \te}(t)\ri|_{\de=0}\, =\, e^{i 2\te_{23}
\,Q_{m,7}(t)} \, e^{i 2\te_{13} \,Q_{m,5}(t)} \, e^{i 2\te_{12}
\,Q_{m,2}(t)} \eea

We can now perform the $SU(3)$ transformations on the flavor
triplet $\Psi_f$ and obtain another set of currents for the flavor
fields:
\bea \Psi_f'(x) \, =\, e^{i  \al_j F_j }\, \Psi_f (x) \qquad
,\qquad j \,=\, 1, 2,..., 8, \eea which leads to
\bea \non &&\de {\cal L}(x)\,= \,   i \al_j\,{\bar \Psi_f}(x)\,
[F_j, \textsf{M}]\, \Psi_f(x)\, =\, - \al_j \,\pa_\mu
J_{f,j}^{\mu}(x)\, ,
\\ [3mm] \label{fermacu2}
&&J^\mu_{f,j}(x) \,=\,  {\bar \Psi_f}(x)\, \ga^\mu\, F_j\,
\Psi_f(x)\qquad ,\qquad j \,=\, 1, 2,..., 8. \eea

Alternatively, the same currents can be obtained  by applying on
the $ J_{m,j}^{\mu}(x)$ the mixing generator
Eq.(\ref{generator}):
\bea J_{f,j}^{\mu}(x) \, = \, G_\te^{-1}(t)\,
J_{m,j}^{\mu}(x)\,G_\te(t)\qquad ,\qquad j \,=\, 1, 2,..., 8. \eea
The related charges  $Q_{f,j}(t)$ $\equiv$  $\int d^3 {\bf x}
\,J^0_{f,j}(x) $ still close the $su(3)$ algebra. Due to the
off--diagonal (mixing) terms in the mass matrix $\textsf{M}$,
$Q_{f,3}(t)$ and $Q_{f,8}(t)$ are time--dependent. This implies an
exchange of charge between $\nu_e$, $\nu_\mu$ and $\nu_\tau$,
resulting in the flavor oscillations.

In accordance with Eqs.(\ref{noether2}) we define the {\em flavor
charges} for mixed fields as
\bea\non Q_e(t) & \equiv & \frac{1}{3}Q \, + \, Q_{f,3}(t)\, +
\,\frac{1}{\sqrt{3}} Q_{f,8}(t),
\\
Q_\mu(t) & \equiv & \frac{1}{3}Q \, - \, Q_{f,3}(t)+
\,\frac{1}{\sqrt{3}} Q_{f,8}(t),
\\\non
Q_\tau(t) & \equiv & \frac{1}{3}Q \, -  \, \frac{2}{\sqrt{3}}
Q_{f,8}(t). \eea
with \bea Q_e(t) \, + \,Q_\mu(t) \,+ \,Q_\tau(t) \, = \, Q.\eea
These charges have a simple expression in terms of the flavor
ladder operators:
\bea\lab{flavchar} Q_\si(t) & = & \sum_{r} \int d^3 {\bf k}\lf(
\al^{r\dag}_{{\bf k},\si}(t) \al^{r}_{{\bf k},\si}(t)\, -\,
\bt^{r\dag}_{-{\bf k},\si}(t)\bt^{r}_{-{\bf k},\si}(t)\ri)\eea
with $\si= e,\mu,\tau$, because of the connection with the Noether
charges of Eq.(\ref{charge}) via the mixing generator: \bea
Q_\si(t) = G^{-1}_\te(t)\;Q_i\; G_\te(t).\eea

Notice also that the operator $\Delta Q_\si(t)\equiv  Q_\si(t) -
Q_i$ with $(\si,i)=(e,1) , (\mu,2) , (\tau,3)$\,, describes how
much the mixing violates the (lepton) charge conservation for a
given generation.

\vspace{0.3cm}

\section{Neutrino Oscillations}

 The oscillation formulas are
obtained by taking expectation values of the above charges on the
(flavor) neutrino state. Consider for example an initial electron
neutrino state defined as $|\nu_e\ran \equiv \al_{{\bf k},e}^{r
\dag}(0) |0\ran_{f}$ \cite{BHV98,Blasone:1999jb,BCRV01}. Working
in the Heisenberg picture, we obtain
\bea\non \label{charge1} {\cal Q}^\rho_{{\bf k},\si}(t) &\equiv&
\langle \nu_\rho|Q_\si(t)| \nu_\rho\ran - \, {}_f\lan 0 |Q_\si(t)|
0\ran_f =
\\
&=& \lf|\lf \{\al^{r}_{{\bf k},\si}(t), \al^{r \dag}_{{\bf
k},\rho}(0) \ri\}\ri|^{2} \;+ \;\lf|\lf\{\bt_{{-\bf k},\si}^{r
\dag}(t), \al^{r \dag}_{{\bf k},\rho}(0) \ri\}\ri|^{2},\eea
\bea\non \label{charge2} {\cal Q}^{\bar \rho}_{{\bf k},\si}(t)
&\equiv& \langle {\bar \nu}_\rho|Q_\si(t)| {\bar \nu}_\rho\ran -
\, {}_f\lan 0 |Q_\si(t)| 0\ran_f =\\&=& - \lf|\lf \{\bt^{r}_{{\bf
k},\si}(t), \bt^{r \dag}_{{\bf k},\rho}(0) \ri\}\ri|^{2} \;-
\;\lf|\lf\{\al_{{-\bf k},\si}^{r \dag}(t), \bt^{r \dag}_{{\bf
k},\rho}(0) \ri\}\ri|^{2}, \eea
where $|0\ran_{f}\equiv |0(0)\ran_{f}$. Overall charge
conservation is obviously ensured at any time: \bea {\cal
Q}^{\rho}_{{\bf k},e}(t) + {\cal Q}^{\rho}_{{\bf k},\mu}(t)+{\cal
Q}^{\rho}_{{\bf k},\tau}(t) \; = \; 1,\qquad\;\quad {\rho}=e, \mu,
\tau.\eea We remark that the expectation value of $Q_\si$ cannot
be taken on vectors of the Fock space built on $|0\ran_{m}$,
\cite{BHV98,Blasone:1999jb,BCRV01}. Also we observe that ${}_f\lan
0 |Q_\si(t)| 0\ran_f  \neq 0$, in contrast with the two flavor
case \cite{Blasone:1999jb,yBCV02,comment}. We introduce the
following notation:
\bea\non \De_{ij}^{\bf k}\equiv \frac{\om_{k,j} - \om_{k,i}}{2}
\quad , \quad \Om_{ij}^{\bf k}\equiv \frac{\om_{k,i} +
\om_{k,j}}{2} \eea
Then the  oscillation (in time) formulae for the flavor charges,
on an initial electron neutrino state, follow as \cite{yBCV02}:
\bea \non {\cal Q}^e_{{\bf k},e}(t) \, &=& \,1 \,-\, \sin^{2}( 2
\te_{12})\cos^{4}\te_{13} \, \Big[|U_{12}^{\bf k}|^2\, \sin^{2}
\lf(  \De_{12}^{\bf k}  t \ri)
 +\,|V_{12}^{\bf k}|^2 \, \sin^{2}
\lf(\Om_{12}^{\bf k}t \ri)\Big] \
\\ \non && -\, \sin^{2}(2 \te_{13})\cos^{2}\te_{12}
\, \Big[|U_{13}^{\bf k}|^2\,
 \sin^{2} \lf( \De_{13}^{\bf k}  t \ri)
 +\,|V_{13}^{\bf k}|^2\, \sin^{2} \lf( \Om_{13}^{\bf k} t \ri)\Big] \
\\ &&
-\, \sin^{2}(2 \te_{13})\sin^{2}\te_{12} \, \Big[|U_{23}^{\bf
k}|^2 \, \sin^{2} \lf( \De_{23}^{\bf k} t \ri) +\,|V_{23}^{\bf
k}|^2\, \sin^{2} \lf( \Om_{23}^{\bf k} t \ri)\Big]\,, \eea
\bea\non {\cal Q}^e_{{\bf k},\mu}(t)
 &=& 2 J_{\CP}
 \Big[|U_{12}^{\bf k}|^2\, \sin(2\De_{12}^{\bf k}t)
- |V_{12}^{\bf k}|^2\, \sin(2\Om_{12}^{\bf k} t) + (|U_{12}^{\bf
k}|^2 - |V_{13}^{\bf k}|^2 ) \sin(2\De_{23}^{\bf k}t)
\\ \non
&+& (|V_{12}^{\bf k}|^2 - |V_{13}^{\bf k}|^2 ) \sin(2\Om_{23}^{\bf
k}t)
  - |U_{13}^{\bf k}|^2\,
\sin(2\De_{13}^{\bf k}t)+ |V_{13}^{\bf k}|^2\, \sin(2\Om_{13}^{\bf
k}t)\Big]
\\ \non
&+&\, \cos^{2}\te_{13} \sin\te_{13}
\Big[\cos\de\sin(2\te_{12})\sin(2\te_{23}) + 4
\cos^2\te_{12}\sin\te_{13}\sin^2\te_{23}\Big]\times\\\non
&\times&\Big[|U_{13}^{\bf k}|^2\sin^{2} \lf(\De_{13}^{\bf k} t
\ri) + |V_{13}^{\bf k}|^2\ \sin^{2} \lf( \Om_{13}^{\bf k} t
\ri)\Big]
\\ \non
& -& \cos^{2}\te_{13}\sin\te_{13}
 \Big[\cos\de\sin(2\te_{12})\sin(2\te_{23}) -
4 \sin^2\te_{12}\sin\te_{13}\sin^2\te_{23}\Big]\times
\\\non &\times& \Big[|U_{23}^{\bf k}|^2\ \sin^{2} \lf( \De_{23}^{\bf
k} t \ri)
 + |V_{23}^{\bf k}|^2\
\sin^{2} \lf( \Om_{23}^{\bf k} t \ri)\Big]
\\\non
& +&\cos^{2}\te_{13} \sin(2\te_{12}) \Big[ (\cos^2\te_{23} -
\sin^2\te_{23}\sin^2\te_{13})\sin(2\te_{12})
\\\non &+&\cos\de\cos(2\te_{12})\sin\te_{13}\sin(2\te_{23})\Big]
 \Big[|U_{12}^{\bf k}|^2\ \sin^{2} \lf(\De_{12}^{\bf k} t
\ri) + |V_{12}^{\bf k}|^2\ \sin^{2} \lf( \Om_{12}^{\bf k} t
\ri)\Big]\,,\\ \eea
\bea\non {\cal Q}^e_{{\bf k},\tau}(t)&=& - 2 J_{\CP}
 \Big[|U_{12}^{\bf k}|^2\ \sin(2\De_{12}^{\bf k}t)
- |V_{12}^{\bf k}|^2\, \sin(2\Om_{12}^{\bf k} t) + (|U_{12}^{\bf
k}|^2\, - |V_{13}^{\bf k}|^2 ) \sin(2\De_{23}^{\bf k}t)
\\ \non
&+& (|V_{12}^{\bf k}|^2\, - |V_{13}^{\bf k}|^2 )
\sin(2\Om_{23}^{\bf k}t)
 \, - |U_{13}^{\bf k}|^2\,
\sin(2\De_{13}^{\bf k}t)+ |V_{13}^{\bf k}|^2\, \sin(2\Om_{13}^{\bf
k}t)\Big]
\\ \non
&-& \cos^{2}\te_{13}\sin\te_{13}
\Big[\cos\de\sin(2\te_{12})\sin(2\te_{23}) -4
\cos^2\te_{12}\sin\te_{13}\cos^2\te_{23}\Big]\times\\ \non
&\times& \Big[|U_{13}^{\bf k}|^2\, \sin^{2} \lf( \De_{13}^{\bf k}
t \ri) + |V_{13}^{\bf k}|^2\, \sin^{2}\lf(\Om_{13}^{\bf k}
t\ri)\Big]
\\ \non
& +&\cos^{2}\te_{13}\sin\te_{13}
 \Big[\cos\de\sin(2\te_{12})\sin(2\te_{23}) +
  4 \sin^2\te_{12}\sin\te_{13}\cos^2\te_{23}\Big]\times\\ \non
&\times& \Big[|U_{23}^{\bf k}|^2 \, \sin^{2} \lf(\De_{23}^{\bf k}
t \ri) +
 |V_{23}^{\bf k}|^2\,\sin^{2} \lf(\Om_{23}^{\bf k}  t \ri)\Big]
\\ \non
&+& \cos^{2}\te_{13} \sin(2\te_{12}) \Big[ (\sin^2\te_{23} -
\sin^2\te_{13}\cos^2\te_{23})\sin(2\te_{12})
\\ \non &-&\cos\de\cos(2\te_{12})\sin\te_{13}\sin(2\te_{23})\Big]
 \Big[|U_{12}^{\bf k}|^2\, \sin^{2} \lf( \De_{12}^{\bf k} t \ri)
+ |V_{12}^{\bf k}|^2\, \sin^{2} \lf(\Om_{12}^{\bf k} t
\ri)\Big]\,,\\ \eea
where we used the relations Eqs.(\ref{ident1})-(\ref{ident4}).
We also introduced the Jarlskog factor $J_{\CP}$ defined as
\cite{Jarlskog}
\bea J_{\CP} \equiv Im (u_{i\alpha}u_{j \beta}u^{*}_{i
\beta}u^{*}_{j\alpha}), \eea
where the $u_{ij}$ are the elements of mixing matrix ${\cal U}$
and $i \neq j$, $\alpha \neq \beta$. In the parameterization
Eq.(\ref{fermix123}), $J_{\CP}$ is given by
 \bea
J_{\CP} = \frac{1}{8}\, \sin \de\, \sin(2\te_{12})
\sin(2\te_{13})\cos\te_{13} \sin(2\te_{23})  . \eea
Evidently, $J_{\CP}$ vanishes if $ \theta_{ij}=0, \pi/2$ and/or
$\de=0,\pi$: all CP--violating effects are proportional to it.

The above oscillation formulas are exact. The differences with
respect to the usual formulas for neutrino oscillations are in the
energy dependence of the amplitudes and in the additional
oscillating terms. For $|{\bf k}|\gg\sqrt{m_1m_2}$, we have
$|U_{ij}^{{\bf k}}|^{2}\rar 1$ and $|V_{ij}^{{\bf k}}|^{2}\rar 0$
and the traditional (Pontecorvo) oscillation formulas are
approximately recovered. Indeed, for sufficiently small time
arguments, a correction to the Pontecorvo formula is present even
in the relativistic limit.

In Appendix G the oscillation formulas for the flavor charges on
an initial electron anti-neutrino state are given.

We plot in Figures 3.2 and 3.4 the QFT oscillation formulas ${\cal
Q}^e_{{\bf k},e}(t)$ and ${\cal Q}^e_{{\bf k},\mu}(t)$ as a
function of time, and in Figures 3.3 and 3.5 the corresponding
Pontecorvo oscillation formulas $P^{\bf k}_{e\rightarrow e}(t)$
and $P^{\bf k}_{e\rightarrow \mu}(t)$. The time scale is in
$T_{12}$ units, where $T_{12}= \pi/\De_{12}^{\bf k}$ is, for the
values of parameters of Tab.(1), the largest oscillation period.
\vspace{0.5cm}
\begin{figure}
\vspace{1cm} \centerline{\epsfysize=3.0truein\epsfbox{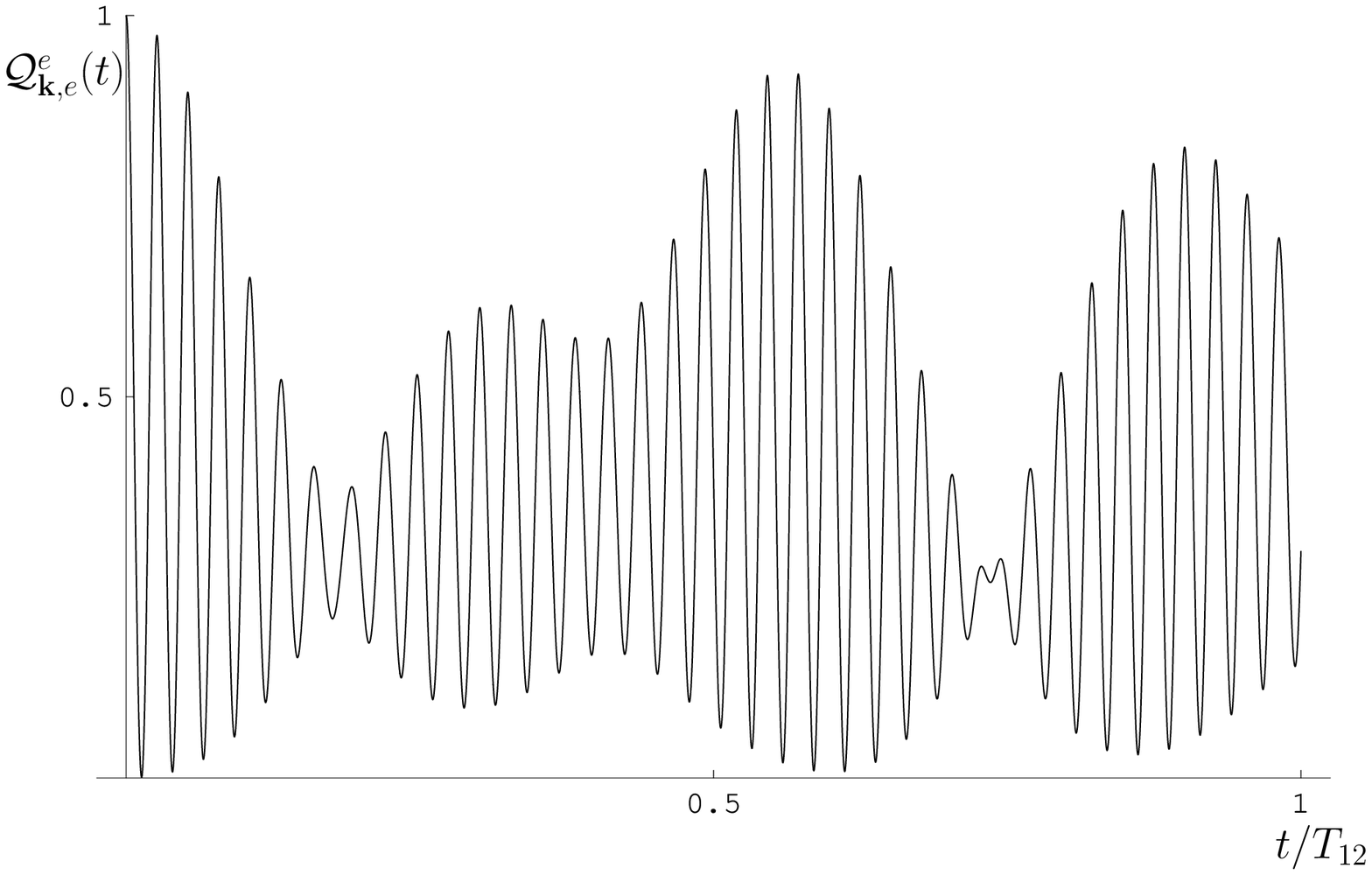}}
\vspace{.2cm} \caption{Plot of QFT oscillation formula: ${\cal
Q}^e_{{\bf k},e}(t)$ in function of time for $k =55$ and
parameters as in Tab.(1). }

\vspace{0.5cm}

\hrule
\end{figure}

\begin{figure}
\centerline{\epsfysize=3.0truein\epsfbox{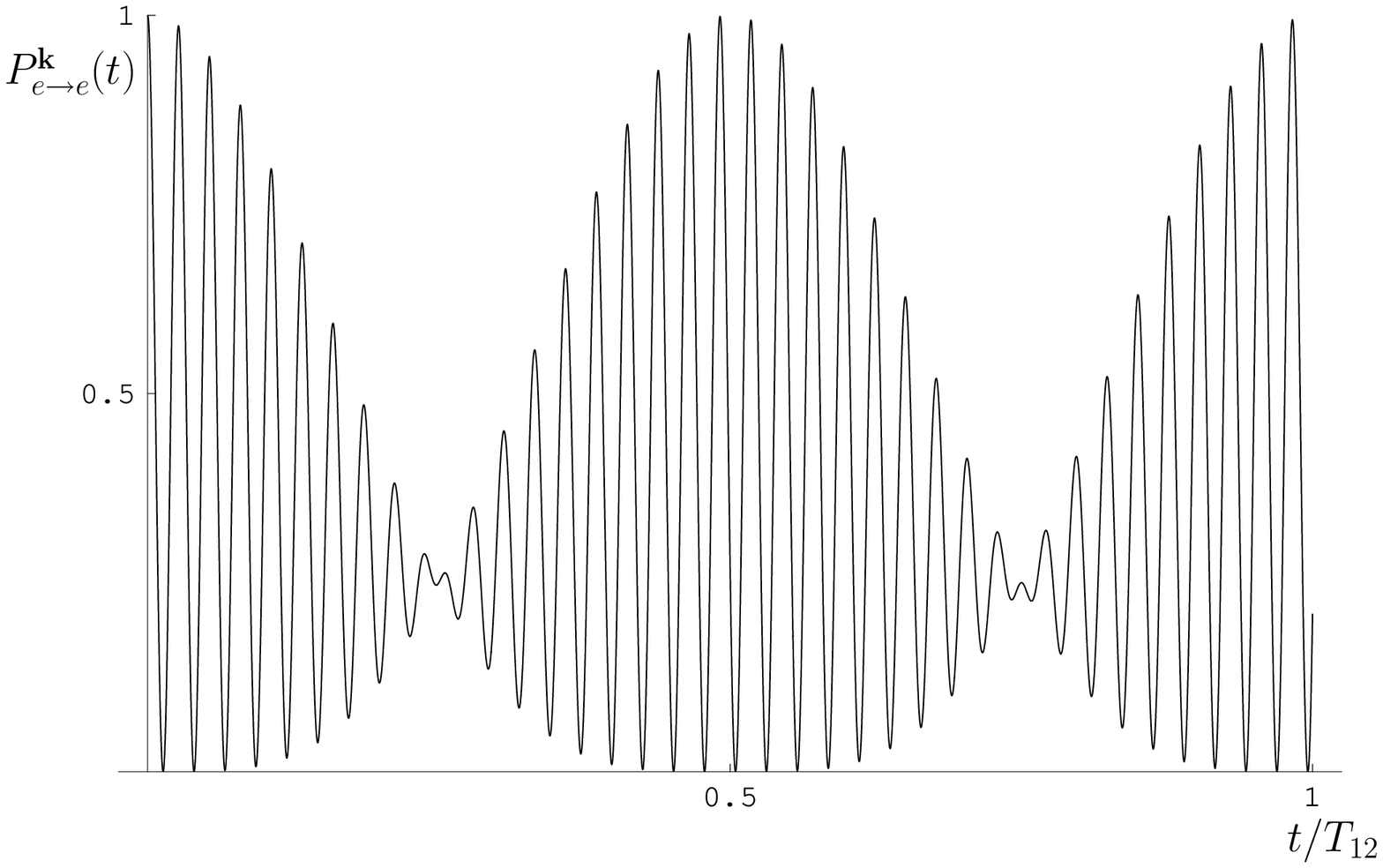}} \vspace{.2cm}
\caption{ Plot of QM oscillation formula: $P^{\bf k}_{e\rightarrow
e}(t)$ in function of time for $k =55$ and parameters as in
Tab.(1). }

\vspace{0.5cm}

\hrule
\end{figure}

\begin{figure}

\centerline{\epsfysize=3.0truein\epsfbox{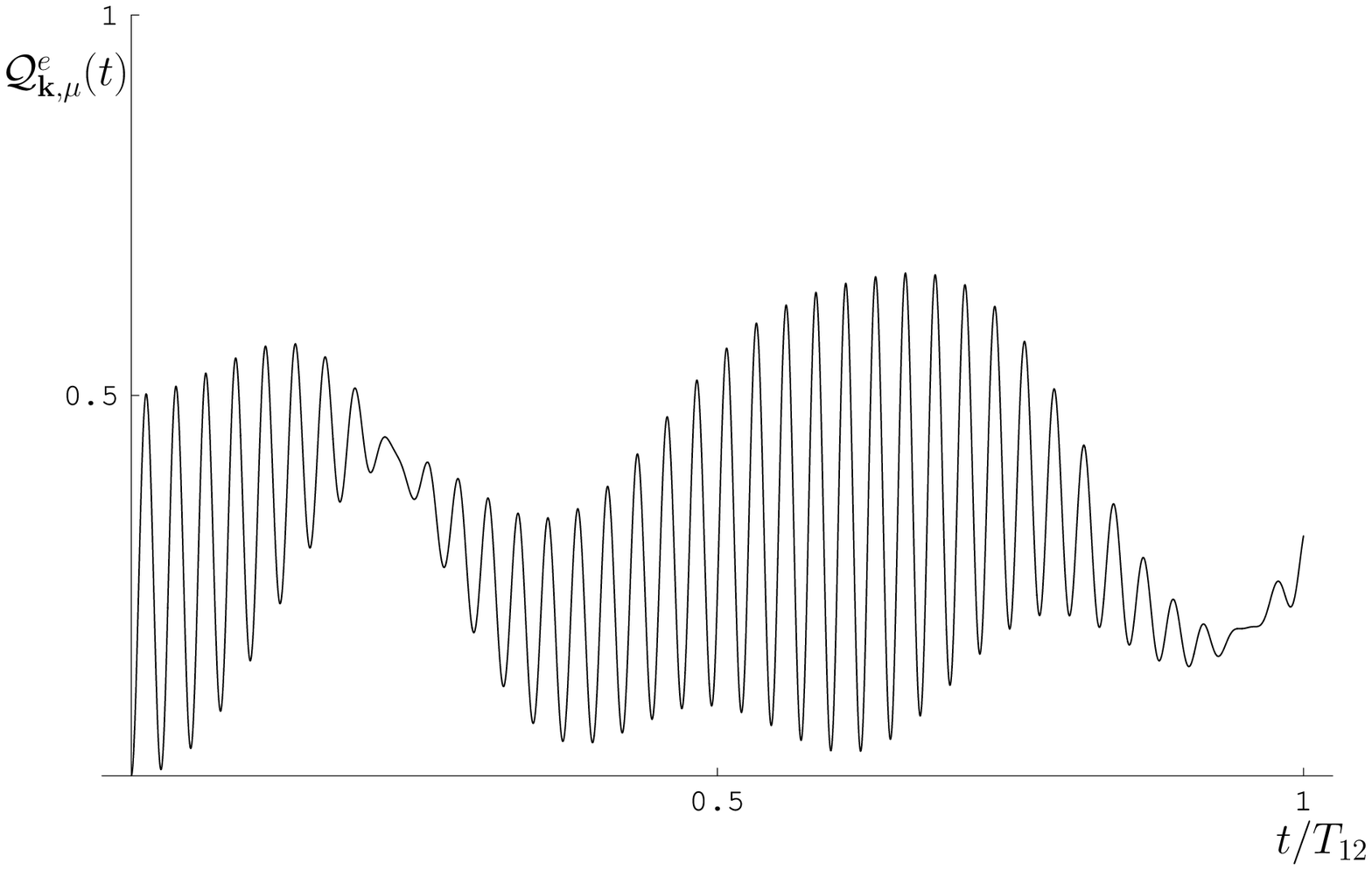}} \vspace{.2cm}
\caption{ Plot of QFT oscillation formula: ${\cal Q}^e_{{\bf
k},\mu}(t)$ in function of time for $k =55$ and parameters as in
Tab(1). }

\vspace{0.5cm}
 \hrule
\end{figure}

\begin{figure}
\centerline{\epsfysize=3.0truein\epsfbox{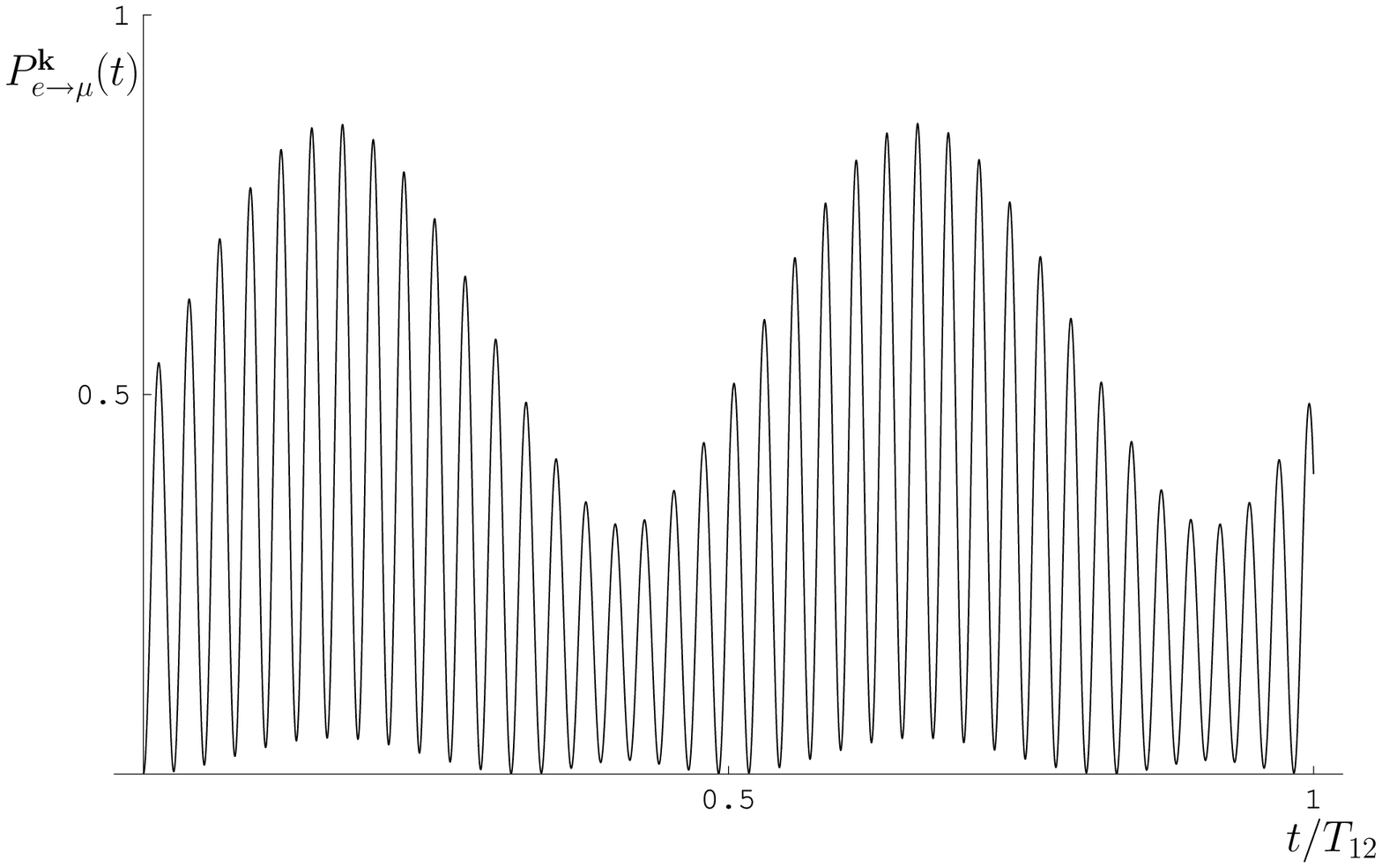}} \vspace{.2cm}
\caption{ Plot of QM oscillation formula: $P^{\bf k}_{e\rightarrow
\mu}(t)$ in function of time for $k =55$ and parameters as in
Tab.(1). }

\vspace{0.5cm}

 \hrule
\end{figure}

\section{CP and T violations in neutrino oscillations}

In this Section we consider the oscillation induced CP and T
violation in the context of the present QFT framework. Let us
first briefly  recall the situation in QM\footnote{We use here the
"hat"  for QM quantities. For notational simplicity, we also
suppress momentum indices where unnecessary.}: there, the CP
asymmetry between the probabilities of two conjugate neutrino
transitions, due to CPT invariance and unitarity of the mixing
matrix, is given as \cite{Fritzsch}
\bea {\hat \Delta}_{\CP}^{\rho\si}(t)\equiv P_ {\nu_{\sigma}\rar
\nu_{\rho}}(t) - P_ {\overline{\nu}_{\sigma}\rar
\overline{\nu}_{\rho}}(t),
 \eea
 where
$ \sigma, \rho = e, \mu, \tau.$ The T violating asymmetry can be
obtained in similar way as \cite{Fritzsch}
\bea\lab{DeT} {\hat \Delta}_{\T}^{\rho\si}(t)\,\equiv\, P_
{\nu_{\sigma}\rar \nu_{\rho}}(t) - P_ {\nu_{\rho}\rar
\nu_{\sigma}}(t)\, = \,P_ {\nu_{\sigma}\rar \nu_{\rho}}(t) - P_
{\nu_{\sigma}\rar \nu_{\rho}}(-t)\,. \eea
The relationship $ {\hat \Delta}_{\CP}^{\rho\si}(t)= {\hat
\Delta}_{\T}^{\rho\si}(t)$ is a consequence of CPT invariance.

The corresponding quantities in QFT have to be defined in the
framework of the previous Section, i.e. as expectation values of
the flavor charges on states belonging to the flavor Hilbert
space. We thus have for the CP violation:
\bea\non\lab{DECP} \De^{\rho\si}_\CP(t) & \equiv &
 {\cal Q}^\rho_{{\bf k},\si}(t)  \, +\,
{\cal Q}^{\bar \rho}_{{\bf k},\si}(t)
\\\non
&=& \lf|\lf \{\al^{r}_{{\bf k},\si}(t), \al^{r \dag}_{{\bf
k},\rho}(0) \ri\}\ri|^{2} \,+ \,\lf|\lf\{\bt_{{-\bf k},\si}^{r
\dag}(t), \al^{r \dag}_{{\bf k},\rho}(0) \ri\}\ri|^{2}- \\&-&
\lf|\lf \{\al^{r\dag}_{-{\bf k},\si}(t), \bt^{r \dag}_{{\bf
k},\rho}(0) \ri\}\ri|^{2} \,- \,\lf|\lf\{\bt_{{\bf k},\si}^{r}(t),
\bt^{r \dag}_{{\bf k},\rho}(0) \ri\}\ri|^{2}\,. \eea

We have
\bea &&\sum_\si \De^{\rho\si}_\CP \, = \,0 \qquad, \quad
\rho,\si=e,\mu,\tau, \eea
which follows from the fact that \bea \sum_\si Q_\si(t)
=Q,\;\;\;\;\; \langle \nu_\rho|Q| \nu_\rho\ran\,=\,1\;\;\;\;\;
\text {and}\;\;\;\;\; \langle {\bar \nu}_\rho|Q| {\bar
\nu}_\rho\ran \,=\,-1.\eea

\vspace{0.2cm}

We can calculate the CP asymmetry Eq.(\ref{DECP}) for a specific
case, namely for the transition $\nu_e \longrightarrow \nu_\mu $.
We obtain
\bea\non \De^{e\mu}_\CP(t) &=&4 J_{\CP}
 \Big[|U_{12}^{\bf k}|^2\, \sin(2 \De_{12}^{\bf k}t)
- |V_{12}^{\bf k}|^2\, \sin(2 \Om_{12}^{\bf k} t) + (|U_{12}^{\bf
k}|^2 - |V_{13}^{\bf k}|^2 ) \sin(2 \De_{23}^{\bf k}t)
\\\non \lab{Demt}
&& + (|V_{12}^{\bf k}|^2 - |V_{13}^{\bf k}|^2 ) \sin(2
\Om_{23}^{\bf k}t)
  - |U_{13}^{\bf k}|^2\,
\sin(2 \De_{13}^{\bf k}t)+ |V_{13}^{\bf k}|^2\, \sin(2
\Om_{13}^{\bf k}t)\Big]\,,\\ \eea
and $ \De^{e\tau}_\CP(t)= - \De^{e\mu}_\CP (t)$. As already
observed for oscillation formulas, high-frequency oscillating
terms and Bogoliubov coefficients in the oscillation amplitudes
appear in Eq.(\ref{Demt}) as a QFT correction to the QM formula.

The definition of the QFT analogue of the T-violating quantity
Eq.(\ref{DeT}) is more delicate. Indeed,
 defining  $\De_\T$ as
$ \De^{e\mu}_\T \equiv {\cal Q}^e_{\mu}(t)  \, -\, {\cal
Q}^\mu_{e}(t) $ does not seem to work, since we obtain
$\De^{e\mu}_\T \, - \,\De^{e\mu}_\CP \, \neq \, 0 $ in contrast
with CPT conservation.

A more consistent definition of the time-reversal violation in QFT
is then:
\bea
 \De^{\rho\si}_\T (t) \, \equiv
\, {\cal Q}^\rho_{{\bf k},\si}(t)  \, -\, {\cal Q}^\rho_{{\bf
k},\si}(-t)\qquad, \quad \rho,\si=e,\mu,\tau\,. \eea
With such definition, the equality $\De^{\rho\si}_\T(t)
\,=\,\De^{\rho\si}_\CP(t)$ follows from ${\cal Q}^\rho_{{\bf
k},\si}(-t)  \, =\, - {\cal Q}^{\bar \rho}_{{\bf k},\si}(t)$.

We plot in  Fig. 3.6  the CP asymmetry Eq.(\ref{Demt}) for sample
values of the parameters as in Tab.(1). In Fig. 3.7 the
corresponding standard QM quantity is plotted for the same values
of parameters. \vspace{2cm}
\begin{figure}
\centerline{\epsfysize=3.0truein\epsfbox{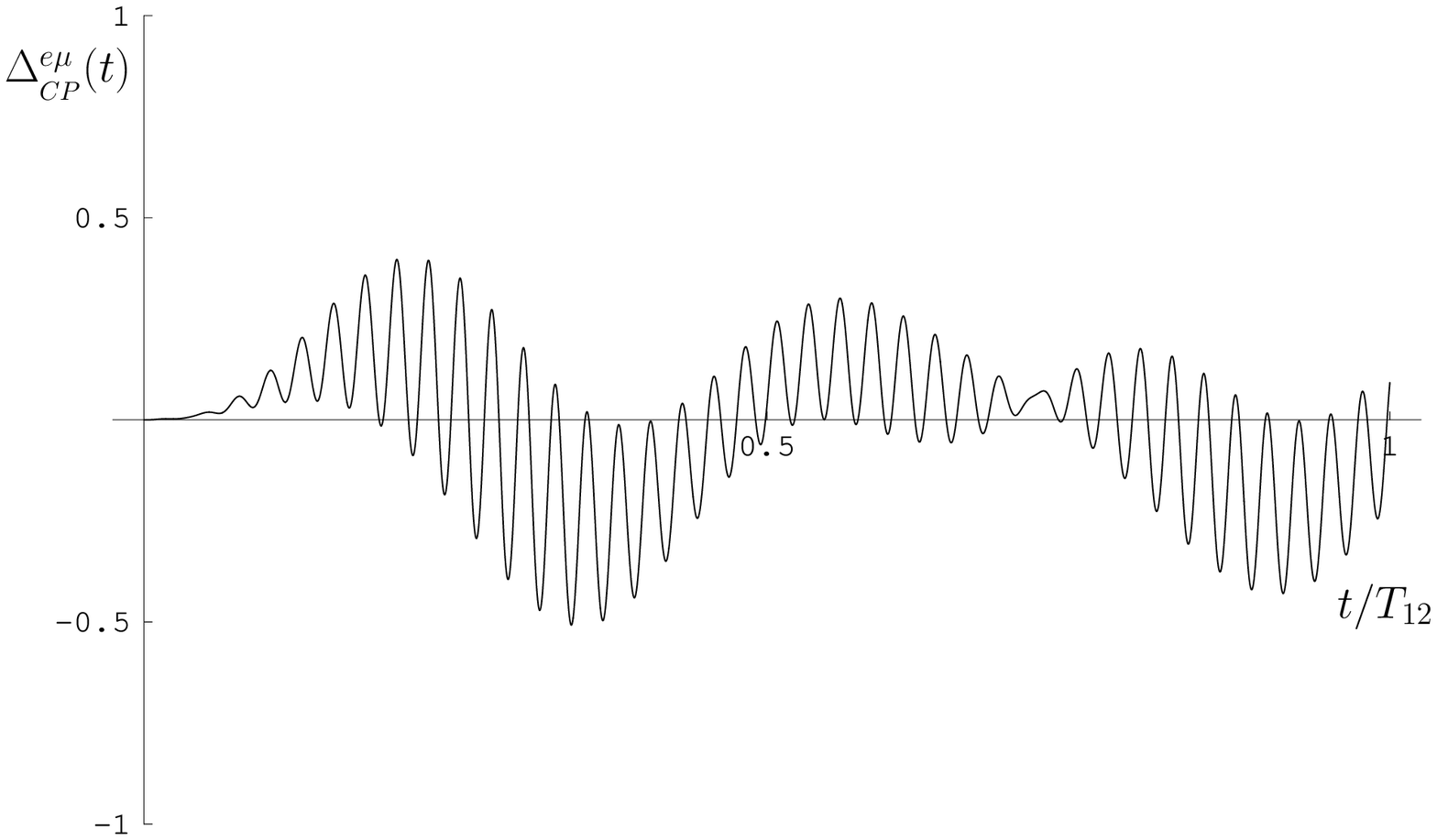}} \vspace{.2cm}
\caption{ Plot of the QFT CP asymmetry $\Delta^{e\mu}_{\CP}(t)$,
in function of time for $k =55$ and parameters as in Tab.(1). }

\vspace{0.5cm}

 \hrule
\end{figure}
\vspace{2cm}
\begin{figure} \centerline{\epsfysize=3.0truein\epsfbox{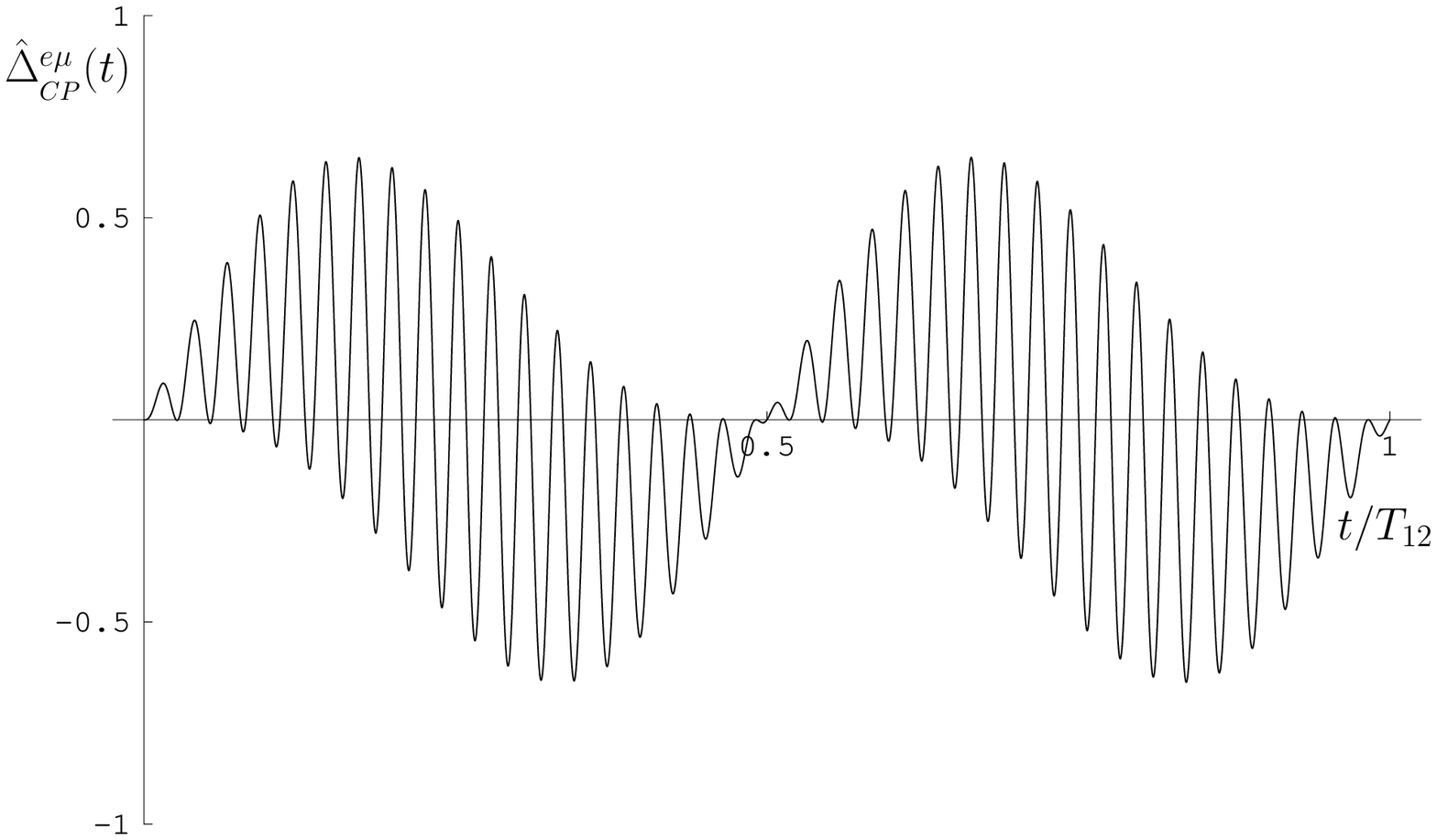}}
\vspace{.2cm} \caption{ Plot of the QM CP asymmetry ${\hat
\Delta}^{e\mu}_{\CP}(t)$, in function of time for $k =55$
 and parameters as in Tab.(1). }

\vspace{0.5cm}

 \hrule
\end{figure}

\section{Conclusions}

In this chapter, we have discussed the mixing of (Dirac) fermionic
fields in Quantum Field Theory for the case of three flavors with
CP violation. We constructed the flavor Hilbert space and studied
the currents and charges for mixed fields (neutrinos). We will
show in the next chapter that the algebraic structure associated
with the mixing for the case of three generation  turned out to be
that of a deformed $su(3)$ algebra, when a CP violating phase is
present.

We have derived all the known parameterization of the three-flavor
mixing matrix and a number of new ones. We have shown that these
parameterizations actually reflect the group theoretical structure
of the generator of the mixing transformations.

By use of the flavor Hilbert space, we have calculated the exact
QFT oscillation formulas, a generalization of the usual QM
Pontecorvo formulas. The comparison between the exact oscillation
formulas and the usual ones has been explicitly exhibited for
sample values of the neutrino masses and mixings. CP and T
violation induced by neutrino oscillations have also been
discussed.

As already remarked in the Introduction, the corrections
introduced by the present formalism to the usual Pontecorvo
formulas are in principle experimentally testable. The fact that
these corrections may be quantitatively below the experimental
accuracy reachable at the present state of the art in the
detection of the neutrino oscillations, does not justify
neglecting them in the analysis of the particle mixing and
oscillation mechanism. The exact oscillation formulas here derived
are the result of a mathematically consistent analysis which
cannot be ignored in a correct treatment of the field mixing
phenomenon. As we have seen above, the QFT formalism accounts for
all the known parameterizations of the mixing matrix and explains
their origin and their reciprocal relations, thus unifying the
phenomenological proposals scattered in the literature where such
parameterizations have been presented. Moreover, the QFT formalism
clearly points to the truly non-perturbative character of the
particle mixing phenomenon. A lot of Physics must be there waiting
to be discovered.

\chapter{Group theoretical aspects of neutrino mixing in Quantum Field Theory}

We analyze some aspects of three flavor neutrino mixing.
Particular emphasis is given to the related algebraic structures
and their deformation in the presence of CP violation. A novel
geometric phase related to CP violation is introduced.

\section{Introduction}

In this chapter we discuss the group structure involved in the
mixing and the related representations, both for two- and
three-flavor mixing. The deformation of the associated algebra as
well as the geometric phase due to CP violation are also discussed
\cite{yBCV02,BCV2003}. In Section 4.2 we study the deformation of
SU(3) algebra induced by CP violation; in Section 4.3 we analyze
the group representations, considering the case of two and three
generations. The Section 4.4 is devoted to conclusions.

\section{SU(3) deformed algebra}

We investigate the algebraic structures associated with the mixing
generator of three Dirac neutrinos fields Eq.(\ref{generator}). To
this end, we consider the Lagrangian density describing three
Dirac neutrinos fields:
\bea \label{lag123} {\cal L}(x)&=&  {\bar \Psi_m}(x) \lf( i
\not\!\partial -  \textsf{M}_d\ri) \Psi_m(x) \;=\; {\bar
\Psi_f}(x) \lf( i \not\!\partial - \textsf{M} \ri) \Psi_f(x)\, ,
\eea
where $\textsf{M}_d = diag(m_1,m_2,m_3)$ and the matrix $
\textsf{M}$ is non-diagonal.

The above Lagrangian is invariant under global $U(1)$ phase
transformations, leading to a conserved (total) charge $Q= \int
d^3{\bf x}\,\Psi^\dag_m(x)\,\Psi_m(x)\,=\, \int  d^3{\bf
x}\,\Psi^\dag_f(x)\,\Psi_f(x)$.

We then study the invariance of ${\cal L}$ under global phase
transformations of the kind:
\bea \label{masssu3} \Psi_m'(x) \, =\, e^{i \al_j  {\ti F}_j}\,
\Psi_m (x) \, \qquad, \qquad
 j=1, 2,..., 8.
\eea
where  \bea\label{defgen} &&{\ti F}_{j}=\frac{1}{2}{\ti
\lambda}_{j}\quad,\quad j=1,..,8 \eea and the ${\ti \lambda}_{j}$
are a generalization of the usual Gell-Mann matrices
$\lambda_{j}$:
\bea\label{gelm} \non &&{\ti \lambda}_{1}=\begin{pmatrix}
  0 & e^{i\de_2} & 0 \\
 e^{-i\de_2} & 0 & 0 \\
  0 & 0 & 0
\end{pmatrix} \quad,\quad\quad\quad
{\ti \lambda}_{2}=\begin{pmatrix}
  0 & -i e^{i\de_2} & 0 \\
  i e^{-i\de_2}& 0 & 0 \\
  0 & 0 & 0
\end{pmatrix} \\[3mm]\non &&{\ti \lambda}_{3}=
\begin{pmatrix}
  1 & 0 & 0 \\
  0 & -1 & 0 \\
  0 & 0 & 0
\end{pmatrix}
\;, \;\;\;\;\;\;\;\;\;\;\;\;\;\quad\quad {\ti \lambda}_{4}=
\begin{pmatrix}
  0 & 0 & e^{-i\de_5}\\
  0 & 0 & 0 \\
  e^{i\de_5} & 0 & 0
\end{pmatrix}
\\[3mm]\non
&&{\ti \lambda}_{5}=\begin{pmatrix}
  0 & 0 & -ie^{-i\de_5} \\
  0 & 0 & 0 \\
  ie^{i\de_5} & 0 & 0
\end{pmatrix}\;, \;\quad\quad
{\ti \lambda}_{6}=\begin{pmatrix}
  0 & 0 & 0 \\
  0 & 0 & e^{i\de_7} \\
  0 & e^{-i\de_7} & 0
\end{pmatrix} \\[3mm] &&{\ti \lambda}_{7}=\begin{pmatrix}
  0 & 0 & 0 \\
  0 & 0 & -i e^{i\de_7}\\
  0 & ie^{-i\de_7} & 0
\end{pmatrix}\;, \;\quad\quad{\ti \lambda_{8}}=\frac{1}{\sqrt{3}}\begin{pmatrix}
  1 & 0 & 0 \\
  0 & 1 & 0 \\
  0 & 0 & -2
\end{pmatrix}.
\eea
These are normalized as the  Gell-Mann matrices: $tr(  \lambda_{j}
\lambda_{k}) =2\delta_{jk}$ . One then obtains the following set
of charges \cite{BCV2003}:
\bea\label{su3charges} &&{}\hspace{-.5cm} {\ti Q}_{m,j}(t)\, =\,
\int  d^3{\bf x}\,\Psi^\dag_m(x)\, {\ti F}_j\, \Psi_m(x) \;,
\qquad j=1, 2,..., 8. \eea Thus the CKM matrix
Eq.(\ref{fermix123}) is generated by ${\ti Q}_{m,2}(t)$, ${\ti
Q}_{m,5}(t)$ and ${\ti Q}_{m,7}(t)$, with
$\{\de_2,\de_5,\de_7\}\rar \{0,\de,0\}$. An interesting point is
that the algebra generated by the matrices Eqs.(\ref{gelm}) {\em
is not} $su(3)$ unless the condition $\De \equiv \de_2+\de_5
+\de_7 =0$ is imposed: such a condition is however incompatible
with the presence of a CP violating phase. In the
parameterizations of the mixing matrices of Section 3.2 and
Section 3.3 (cf. e.g. the discussion after Eq.(\ref{matrix})); we
have the correspondence $\{\de_2,\de_5,\de_7\}\lrar
\{\de_{12},\de_{13},\de_{23}\}$.

The ${\ti F}_{j}$ satisfy a deformed $su(3)$ algebra with deformed
commutation relations given by \cite{yBCV02,BCV2003}:
\bea \label{com}\non &&[{\ti F}_2,{\ti F}_5]\, =\, \frac{i}{2} \,
{\ti F}_7 \, e^{-i \,\De\, ({\ti F}_3-\sqrt{3}{\ti F}_8)}
 \quad,\quad\quad [{\ti F}_1,{\ti F}_4]\, =\, \frac{i}{2} \, {\ti F}_7 \, e^{-i
\,\De\,({\ti F}_3-\sqrt{3}{\ti F}_8) }\quad,
\\ \non
&&[{\ti F}_1,{\ti F}_5]\, =\,- \frac{i}{2} \, {\ti F}_6 \, e^{-i
\,\De\,({\ti F}_3-\sqrt{3}{\ti F}_8)}\quad,\quad [{\ti F}_2,{\ti
F}_4]\, =\, \frac{i}{2} \, {\ti F}_6\, e^{-i \,\De\, ({\ti
F}_3-\sqrt{3}{\ti F}_8)}\quad,
\\\non
&& [{\ti F}_2,{\ti F}_7]\, =\, -\frac{i}{2} \, {\ti F}_5 \, e^{-i
\,\De\,({\ti F}_3+\sqrt{3}{\ti F}_8) }
 \quad , \quad [{\ti F}_1,{\ti F}_7]\, =-\, \frac{i}{2} \, {\ti F}_4 \, e^{-
i \,\De\,({\ti F}_3+\sqrt{3}{\ti F}_8)}\quad,
\\\non&& [{\ti F}_1,{\ti F}_6]\, =\, \frac{i}{2} \, {\ti F}_5
\, e^{-i \,\De\, ({\ti F}_3+\sqrt{3}{\ti F}_8)} \quad ,\quad \quad
[{\ti F}_2,{\ti F}_6]\, =\, - \frac{i}{2} \, {\ti F}_4\, e^{-i
\,\De\,({\ti F}_3+\sqrt{3}{\ti F}_8)}\quad, \\\non &&
 [{\ti F}_5,{\ti F}_7]\, =\, \frac{i}{2} \, {\ti F}_2\,
 e^{ 2i\,\De\, {\ti F}_3}
 \quad , \quad \quad \quad\quad \quad[{\ti F}_4,{\ti F}_7]\, =\, \frac{i}{2} \, {\ti
F}_1\, e^{2 i \,\De\,{\ti F}_3}\quad,
\\
&& [{\ti F}_5,{\ti F}_6]\, =\,- \frac{i}{2} \, {\ti F}_1 \, e^{2i
\,\De\,{\ti F}_3} \quad , \quad  \quad \quad \quad [{\ti F}_4,{\ti
F}_6]\, =\, \frac{i}{2} \, {\ti F}_2\, e^{2 i \,\De\,{\ti F}_3},
\eea
where  $\De \equiv \de_2+\de_5 +\de_7$. The other commutators are
the usual $su(3)$ ones. For $\De=0$, the $su(3)$ algebra is
recovered.

When CP violation is allowed, then $\De \neq 0$ and the $su(3)$
algebra is deformed. Let us introduce the raising and lowering
operators, defined as \cite{mesons}:
\bea && {\ti T}_\pm \equiv {\ti F}_1 \pm i {\ti F}_2 \quad, \quad
{\ti U}_\pm \equiv {\ti F}_6 \pm i {\ti F}_7 \quad, \quad {\ti
V}_\pm \equiv {\ti F}_4 \pm i {\ti F}_5 \eea
We also define:
\bea\non && {\ti Y}= \frac{2}{\sqrt{3}}{\ti F}_8 \quad, \quad {\ti
T}_3 \equiv {\ti F}_3 \quad, \quad {\ti U}_3 \equiv
\frac{1}{2}\lf(\sqrt{3} {\ti F}_8 - {\ti F}_3 \ri) \quad, \quad
{\ti V}_3 \equiv \frac{1}{2}\lf(\sqrt{3} {\ti F}_8 + {\ti F}_3
\ri)\\ \eea The commutation relations are
 \bea\non \lab{co1}
&& [{\ti T}_3,{\ti T}_\pm] \,=\, \pm {\ti T}_\pm \quad, \quad
[{\ti T}_3,{\ti U}_\pm] \,=\, \mp \frac{1}{2} {\ti U}_\pm \quad,
\quad [{\ti T}_3,{\ti V}_\pm] \,=\, \pm \frac{1}{2} {\ti V}_\pm
\quad,
\quad [{\ti T}_3,{\ti Y}] \,=\, 0, \\
\\
&& [{\ti Y},{\ti T}_\pm] = 0 \quad,\quad \quad \quad [{\ti Y},{\ti
U}_\pm] \,=\, \pm  {\ti U}_\pm \quad,\quad \quad [{\ti Y},{\ti
V}_\pm] \,=\, \pm {\ti V}_\pm \quad,\quad
\\
&& [{\ti T}_+,{\ti T}_-] = 2 {\ti T}_3 \quad,\quad \quad [{\ti
U}_+,{\ti U}_-] \,=\, 2 {\ti U}_3 \quad, \quad \quad [{\ti
V}_+,{\ti V}_-] \,=\, 2 {\ti V}_3 \quad, \quad
\\
&& [{\ti T}_+,{\ti V}_+] = [{\ti T}_+,{\ti U}_-] \,=\,[{\ti
U}_+,{\ti V}_+] = 0,
 \eea
 that are similar to the standard SU(3) commutation relations.
 However, the following commutators are deformed:
  \bea\non \lab{co2}  [{\ti T}_+,{\ti V}_-] \,=\, -
{\ti U}_- \,e^{2 i \De {\ti U}_3} \quad, \quad [{\ti T}_+,{\ti
U}_+] \,=\, {\ti V}_+ \,e^{-2 i \De {\ti V}_3} \quad, \quad [{\ti
U}_+,{\ti V}_-] \,=\, {\ti T}_- \,e^{2 i \De {\ti T}_3}.\\ \eea

In a similar way with the above derivation, we can study the
invariance properties of the Lagrangian Eq.(\ref{lag123}) under
the transformations:
\bea \label{flavsu3} \Psi_f'(x) \, =\, e^{i \al_j  {\ti F}_j}\,
\Psi_f (x) \, , \qquad  j=1, 2,..., 8. \eea
Then the following charges are obtained
\bea\label{su3fcharges} &&{}\hspace{-.5cm} {\ti Q}_{f,j}(t)\, =\,
\int  d^3{\bf x}\,\Psi^\dag_f(x)\, {\ti F}_j\, \Psi_f(x) \;,
\qquad j=1, 2,..., 8. \eea

\section{Group representations and the oscillation formula}

We now study the group representations. Let us first consider the
simple case of two generations and then discuss the three flavor
case.

\subsection{Two flavors}

In this case, the group is $SU(2)$ and the charges in the mass
basis read: \bea Q_{m,j}(t)\,=\,\frac{1}{2}\int d^{3}{\bf x}\,
\Psi_m^\dag(x) \, \tau_j\, \Psi_m(x),\quad \qquad j \,=\, 1, 2, 3,
\eea where $\Psi_m^T=(\nu_1,\nu_2)$ and $\tau_j=\si_j/2$ with
$\si_j$ being the Pauli matrices.

The states with definite masses can then be defined as eigenstates
of $Q_{m,3}$:
\bea \label{eigsu2a}
Q_{m,3}|\nu_{1}\rangle=\frac{1}{2}|\nu_{1}\rangle\qquad ;\qquad
Q_{m,3}|\nu_{2}\rangle=-\frac{1}{2}|\nu_{2}\rangle \eea
 and similar ones for antiparticles, where \bea Q_{m,3}=\frac{1}{2}\int
d^{3}{\bf x} \lf[\nu_{1}^{\dag}(x)\nu_{1}(x)-
 \nu_{2}^{\dag}(x)\nu_{2}(x)\ri].\eea

\vspace{0.5cm}
\centerline{\epsfysize=0.6truein\epsfbox{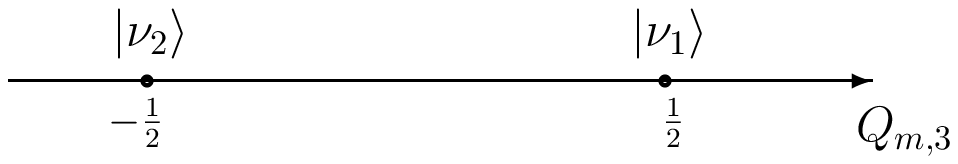}}
\vspace{.2cm} \centerline{\small Figure 4.1:  Plot of the mass
doublet $SU(2)$.}

\vspace{0.5cm}

 We have $|\nu_i\ran = \al^{r\dag}_{{\bf k},i}
|0\ran_m$, $i=1,2$.

The Eq.(\ref{eigsu2a}) expresses the obvious fact that the mass
eigenstates, treated as free particle states, are eigenstates of
the conserved $U(1)$ charges associated to $\nu_1$ and $\nu_2$:
\bea\label{su2noether} Q_{1}\, \equiv \,\frac{1}{2}Q \,+ \,Q_{m,3}
\qquad; \qquad Q_{2}\, \equiv \,\frac{1}{2}Q \,- \,Q_{m,3}. \eea

The next step is to define flavor states using a similar
procedure. We need to be careful here since the diagonal $SU(2)$
generator $Q_{f,3}$ is time-dependent in the flavor basis. Thus we
define states (Hilbert space) at a reference time $t=0$ from:
\bea
Q_{f,3}(0)|\nu_{e}\rangle=\frac{1}{2}|\nu_{e}\rangle\quad;\quad
Q_{f,3}(0)|\nu_{\mu}\rangle=-\frac{1}{2}|\nu_{\mu}\rangle. \eea
with $|\nu_\si\ran = \al^{r\dag}_{{\bf k},\si}(0) |0(0)\ran_f$,
$\si = e,\mu$ and similar ones for antiparticles.

\vspace{0.5cm}
\centerline{\epsfysize=0.6truein\epsfbox{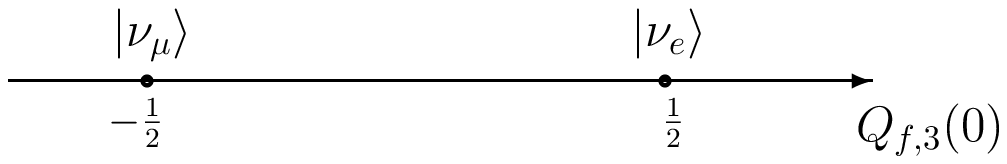}}
 \vspace{.2cm} \centerline{\small Figure 4.2: Plot of the flavor doublet $SU(2)$.}

\vspace{0.5cm}

The flavor states so defined are  eigenstates of the flavor
charges at time $t=0$:
\bea &&Q_{e}(t)=\frac{1}{2}Q + Q_{f,3}(t)\qquad; \qquad
Q_{\mu}(t)=\frac{1}{2}Q - Q_{f,3}(t),
\\ [2mm] \label{su2flavstates}
&& Q_{e}(0)|\nu_{e}\rangle=|\nu_{e}\rangle\qquad ;\qquad
Q_{\mu}(0)|\nu_{\mu}\rangle=|\nu_{\mu}\rangle. \eea
and $Q_{e}(0)|\nu_\mu\rangle=Q_{\mu}(0)|\nu_e\rangle=0$, with
 $Q_{e}(t)$ given by \bea\non
Q_{e}(t)&=& \int d^{3}{\bf x} \lf[\nu_{e}^{\dag}(x)\nu_{e}(x)\ri]=
\int d^{3}{\bf x} \Big[\nu_{1}^{\dag}(x)\nu_{1}(x)\cos^{2}\theta+
\\ &+&
 \nu_{2}^{\dag}(x)\nu_{2}(x)\sin^{2}\theta + \lf(\nu_{1}^{\dag}(x)\nu_{2}(x)+
 \nu_{2}^{\dag}(x)\nu_{1}(x)\ri)\sin\theta \cos\theta \Big] \eea
and similar equation for $Q_{\mu}(t)$.

This result is far from being trivial since the usual Pontecorvo
states \cite{Pontecorvo}:
\begin{eqnarray} \label{nue0a}
|\nu_{e}\rangle_P &=& \cos\theta\;|\nu_{1}\rangle \;+\;
\sin\theta\; |\nu_{2}\rangle \;
\\ [2mm] \label{nue0b}
|\nu_{\mu}\rangle_P &=& -\sin\theta\;|\nu_{1}\rangle \;+\;
\cos\theta\; |\nu_{2}\rangle \; ,
\end{eqnarray}
are {\em not} eigenstates of the flavor charges, indeed, for
example, we have \bea\non Q_{e}(0) |\nu_{e}\rangle_P =
\cos^{3}\theta |\nu_{1}\rangle + \sin^{3}\theta |\nu_{2}\rangle+
\sin\theta \cos\theta \lf[U^{*}_{\textbf{k}} |\nu_{1}\rangle
\sin\theta+ U_{\textbf{k}} |\nu_{2}\rangle \cos\theta- \ri. \\-
\non \sum_{r}\int d^{3}{\bf k} \lf(\cos\theta
u_{\textbf{k},1}^{r\dag}v_{-\textbf{q},2}^{r}\beta_{-\textbf{k},2}^{r\dag}
\alpha_{\textbf{k},1}^{r\dag}\alpha_{\textbf{q},1}^{s\dag}|0\rangle_{1,2}
+ \sin\theta
u_{\textbf{k},1}^{r\dag}v_{-\textbf{q},2}^{r}\beta_{-\textbf{k},2}^{r\dag}
\alpha_{\textbf{k},1}^{r\dag}\alpha_{\textbf{q},2}^{s\dag}|0\rangle_{1,2}\ri)
-\\ -\non  \sum_{r}\int d^{3}{\bf k} \lf(\cos\theta
u_{\textbf{k},2}^{r\dag}v_{-\textbf{q},1}^{s}\beta_{-\textbf{k},1}^{r\dag}
\alpha_{\textbf{k},2}^{r\dag}\alpha_{\textbf{q},1}^{s\dag}|0\rangle_{1,2}
+ \sin\theta
u_{\textbf{k},2}^{r\dag}v_{-\textbf{q},1}^{s}\beta_{-\textbf{k},1}^{s\dag}
\alpha_{\textbf{k},2}^{r\dag}\alpha_{\textbf{q},2}^{s\dag}|0\rangle_{1,2}
\ri).\\\eea

At a  time $t\neq 0$, oscillation formulae can be derived
 for the flavor charges from the following relation
\bea\non \langle\nu_{e}|Q_{f,3}(t)|\nu_{e}\rangle &=& \frac{1}{2}
- |U^{\bf k}_{12}|^{2}\sin^{2}(2\theta)
\sin^{2}\lf(\frac{\omega_{k,2}-\omega_{k,1}}{2}t\ri)-
\\
&-&|V^{\bf k}_{12}|^{2}\sin^{2}(2\theta)
\sin^{2}\lf(\frac{\omega_{k,2}+\omega_{k,1}}{2}t\ri) \eea
where the non-standard oscillation term do appear and $Q_{f,3}(t)$
is \bea Q_{f,3}(t) =\frac{1}{2}[Q_{e}(t)-Q_{\mu}(t)]. \eea

\subsection{Three flavors}
%
%%%%%%%%%%%%%%%%%%%%%%%%%%%%%%%%%%%%%%%%%%%%%%%%%%%%%%%%%%%%%
%\begin{figure}[t]
%\setlength{\unitlength}{1mm} \vspace*{70mm} %%
%% dvips
%\special{psfile="tripl3b.eps"
         %hscale=65 vscale=65
         %hoffset=100pt voffset=-0pt}
%\caption{The triplet.} \mlab{fig4}
%\end{figure}
%%%%%%%%%%%%%%%%%%%%%%%%%%%%%%%%%%%%%%%%%%%%%%%%%%%%%%%%%%%%%%%
%
Having discussed above the  procedure for the definition of flavor
states in the case of two flavors, we can directly write the {\em
flavor charges} for three flavor mixed fields as
\bea\non\label{char123} Q_e(t) & \equiv & \frac{1}{3}Q \, + \,
Q_{f,3}(t)\, + \,\frac{1}{\sqrt{3}} Q_{f,8}(t),
\\
Q_\mu(t) & \equiv & \frac{1}{3}Q \, - \, Q_{f,3}(t)+
\,\frac{1}{\sqrt{3}} Q_{f,8}(t),
\\\non
Q_\tau(t) & \equiv & \frac{1}{3}Q \, -  \, \frac{2}{\sqrt{3}}
Q_{f,8}(t). \eea

\bea Q_{\si}(0)|\nu_{\si}\rangle=|\nu_{\si}\rangle \quad, \qquad
Q_{\si}(0)|{\bar \nu}_{\si}\rangle=-|{\bar \nu}_{\si}\rangle \quad
, \qquad \si=e,\mu,\tau, \eea
leading to
\bea |\nu_\si\ran \equiv \al^{r\dag}_{{\bf k},\si}(0)
|0(0)\ran_f\quad, \qquad |{\bar \nu}_\si\ran \equiv
\bt^{r\dag}_{{\bf k},\si}(0) |0(0)\ran_f\quad, \qquad
\si=e,\mu,\tau. \eea

These neutrino and antineutrino states can be related to the
fundamental representation  ${\bf 3}$ and ${\bf 3^*}$ of the
(deformed) $SU(3)$ mixing group above introduced, as shown in
Fig.4.3 for neutrinos.

\vspace{0.5cm}
\centerline{\epsfysize=3truein\epsfbox{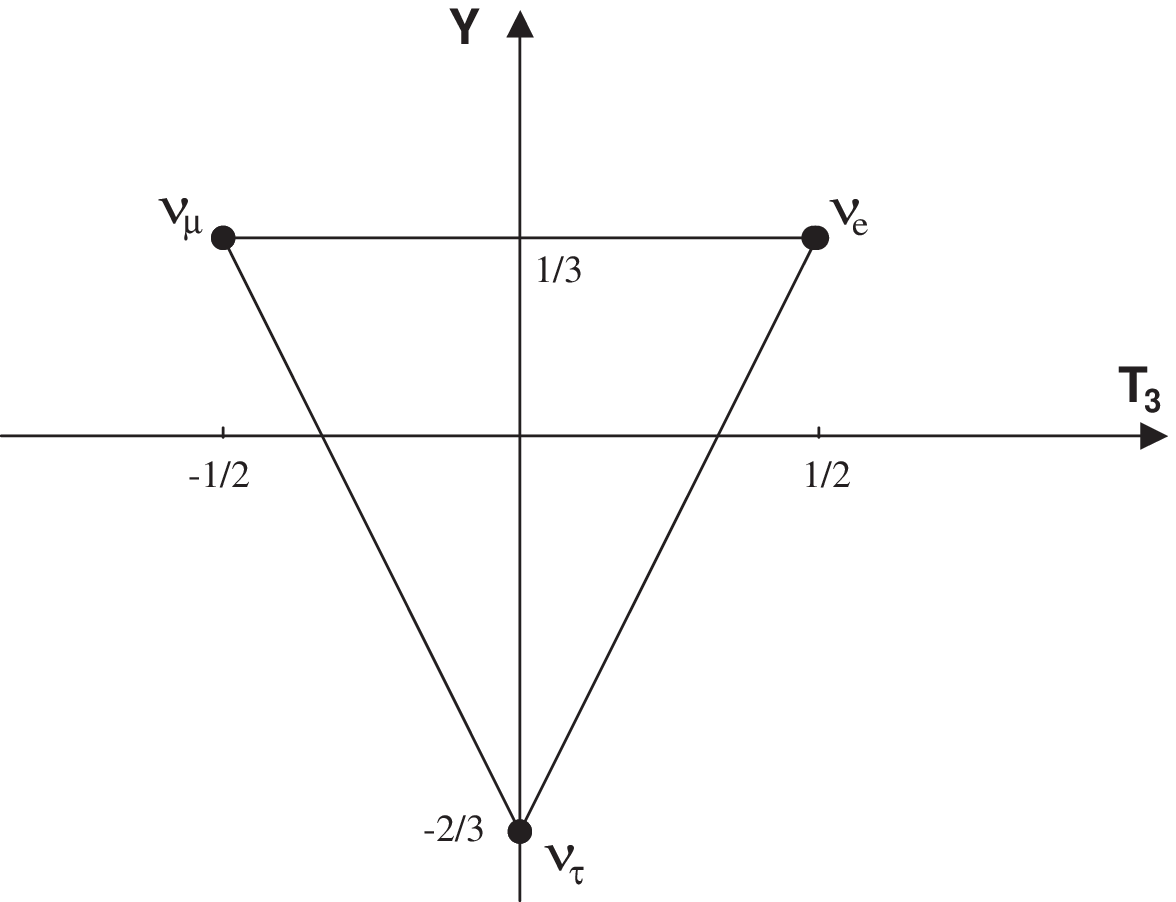}}
 \vspace{.2cm} \centerline{\small Figure 4.3: Plot of the flavor triplet $SU(3)$.}

\vspace{0.5cm}

To have the Fig.4.3, we have compared the generalized
Gell-Mann--Nishijima relation that defines the charge operator as
a function of $SU(3)$ generators:
 \bea q =
\frac{1}{3}Q+(T_{3}+\frac{Y}{2}), \eea to the Eqs.(\ref{char123}).

Note that the position of the points in the ${\ti Y}-{\ti T}_{3}$
is the same as for the ordinary $SU(3)$, since the diagonal
matrices ${\ti \lambda_{3}}$, ${\ti \lambda_{8}}$ do not contain
phases. However, a closed loop around the triangle gives a
non-zero phase which is of geometrical origin \cite{Anandan} and
only depends on the CP phase. A similar situation is valid for
antineutrinos.

\vspace{0.5cm}

To see this more in detail, let us consider the octet
representation as in Fig.4.4 and define the normalized state
$|A\rangle$: $ \langle A |A\rangle =1$. Then all the other states
are also normalized, except for  $|G\rangle$: $|G\rangle =
\frac{1}{\sqrt{2}}\ti{T}_{-}|A\rangle$.

\vspace{0.5cm} \centerline{\epsfysize=3truein\epsfbox{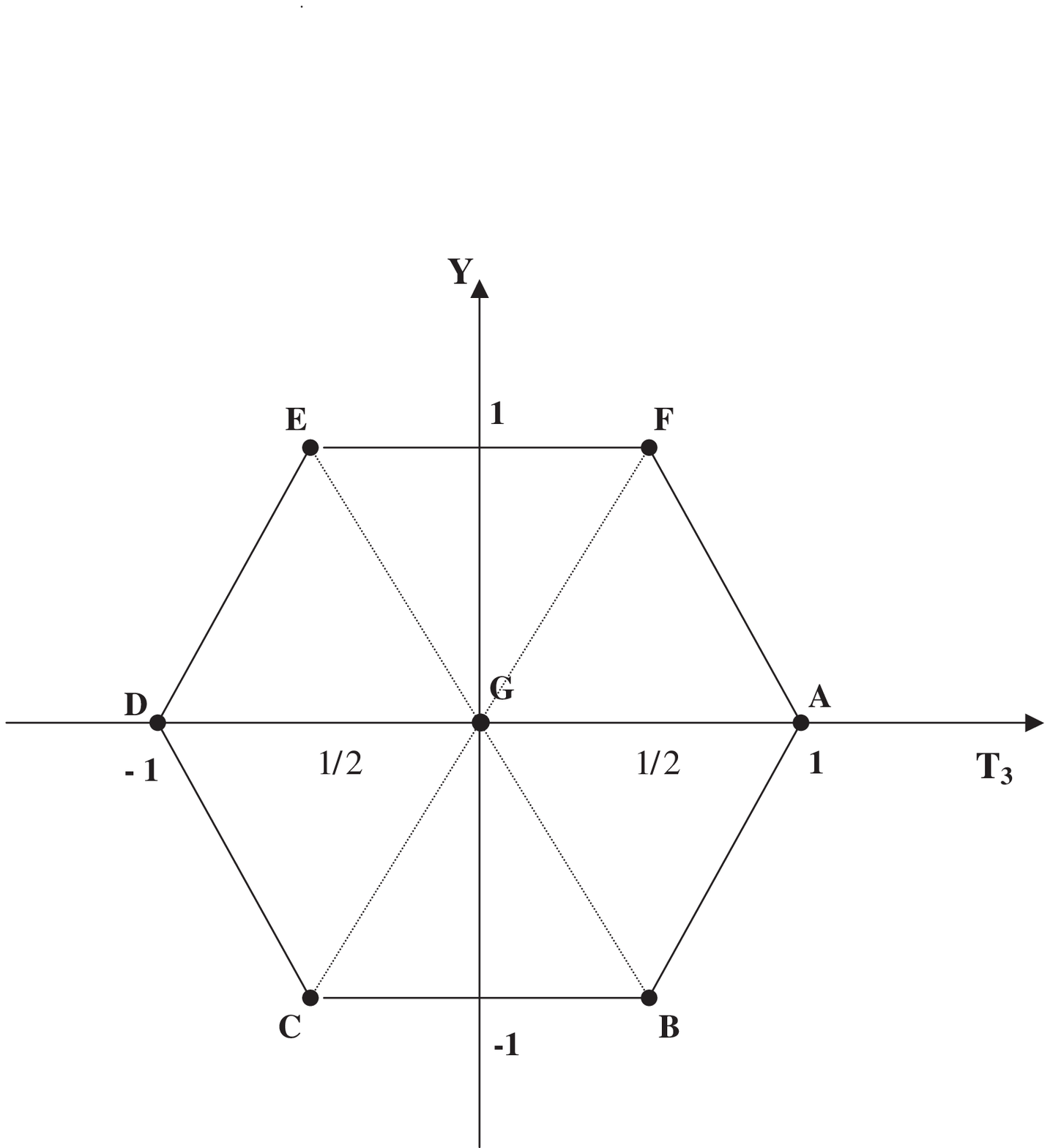}}
 \vspace{.2cm} \centerline{\small Figure 4.4: Plot of the $SU(3)$ octet.}

\vspace{0.5cm}
%
%%%%%%%%%%%%%%%%%%%%%%%%%%%%%%%%%%%%%%%%%%%%%%%%%%%%%%%%%%%%%
%\begin{figure}[t]
%\setlength{\unitlength}{1mm} \vspace*{70mm} %%
%% dvips
%\special{psfile="octetb.eps"
 %        hscale=60 vscale=60
  %       hoffset=100pt voffset=-0pt}
%\caption{The octet.}
%\end{figure}
%%%%%%%%%%%%%%%%%%%%%%%%%%%%%%%%%%%%%%%%%%%%%%%%%%%%%%%%%%%%%%%
We obtain the following paths
\bea\non
 &(AGBA): &\ti{V}_{+}\ti{U}_{-}\ti{T}_{-}|A\rangle =
\ti{V}_{+}([\ti{U}_{-},\ti{T}_{-}] +
\ti{T}_{-}\ti{U}_{-})|A\rangle =\ti V_{+}\ti V_{-}e^{2i\Delta \ti
V_{3}}|A\rangle = e^{i\Delta}|A\rangle
\\[2mm] \non
&(ABGA): &\ti T_{+}\ti U_{+}\ti V_{-}|A\rangle =
e^{-i\Delta}|A\rangle
\\[2mm] \non
&(AFGA): &\ti T_{+}\ti V_{-}\ti U_{+}|A\rangle =
-e^{-i\Delta}|A\rangle
\\[2mm] \non
&(AGFA): &\ti U_{-}\ti V_{+}\ti T_{-}|A\rangle =
-e^{i\Delta}|A\rangle
\\[2mm] \non
&(AFGBA): &\ti V_{+}\ti U_{-}\ti V_{-}\ti U_{+}|A\rangle =
|A\rangle
\\[2mm]
&(AFEDCBA): &\ti V_{+}\ti T_{+} \ti U_{-}\ti V_{-}\ti T_{-}\ti
U_{+}|A\rangle = |A\rangle \eea where we have used \bea\non
&&{}\hspace{-1.5cm} \ti U_{-}|A\rangle = \ti T_{+}|A\rangle = \ti
V_{+}|A\rangle = 0, \quad\ti T_{3}|A\rangle = |A\rangle \;, \quad
\ti V_{3}|A\rangle = \frac{1}{2}|A\rangle \;, \quad \ti
U_{3}|A\rangle = - \frac{1}{2}|A\rangle\\ \eea and the commutation
relations.
%
%%%%%%%%%%%%%%%%%%%%%%%%%%%%%%%%%%%%%%%%%%%%%%%%%%%%%%%%%%%%%%
%\begin{figure}[t]
%\setlength{\unitlength}{1mm} \vspace*{70mm} %%
%%% dvips
%\special{psfile="antitripl3b.eps"
%         hscale=60 vscale=60
%         hoffset=100pt voffset=-0pt}
%\caption{Plot of the triplet $3^{*}, (p,q)=(0,1)$.}
%
%\mlab{fig5}
%\end{figure}
%%%%%%%%%%%%%%%%%%%%%%%%%%%%%%%%%%%%%%%%%%%%%%%%%%%%%%%%%%%%%%%%

We thus see  that the phase sign change if we change the versus of
the path on the triangles; the paths on two opposite triangles and
around the hexagon bring no phase.

\section{Conclusions}
In this chapter, we have discussed some aspects of the
quantization of mixed fermions (neutrinos) in the context of
Quantum Field Theory.

In particular, we have analyzed the algebraic structures arising
in connection with field mixing and their deformation due to the
presence of CP violating phase, in the case of neutrino mixing
among three generations.

We have defined flavor states in terms of the  representations of
the group associated with field mixing. A new geometric phase
arising from  CP violation was discovered.

\vspace{0.5cm}

\chapter{Mixing and oscillations of Majorana fermions}

We study the mixing of Majorana fermions in Quantum Field Theory:
Majorana field are treated in the case of mixing among two
generations. We show how to consistently calculate oscillation
formulas, which agree with previous results for Dirac fields.

\section{Introduction}

In the derivation of the oscillation formulas by use of the flavor
Hilbert space a central role is played by the flavor charges
\cite{currents} and indeed it was found that these operators
satisfy very specifical physical requirements
\cite{Blasone:1999jb,fujii1}.
However, these charges vanish identically in the case of neutral
fields. In this chapter we will provide a consistent treatment of
Majorana fermions. In order to keep the discussion transparent, we
limit ourselves to the case of two generations.

Apart from the explicit quantization of the neutral mixed fields,
the main point of this chapter is the study of the momentum
operator (and more in general of the energy-momentum tensor) for
those fields \cite{Blaspalm}. We show how to define it in a
consistent way and by its use we then derive the oscillation
formulas, which match the ones already obtained for Dirac fields.
We also comment on its relevance for the study of Dirac mixed
fields, where, when CP violation is present, the charge
interpretation requires a further effort
\cite{fujii2,yBCV02,comment}.

In Section 5.2 we treat Majorana fields. In Section 5.3 we discuss
some general consequences of the results presented in this chapter
and draw conclusions.

\section{Majorana fermions}

We consider the case of mixing of two Majorana fermion fields. The
charge-conjugation operator $C$ is defined as satisfying the
relations
\bea \label{chargeconjoperator}
 {C}^{-1}\gamma_\mu\;{C} =
-\gamma_\mu^T \quad,\quad {C}^{\dag} = {C}^{-1} \quad,\quad
{C}^{T} = -{C} \ . \eea
from which we define the charge conjugate $\psi^c$ of $\psi$ as
\bea \psi^c(x) \equiv \gamma_0\;{C}\;\psi^*(x) \ . \eea
Now we define a Majorana fermion as the field that satisfies the
Dirac equation
\bea \label{eq:DiracEquation} (i \!\not\!\partial - m) \psi = 0
\eea
and the self-conjugation relation
\bea \label{eq:MajoranaCondition} \psi = \psi^c \;. \eea
Thus the two equations (\ref{eq:DiracEquation}) and
(\ref{eq:MajoranaCondition}) ensure that the Majorana field is a
neutral fermion field.

We now proceed by introducing the following Lagrangian:
\bea {\cal L} (x) &=& {\bar \psi_f}(x) ( i\!\not\!\partial - M )
\psi_f(x) \, =\,  {\bar \psi_m}(x) ( i\!\not\!\partial - M_d )
\psi_m(x)\,, \eea
with $\psi_f^T = (\nu_e, \nu_\mu)$ being the flavor fields and $M
= \left( \brr{cc} m_e & m_{e\mu}
\\ m_{e\mu} & m_\mu \err \right)$.
The flavor fields are connected to the free fields $\psi_m^T =
(\nu_1, \nu_2)$ with $M_d=diag(m_1, m_2)$ by the rotation:
\bea \label{eq:SimpleMajoranaMixing} &&\nu_{e}(x) = \nu_{1}(x) \;
\cos\te   + \nu_{2}(x) \; \sin\te\; ,
\\ [2mm] \label{eq:SimpleMajoranaMixing2}
 &&\nu_{\mu}(x) =- \nu_{1}(x) \; \sin\te   +
\nu_{2}(x)\; \cos\te\;. \eea
The quantization of the free fields is given by \cite{solar}
\be \nu_i (x) = \sum_{r=1,2} \int \frac {d^3 {\bf k}}
{(2\pi)^{\frac{3}{2}}} \; e^{i{\bf k}\cdot\mathbf{x}}
\left[u^r_{{\bf k},i}(t)\al^r_{{\bf k},i} +  v^r_{-{\bf k},i}(t)
\al^{r{\dag}}_{-{\bf k},i}\right], \quad i = 1,2.\ee
where $u^r_{{\bf k},i}(t) = e^{-i\omega_{k,i}t}u^r_{{\bf k},i}$,
$v^r_{{\bf k},i}(t) = e^{i\omega_{k,i}t}v^r_{{\bf k},i}$, with
$\omega_{k,i} = \sqrt{{\bf k}^2 + m^2_i}$. In order for the
Majorana condition (\ref{eq:MajoranaCondition}) to be satisfied,
the four spinors must also satisfy the following condition:
\bea \label{eq:MajoranaSpinorCondition} v^s_{{\bf k},i} = \gamma_0
\;{C} (u^{s}_{{\bf k},i})^*\,\quad;\quad u^{s}_{{\bf k},i} =
\gamma_0\; {C} (v^{s}_{{\bf k},i})^*\,. \eea
The equal time anticommutation relations are: \bea\non
\{\nu^{\al}_{i}(x), \nu^{\bt{\dag} }_{j}(y)\}_{t=t'} &=
\de^{3}{(\bf x-y)} \de_{\al\bt} \de_{ij} \quad, \quad
\{\nu^{\al}_{i}(x), \nu^{\bt}_{j}(y)\}_{t=t'} &= \de^{3}{(\bf
x-y)} (\gamma_0 {C})^{\al\bt} \de_{ij},\\ \eea with
$\al,\bt=1,..,4$
and
\be \{\al^{r}_{{\bf k},i}, \al^{s{\dag} }_{{\bf q},j}\} =
\de^{3}{(\bf k-q)}\de _{rs}\de_{ij}   \;,\qquad i,j=1,2\;. \ee
All other anticommutators are zero. The orthonormality and
completeness relations are:
\be u^{r{\dag}}_{{\bf k},i} u^{s}_{{\bf k},i} = v^{r{\dag}}_{{\bf
k},i} v^{s}_{{\bf k},i} = \delta_{rs} \;,\quad  u^{r{\dag}}_{{\bf
k},i} v^{s}_{-{\bf k},i} = v^{r{\dag}}_{-{\bf k},i} u^{s}_{{\bf
k},i} = 0\;,\quad \sum_{r=1,2}(u^{r}_{{\bf k},i} u^{r{\dag}}_{{\bf
k},i} + v^{r}_{-{\bf k},i} v^{r{\dag}}_{-{\bf k},i}) = I \;. \ee

We can recast
Eqs.(\ref{eq:SimpleMajoranaMixing}),(\ref{eq:SimpleMajoranaMixing2})
into the form:
\bea
&&\nu_e^{\al}(x) = G^{-1}_{\te}(t)\; \nu_{1}^{\al}(x)\; G_{\te}(t)\;,\\[2mm]
&&\nu_\mu^{\al}(x) = G^{-1}_{\te}(t)\; \nu_{2}^{\al}(x)\;
G_{\te}(t) \;, \eea
where $G_{\te}(t)$ is given by
\be G_{\te}(t) = \exp\lf[\frac {\te}{2} \int d^{3}{\bf x}
\lf(\nu_{1}^{\dag}(x) \nu_{2}(x) - \nu_{2}^{\dag}(x) \nu_{1}(x)
\ri)\ri]\;. \ee

We have $G_{\te}(t)=\prod_{\bf k} G_{\te}^{\bf k}(t)$. Moreover
for a given ${\bf k}$, in the reference frame where ${\bf
k}=(0,0,|{\bf k}|)$, the spins decouple \cite{BV95} and one has
$G_{\te}^{\bf k}(t)=\prod_r G_{\te}^{{\bf k},r}(t)$ with
\bea\non G_{\te}^{{\bf k},r}(t) = \exp\Big\{\te  \Big[ U_{\bf
k}^*(t)\; \al^{r{\dag}}_{{\bf k},1} \al^{r}_{{\bf k},2} - U_{\bf
k}(t)\; \al^{r{\dag}}_{-{\bf k},2} \al^{r}_{-{\bf k},1}-\\ - \ep^r
V_{\bf k}^*(t) \al^{r}_{-{\bf k},1} \al^{r}_{{\bf k},2}+  \ep^r
V_{\bf k}(t) \al^{r{\dag}}_{{\bf k},1} \al^{r{\dag}}_{{-\bf
k},2}\Big]\Big\}\,, \eea
where $U_{{\bf k}}(t)$ and $V_{{\bf k}}(t)$ are Bogoliubov
coefficients given by
\be U_{{\bf k}}(t)\equiv |U_{{\bf
k}}|\;e^{i(\om_{k,2}-\om_{k,1})t} \quad,\quad V_{{\bf
k}}(t)\equiv|V_{{\bf k}}|\;e^{i(\om_{k,2}+\om_{k,1})t}\;, \ee
\bea &&|U_{{\bf
k}}|\equiv\lf(\frac{\om_{k,1}+m_{1}}{2\om_{k,1}}\ri)^{\frac{1}{2}}
\lf(\frac{\om_{k,2}+m_{2}}{2\om_{k,2}}\ri)^{\frac{1}{2}}
\lf(1+\frac{|{\bf
k}|^{2}}{(\om_{k,1}+m_{1})(\om_{k,2}+m_{2})}\ri)\;,
\\[2mm]
 &&|V_{{\bf k}}|\equiv\lf(\frac{\om_{k,1}+m_{1}}{2\om_{k,1}}
\ri)^{\frac{1}{2}}
\lf(\frac{\om_{k,2}+m_{2}}{2\om_{k,2}}\ri)^{\frac{1}{2}}
\lf(\frac{|\bf k|}{(\om_{k,2}+m_{2})}- \frac{|\bf
k|}{(\om_{k,1}+m_{1})}\ri)\;, \eea
\be |U_{{\bf k}}|^{2}+|V_{{\bf k}}|^{2}=1 \;. \ee

The flavor fields can be thus expanded as:
\be \nu_\si (x) = \sum_{r=1,2} \int \frac {d^3 {\bf k}}
{(2\pi)^{\frac{3}{2}}} e^{i{\bf k}\cdot\mathbf{x}} \left[u^r_{{\bf
k},j}(t)\al^r_{{\bf k},\sigma}(t) + v^r_{-{\bf k},j}(t)
\al^{r{\dag}}_{-{\bf k},\sigma}(t)\right], \ee
with $\sigma,j = (e,1),(\mu,2)$ and the flavor annihilation
operators given by (for ${\bf k}=(0,0,|{\bf k}|)$):
\bea\non &&\al^{r}_{{\bf k},e}(t) \equiv G^{-1}_{\te}(t) \;
\al^{r}_{{\bf k},1}\; G_{\te}(t) = \cos \theta \, \al^{r}_{{\bf
k},1} + \sin \theta\,\lf(U_{\bf k}^*(t) \,\al^{r}_{{\bf k},2}+
\ep^r V_{\bf k}(t)\, \al^{r{\dag}}_{{-\bf k},2}\ri)\;,
\\[2mm] \non
&&\al^{r}_{{\bf k},\mu}(t) \equiv G^{-1}_{\te}(t)\; \al^{r}_{{\bf
k},2}\; G_{\te}(t) = \cos \theta\, \al^{r}_{{\bf k},2} -\sin
\theta\, \lf(U_{\bf k}(t) \,\al^{r}_{{\bf k},1} -\ep^r V_{\bf
k}(t) \,\al^{r{\dag}}_{{-\bf k},1}\ri)\;.\\ \eea

We now consider the action of the generator of the mixing
transformations on the vacuum $|0\rangle_\mass$. The flavor vacuum
is defined as:
\be |0(\theta,t)\rangle_{e,\mu} \equiv
G_\theta^{-1}(t)|0\rangle_\mass\;.\ \ee

We define the state for a mixed particle with definite flavor,
spin and momentum as:
\be |\alpha^r_{{\bf k},e}(t)\rangle \equiv \alpha^{r
\dagger}_{{\bf k},e}(t)|0(t)\rangle_{e,\mu} = G_\theta^{-1}(t)
\alpha^{r \dagger}_{{\bf k},1}|0\rangle_\mass\;. \ee

The anticommutators of the flavor ladder operators at different
times are:
\bea\non \left\{ \alpha^r_{{\bf k},e}(t),\alpha^{r \dagger}_{{\bf
k},e}(t') \right\} &=& \cos^2\theta + \sin^2\theta \left( |U_{\bf
k}|^2 e^{-i(\omega_2-\omega_1)(t-t')} + |V_{\bf k}|^2
e^{i(\omega_2+\omega_1)(t-t')} \right) \;,\\
\\[2mm]
\left\{\alpha^{r \dagger}_{{-\bf k},e}(t), \alpha^{r
\dagger}_{{\bf k},e}(t')\right\} &=& \epsilon^r\, \sin^2\theta
\,|U_{\bf k}| |V_{\bf k}|\, \left(e^{i\omega_2(t-t')} -
e^{-i\omega_2(t-t')}\right) e^{-i\omega_1 (t+t' )} \;,
\\[2mm]
\left\{ \alpha^r_{{\bf k},\mu}(t),\alpha^{r \dagger}_{{\bf
k},e}(t') \right\} &=& \cos\theta \sin\theta\,|U_{\bf k}| \,\left(
e^{i(\omega_2-\omega_1)t'} - e^{i(\omega_2-\omega_1)t} \right) \;,
\\[2mm]
\left\{\alpha^{r \dagger}_{{-\bf k},\mu}(t),\alpha^{r
\dagger}_{{\bf k},e}(t')\right\} &=& \epsilon^r\, \cos\theta
\sin\theta \, |V_{\bf k}| \, \left( e^{-i(\omega_2+\omega_1)t'} -
e^{-i(\omega_2+\omega_1)t} \right) \;. \eea
The following quantity is constant in time:
\bea\non\label{one1} &&\left|\left\{ \alpha^r_{{\bf k},e}(t),
\alpha^{r \dagger}_{{\bf k},e}(t') \right\}\right|^2 +
\left|\left\{\alpha^{r \dagger}_{{-\bf k},e}(t), \alpha^{r
\dagger}_{{\bf k},e}(t')\right\}\right|^2 +\\&&+ \left|\left\{
\alpha^r_{{\bf k},\mu}(t), \alpha^{r \dagger}_{{\bf k},e}(t')
\right\}\right|^2 + \left|\left\{\alpha^{r \dagger}_{{-\bf
k},\mu}(t), \alpha^{r \dagger}_{{\bf k},e}(t')\right\}\right|^2 =
1\;. \eea

The corresponding of Eq.(\ref{one1}) for Dirac fields, was
consistently interpreted as expressing the conservation of total
charge. In the present case we are dealing with a neutral field
and thus the charge operator vanishes identically. Nevertheless
the quantities in Eq.(\ref{one1}) are well defined and are the
Majorana field counterpart of the corresponding ones for the case
of Dirac fields. Thus we look for a physical interpretation of
such oscillating quantities.

Let us consider the momentum operator defined as \bea P^j \equiv
\int d^3{\bf x}\; {T}^{0j}(x),\eea where the energy-momentum
tensor for the fermion field, ${T}^{\mu\nu}$, is defined by
\bea{T}^{\mu\nu} \equiv i \bar{\psi}\gamma^\nu
\partial_\mu  \psi.\eea For the free fields $\psi_i$ we have:
\bea\non {\bf P}_i = \int d^3{\bf x}\, \psi^\dagger_i(x)
(-i\nabla)\psi_i(x) = \int d^3 {\bf k} \sum_{r=1,2} {\bf k}
 \left( \alpha^{r\dagger}_{{\bf k},i}\alpha^{r}_{{\bf k},i} -
 \alpha^{r\dagger}_{{-\bf
k},i}\alpha^{r}_{{-\bf k},i} \right), \quad i = 1,2 \;.\\ \eea
We then define the momentum operator for mixed fields:
\bea\non {\bf P}_\sigma(t) = \int d^3{\bf x}
\,\psi^\dagger_\sigma(x) (-i\nabla)\psi_\sigma(x) = \int d^3 {\bf
k} \sum_{r=1,2} {\bf k} \left( \alpha^{r\dagger}_{{\bf
k},\sigma}(t) \alpha^{r}_{{\bf k},\sigma}(t) -
\alpha^{r\dagger}_{{-\bf k},\sigma}(t)\alpha^{r}_{{-\bf
k},\sigma}(t) \right),\\\eea with $\sigma = e,\mu$.
We have \bea{\bf P}_\sigma(t)= G_\theta^{-1}(t) {\bf P}_i
G_\theta(t)\eea and the conservation of total momentum as a
consequence of
\bea && {\bf P}_e(t) + {\bf P}_\mu(t) = {\bf P}_1 + {\bf P}_2
\equiv {\bf P}\quad ,\quad [{\bf P},G_\theta(t)] = 0 \quad , \quad
[{\bf P}, H] = 0\;. \eea

We now consider the expectation values on the flavor state
$|\alpha_{{\bf k},e}^r\rangle\equiv |\alpha_{{\bf
k},e}^r(0)\rangle$. At time $t=0$, this state is an eigenstate of
the momentum operator ${\bf P}_e(0)$:
\be {\bf P}_e(0)\, |\alpha_{{\bf k},e}^r\rangle \, =\, {\bf k}
\,|\alpha_{{\bf k},e}^r\rangle \;. \ee

At $t \neq 0$ the expectation value for the momentum (normalized
to initial value) gives:
\bea {\cal P}^e_{{\bf k},\sigma}(t) \equiv \frac{\langle
\alpha^r_{{\bf k},e}|{\bf P}_\sigma(t)|\alpha^r_{{\bf
k},e}\rangle} {\langle\alpha^r_{{\bf k},e}|{\bf
P}_\sigma(0)|\alpha^r_{{\bf k},e}\rangle} = \left|\left\{
\alpha^r_{{\bf k},\sigma}(t),\alpha^{r \dagger}_{{\bf k},e}(0)
\right\}\right|^2 + \left|\left\{\alpha^{r \dagger}_{{-\bf
k},\sigma}(t),\alpha^{r \dagger}_{{\bf
k},e}(0)\right\}\right|^2,\eea with $\sigma = e,\mu$,
which is the same form of the expression one obtains for the
expectation values of the flavor charges in the case of Dirac
fields \cite{BHV98}.
The flavor vacuum expectation value of the momentum operator ${\bf
P}_\sigma(t)$ vanishes at all times:
\be _{e,\mu}\langle 0|{\bf P}_\sigma(t)|0\rangle_{e,\mu} = 0 \quad
, \quad \sigma = e,\mu \;. \ee

The explicit calculation of the oscillating quantities ${\cal
P}^e_{{\bf k},\sigma}(t)$ gives:
\bea\non {\cal P}^e_{{\bf k},e} (t) &=& 1 - \sin^2 2\theta
\left[|U_{\bf k}|^2\, \sin^2 \lf( \frac{\omega_{k,2} -
\omega_{k,1}}{2} t \ri)  + |V_{\bf k}|^2\, \sin^2\lf(
\frac{\omega_{k,2} + \omega_{k,1}}{2} t \ri)\right],\\
\\[2mm]
{\cal P}^e_{{\bf k},\mu} (t) &=& \sin^2 2\theta \left[|U_{\bf
k}|^2\, \sin^2 \lf( \frac{\omega_{k,2} - \omega_{k,1}}{2} t \ri) +
|V_{\bf k}|^2\, \sin^2 \lf( \frac{\omega_{k,2} + \omega_{k,1}}{2}
t \ri)\right], \eea
in complete agreement with the Dirac  field case \cite{BHV98}.

\section{Discussion and Conclusions}

In this chapter, we have studied the mixing among two generations
in the case of Majorana in the context of Quantum Field Theory.

The main point is the calculation of oscillation formulas, which
we obtained by use of the momentum operator, that is well defined
for (mixed) neutral fields, whereas the charge operator vanishes
identically. The results confirm the oscillation formulas already
obtained in the case of Dirac fields by use of the flavor charges,
and it also reveals to be useful in the case of three flavor
mixing, where the presence of the CP violating phase introduces
ambiguities in the treatment based on flavor charges
\cite{yBCV02}.

It is indeed interesting to comment on this point: for Dirac
fields, the momentum operator is given as
\bea\non {\bf P}_\sigma(t) =\, \int d^3 {\bf k} \sum_r \frac {\bf
k} {2} \Big( \alpha^{r\dagger}_{{\bf k},\sigma}(t)
\alpha^{r}_{{\bf k},\sigma}(t)  - \alpha^{r\dagger}_{{-\bf
k},\sigma}(t)\alpha^{r}_{{-\bf k},\sigma}(t)+ \\+
\beta^{r\dagger}_{{\bf k},\sigma}(t)\beta^{r}_{{\bf k},\sigma}(t)
-\beta^{r\dagger}_{-{\bf k},\sigma}(t)\beta^{r}_{-{\bf
k},\sigma}(t) \Big)\quad , \quad \sigma = e,\mu,\tau\;. \eea
This operator can be used for the calculation of the oscillation
formulas for Dirac neutrinos in analogy with what done above in
the Majorana case. Although in this case the charge operator is
available and it has been used successfully for deriving the
oscillation formula \cite{BHV98} in the two-flavor case, it has
emerged that for three-flavor mixing, the CP violating phase
introduces complications in the identification of the observables
and indeed the matter is still object of discussion
\cite{fujii2,yBCV02,comment}. The main problem there  is that the
flavor charges at time $t$ do not annihilate the flavor vacuum:
${}_f\lan 0 | Q_\si(t) | 0\ran_f \, \neq \, 0$ and this
expectation value needs to be subtracted by hand in order to get
the correct oscillation formulas \cite{yBCV02}.

However, we see easily how the use of the momentum operator
confirms the results of Ref.\cite{yBCV02}, without presenting any
ambiguity. We have indeed:
\bea &&{}_f\lan 0 | {\bf P}_\si(t) | 0\ran_f\, =\, 0
\\ [2mm]
&&\frac{\langle \nu_\rho|{\bf P}_\si(t)| \nu_\rho\ran} {\langle
\nu_\rho|{\bf P}_\si(0)| \nu_\rho\ran}
 \;=\, \lf|\lf \{\al^{r}_{{\bf k},\si}(t), \al^{r
{\dag}}_{{\bf k},\rho}(0) \ri\}\ri|^{2} \;+ \;\lf|\lf\{\bt_{{-\bf
k},\si}^{r {\dag}}(t), \al^{r {\dag}}_{{\bf k},\rho}(0)
\ri\}\ri|^{2} \,, \eea
with $\si,\rho = e,\mu,\tau$ and $|\nu_\rho\ran \equiv \al^{r
{\dag}}_{{\bf k},\rho}(0) |0\ran_f$. This follows from the
following relations:
\bea && {}_f\lan 0| \alpha^{r\dagger}_{{\bf k},\sigma}(t)
\alpha^{r}_{{\bf k},\sigma}(t)  | 0\ran_f \, =\, {}_f\lan
0|\alpha^{r\dagger}_{{-\bf k},\sigma}(t) \alpha^{r}_{{-\bf
k},\sigma}(t) | 0\ran_f
\\ [2mm]
&& {}_f\lan 0| \bt^{r\dagger}_{{\bf k},\sigma}(t) \bt^{r}_{{\bf
k},\sigma}(t)  | 0\ran_f\, =\, {}_f\lan 0|\bt^{r\dagger}_{{-\bf
k},\sigma}(t) \bt^{r}_{{-\bf k},\sigma}(t) | 0\ran_f \eea
which are valid even in presence of CP violation, when ${}_f\lan
0| \alpha^{r\dagger}_{{\bf k},\sigma}(t) \alpha^{r}_{{\bf
k},\sigma}(t)  | 0\ran_f \neq {}_f\lan 0| \bt^{r\dagger}_{-{\bf
k},\sigma}(t) \bt^{r}_{-{\bf k},\sigma}(t)  | 0\ran_f $.

These results seem to suggest that perhaps a redefinition of the
flavor charge operators is necessary in presence of CP violation
and
further study in this direction is in progress.

\chapter{Neutrino mixing and cosmological constant}

We show that the non--perturbative vacuum structure associated
with neutrino mixing leads to a non--zero contribution to the
value of the cosmological constant. Such a contribution comes from
the specific nature of the mixing phenomenon. Its origin is
completely different from the one of the ordinary contribution of
a massive spinor field. We estimate this neutrino mixing
contribution by using the natural cut--off appearing in the
quantum field theory formalism for neutrino mixing and
oscillation.

\vspace{8mm}

\section{Introduction}

By resorting to the recent discovery of the unitary inequivalence
between the mass and the flavor vacua for neutrino fields in
quantum field theory (QFT)
\cite{BHV98,Blasone:1999jb,BV95,BCRV01,JM01,fujii1,fujii2,comment,JM011,binger,CJMV2003},
we show that the non--perturbative vacuum structure associated
with neutrino mixing may lead to a non--zero contribution to the
value of the cosmological constant \cite{cosmolog2003}.

The contribution we find comes from the specific nature of the
field mixing and is therefore of different origin with respect to
the ordinary well known perturbative vacuum energy contribution of
a massive spinor field.

The nature of the cosmological constant, say $\Lambda$, is one of
the most intriguing issues in  modern theoretical physics and
cosmology. Data coming from observations indicate that not only
$\Lambda$ is different from zero, but it also dominates the
universe dynamics driving an accelerated expansion (see for
example \cite{WMAP}).

In the classical framework, $\Lambda$ can be considered as an
intrinsic (i.e. not induced by matter) curvature of  space-time or
a sort of shift in the matter Lagrangian. In the latter case
$\Lambda$ can be considered as an unclustered, non interacting
component of the cosmic fluid with a constant energy density
$\rho$ and the equation of state $p=-\rho$. These properties and
the fact that the presence of a cosmological constant fluid has to
be compatible with the structure formation, allow to set the upper
bound $\Lambda < 10^{-56} cm^{-2}$ \cite{Zeldovic1}.

In the quantum framework, the standard approach is to consider the
cosmological constant as a gravitational effect of vacuum energy
\cite{Zeldovic2,weinberg,sahni}. A common problem of all these
approaches is that they do not provide a value of $\Lambda$ in the
bound given above. This is known as the {\it cosmological constant
problem} \cite{sahni} and it has been tried to address it in
different ways (see for example \cite{Cap}).

In this chapter, we show  that the vacuum energy induced by  the
neutrino mixing may contribute to the value of cosmological
constant in a  fundamentally different way from the usual
zero-point energy contribution, as already stated above.

Indeed,  it has been realized \cite{BHV98,Blasone:1999jb,BV95}
that the mixing of massive neutrino fields is a highly non-trivial
transformation in QFT. The vacuum for neutrinos with definite mass
is not invariant under the mixing transformation and in the
infinite volume limit it is unitarily inequivalent to the vacuum
for the neutrino fields with definite flavor number. This affects
the oscillation formula which turns out to be different from the
usual Pontecorvo formula \cite{Pontecorvo} and a number of
consequences have been already discussed \cite{Blasone:1999jb}.

The existence of the two inequivalent vacua for the flavor and the
mass eigenstate neutrino fields, respectively, is crucial in order
to obtain a non--zero contribution to the cosmological constant as
we show below.

In Section 6.2, we show that the neutrino contribution to the
value of the cosmological constant is non--zero and then we
estimate its value by using the natural scale of neutrino mixing
as cut--off. The result turns out to be compatible with the above
mentioned upper bound on $\Lambda$. Section 6.3 is devoted to the
conclusions. The Appendix J is devoted to the tetradic formalism
in the Friedmann Robertson Walker (FRW) space--time.

\section{Neutrino mixing contribution to the cosmological
constant}

The connection between the vacuum energy density
$\lan\rho_{vac}\ran$ and the cosmological constant $\Lambda$ is
provided by the well known relation
\bea\label{ed} \lan\rho_{vac}\ran= \frac{\Lambda}{4\pi G} , \eea
where $G$ is the gravitational constant.

In order to calculate $\lan\rho_{vac}\ran$ we have to consider
tetrads and  spinorial connection. The symmetries of the
cosmological metric make this task easier.

Tetrads are defined as a local inertial coordinate system at every
space-time point defined by \bea g_{\mu
\nu}=e^{a}_{\mu}\;e^{b}_{\nu}\;\eta_{a b},\eea where $g_{\mu \nu}$
it the curved space-time metric and $\eta_{a b}$ is the Minkowski
metric. In the tetrads framework the parallel transport in a
torsion free space-time is defined by the affine spin connection
one form \bea \omega^{a}_{b}=\omega^{a}_{b \mu}d x^{\mu},\eea
which satisfies the Cartan Structure equations
\begin{equation}\label{eq forma di connessione}
  d e^{a} + \omega^{a}_{b}\wedge e^{b}=0,
\end{equation}
where $e^{a}$ is the tetrad one-form defined by \bea
e^{a}=e^{a}_{\nu}dx^{\nu}.\eea The energy--momentum tensor density
${\cal T}_{\mu \nu}$ is obtained  by varying the action with
respect to the metric $g_{\mu\nu}$:

 \bea\ \label{impulsoenergia}
{\cal T}_{\mu\nu}=\frac{2}{\sqrt{-g}}\frac{\delta S}{\delta
g^{\mu\nu}(x)} ,
 \eea
where the action is
 \bea\
S=\int \sqrt{-g}{\cal L}(x) d^{4}x.
 \eea%

In the present case, the energy momentum tensor density is given
by \bea\label{tmn}
 {\cal T}_{\mu\nu}(x) = \frac{i}{2}\left({\bar
\Psi}_{m}(x)\gamma_{\mu} \overleftrightarrow{D}_{\nu}
\Psi_{m}(x)\right)\ ,
 \eea
where $\overleftrightarrow{D}_{\nu}$ is the covariant derivative:
\begin{equation}\label{der spin}
D_{\nu}=\partial_{\nu}+\Gamma_{\nu},
 \quad\;\quad\;\quad  \Gamma_{\nu}=\frac{1}{8} \omega^{a
b}_{\nu}[\gamma_{a},\gamma_{b}], \quad\;\quad\;\quad
  \gamma_{\mu}(x)=\gamma^{c}e_{c \mu}(x),
\end{equation}
 being $\gamma^{c}$ the standard Dirac matrices, and
 $\bar{\Psi}\overleftrightarrow{D}_{\nu}\Psi=\bar{\Psi}D\Psi-(D\bar{\Psi})
\Psi$. Let us consider the FRW metric of the form
\begin{equation}\label{metrica di friedmann}
  ds^{2}=dt^{2}-a^{2}(t)\left(\frac{dr^{2}}{\rho}+r^{2}d\theta^{2}+r^{2}\sin^{2}
(\theta)d\phi^{2}\right),
\end{equation}
with $\rho=1-\kappa r^{2}$, and $\kappa=0,1,-1$.

In the FRW metric Eq.(\ref{metrica di friedmann}), the most
natural choice for tetrads is \bea\non\label{tetradi di Friedmann}
e^{0}&=&e^{0}_{0}\;d x^{0}=
dt,\qquad\qquad\quad e^{1}=e^{1}_{1}\;d x^{1}=\frac{a(t)}{\rho}\;d r,\\
e^{2}&=&e^{2}_{2}\;d x^{2}=a(t)\;r\; d \theta, \qquad
e^{3}=e^{3}_{3}\;d x^{3}=a(t)\;r\; \sin(\theta) d r. \eea Using
Eq.(\ref{eq forma di connessione}) and the definitions in
Eq.(\ref{der spin}) we have (see the Appendix J)

\bea\label{derivata covariante spinoriale componente zero}
  D_{0}=\partial_{0}+\frac{1}{4}[\gamma_{i},\gamma_{0}]\;{\cal H}\;e^{i}_{0}+\frac{1}{4}
  [\gamma_{1},\gamma_{j}]\;\frac{\rho}{a r}\;e^{j}_{0}+\frac{1}{4}
  [\gamma_{2},\gamma_{3}]\;\frac{\tan(\theta)}{a
r}\;e^{3}_{0},\eea with $i=1,2,3$, $j=2,3$.

Since we choose diagonal tetrads every term $e^{k}_{0}$ with
$k=1,2,3$ is null. This implies that all the terms but the first
are null whatever the value of the commutators of the Dirac
matrices. It is worth to stress that this result is independent of
the choice of tetrads because the (0,0) component of the energy
momentum tensor of a fluid is equivalent to the energy density
only if the tetrads (or in general the chosen coordinates) are
time-orthogonal as in Eq.(\ref{tetradi di Friedmann}). If we
choose non time-orthogonal tetrads, i.e. $e^{0}_{0}$ not constant,
${\cal T}_{00}(x)$ does not represent the energy density because
it acquires "pressure components" due to the different orientation
of the tetrad. In our calculation these terms are the second,
third and forth term of the Eq.(\ref{derivata covariante
spinoriale componente zero}).

Thus the temporal component of the spinorial derivative in the FRW
metric is just the standard time derivative \cite{cosmolog2003}:
\begin{equation}
D_{0}=\partial_{0}.
\end{equation}
This is not surprising if we consider the symmetries of the metric
element Eq.(\ref{metrica di friedmann}). Thus, the (0,0) component
of the stress energy tensor density is
\begin{equation}\label{sttress energy tensor esplicito}
  {\cal T}_{00} = {\cal T}_{00}^{Flat}.
\end{equation}

This allows us to use ${\cal T}_{00}^{Flat}$ to compute the
cosmological constant. From Eq.(\ref{tmn}) we thus obtain
\bea\
 {\cal T}_{00}(x) = \frac{i}{2}:\left({\bar \Psi}_{m}(x)\gamma_{0}
\overleftrightarrow{\partial}_{0} \Psi_{m}(x)\right):\eea where
$:...:$ denotes  the customary normal ordering with respect to the
mass vacuum in the flat space-time.

In terms of the annihilation and creation operators of fields
$\nu_{1}$ and $\nu_{2}$, the energy-momentum tensor \bea
T_{00}=\int d^{3}x {\cal T}_{00}(x)\eea is given by
\bea T^{(i)}_{00}= \sum_{r}\int d^{3}{\bf k}\,
\omega_{k,i}\lf(\al_{{\bf k},i}^{r\dag} \al_{{\bf k},i}^{r}+
\beta_{{\bf -k},i}^{r\dag}\beta_{{\bf
-k},i}^{r}\ri),\quad\quad\quad i=1,2, \eea
and we note that $T^{(i)}_{00}$ is time independent.

Next, our task is to compute the expectation value of
$T^{(i)}_{00}$ in the flavor vacuum $| 0\ran_f$, which, as already
recalled, is the one relevant to mixing and oscillations. Thus,
the contribution $\lan\rho_{vac}^{mix}\ran$ of the neutrino mixing
to the vacuum energy density is:
 \bea\
 {}_f\lan 0 |\sum_{i} T^{(i)}_{00}(0)|
0\ran_f = \lan\rho_{vac}^{mix}\ran \eta_{00} ~.
 \eea

We observe that within the above  QFT formalism for neutrino
mixing
 we have \bea {}_f\lan 0 |T^{(i)}_{00}| 0\ran_f={}_f\lan
0(t) |T^{(i)}_{00}| 0(t)\ran_f \eea for any t. We then obtain
\bea {}_f\lan 0 |\sum_{i} T^{(i)}_{00}(0)| 0\ran_f =
\sum_{i,r}\int d^{3}{\bf k} \, \omega_{k,i}\lf({}_f\lan 0
|\al_{{\bf k},i}^{r\dag} \al_{{\bf k},i}^{r}| 0\ran_f + {}_f\lan 0
|\beta_{{\bf k},i}^{r\dag} \beta_{{\bf k},i}^{r}| 0\ran_f \ri) .
\eea
For simplicity we restrict ourselves to the two flavor mixing and
we use Dirac neutrino fields. Since \cite{BV95}
\bea\label{con} {}_f\lan 0 |\al_{{\bf k},i}^{r\dag} \al_{{\bf
k},i}^{r}| 0\ran_f = {}_f\lan 0 |\beta_{{\bf
k},i}^{r\dag}\beta_{{\bf k},i}^{r}| 0\ran_f\, = \,\sin^{2} \theta
|V_{\bf k}|^{2}, \eea
we get
\bea\label{aspT} {}_f\lan 0 |\sum_{i} T^{(i)}_{00}(0)| 0\ran_f
=\,8\sin^{2}\theta \int d^{3}{\bf
k}\lf(\omega_{k,1}+\omega_{k,2}\ri) |V_{\bf k}|^{2}
=\lan\rho_{vac}^{mix}\ran \eta_{00}, \eea
i.e.
\bea\label{cc} \lan\rho_{vac}^{mix}\ran = 32 \pi^{2}\sin^{2}\theta
\int_{0}^{K} dk \, k^{2}(\omega_{k,1}+\omega_{k,2}) |V_{\bf
k}|^{2} , \eea
where the cut-off $K$ has been introduced. Eq.(\ref{cc}) is our
result: it shows that the cosmological constant gets a non-zero
contribution induced from the neutrino mixing \cite{cosmolog2003}.
Notice that such a contribution is indeed zero in the no-mixing
limit when the mixing angle $\theta = 0$ and/or $m_{1} = m_{2}$.
It is to be remarked that the contribution is zero also in the
limit of $V_{\bf k} \rightarrow 0$, namely in the limit of the
traditional phenomenological mixing treatment.

It is  interesting to note that, for high momenta, the function
$|V_{\bf k}|^{2}$ produces a drastic decrease in the degree of
divergency of the above integral, in comparison with the case of a
free field. Thus, if for example we chose $K\gg \sqrt{m_1 m_2}$,
we obtain:
\bea\label{cc2} \lan\rho_{vac}^{mix}\ran \propto
\sin^2\theta\,(m_2 -m_1)^2 \, K^2 ,
 \eea
whereas the usual zero-point energy contribution would be going
like $K^4$.

Of course, we are not in a position to make our result independent
on the cut-off choice. However, this was not our goal. What we
have shown is that a non - zero contribution to the value of the
cosmological constant may come from the mixing of the neutrinos.
We have not solved the cosmological constant problem. Although it
might be unsatisfactory from a general theoretical point of view,
we may try to estimate the neutrino mixing contribution by making
our choice for the cut-off. Since we are dealing with neutrino
mixing, at a first trial it might be reasonable to chose the
cut-off proportional to the natural scale we have in the mixing
phenomenon, namely $\textbf{k}^{2}_{0}\simeq m_{1} m_{2}$
 \cite{BV95}.

With such a choice, using $K\sim k_{0}$, $m_{1}=7 \times
10^{-3}eV$, $m_{2}=5 \times 10^{-2}eV$, $k_{0}=10^{-3}eV$ and
 $\sin^{2}\theta\simeq 1$ in Eq.(\ref{cc}), we obtain
\bea \lan\rho_{vac}^{mix}\ran =1.3 \times 10^{-47}GeV^{4}\eea

Using Eq.(\ref{ed}), we have agreement with the upper bound given
in Section 6.1:
\bea \Lambda \sim 10^{-56}cm^{-2} , \eea
Another possible choice is to use the electro-weak scale cut-off:
$K\approx 100 GeV$. We then have \bea \lan\rho_{vac}^{mix}\ran
=1.5 \times 10^{-15}GeV^{4}\eea and \bea \Lambda \sim
10^{-24}cm^{-2} , \eea which is, however, beyond the accepted
upper bound.

\section{Conclusions}
In this chapter we have shown that the neutrino mixing can give
rise to a non zero contribution to the cosmological constant
\cite{cosmolog2003}. We have shown that this contribution is of a
different nature with respect to that given by the zero-point
energy of free fields and we estimated it by using the cut--off
given by the natural scale of the neutrino mixing phenomenon. The
different origin of the mixing contribution also manifests in the
different ultraviolet divergency order (quadratic rather than
quartic, see Eq.(\ref{cc2})).
 The obtained value is consistent
with the accepted upper bound for the value of $\Lambda$. On the
contrary, by using as a cut off the one related with the
electroweak scale, the $\Lambda$ value is greater than such upper
bound.

It is worth to stress once more that the origin of the present
contribution is completely different from that of the ordinary
contribution to the vacuum zero energy of a massive spinor field.
As we have shown, the effect we find is not originated from a
radiative correction at some perturbative order \cite{Col-Weinb}.
Our effect is exact at any order. It comes from the property of
QFT of being endowed with infinitely many representations of the
canonical (anti-)commutation relations in the infinite volume
limit. Therefore, the new result we find is that it is the {\it
mixing} phenomenon which provides such a vacuum energy
contribution, and this is so since the field mixing involves
unitary inequivalent representations. Indeed, as
Eqs.(\ref{con})-(\ref{cc}) show the contribution vanishes as
$V_{\bf k} \rightarrow 0$, namely in the quantum mechanical limit
where the representations of the (anti-)commutation relations are
all each other unitarily equivalent. Our result thus discloses a
new possible mechanism contributing to the cosmological constant
value.

As a final consideration, we observe that this effect could also
be exploited in the issue of dark energy without introducing
exotic fields like quintessence. In fact, neutrinos constitute a
cosmic background of unclustered components whose mixing and
oscillations could drive the observed accelerated expansion.

\newpage
\mbox{ }

\begin{center}

\chapter*{PART 2. BOSON MIXING}
\addcontentsline{toc}{chapter}{PART 2. BOSON MIXING} \noindent

\end{center}
\newpage
\mbox{ }
\newpage
\vspace{.6in}

\bigskip
\bigskip
\chapter{Boson mixing in Quantum Mechanics}
 \vspace{.4in}

In this chapter we present the theoretical model describing the
meson mixing and oscillations in Quantum Mechanics. In particular,
the boson oscillations will be treated by using the Gell-Mann and
Pais formalism.

\section{Usual representation of boson oscillations: the Gell-Mann
Pais model}

Quark mixing and meson mixing are widely accepted and verified
\cite{mesons}. However, many features of the physics of mixing are
still obscure, for example the issue related to its origin in the
context of Standard Model and the related problem of the
generation of masses.

The problem of the boson mixing is known since $1955$ when
Gell-Mann and Pais predicted the existence of two neutral kaons
\cite{Gell-Mann} : $K^{0}$ of strangeness $S=1$ and $\bar{K}^{0}$
of strangeness $S=-1$. These are particle and antiparticle, and
are connected by the process of charge conjugation, which involves
a reversal of values of $I_{3}$ and a change of strangeness
$\Delta S=2$. Strong interactions conserve $I_{3}$ and $S$, so
that as far as production is concerned, the separate neutral-kaon
eigenstates are $K^{0}-\bar{K}^{0}$.

Now suppose $K^{0}$ and $\bar{K}^{0}$ particles propagate through
empty space. Since both are neutral, both can decay to pions by
weak interaction, with $\Delta S=1$. Thus, mixing can occur via
(virtual) intermediate pion states:

\bea \non K^{0}\leftrightarrows 2\pi \leftrightarrows \bar{K}^{0} \\
K^{0}\leftrightarrows 3\pi \leftrightarrows \bar{K}^{0} \eea

These transitions are $\Delta S=2$ and thus second order weak
interactions. Although extremely weak, this implies that if one
has a pure $K^{0}$-state at $t=0$, at any later time $t>0$ one
will have a superposition of both $K^{0}$ and $\bar{K}^{0}$, so
that the state can be written \bea |K(t)\rangle =\alpha
(t)|K^{0}\rangle +\beta (t)|\bar{K}^{0}\rangle.\eea

The phenomena has been explained by realizing that what we observe
is the mixture of two mass and mean life eigenstates $K_{S}$ and
$K_{L}$ expressed by

\bea\label{Ks} \non |K_{S}\rangle &=&
\frac{1}{\sqrt{2(1+|\varepsilon|^{2})}}
[(1+\varepsilon)|K^{0}\rangle
+(1-\varepsilon)|\bar{K}^{0}\rangle]=\frac{1}{\sqrt{(1+|\varepsilon|^{2})}}
(|K_{1}\rangle+\varepsilon|K_{2}\rangle), \\\non|K_{L}\rangle &=&
\frac{1}{\sqrt{2(1+|\varepsilon|^{2})}}[(1+\varepsilon)|K^{0}\rangle
-(1-\varepsilon)|\bar{K}^{0}\rangle]=\frac{1}{\sqrt{(1+|\varepsilon|^{2})}}
(|K_{2}\rangle+\varepsilon|K_{1}\rangle),\\\eea where
$\varepsilon$ is a small, complex parameter responsible for $CP$
symmetry breaking and $K_{1}$, $K_{2}$ are $CP$ eigenstates: \bea
\non |K_{1}\rangle &=&\frac{1}{\sqrt{2}}(|K^{0}\rangle
+|\bar{K}^{0}\rangle)
\qquad \qquad CP|K_{1}\rangle=|K_{1}\rangle, \\
|K_{2}\rangle &=&\frac{1}{\sqrt{2}}(|K^{0}\rangle -|\bar{K}
^{0}\rangle )\qquad \qquad CP|K_{2}\rangle=-|K_{2}\rangle \eea
with the convention \bea \non CP|K^{0}\rangle &=&|\bar{K}^{0}\rangle\\
CP|\bar{K}^{0}\rangle &=&|K^{0}\rangle. \eea

Unlike $K^{0}$ and $\bar{K}^{0}$, distinguished by their mode of
production (the $K^{0}$ can be produced by nonstrange particles in
association with a hyperon and $\bar{K}^{0}$ can be produced only
in association with a kaon or antihyperon, of strangeness $S=1$),
$K_{S}$ and $K_{L}$ are distinguished by their mode of decay.
Consider $2\pi$ and $3\pi$ decay modes.

Since pions have no spin, angular momentum conservation requires
that the two pions resulting from $K^{0}\longrightarrow 2\pi$
decay carry relative angular momentum equal to the spin of the
kaon. A neutral $2\pi $ state with specified angular momentum $l$
is an eigenstate of $C$ with eigenvalue $C=(-1)^{l}$, since the
action of $C$ is just to exchange the two pions.

Being $K^{0}$ a spinless particle, Gell-Mann e Pais concluded that
only the component $K_{1}$, $CP$ eigenstate with eigenvalue $1$,
would be capable of $2\pi$ decay; while $K_{2}$ would only decay
in $3\pi$ state with $CP=-1$. But, in $1964$ Christenson, Cronin,
Fitch and Turlay \cite{christenson} demonstrate that the $K_{2}$
state could also decay to $\pi^{+}\pi^{-}$ with a branching ratio
of order $10^{-3}$. Then there is a $CP$ violation in the $K^{0}$
decay and the physical component of $K^{0}$ are $K_{S}$ (short
lived component) and $K_{L}$ (long lived component), where $K_{S}$
consists principally of a $CP=+1$ amplitude, but with a little
$CP=-1$, and $K_{L}$ vice versa. Experimentally the mean lives of
$K_{S}$ and $K_{L}$ are $\tau_{S}=(0.8934\mp0.0008)\times
10^{-10}\sec$, $\tau_{L}=(5.17\mp0.04)\times 10^{-8}\sec$
respectively.

An important phenomenon is the $K^{0}$ regeneration
\cite{Perkins}. Suppose we produce a pure $K^{0}$ beam and let it
travel in vacuo for the order of $100K_{S}$ mean lives, so that
all the $K_{S}$ component has decayed and we are left with $K_{L}$
only. Now let the $K_{L}$ beam traverse a slab of material and
interact. Immediately, the strong interactions will pick out the
strangeness $S=+1$ and $S=-1$ components of the beam.

Thus, of the original $K^{0}$ beam intensity, about $50\%$
  has disappeared by $K_{S}$ decay. The remainder, $K_{L}$ upon
traversing a slab where its nuclear interaction can be observed,
should consist of $50\%$ $K^{0}$ and $50\%$ $\bar{K}^{0}$.

The ${K}^{0}$ and $\bar{K}^{0}$ must be absorbed differently;
${K}^{0}$ particles can only undergo elastic and charge exchange
scattering, while $\bar{K}^{0}$ particles can also undergo
strangeness exchange giving hyperons: \bea K^{0}+p\rightarrow
K^{+}+n \\ K^{0}+n\rightarrow K^{0}+n \eea and \bea
\bar{K}^{0}+p\rightarrow \left\{
\begin{array}{c}
\Lambda ^{0} + \pi ^{+} \\
\Sigma ^{+} + \pi ^{0} \\
p + K^{+}+K^{-}
\end{array}
\right. \eea

\bea \bar{K}^{0}+n\rightarrow \Lambda ^{0}+\pi ^{0}. \eea

 With more strong channels open, the $\bar{K}^{0}$ is therefore
 absorbed more strongly than
${K}^{0}$. After emerging from the slab, we shall therefore have a
${K}^{0}$ amplitude $f|{K}^{0}\rangle$ and a $\bar{K}^{0}$
amplitude $\bar{f}|\bar{K}^{0}\rangle$, where $\bar{f}<f<1$. If we
neglect the $CP$ symmetry breaking, the emergent beam will be \bea
\non \frac{1}{2}(f|{K}^{0}\rangle -
\bar{f}|\bar{K}^{0}\rangle)&=&\frac{f+\bar{f}}{2\sqrt{2}}(|{K}^{0}\rangle
-|\bar{K}^{0}\rangle)+ \frac{f-\bar{f}}{2\sqrt{2}}(|{K}^{0}\rangle
+|\bar{K}^{0}\rangle) \\ &=& \frac{1}{2}(f +
\bar{f})|{K}_{L}\rangle+\frac{1}{2}(f - \bar{f})|{K}_{S}\rangle.
\eea

Since $f\neq \bar{f}$, it follows that some of the ${K}_{S}$ state
has been regenerated.

The main prediction of the particle mixture hypothesis is the
possibility of observing the $K_{S}- K_{L}$ interference. If we
make the assumption of exponential decay, each component will have
a time dependence of the form $e^{-(\Gamma_{i} /2\hslash
+iE_{i}/\hslash )t}$ where $E_{i}$ is the total energy of the
particle $i = L, S$ and $\Gamma_{i}=\hslash /\tau _{i}$ is the
width of the state, $\tau _{i}$ being the mean life in the frame
in which the energy $E_{i}$ is defined.

Set $\hslash =c=1$, and measure all times in the rest frame, so
that $\tau _{i}$ is the proper lifetime and $E_{i}=m_{i}$, the
particle rest mass, then the time dependence became
$e^{-(\Gamma_{i} /2+im_{i})t}$.

 Suppose that we produce $K^{0}$ at $t=0$, from
Eqs.(\ref{Ks}), the $K^{0}$ state at $t=0$ is a coherent
superposition of $K_{S}$ and $K_{L}$ \bea |K^{0}\rangle
=\sqrt{\frac{1+|\varepsilon|^{2}}{2(1+\varepsilon)^{2}}}(|K_{S}\rangle
+|K_{L}\rangle).\eea

After time $t$, as $|K_{S}\rangle$ and $|K_{L}\rangle$ states are
definite mass eigenstates, we have \bea |K^{0}(t)\rangle
=\sqrt{\frac{1+|\varepsilon|^{2}}{2(1+\varepsilon)^{2}}}\left[e^{-(\Gamma
_{S}/2 + i m_{S})t} |K_{S}\rangle +e^{-(\Gamma _{L}/2 + i
m_{L})t}|K_{L}\rangle \right]\eea and then after a time $t$ for
free decay in vacuo, the $K^{0}$ fraction of the beam will be \bea
\non P(K^{0},t)&=& \left| \langle K^{0}(t)|K^{0}\rangle\right|
^{2}=\frac{1}{4}\left| e^{-(\Gamma _{S}/2+im_{S})t}+e^{-(\Gamma
_{L}/2+im_{L})t}\right| ^{2} \\ \non &=&\frac{1}{4}\left[
e^{-\Gamma _{S}t}+e^{-\Gamma
_{L}t}+e^{i(m_{L}-m_{S})t}e^{-(\Gamma_{S}+\Gamma
_{L})t/2}+e^{-i(m_{L}-m_{S})t}e^{-(\Gamma _{S}+\Gamma
_{L})t/2}\right] \\  &=& \frac{1}{4}\left[ e^{-\Gamma
_{S}t}+e^{-\Gamma _{L}t}+2e^{-(\Gamma _{S}+\Gamma _{L})t/2}\cos
(\Delta m t)\right], \eea where $\Delta m =
m_{L}-m_{S}=(3.489\mp0.009)\times10^{-12}MeV$.

Similarly, the $\bar{K}^{0}$ fraction of the beam will be \bea
P(\bar{K}^{0},t)=\frac{1}{4}\left[ e^{-\Gamma _{S}t}+e^{-\Gamma
_{L}t}-2e^{-(\Gamma _{S}+\Gamma _{L})t/2}\cos (\Delta m
t)\right].\eea

The neutral $K^{0}-\bar{K}^{0}$ boson system is not the only one
where the quantum mechanical mass mixing can be considered. We can
expect to observe the same phenomenon in other neutral boson
systems: $D^{0}-\bar{D}^{0}$, $B^{0}-\bar{B}^{0}$ and
$\eta-\eta'$. Generally, flavor oscillations of particles can
occur when states produced and detected in a given experiment, are
superpositions of two or more eigenstates with different masses.
The mesons oscillations has been used to place stringent
constraints on physics beyond the Standard Model.

\bigskip
\bigskip
\chapter{Quantum field theory of boson mixing}

\vspace{.4in}

We consider the quantum field theoretical formulation of boson
field mixing and obtain the exact oscillation formula. This
formula does not depend on arbitrary mass parameters. We show that
the space for the mixed field states is unitarily inequivalent to
the state space where the unmixed field operators are defined. We
also study the structure of the currents and charges for the mixed
fields.

\vspace{8mm}

\section{Introduction}

%Also the violation of
%CP symmetry is intimately connected with the structure of the
%Cabibbo--Kobayashi--Maskawa mixing matrix for quarks.

A rich non--perturbative vacuum structure has been discovered to
be associated with the mixing of fermion fields in the context of
Quantum Field Theory \cite{BV95,hannabus}. The careful study of
such a structure \cite{fujii1} has led to the determination of the
exact QFT formula for neutrino oscillations
\cite{BHV98,Blasone:1999jb}, exhibiting new features with respect
to the usual quantum mechanical Pontecorvo formula
\cite{Pontecorvo}. Actually, it turns out
\cite{lathuile,BCRV01,binger,CJMV2003} that the non--trivial
nature of the mixing transformations manifests itself also in the
case of the mixing of boson fields. Of course, in this case the
condensate structure for the ``flavor'' vacuum is very much
different from the fermion case and a careful analysis is
necessary in order to understand which phenomenological
consequences are to be expected for the oscillations of mixed
bosons.

In this chapter, we perform this analysis first at a formal level
and then we study the oscillations of mixed mesons (charged and
neutral systems). We will treat these particles as stable ones, an
approximation which however does not affect the general validity
of our results. In the framework of the QFT analysis of Refs.
\cite{lathuile,BV95}, a study of the meson mixing and oscillations
has been carried out in Ref.\cite{binger}, where modifications to
the usual oscillation formulas, connected with the vacuum
structure, have been presented. However, the results of Ref.
\cite{binger} can be improved in many respects \cite{BCRV01} and
in the present chapter we show that the oscillation formula there
obtained has to be actually replaced with the exact one here
presented.

In Section 8.2 we study the quantum field theory of two mixed
spin-zero charged boson fields. In Section 8.3 we analyze the
structure of currents for mixed fields and we derive the exact
oscillation formula for complex and neutral fields in Sections 8.4
and 8.5, respectively. Section 8.6 is devoted to conclusions. Some
mathematical derivations are given in the Appendix K.

%%%%%%%%%%%%%%%%%%%%%%%%%%%%%%%%%%%%%%%%%%%%%%%%%%%%%%%%%%%%%
\section{Mixing of boson fields in QFT}
%%%%%%%%%%%%%%%%%%%%%%%%%%%%%%%%%%%%%%%%%%%%%%%%%%%%%%%%%%%%%

The observed boson oscillations always involve particles with zero
electrical charge. In the case of $K^0-{\bar K^0}$, $B^0-{\bar
B^0}$, $D^0-{\bar D^0}$, what oscillate are some other quantum
numbers such as the strangeness and the isospin. Therefore, in the
study of boson mixing, for these particles, we can consider
\cite{lurie} complex fields. The charge in question is some
``flavor charge'' (e.g. the strangeness) and thus the complex
fields are ``flavor charged'' fields, referred to as ``flavor
fields'' for simplicity.

We define the mixing relations as:
\bea\non &&\phi_{A}(x) = \phi_{1}(x) \; \cos\te + \phi_{2}(x) \;
\sin\te
\\[2mm] \lab{bosmix12}
&&\phi_{B}(x) =- \phi_{1}(x) \; \sin\te + \phi_{2}(x)\; \cos\te
\eea
where generically we denote the mixed fields with suffixes $A$ and
$B$.
Let the fields $\phi_{i}(x)$, $i=1,2$, be free complex fields with
definite masses. Their conjugate momenta are
$\pi_{i}(x)=\pa_{0}\phi_{i}^{\dag}(x)$ and the commutation
relations are the usual ones:
\bea &&\lf[\phi_{i}(x),\pi_{j}(y)\ri]_{t=t'}=
[\phi_{i}^{\dag}(x),\pi_{j}^{\dag}(y)]_{t=t'}=i\de^{3} ({\bf
x}-{\bf y}) \, \de_{ij} , \qquad i,j=1,2\,. \lab{2.50} \eea
with the other equal--time commutators vanishing. The Fourier
expansions of fields and momenta are:
\bea\lab{bosfield} \phi_{i}(x) = \int \frac{d^3 {\bf
k}}{(2\pi)^{\frac{3}{2}}} \frac{1}{\sqrt{2\om_{k,i}}} \lf( a_{{\bf
k},i}\, e^{-i \om_{k,i} t} + b^{\dag }_{-{\bf k},i}\, e^{i
\om_{k,i} t}  \ri) e^{i {\bf k}\cdot {\bf x}} \eea
\bea\lab{bosmom} \pi_{i}(x) = i\,\int \frac{d^3 {\bf
k}}{(2\pi)^{\frac{3}{2}}} \sqrt{\frac{\om_{k,i}}{2}} \lf( a^{\dag
}_{{\bf k},i}\, e^{i \om_{k,i} t} - b_{-{\bf k},i}\, e^{-i
\om_{k,i} t} \ri) e^{i {\bf k}\cdot {\bf x}}\,, \eea
where $\om_{k,i}=\sqrt{{\bf k}^2 + m_i^2}$ and $[a_{{\bf
k},i},a_{{\bf p},j}^{\dag} ]= [b_{{\bf k},i},b_{{\bf p},j}^{\dag}
]=\de^{3}({\bf k}-{\bf p}) \de_{ij}\, ,$ with $i,j=1,2\,$ and the
other commutators vanishing. We will consider stable particles,
which will not affect the general validity of our results.

We now proceed in a similar way to what has been done \cite{BV95}
for fermions and recast Eqs.(\ref{bosmix12}) into the form
\cite{BCRV01}:
\bea \lab{bosmix12a}
\phi_{A}(x) = G^{-1}_\te(t)\; \phi_{1}(x)\; G_\te(t) \\[2mm]
\lab{bosmix12b} \phi_{B}(x) = G^{-1}_\te(t)\; \phi_{2}(x)\;
G_\te(t) \eea
and similar ones for $\pi_{A}(x)$, $\pi_B(x)$. $G_\te(t)$  denotes
the operator which implements the mixing transformations
(\ref{bosmix12}):
\bea\lab{bosgen}\non G_\te(t) = exp\lf[-i\;\te \int d^{3}{\bf x}
\lf(\pi_{1}(x)\phi_{2}(x) - \phi_{1}^{\dag}(x)\pi_{2}^{\dag}(x)
-\pi_{2}(x)\phi_{1}(x) +
\phi_{2}^{\dag}(x)\pi_{1}^{\dag}(x)\ri)\ri] ,\\ \eea
which is (at finite volume) a unitary operator:
$G^{-1}_\te(t)=G_{-\te}(t)=G^{\dag}_\te(t)$. The generator of the
mixing transformation in the exponent of $G_{\te}(t)$  can also be
written as
\bea\lab{bosgenS} G_\te(t) = \exp[\te(S_{+}(t) - S_{-}(t))] ~.
\eea
The operators
\bea\lab{Spt} S_+(t)= S_{-}^{\dag}(t) \equiv -i\;\int d^3 {\bf x}
\; (\pi_{1}(x)\phi_{2}(x) - \phi_{1}^{\dag}(x)\pi_{2}^{\dag}(x))
\,, \eea
together with
\bea\lab{2.56} S_{3} \equiv \frac{-i}{2} \int d^3 {\bf x}
\lf(\pi_{1}(x)\phi_{1}(x) - \phi_{1}^{\dag}(x)\pi_{1}^{\dag}(x) -
\pi_{2}(x)\phi_{2}(x) +\phi_{2}^{\dag}(x) \pi_{2}^{\dag}(x) \ri)
\eea
\bea\lab{S0} S_{0} =\frac{Q}{2}\equiv \frac{-i}{2} \int d^3 {\bf
x} \lf( \pi_{1}(x)\phi_{1}(x) -\phi_{1}^{\dag}(x)\pi_{1}^{\dag}(x)
+\pi_{2}(x)\phi_{2}(x) - \phi_{2}^{\dag}(x)\pi_{2}^{\dag}(x) \ri)
, \eea
close the $su(2)$ algebra (at each time $t$): \bea[S_{+}(t) ,
S_{-}(t)]=2S_{3},\quad  [S_{3} , S_{\pm}(t) ] = \pm S_{\pm}(t)
,\quad [S_{0} , S_{3}]= [S_{0} , S_{\pm}(t) ] = 0.\eea Note that
$S_{3}$ and $S_{0}$ are time independent. It is useful to write
down explicitly the expansions of the above generators in terms of
annihilation and creation operators:
\bea\non\lab{bosSp} S_{+}(t)=\int d^3 {\bf k} \lf( U^*_{{\bf
k}}(t) \, a_{{\bf k},1}^{\dag}a_{{\bf k},2} - V_{{\bf k}}^{*}(t)
\, b_{-{\bf k},1}a_{{\bf k},2} + V_{{\bf k}}(t) \, a_{{\bf
k},1}^{\dag}b_{-{\bf k},2}^{\dag} - U_{{\bf k}}(t) \, b_{-{\bf
k},1}b_{-{\bf k},2}^{\dag} \ri),\\ \eea
\bea\non\lab{bosSm} S_{-}(t)=\int d^3 {\bf k} \lf( U_{{\bf k}}(t)
\, a_{{\bf k},2}^{\dag}a_{{\bf k},1} - V_{{\bf k}}(t) \, a_{{\bf
k},2}^{\dag}b_{-{\bf k},1}^{\dag} + V_{{\bf k}}^{*}(t) \, b_{-{\bf
k},2}a_{{\bf k},1} - U^*_{{\bf k}}(t) \, b_{-{\bf k},2}b_{-{\bf
k},1}^{\dag} \ri),\\ \eea
\bea\lab{bosS3} S_{3}=\frac{1}{2} \int d^3 {\bf k} \lf( \, a_{{\bf
k},1}^{\dag}a_{{\bf k},1} \,- \, b_{-{\bf k},1}^\dag b_{-{\bf
k},1}\, - \, a_{{\bf k},2}^{\dag}a_{{\bf k},2} \,+ \, b_{-{\bf
k},2}^\dag b_{-{\bf k},2} \ri), \eea
\bea\lab{bosS0} S_{0}=\frac{1}{2} \int d^3 {\bf k} \lf( \, a_{{\bf
k},1}^{\dag}a_{{\bf k},1} \,- \, b_{-{\bf k},1}^\dag b_{-{\bf
k},1}\, + \, a_{{\bf k},2}^{\dag}a_{{\bf k},2} \,- \, b_{-{\bf
k},2}^\dag b_{-{\bf k},2} \ri)\,. \eea

As for the case of the fermion mixing, the structure of  the
generator Eq.(\ref{bosgen}) is recognized to be the one of a
rotation combined with a Bogoliubov transformation (see below
Eqs.(\ref{bosanna})-(\ref{bosannd})). Indeed, in the above
equations, the coefficients \bea U_{{\bf k}}(t)\equiv |U_{{\bf
k}}| \; e^{i(\om_{k,2}- \om_{k,1})t}~, \;\;\;\;\;\;  V_{{\bf
k}}(t)\equiv |V_{{\bf k}}| \; e^{i(\om_{k,1}+ \om_{k,2})t}\eea
appear to be the Bogoliubov coefficients. They are defined as
\bea &&|U_{{\bf k}}|\equiv \frac{1}{2} \lf(
\sqrt{\frac{\om_{k,1}}{\om_{k,2}}} +
\sqrt{\frac{\om_{k,2}}{\om_{k,1}}}\ri) ~, \;\;\;\;\;\; |V_{{\bf
k}}|\equiv  \frac{1}{2} \lf( \sqrt{\frac{\om_{k,1}}{\om_{k,2}}} -
\sqrt{\frac{\om_{k,2}}{\om_{k,1}}} \ri)  \eea
and satisfy the relation
\bea\lab{2.60} &&|U_{{\bf k}}|^{2}-|V_{{\bf k}}|^{2}=1\,, \eea
which is in fact to be expected in  the boson case (note the
difference with respect to the fermion case \cite{BV95}). We can
thus  put \bea |U_{{\bf k}}|\equiv \cosh \xi^{\bf k}_{1,2} ~,
\;\;\;\;\;\; |V_{{\bf k}}|\equiv \sinh \xi^{\bf k}_{1,2}, \quad
\text{with}\quad \xi^{\bf k}_{1,2}= \frac{1}{2} \ln
\lf(\frac{\om_{k,1}}{\om_{k,2}}\ri).\eea

We now consider the action of the generator of the mixing
transformations on the vacuum $|0 \ran_{1,2}$ for the fields
$\phi_{1,2}(x)$:  $a_{{\bf k},i}|0 \ran_{1,2} = 0, ~ i=1,2$ . The
generator induces an $SU(2)$ coherent state structure on such
state \cite{Per}:
\bea\label{2.61} |0(\te, t) \ran_\AB \equiv G^{-1}_\te(t)\; |0
\ran_{1,2}\,. \eea

From now on we will refer to the state $|0(\te, t) \ran_\AB$ as to
the ``flavor'' vacuum for bosons. The suffixes $A$ and $B$ label
the flavor charge content of the state. We have \bea\,_\AB\lan
0(\te,t)|0 (\te,t)\ran_\AB \,= \,1.\eea

In the following, we will consider the Hilbert space for flavor
fields at a given time $t$, say $t=0$, and it is useful to define
\bea|0 (t)\ran_\AB\equiv|0(\te,t)\ran_\AB \quad \text{and}\quad
|0\ran_\AB\equiv|0(\te,t=0)\ran_\AB\eea for future reference. A
crucial point is that the flavor and the mass vacua are orthogonal
in the infinite volume  limit \cite{BCRV01}. We indeed have (see
Appendix K):
\bea\label{2.61a} \,_{1,2}\lan 0|0 (t)\ran_\AB \,= \,
\prod\limits_{\bf k} \, _{1,2}\lan 0| G_{{\bf k},\te}^{-1}(t)
|0\ran_{1,2}\,=\, \prod\limits_{\bf k} \,f_0^{\bf k}(\te) ~,
~~~\text{for~any}~t, \eea
where we have used $G^{-1}_\te(t) = \prod\limits_{\bf k} \,
G_{{\bf k},\te}^{-1}(t)$ (see Eqs.(\ref{bosgenS}), (\ref{bosSp})
and (\ref{bosSm})). In the infinite volume limit, we obtain
\bea\label{2.61b} \lim\limits_{V\rar \infty}\,_{1,2}\lan 0|0
(t)\ran_\AB = \lim\limits_{V\rar \infty}\, e^{\frac{V}{(2\pi)^3}
 \int d^3 {\bf k} \, \ln f_0^{\bf k}(\te) } \, = \, 0 ~,
~~~\text{for~any}~t. \eea

From the Appendix K, Eq.(\ref{A74}), we see that $\ln f_0^{\bf
k}(\te)$ is indeed negative for any values of ${\bf k}$, $\te$ and
$m_1, m_2$ (note that $0\le \te \le \pi/4$). We also observe that
the orthogonality disappears when   $\te=0$ and/or $m_1=m_2$,
consistently with the fact that in both cases  there is no mixing.
These features are similar to the case of fermion  mixing
\cite{BV95}: the orthogonality is essentially due to the infinite
number of degrees of freedom \cite{Itz,Um1}.
%It is important to note that the orthogonality is
%also between the states $|0 (\te, t)\ran_\AB $ and $|0 (\te',
%t')\ran_\AB$, for $\te \neq \te'$ and/or $t\neq t'$
%(see Appendix B).

%
We can define annihilation operators  for the vacuum $|0(t)
\ran_\AB$ as \bea a_{{\bf k},A}(\te ,t) &\equiv& G^{-1}_\te(t) \;
a_{{\bf k},1}\;G_\te(t),\\a_{{\bf k},B}(\te ,t) &\equiv&
G^{-1}_\te(t) \; a_{{\bf k},2}\;G_\te(t),\\b_{-{\bf k},A}(\te ,t)
&\equiv& G^{-1}_\te(t) \; b_{-{\bf k},1}\;G_\te(t),\\b_{-{\bf
k},B}(\te ,t) &\equiv& G^{-1}_\te(t) \; b_{-{\bf k},2}\;G_\te(t),
\eea with $a_{{\bf k},A}(\te ,t) |0(t) \ran_\AB = 0$. For
simplicity we will use the notation $ a_{{\bf k},A}(t) \equiv
a_{{\bf k},A}(\te ,t)$. Explicitly, we have:
\bea \label{bosanna} a_{{\bf k},A}(t)&=&\cos\te\;a_{{\bf
k},1}\;+\;\sin\te\;\lf( U^*_{{\bf k}}(t)\; a_{{\bf k},2}\;+\;
V_{{\bf k}}(t)\; b^{\dag}_{-{\bf k},2}\ri)\, ,
\\
a_{{\bf k},B}(t)&=&\cos\te\;a_{{\bf k},2}\;-\;\sin\te\;\lf(
U_{{\bf k}}(t)\; a_{{\bf k},1}\;- \; V_{{\bf k}}(t)\;
b^{\dag}_{-{\bf k},1}\ri)\, ,
\\
b_{-{\bf k},A}(t)&=&\cos\te\;b_{-{\bf k},1}\;+\;\sin\te\;\lf(
U^*_{{\bf k}}(t)\; b_{-{\bf k},2}\;+ \; V_{{\bf k}}(t)\;
a^{\dag}_{{\bf k},2}\ri)\, ,
\\ \label{bosannd}
b_{-{\bf k},B}(t)&=&\cos\te\;b_{-{\bf k},2}\;-\;\sin\te\;\lf(
U_{{\bf k}}(t)\; b_{-{\bf k},1}\;- \; V_{{\bf k}}(t)\;
a^{\dag}_{{\bf k},1}\ri) ~. \eea
These operators satisfy the canonical commutation relations (at
equal times). In their expressions the Bogoliubov transformation
part is evidently characterized by the terms with the $U$ and $V$
coefficients. The condensation density of the flavor vacuum is
given for any $t$ by
\bea\label{2.63} {}_\AB\lan 0(t)| a_{{\bf k},i}^{\dag} a_{{\bf
k},i} |0(t)\ran_\AB= {}_\AB\lan 0(t)| b_{-{\bf k},i}^{\dag}
b_{-{\bf k},i} |0(t)\ran_\AB=\sin^{2}\te\;  |V_{{\bf k}}|^{2} ,
  \eea with $i=1,2.$
%

%%%%%%%%%%%%%%%%%%%%%%%%%%%
\subsection{Arbitrary mass parameterization}
%%%%%%%%%%%%%%%%%%%%%%%%%%%

Above we have expanded the mixed fields $\phi_{A,B}$ in the same
basis as the free fields $\phi_{1,2}$. However, as noticed in
 the case of fermion mixing \cite{fujii1}, this is not the most
general possibility. Indeed, one could as well expand the flavor
fields in a basis of fields with arbitrary masses. Of course,
these arbitrary mass parameters should not appear in the
physically observable quantities. Thus, as a check for the
validity of the oscillation formula we are going to derive in
Section 8.4, it is  important to consider this generalization. Let
us first rewrite the free fields $\phi_{1,2}$ in the form
\cite{BCRV01}
\bea\label{gener1} \phi_{i}(x) &=& \int \frac{d^3 {\bf
k}}{(2\pi)^{\frac{3}{2}}} \lf(u^\phi_{{\bf k},i}(t)\, a_{{\bf
k},i}\, + \, v^\phi_{-{\bf k},i}(t) b^{\dag }_{-{\bf k},i} \ri)
e^{i {\bf k}\cdot {\bf x}} \, ,
\\[3mm] \label{gener2}
\pi_{i}(x) & =& i\,\int \frac{d^3 {\bf k}}{(2\pi)^{\frac{3}{2}}}
\lf( u^\pi_{{\bf k},i}(t)\,a^{\dag }_{{\bf k},i}\: -
\,v^\pi_{-{\bf k},i}(t)\, b_{-{\bf k},i} \ri)e^{i {\bf k}\cdot
{\bf x}} \quad , \qquad i=1,2, \eea
where we have introduced the notation
\bea &&u^\phi_{{\bf k},i}(t)\, \equiv \,\frac{1}{\sqrt{2
\om_{k,i}}} e^{-i \om_{k,i}t} \quad ,\quad v^\phi_{-{\bf
k},i}(t)\,\equiv \, \frac{1}{\sqrt{2 \om_{k,i}}}e^{i \om_{k,i}t}
~,
\\ [2mm]
&&u^\pi_{{\bf k},i}(t)\,\equiv\,\sqrt{\frac{\om_{k,i}}{2}} e^{i
\om_{k,i}t}\qquad, \quad v^\pi_{-{\bf k},i}(t)\,\equiv \,
\sqrt{\frac{\om_{k,i}}{2}} e^{-i \om_{k,i}t} , \qquad i=1,2~. \eea

We now define
\bea\label{rho} \rho^{\bf k *}_{\al\bt}(t)& \equiv& u^\pi_{{\bf
k},\al}(t) u^\phi_{{\bf k},\bt}(t) \,+\, v^\phi_{-{\bf
k},\al}(t)v^\pi_{-{\bf k},\bt}(t) \,=\,
%\frac{1}{2}\lf[\sqrt{\frac{\om_{k,\al}}{\om_{k,\bt}}}
%+\sqrt{\frac{\om_{k,\bt}}{\om_{k,\al}}} \,\ri]
e^{i (\om_{k,\al} - \om_{k,\bt}) t}\,\cosh\,\xi_{\al,\bt}^{\bf k}
~,
\\ [2mm]
\la^{\bf k *}_{\al\bt}(t)& \equiv& v^\pi_{-{\bf k},\al}(t)
u^\phi_{{\bf k},\bt}(t) \,-\, u^\phi_{{\bf k},\al}(t)v^\pi_{-{\bf
k},\bt}(t)\,= \,
%\frac{1}{2}\lf[\sqrt{\frac{\om_{k,\al}}{\om_{k,\bt}}}
%-\sqrt{\frac{\om_{k,\bt}}{\om_{k,\al}}} \,\ri]
e^{-i (\om_{k,\al} + \om_{k,\bt}) t} \, \sinh\,\xi_{\al,\bt}^{\bf
k}~,
\\ [2mm]
\xi_{\al,\bt}^{\bf k}&\equiv & \frac{1}{2}
\ln\lf(\frac{\om_{k,\al}}{\om_{k,\bt}}\ri) \qquad , \qquad \al,\bt
= 1,2,A,B , \eea
where $\om_{k,\al}\equiv\sqrt{k^2 + \mu_\al^2}$. We denote with
$\mu_A$ and $\mu_B$ the arbitrary mass parameters while
$\mu_1\equiv m_1$ and $\mu_2\equiv m_2$ are the  physical masses.
Note that $\rho^{\bf k }_{1 2}(t)=U_{\bf k}(t)$ and $\la^{\bf k
}_{1 2}(t)=V_{\bf k }(t)$. We can now write the expansion of the
flavor fields in the general form (we use a tilde to denote the
generalized ladder operators):
\bea\label{gener7} \phi_{\si}(x) = \int \frac{d^3 {\bf
k}}{(2\pi)^{\frac{3}{2}}} \lf(u^\phi_{{\bf k},\si}(t)\, {\wti
a}_{{\bf k},\si}(t) \, + \, v^\phi_{-{\bf k},\si}(t) \, {\wti
b}^{\dag }_{-{\bf k},\si}(t) \ri) e^{i {\bf k}\cdot {\bf x}}, \eea
with $ \si = A,B$, which is to be compared with the expansion in
the free field basis as given in Eqs.(\ref{bosmix12a}),
(\ref{bosmix12b}):
\bea\label{gener8} \phi_{\si}(x) = \int \frac{d^3 {\bf
k}}{(2\pi)^{\frac{3}{2}}} \lf(u^\phi_{{\bf k},i}(t)\, a_{{\bf
k},\si}(t) \, + \, v^\phi_{-{\bf k},i}(t) \,b^{\dag }_{-{\bf
k},\si}(t) \ri) e^{i {\bf k}\cdot {\bf x}}~, \eea
where $(\si,i)=(A,1),(B,2)$.

The relation between the two sets of flavor operators is given as
\cite{BCRV01}

\bea\non \label{bogol}\left(
\begin{array}{c}{\wti a}_{{\bf k},\sigma}(t)
\\ [2mm] {\wti b}^{\dag }_{-{\bf k},\sigma}(t)\end{array}
\right) & =& J^{-1}(t) \left(
\begin{array}{c} a_{{\bf k},\sigma}(t)
\\ [2mm] b^{\dag }_{-{\bf k},\sigma}(t)\end{array}
\right) J(t) \, =\,\left(\begin{array}{cc} \rho^{\bf k *}_{\si
i}(t) & \lambda^{\bf k}_{\sigma i}(t)
\\ [2mm]\lambda^{\bf k *}_{\sigma i}(t) &
\rho^{\bf k}_{\sigma i}(t)  \end{array} \right) \,\left(
\begin{array}{c}
a_{{\bf k},\sigma}(t)
\\ [2mm]
b^{\dag }_{-{\bf k},\sigma}(t) \end{array} \right),\\
\\ [2mm]
J(t)&=& \exp\left\{ \int d^3 {\bf k} \, \xi_{\sigma,i}^{\bf
k}\left[ a^{\dag}_{{\bf k},\sigma}(t)b^{\dag}_{{-\bf k},\sigma}(t)
- b_{{-\bf k},\sigma}(t)a_{{\bf k},\sigma}(t)\right]\right\}\, ,
\eea
with \bea \xi_{\sigma,i}^{\bf k}\equiv \frac{1}{2}
\ln\lf(\frac{\om_{k,\si}}{\om_{k,i}}\ri).\eea

For $\mu_A=m_1$ and $\mu_B=m_2$ one has $J = 1$. Note that the
transformation Eq.(\ref{bogol}) is in fact a Bogoliubov
transformation which leaves invariant the form $a^{\dag}_{{\bf
k},\sigma}(t)a_{{\bf k},\sigma}(t) - b^{\dag}_{{-\bf
k},\sigma}(t)b_{{-\bf k},\sigma}(t)$.

%%%%%%%%%%%%%%%%%%%%%%%%%%%%%%%%%%%%%%%%%%%%%%%%%%%%%%%%%%%%
\section{The currents for mixed boson fields}
%%%%%%%%%%%%%%%%%%%%%%%%%%%%%%%%%%%%%%%%%%%%%%%%%%%%%%%%%%%%

Before presenting the exact oscillation formula, let us
investigate in this Section the structure of currents and charges
for the mixed fields \cite{BCRV01}. This will enable us to
identify the relevant physical observables to look at for flavor
oscillations. Since we are here interested in vacuum oscillations,
in the following we neglect interaction terms and only consider
the free field Lagrangian for two charged scalar fields with a
mixed mass term:
\bea\label{boslagAB} {\cal L}(x)&=& \pa_\mu \Phi_f^\dag(x)\,
\pa^\mu \Phi_f (x)\, - \, \Phi_f^\dag(x)  M \Phi_f(x) \, , \eea
with $\Phi_f^T=(\phi_A,\phi_B)$, \bea M = \left(\begin{array}{cc}
m_A^2 & m_{AB}^2 \\ m_{AB}^2 & m_B^2\end{array} \right).\eea By
means of Eq.(\ref{bosmix12}),
\bea\label{bosmix}
\Phi_f(x) \, =\, \left(\begin{array}{cc} \cos \te & \sin \te \\[2mm]
-\sin \te & \cos \te \end{array} \right) \Phi_m (x) ~, \eea
${\cal L}$ becomes diagonal in the basis
$\Phi_m^T=(\phi_1,\phi_2)$:
\bea \label{boslag12} {\cal L}(x)&=&\pa_\mu \Phi_m^\dag(x)
\,\pa^\mu \Phi_m(x) \, - \, \Phi_m^\dag(x)  M_d \Phi_m (x) ~, \eea
where $M_d = diag(m_1^2,m_2^2)$ and \bea m_A^2 = m_1^2\cos^{2}\te
+ m_2^2 \sin^{2}\te\, \\ m_B^2 = m_1^2\sin^{2}\te + m_2^2
\cos^{2}\te\, \\ m_{AB}^2 =(m_2^2-m_1^2)\sin\te \cos\te.\eea

The Lagrangian  ${\cal L}$ is invariant under the global $U(1)$
phase transformations  \bea\Phi_m' \, =\, e^{i \al }\, \Phi_m\eea
as a result, we have the conservation of the Noether  charge \bea
Q=\int d^3{\bf x} \, I^0(x),\eea which is indeed the total charge
of the system (we have $I^\mu(x)= i \Phi_m^\dag(x)
\stackrel{\lrar}{\pa^\mu} \Phi_m (x) $ with $
\stackrel{\lrar}{\pa^\mu}  \,\equiv \,\stackrel{\rar}{\pa^\mu} -
\stackrel{\lar}{\pa^\mu}   $).

Let us now consider the $SU(2)$ transformation
\bea \label{masssu2} \Phi_m'(x) \, =\, e^{i \al_j \tau_j}\,
\Phi_m(x) \qquad, \qquad j= 1,2,3 \, , \eea
with $\al_j$ real constants, $\tau_j=\si_j/2$ and $\sigma_j$ being
the Pauli matrices. For $m_1\neq m_2$, the Lagrangian is not
generally invariant under (\ref{masssu2}) and we obtain, by use of
the equations of motion,
\bea \label{boscu1} \de {\cal L}(x)&= &  - i \,\al_j\,
\Phi_m^\dag(x) \, [M_d\,, \,\tau_j] \, \Phi_m(x) \, = \, -\al_j
\,\pa_\mu\, J^\mu_{m,j}(x)\, , \eea where the currents are
 \bea J^\mu_{m}(x)=\frac{\partial {\cal L}}{\partial\left(\partial _{\mu }\Phi_m(x)
\right)}\delta\Phi_m(x)+\delta\Phi_m^\dag(x)\frac{\partial {\cal
L}}{\partial \left(\partial_{\mu}\Phi_m^\dag(x)\right)}, \eea and
then \bea J^\mu_{m,j}(x) &=& i\, \Phi_m^\dag(x) \, \tau_j\,
\stackrel{\lrar}{\pa^\mu}\, \Phi_m (x) \quad , \qquad j= 1,2,3  .
\eea

We thus obtain the currents:

\bea J_{m,1}^{\mu }=\frac{i}{2}\left[(\partial^{\mu }\phi
_{1}^{\dagger}) \phi_{2}+ (\partial^{\mu }\phi _{2}^{\dagger})
\phi_{1}-\phi_{1}^{\dagger}(\partial ^{\mu }\phi_{2})
-\phi_{2}^{\dagger}(\partial^{\mu}\phi _{1}) \right],\eea \bea
J_{m,2}^{\mu }=\frac{1}{2}\left[(
\partial ^{\mu }\phi _{1}^{\dagger }) \phi _{2}-( \partial ^{\mu
}\phi _{2}^{\dagger }) \phi _{1}-\phi _{1}^{\dagger }(\partial
^{\mu }\phi _{2}) +\phi _{2}^{\dagger }(\partial ^{\mu }\phi _{1})
\right] ,\eea \bea J_{m,3}^{\mu }=\frac{i}{2}\left[(
\partial ^{\mu }\phi _{1}^{\dagger }) \phi _{1}-(
\partial ^{\mu }\phi _{2}^{\dagger }) \phi _{2}-\phi
_{1}^{\dagger }(\partial ^{\mu }\phi _{1}) +\phi _{2}^{\dagger
}(\partial ^{\mu }\phi _{2}) \right] . \eea

 The  corresponding charges, \bea Q_{m,j}(t)\equiv \int d^3 {\bf
 x}
\,J^0_{m,j}(x),\eea close  the $su(2)$ algebra (at each time $t$).
The Casimir operator $C_m$ is proportional to the total charge:
\bea C_m \equiv \Big[ \sum\limits_{j=1}^3
Q_{m,j}^2(t)\Big]^{\frac{1}{2}}=\,\frac{1}{2}Q.\eea Observe also
that the transformation induced by $Q_{m,2}(t)$,
\bea \Phi_f(x) &=& e^{-2 i\te Q_{m,2}(t)} \Phi_m(x) e^{2i\te
Q_{m,2}(t)} \eea
is just the mixing transformation Eq.(\ref{bosmix}). Thus $ 2
Q_{m,2}(t)$ is the generator of the mixing transformations.
Moreover, \bea Q_{m, \pm}(t) \equiv {\frac{1}{2}}\left[Q_{m,1}(t)
\pm iQ_{m,2}(t)\right],\eea $Q_{m,3}$, and $C_{m}$ are nothing but
$S_\pm(t)$, $S_{3}$, and $S_{0}$, respectively, as introduced in
Eqs.(\ref{Spt})-(\ref{S0}). From Eq.(\ref{boscu1}) we also see
that $Q_{m,3}$ and $C_{m}$ are conserved, consistently with
Eqs.(\ref{bosS3}), (\ref{bosS0}). Observe that the combinations
\bea Q_{1,2}&\equiv& \frac{1}{2}Q \pm Q_{m,3}
\\[2mm] \label{noether1}
Q_i & = & \int d^3 {\bf k}  \lf( a^{\dag}_{{\bf k},i} a_{{\bf
k},i}\, -\, b^{\dag}_{-{\bf k},i}b_{-{\bf k},i}\ri) \quad ,\qquad
i=1,2, \eea
are simply the conserved\footnote{Note that, in absence of mixing,
these charges would indeed be the flavor charges, being the flavor
conserved for each generation.} (Noether) charges for the free
fields $\phi_1$ and $\phi_2$ with \bea Q_1 + Q_2 = Q.\eea

We now perform the $SU(2)$ transformations on the flavor doublet
$\Phi_f$:
\bea \Phi_f'(x) \, =\, e^{ i  \al_j \tau_j }\, \Phi_f (x) \quad
,\qquad j \,=\, 1, 2, 3, \eea
and obtain:
\bea \de {\cal L}(x)&= & -i \,\al_j\,\Phi_f^\dag(x)\, [M,\tau_j]\,
\Phi_f(x) \, = \, -\al_j \,\pa_\mu  J_{f,j}^{\mu}(x) ~,
\\ [3mm]
J^\mu_{f,j}(x) &=&  i \,\Phi_f^\dag(x) \, \tau_j\,
\stackrel{\lrar}{\pa^\mu}\, \Phi_f (x) \quad ,\qquad j \,=\, 1, 2,
3. \eea

The related charges, \bea Q_{f,j}(t) \equiv \int d^3 {\bf x}
\,J^0_{f,j}(x)\eea still fulfil the $su(2)$ algebra and \bea
C_f=C_m =\frac{1}{2}Q.\eea Due to the off--diagonal (mixing) terms
in the mass matrix $M$, $Q_{f,3}(t)$ is time--dependent. This
implies an exchange of charge between $\phi_A$ and $\phi_B$,
resulting in the flavor oscillations. This suggests to us to
define indeed the {\em flavor charges} as
\bea \label{flacha} Q_A(t) & \equiv &\frac{1}{2}Q  \, + \,
Q_{f,3}(t)~,
\\[2mm]
Q_B(t) & \equiv & \frac{1}{2}Q \, -  \, Q_{f,3}(t)~, \eea
with \bea Q_A(t) \, + \,Q_B(t) \, = \, Q.\eea

These charges have a simple expression in terms of the flavor
ladder operators:
\bea\label{flacha12}
 Q_\si(t) & = & \int d^3 {\bf k}  \lf( a^{\dag}_{{\bf k},\si}(t)
a_{{\bf k},\si}(t)\, -\, b^{\dag}_{-{\bf k},\si}(t)b_{-{\bf
k},\si}(t)\ri) \quad ,\qquad\si= A, B ~. \eea

This is because they are connected to the Noether charges $Q_i$ of
Eq.(\ref{noether1}) via the mixing generator: \bea Q_\si(t) =
G^{-1}_{\te}(t)\;Q_i \; G_{\te}(t),\eea with
$(\si,i)=(A,1),(B,2)$. Note that the flavor charges are invariant
under the transformation Eq.(\ref{bogol}).

%%%%%%%%%%%%%%%%%%%%%%%%%%%%%%%%%%%%%%%%%%%%%%%%%%%%%%%%%%%
\section{The oscillation formula for mixed bosons}
%%%%%%%%%%%%%%%%%%%%%%%%%%%%%%%%%%%%%%%%%%%%%%%%%%%%%%%%%%%

Let us now calculate the oscillation formula for mixed bosons
\cite{BCRV01}. We will first follow the approach of Binger and Ji
\cite{binger} and show that the oscillation formulas there
presented  exhibit a dependence on the arbitrary mass parameters
$\mu_\si$, a feature which is not physically acceptable. We will
do this by using the generalized operators introduced above. Then
we will show how to cure this pathology \cite{BCRV01}, in analogy
to what was done in Chapter 2 for the case of fermion mixing
\cite{BHV98,Blasone:1999jb}, where the exact formula for neutrino
oscillations was derived and it was shown to be independent from
the arbitrary mass parameters that can be introduced in the
expansions of the flavor fields.

\subsection{The oscillation formula of Binger and Ji}

Following Ref.\cite{binger}, let us define the (generalized)
flavor state by acting on the {\em mass vacuum} $|0\ran_{1,2}$
with the flavor creation operators (we omit momentum indices):
\bea\label{BJstate}
 |{\wti a}_A \ran \equiv {\wti a}_A^\dag |0\ran_{1,2} \,=\,
\rho_{A 1} \cos\te |a_1 \ran \, + \, \rho_{A 2} \sin\te |a_2 \ran
~ \eea
with $|a_i \ran =a^\dag_i |0\ran_{12} $.

As already discussed in Ref.\cite{binger}, the flavor state so
defined is not normalized and the normalization factor has to be
introduced as
\bea\label{normaliz} {\wti {\cal N}}_A \equiv \lan {\wti a}_A |
{\wti a}_A \ran \,=\, \rho^2_{A 1} \cos^2\te + \rho^2_{A 2}
\sin^2\te  ~. \eea
We have
\bea \label{na} \lan {\wti a}_A | {\wti N}_A |{\wti a}_A \ran &=&(
\rho^2_{A 1} + \la^2_{A 1} ) \cos^2 \te + ( \rho^2_{A 2} +
\la^2_{A 2} ) \sin^2 \te  ~. \eea

The oscillation formula then follow as:
\bea \label{nat} \lan {\wti a}_A(t) | {\wti N}_A |{\wti a}_A(t)
\ran &=& \lan {\wti a}_A | {\wti N}_A |{\wti a}_A \ran - 4 \frac{
\rho^2_{A 1} \rho^2_{A 2} }{ {\wti {\cal N}}_A } \sin^2 \te \cos^2
\te \sin^2 \lf( \frac{\De E}{2} t \ri) \eea
and a similar one for the expectation value of ${\wti N}_B$. From
the above, as announced, it is evident that these formulas
explicitly depend on the (arbitrary) parameters $\mu_\si$ (see
Eq.(\ref{rho})). We also note that, for $\mu_A=m_1$ and
$\mu_B=m_2$, one has $\rho_{A 1}=1$, $\la_{A 1} =0 $, $\rho_{A
2}=U$ and $\la_{A 2} =V$. Consequently, Eqs.(\ref{na}),
(\ref{nat}) reduce respectively to Eqs.(18) and (20) of
Ref.\cite{binger}.

\subsection{The exact oscillation formula}

We now show how a consistent treatment of the flavor oscillation
for bosons in QFT can be given which does not exhibit the above
pathological dependence on arbitrary parameters \cite{BCRV01}.

There are two key points to be remarked. A general feature of
field mixing is  that the number operator  for mixed particles is
not a well-defined operator. It is so because the mixing
transformations mix creation and annihilation operators and then
the annihilation (creation) operators for flavor particles  and
antiparticles do not commute at different times (see
Eqs.(\ref{bosanna})-(\ref{bosannd})). Moreover, the number
operator does depend on the arbitrary mass parameters. Much care
is therefore required in the use of the  number operator. A second
remark is that the flavor states are not to be defined by using
the vacuum $|0\ran_{1,2}$: the flavor states so defined are in
fact not normalized and the normalization factor
Eq.(\ref{normaliz}) depends on the arbitrary mass parameters.

These two difficulties can be bypassed by using the remedy already
adopted  in chapter 2 \cite{BHV98,Blasone:1999jb} for the case of
fermions: the flavor states have shown to be consistently defined
by acting with the flavor creation operators on the {\em flavor
vacuum}. The observable quantities are then the expectation values
of the {\em flavor charges} on the flavor states: the oscillation
formulas thus obtained do not depend on the arbitrary mass
parameters.

Let us now define\footnote{In the following, we will work in the
Heisenberg picture: this is particularly convenient in the present
context since special care has to be taken with the time
dependence of flavor states (see the discussion in
Ref.\cite{BHV98}).} the state of the $a_A$ particle as \bea |{\wti
a}_{{\bf k},A}\ran_\AB\equiv {\wti a}_{{\bf k},A}^\dag(0) |{\wti
0}\ran_\AB \eea with \bea |{\wti 0}\ran_\AB= J^{-1}|0\ran_\AB,\eea
and consider the expectation values of the flavor charges
Eq.(\ref{flacha12}) on it (analogous results follow if one
considers $|{\wti a}_{{\bf k},B}\ran_\AB$). We obtain:
\bea\non\label{sicharges12}
 {\wti {\cal Q}}^{A}_{{\bf k},\si}(t)&\equiv& {}_\AB\lan {\wti
a}_{{\bf k},A} | {\wti Q}_\si(t) | {\wti a}_{{\bf k},A}\ran_\AB \,
=\, \lf|\lf[{\wti a}_{{\bf k},\si}(t), {\wti a}^{\dag}_{{\bf
k},A}(0) \ri]\ri|^2 \; - \; \lf|\lf[{\wti b}^\dag_{-{\bf
k},\si}(t), {\wti a}^{\dag}_{{\bf k},A}(0) \ri]\ri|^2 ,\\\eea with
$\si=A,B$.

We also have $\,_\AB\langle {\wti 0}| {\wti Q}_{{\bf k},\si}(t)|
{\wti 0}\rangle_\AB =0$ and ${\wti {\cal Q}}^{A}_{{\bf k},A}(t) +
{\wti {\cal Q}}^{A}_{{\bf k},B}(t)= 1$.

A straightforward direct calculation shows that the above
quantities {\em do not depend} on $\mu_A$ and $\mu_B$, i.e.:
\bea \label{independence} {}_\AB\lan {\wti a}_{{\bf k},A} | {\wti
Q}_{{\bf k},\si}(t) | {\wti a}_{{\bf k},A}\ran_\AB \,
=\,{}_\AB\lan a_{{\bf k},A} | Q_{{\bf k},\si}(t) | a_{{\bf
k},A}\ran_\AB \quad, \qquad \si=A,B\, , \eea
and similar one for the expectation values on $|{\ti a}_{{\bf
k},B}\ran_\AB$. Eq.(\ref{independence}) is a central result of
this chapter \cite{BCRV01}: it confirms that the only physically
relevant quantities are the above expectation values of flavor
charges. Note that expectation values of the number operator, of
the kind \bea{}_\AB\lan {\wti a}_{{\bf k},A} | {\wti N}_\si(t) |
{\wti a}_{{\bf k},A}\ran_\AB \, =\, \lf|\lf[{\wti a}_{{\bf
k},\si}(t), {\wti a}^{\dag}_{{\bf k},A}(0) \ri]\ri|^2\eea and
similar ones, do indeed depend on the arbitrary mass parameters,
although the flavor states are properly defined (i.e. on the
flavor Hilbert space). The cancellation of these parameters
happens {\em only} when considering the combination of squared
modula of commutators of the form
Eq.(\ref{sicharges12})\footnote{One may think it could make sense
to take the expectation value of the flavor charges on states
defined on the mass Hilbert space, as the ones defined in
Eq.(\ref{BJstate}). A direct calculation however shows that this
is not the case and these expectation values depends on the mass
parameters: the conclusion is that one must use the flavor Hilbert
space.}.
 A similar cancellation
occurs for fermions \cite{Blasone:1999jb} with the sum of the
squared modula of anticommutators.

Finally, the explicit calculation \cite{BCRV01} gives
\bea  \non
 {\cal Q}^{A}_{{\bf k},A}(t)&=&
\lf|\lf[a_{{\bf k},A}(t), a^{\dag}_{{\bf k},A}(0) \ri]\ri|^2 \; -
\; \lf|\lf[b^\dag_{-{\bf k},A}(t), a^{\dag}_{{\bf k},A}(0)
\ri]\ri|^2
\\[2mm]\non \label{Acharge}
&=& 1 - \sin^{2}( 2 \theta) \lf[ |U_{{\bf k}}|^{2} \; \sin^{2}
\lf( \frac{\omega_{k,2} - \omega_{k,1}}{2} t \ri) -|V_{{\bf
k}}|^{2} \; \sin^{2} \lf( \frac{\omega_{k,2} + \omega_{k,1}}{2} t
\ri) \ri] \, ,\\
\\[4mm]  \non
{\cal Q}^{A}_{{\bf k},B}(t)&=& \lf|\lf[a_{{\bf k},B}(t),
a^{\dag}_{{\bf k},A}(0) \ri]\ri|^2 \; - \; \lf|\lf[b^\dag_{-{\bf
k},B}(t), a^{\dag}_{{\bf k},A}(0) \ri]\ri|^2
\\[2mm] \label{Bcharge}
&=& \sin^{2}( 2 \theta)\lf[ |U_{{\bf k}}|^{2} \; \sin^{2} \lf(
\frac{\omega_{k,2} - \omega_{k,1}}{2} t \ri) -|V_{{\bf k}}|^{2} \;
\sin^{2} \lf( \frac{\omega_{k,2} + \omega_{k,1}}{2} t \ri) \ri] \,
. \eea

Notice the negative sign in front of the $|V_{{\bf k}}|^{2}$ terms
in these formulas, in contrast with the fermion case
\cite{BHV98,Blasone:1999jb}: the boson flavor charge can assume
also negative values. This fact  points to the statistical nature
of the phenomenon: it means that when dealing with mixed fields,
one intrinsically deals with a many--particle system, i.e. a
genuine field theory phenomenon. This situation has a strong
analogy with Thermal Field Theory (i.e. QFT at finite temperature)
\cite{Um1}, where quasi--particle states are ill defined and only
statistical averages make sense. Of course, there is no violation
of charge conservation for the overall system of two mixed fields.

The above formulas are obviously different from the usual quantum
mechanical oscillation formulas, which however are recovered in
the relativistic limit (i.e. for $|{\bf k}|^2\gg \frac{m_1^2
+m_2^2}{2}$). Apart from the extra oscillating term (the one
proportional to $|V_{{\bf k}}|^{2}$) and the momentum dependent
amplitudes, the QFT formulas carry the remarkable information
about the statistics of the oscillating particles: for bosons and
fermions the amplitudes (Bogoliubov coefficients) are drastically
different according to the two different statistics ($|U_{\bf k}|$
and $|V_{\bf k}|$ are circular functions in the fermion case and
hyperbolic functions in the boson case). This fact also fits with
the above mentioned statistical nature of the oscillation
phenomenon in QFT. Note also that our treatment is essentially
 non--perturbative \cite{BCRV01}.

In order to better appreciate the features of the QFT formulas, it
is useful to plot the oscillating charge in time for sample values
of the masses and for different values of the momentum (we use
same units for masses and momentum). It is evident how the effect
of the extra oscillating term is maximal at lower momenta (see
Figures 8.1 and 8.2) and disappears for large $k$ (see Figure 8.3)
where the standard oscillation pattern is recovered. In the
following plots we use $T_k=\frac{4 \pi}{\omega_{k,2} -
\omega_{k,1}}$ and assume maximal mixing.

\begin{figure}
\vspace{.5cm}
\centerline{\epsfysize=3.0truein\epsfbox{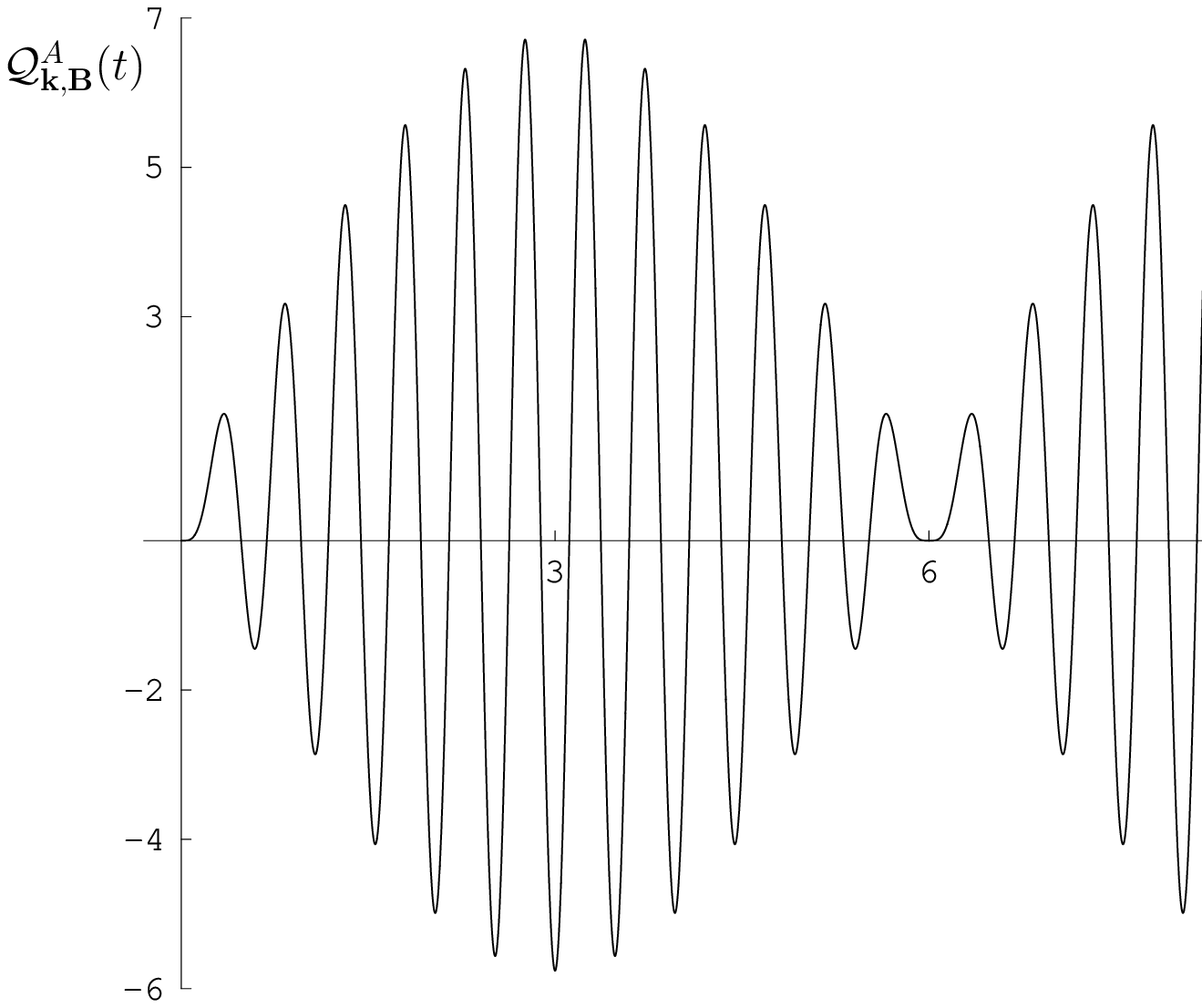}}
\vspace{.2cm} \caption{Plot of ${\cal Q}^{A}_{{\bf k},B}(t)$ in
function of time for $k =0$, $m_1=2$, $m_2=50$ and $\te=\pi/4$.}
\vspace{.1cm}

\hrule
\end{figure}

\begin{figure}
\vspace{0.5cm}
\centerline{\epsfysize=3.0truein\epsfbox{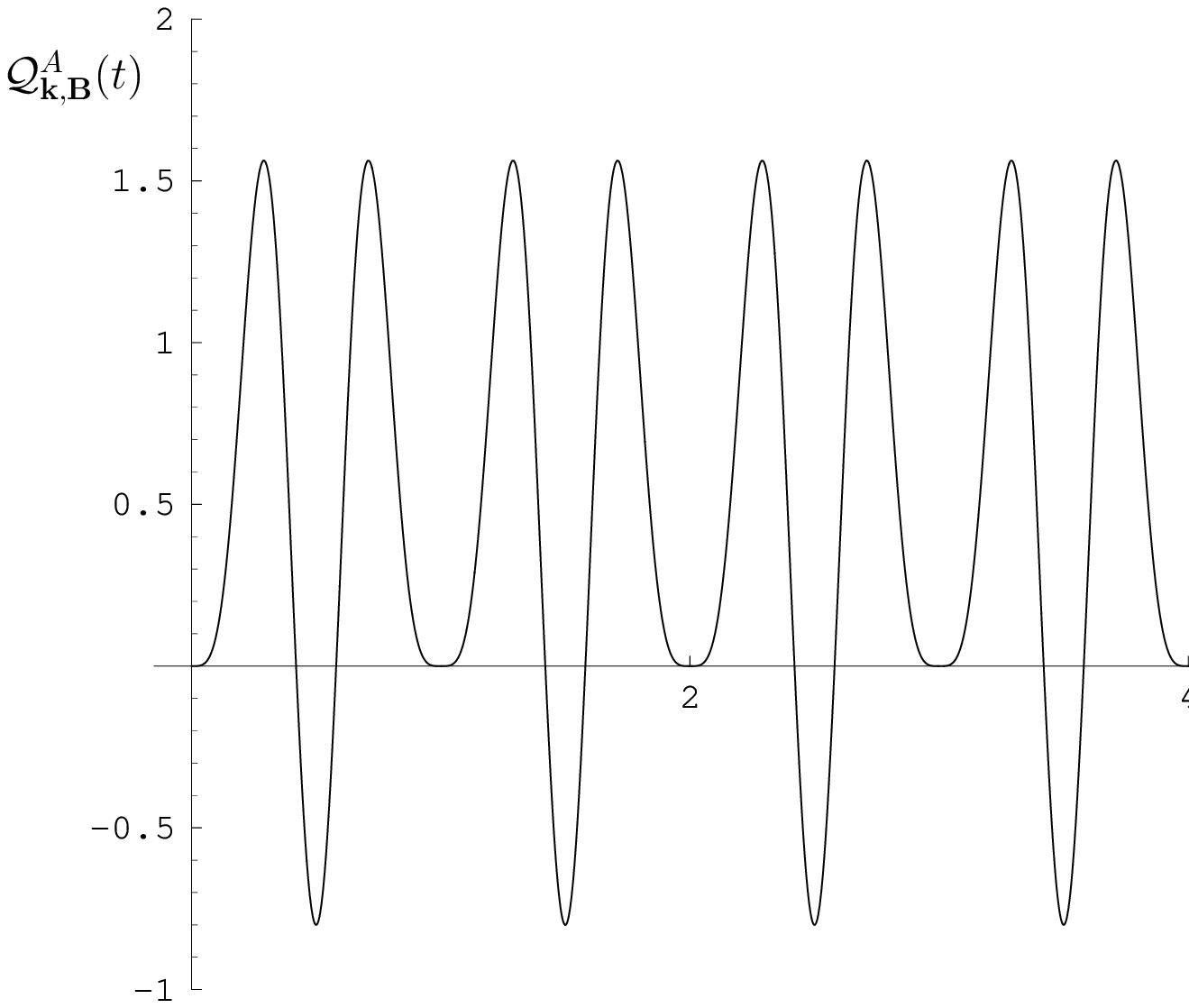}}
\vspace{.2cm} \caption{ Plot of ${\cal Q}^{A}_{{\bf k},B}(t)$ in
function of time for $k =10$, $m_1=2$, $m_2=50$ and $\te=\pi/4$. }
\vspace{.1cm}

\hrule
\end{figure}

\begin{figure}
\vspace{1cm} \centerline{\epsfysize=3.0truein\epsfbox{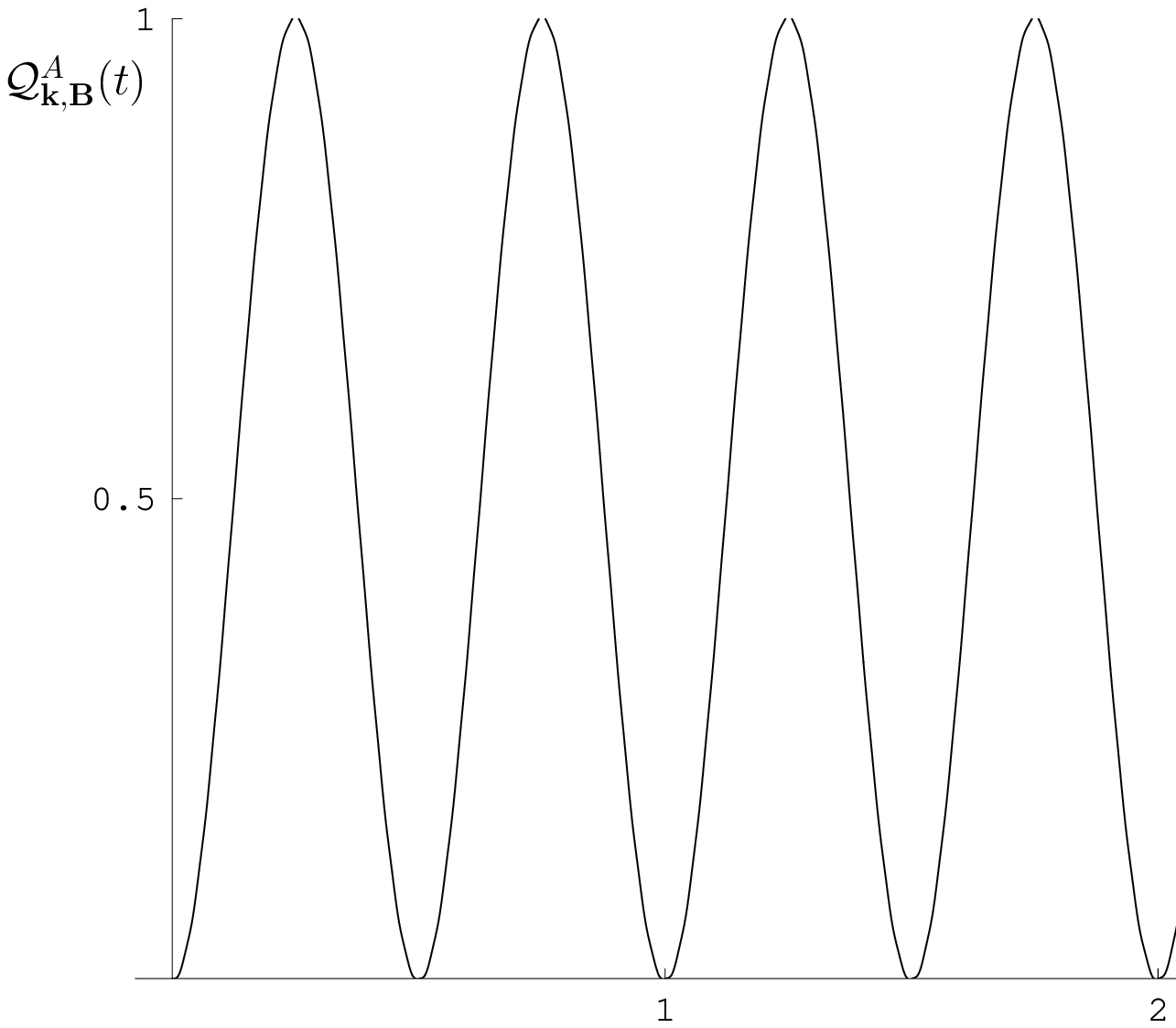}}
\vspace{.2cm} \caption{ Plot of ${\cal Q}^{A}_{{\bf k},B}(t)$ in
function of time for $k =100$, $m_1=2$, $m_2=50$ and $\te=\pi/4$.
} \vspace{.1cm}

\hrule
\end{figure}

It is also interesting  to plot  the time average of the
oscillating charge,  \bea{\bar {\cal Q}}^{A}_{{\bf k},B} =
\frac{1}{n T_k}\int \limits_0^{n T_k} dt \,{\cal Q}^{A}_{{\bf
k},B}(t),\eea as a function of the momentum. In Figure 8.4 we plot
${\cal Q}^{A}_{{\bf k},B}(t)$ averaged over two different time
intervals, i.e. for $n=10$ and $n=100$: it is interesting to
observe how the larger is the time interval, the more the curve
converges to the average of the standard formula, which has the
value $\frac{1}{2}$. The behavior for large k is due to the fact
that, as already observed, the exact oscillation formula reduces
to the quantum mechanical oscillation one in the large momentum
limit (i.e. for $|{\bf k} |^2\gg \frac{m_1^2 +m_2^2}{2}$).

\begin{figure}
\vspace{.5cm}
\centerline{\epsfysize=3.0truein\epsfbox{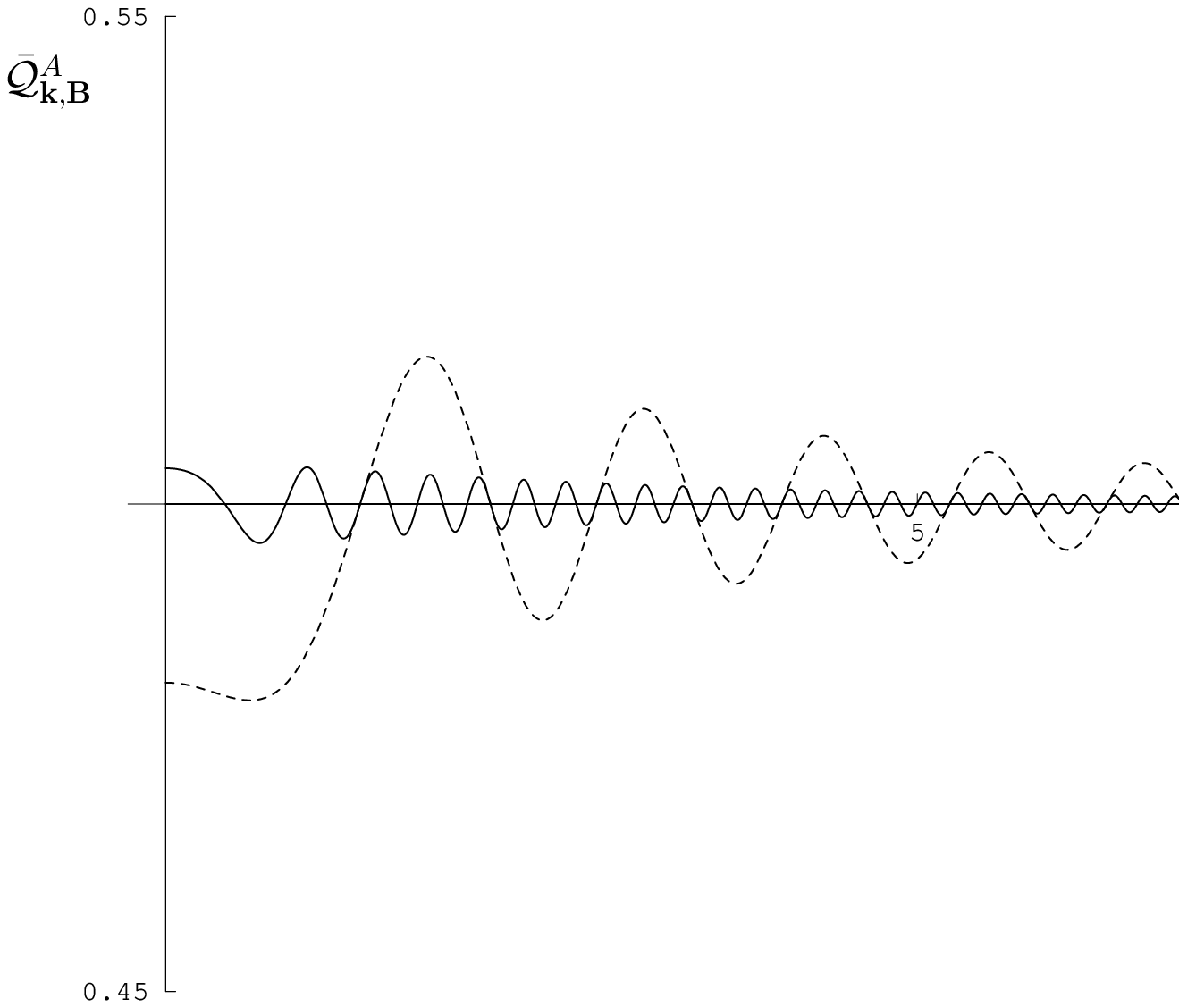}}
\vspace{.2cm} \caption{ The time average of ${\cal Q}^{A}_{{\bf
k},B}(t)$ over 10 $T_k$  (dashed line) and over 100 $T_k$ (solid
line) with respect to the average of the standard oscillation
formula (horizontal axis) as a  function of $k$ for the values
$m_1=2$, $m_2=50$ and  $\te=\pi/4$.} \vspace{.1cm}

\hrule
\end{figure}

\vspace{3cm}
\section{Mixing of neutral particles}

Analyzing the mixing phenomena for the meson sector, we should
also keep in mind another important point. For example, the
$\eta-\eta'$ mixing (which is one of the most interesting systems
due to the large mass difference of the mixed components) involves
the two {\em neutral} particles so that the mixing formulas for
the charged particles do not immediately apply. With this in mind,
we consider here the case of mixing of two spin zero neutral boson
fields \cite{Blaspalm,CJMV2003} . We follow the case of charged
fields as discussed above \cite{lathuile,BCRV01,binger} and
introduce the following lagrangian:
 \bea\label{boslagAB}
 {\cal L}(x)&=& \pa_\mu \Phi_f^T(x)
\pa^\mu \Phi_f (x)\, - \, \Phi_f^T(x)\;{M}\; \Phi_f(x)
\\[2mm]  \label{boslag12}
&=&\pa_\mu \Phi_m^T(x) \pa^\mu \Phi_m(x) \, - \, \Phi_m^T(x)\;
{M}_d\; \Phi_m (x) \eea
with $\Phi_f^T=(\phi_A,\phi_B)$ being the flavor fields and ${M} =
\lf(\brr{cc} m_A^2 & m_{AB}^2 \\ m_{AB}^2 & m_B^2\err \ri)$. Those
are connected to the  free fields $\Phi_m^T=(\phi_1,\phi_2)$ with
${M}_d = diag(m_1^2,m_2^2)$ by a rotation:
\bea\lab{2.53a} &&\phi_{A}(x) = \phi_{1}(x) \; \cos\te +
\phi_{2}(x) \; \sin\te
\\[2mm] \lab{2.53bb}
&&\phi_{B}(x) =- \phi_{1}(x) \; \sin\te + \phi_{2}(x)\; \cos\te
\eea
and a similar one for  the conjugate momenta $\pi_i=\pa_0\phi_i$.
The free fields $\phi_i$ can be quantized in the usual way (we use
$x_0\equiv t$):
\bea\label{phimass} \phi_{i}(x) &=& \int \frac{d^3{\bf
k}}{(2\pi)^{\frac{3}{2}}} \frac{1}{\sqrt{2\om_{k,i}}} \lf( a_{{\bf
k},i}\ e^{-i \om_{k,i} t } + a^{{\dag} }_{-{\bf k},i}\ e^{i
\om_{k,i} t} \ri)e^{i {\bf k x}}
\\ [2mm] \label{pimass}
\pi_{i}(x) &=& -i\ \int \frac{d^3{\bf k}}{(2\pi)^{\frac{3}{2}}}
\sqrt{\frac{\om_{k,i}}{2}} \lf(a_{{\bf k},i}\ e^{-i \om_{k,i} t} -
a^{{\dag} }_{-{\bf k},i}\ e^{i \om_{k,i} t} \ri)e^{i {\bf k x}},
 \eea
with $i=1,2$ and $\om_{k,i}=\sqrt{{\bf k}^2 + m_i^2}$.

The commutation relations are:
\bea\label{commutators} \lf[ \phi_{i}(x),\pi_{j}(y)\ri]_{x_0=y_0}=
 i\, \de_{ij}\,\delta^{3}({\bf x}-{\bf y}) ~,~~~~
\lf[ a_{{\bf k},i}, a_{{\bf p},j}^{\dag} \ri]= \,
\de_{ij}\,\delta^{3}({\bf k}-{\bf p}). \eea

We now recast Eqs.(\ref{2.53a}),(\ref{2.53bb}) into the form:
\bea\lab{2.53c}
\phi_{A}(x) = G^{-1}_\te(t)\; \phi_{1}(x)\; G_\te(t) \\[2mm]
\lab{2.53d} \phi_{B}(x) = G^{-1}_\te(t)\; \phi_{2}(x)\; G_\te(t)
\eea
and similar ones for $\pi_{A}(x)$, $\pi_B(x)$, where $G_\te(t)$
is the generator of the mixing transformations
(\ref{2.53a}),(\ref{2.53bb}):
\bea\lab{neutr} G_\te(t) = \exp\lf[-i\;\te \int d^{3}{\bf x}
\lf(\pi_{1}(x)\phi_{2}(x)  -\pi_{2}(x)\phi_{1}(x) \ri)\ri]\, ,
\eea
which is (at finite volume) a unitary operator:
$G^{-1}_\te(t)=G_{-\te}(t)=G^{\dag}_\te(t)$.

The mixing generator is given by \bea G_{\te}(t) =
\exp[\te(S_{+}(t) - S_{-}(t))]\eea and the $su(2)$ operators are
now realized as
\bea \label{su2charges1} && S_{+}(t) \equiv -i\;\int d^{3}{\bf x}
\; \pi_{1}(x)\phi_{2}(x) \\[2mm] && S_{-} (t)\equiv -i\;\int d^{3}{\bf
x} \;\pi_{2}(x)\phi_{1}(x)
\\ [2mm] \label{su2charges2}
&& S_{3} \equiv \frac{-i}{2} \int d^{3}{\bf x}
\lf(\pi_{1}(x)\phi_{1}(x) - \pi_{2}(x)\phi_{2}(x)\ri) \\[2mm] &&
S_{0} \equiv \frac{-i}{2} \int d^{3}{\bf x} \lf(
\pi_{1}(x)\phi_{1}(x) +\pi_{2}(x)\phi_{2}(x) \ri)\, . \eea

We have, explicitly
\bea\non S_{+}(t)-S_{-}(t)&=& \int d^{3}{\bf k} \Big( U_{\bf
k}^{*}(t) \; a_{{\bf k},1}^{\dag}a_{{\bf k},2}-  V_{{\bf
k}}^{*}(t) \; a_{-{\bf k},1}a_{{\bf k},2}+  V_{{\bf k}}(t) \;
a_{{\bf k},2}^{\dag}a_{-{\bf k},1}^{\dag} \\ &-& U_{{\bf k}}(t) \;
a_{{\bf k},2}^{\dag}a_{{\bf k},1} \Big) \eea
where $ U_{{\bf k}}(t)$ and $ V_{{\bf k}}(t)$ are Bogoliubov
coefficients given by
\bea \lab{bog1}  &&U_{{\bf k}}(t)\equiv | U_{{\bf k}}| \;
e^{i(\om_{k,2}- \om_{k,1})t} \quad , \quad  V_{{\bf k}}(t)\equiv |
V_{{\bf k}}| \; e^{i(\om_{k,1}+ \om_{k,2})t}
\\ [2mm]  \lab{bog2}
&&| U_{{\bf k}}|\equiv \frac{1}{2} \lf(
\sqrt{\frac{\om_{k,1}}{\om_{k,2}}} +
\sqrt{\frac{\om_{k,2}}{\om_{k,1}}}\ri) \quad , \quad | V_{{\bf
k}}|\equiv  \frac{1}{2} \lf( \sqrt{\frac{\om_{k,1}}{\om_{k,2}}} -
\sqrt{\frac{\om_{k,2}}{\om_{k,1}}} \ri)\\[2mm]&& | U_{{\bf k}}|^{2}-|
V_{{\bf k}}|^{2}=1\,, \eea

The flavor fields can be expanded as:
\bea\label{phiflav} \phi_{\si}(x) &=& \int \frac{d^3{\bf
k}}{(2\pi)^{\frac{3}{2}}} \frac{1}{\sqrt{2\om_{k,j}}} \lf( a_{{\bf
k},\si}(t)\ e^{-i \om_{k,j} t } + a^{{\dag} }_{-{\bf k},\si}(t)\
e^{i \om_{k,j} t} \ri)e^{i {\bf k\cdot x}}
\\ [2mm] \label{piflav}
\pi_{\si}(x) &=& -i\ \int \frac{d^3{\bf k}}{(2\pi)^{\frac{3}{2}}}
\sqrt{\frac{\om_{k,j}}{2}} \lf(a_{{\bf k},\si}(t)\ e^{-i \om_{k,j}
t} - a^{{\dag} }_{-{\bf k},\si}(t)\ e^{i \om_{k,j} t} \ri)e^{i
{\bf k \cdot x}} \, , \quad \eea
with $\si,j=(A,1),(B,2)$ and the flavor annihilation operators
given by:
\bea\non a_{{\bf k},A}(t)&\equiv&  G^{-1}_\te(t)\; a_{{\bf k},1}\;
G_\te(t) \,=\, \cos\te\;a_{{\bf k},1}\;+\;\sin\te\; \lf( U_{{\bf
k}}^*(t)\;a_{{\bf k},2}\;+ \;  V_{{\bf k}}(t)\;
a^{\dag}_{-{\bf k},2}\ri),\\
\\ \mlab{2.69}
\non a_{{\bf k},B}(t)&\equiv&  G^{-1}_\te(t)\; a_{{\bf k},2}\;
G_\te(t) \,=\, \cos\te\;a_{{\bf k},2}\;-\;\sin\te\; \lf( U_{{\bf
k}}(t)\;a_{{\bf k},1}\;- \;  V_{{\bf k}}(t)\; a^{\dag}_{-{\bf
k},1}\ri).\\ \eea

We now consider the action of the generator of the mixing
transformations on the vacuum $|0 \ran_\mass$ for the fields
$\phi_{i}(x)$:  $a_{{\bf k},i}|0 \ran_\mass = 0, ~ i=1,2$ . The
generator induces an $SU(2)$ coherent state structure on such
state \cite{Per}:
\bea\label{2.61} |0(\te, t) \ran_\AB \equiv G^{-1}_\te(t)\; |0
\ran_\mass\,. \eea
The state $|0(\te, t) \ran_\AB$ is to the {\em flavor vacuum} for
neutral bosons.

As in the case of complex boson fields, we define the state for a
mixed particle with ``flavor'' $A$ and momentum ${\bf k} $ as:
\bea |  a_{{\bf k},A}(t)\ran &\equiv& a^{\dag}_{{\bf k},A}(t)|0
(t)\ran_\AB \, =\, G^{-1}_\te(t)a^{\dag}_{{\bf k},1}|0\ran_\mass
\eea
In the following we work in the Heisenberg picture, flavor states
will be taken at reference time $t=0$ (including the flavor
vacuum). We also define $|a_{{\bf k},A}\ran\equiv |a_{{\bf
k},A}(0)\ran$.

Let us now consider the (non-vanishing) commutators of the flavor
ladder operators at different times:
\bea\non \label{comma} \Big[a_{{\bf k},A}(t), a^{\dag}_{{\bf
k},A}(t')\Big] &=& \cos^2 \te + \sin^2\te\,\lf(U_{\bf k}|^2 e^{-i
(\om_{2} -\om_1) (t-t')} - |V_{\bf k}|^2 e^{i (\om_{2} +\om_1)
(t-t')}\ri),\\
\\ [2mm] \label{commb}
\Big[ a^{\dag}_{-{\bf k},A}(t),a^{\dag}_{{\bf k},A}(t')\Big]
&=&\sin^2\te\,|U_{\bf k}||V_{\bf k}| \lf(e^{-i \om_{2} (t-t')}-
e^{i \om_{2}(t-t') }\ri) e^{-i \om_{1}  (t+t')},
\\ [2mm]
\label{commc} \Big[a_{{\bf k},B}(t), a^{\dag}_{{\bf k},A}(t')\Big]
&=& \cos\te\sin\te \, |U_{\bf k}| \lf(e^{i (\om_{2} -\om_1) t'} -
 e^{i (\om_{2} -\om_1) t}\ri),
\\ [2mm] \label{commd}
\Big[a^{\dag}_{-{\bf k},B}(t), a^{\dag}_{{\bf k},A}(t')\Big]
&=&\cos\te\sin\te \, |V_{\bf k}| \lf(e^{-i (\om_{2} +\om_1) t} -
 e^{-i (\om_{2} +\om_1) t'}\ri).
\eea

We observe that the following quantity is constant in time:
\bea\non \label{one} \lf|\lf[a_{{\bf k},A}(t), a^{\dag}_{{\bf
k},A}(t') \ri]\ri|^2 \, &-& \,\lf|\lf[a^{\dag}_{-{\bf k},A}(t),
a^{\dag}_{{\bf k},A}(t') \ri]\ri|^2 \,+ \\ + \,\lf|\lf[a_{{\bf
k},B}(t),a^{\dag}_{{\bf k},A}(t') \ri]\ri|^2 \, &-& \,
\lf|\lf[a^{\dag}_{-{\bf k},B}(t), a^{\dag}_{{\bf k},A}(t')
\ri]\ri|^2 \, =\, 1\,. \eea

In the case of the complex boson fields \cite{BCRV01}, the
corresponding of Eq.(\ref{one}) was consistently interpreted as
expressing the conservation of total charge\footnote{We have, for
charged bosonic fields:
\bea\non
 && \lan a_{{\bf k},A} |  Q_\si(t) |  a_{{\bf k},A}\ran\,
=\, \lf|\lf[ a_{{\bf k},\si}(t),  a^{\dag}_{{\bf k},A}(0)
\ri]\ri|^2 \; - \; \lf|\lf[b^{\dag}_{-{\bf k},\si}(t),
a^{\dag}_{{\bf k},A}(0) \ri]\ri|^2 \quad, \qquad \si=A,B\, . \eea
together with $\,_\AB\langle  0| Q_{\si}(t)|
 0\rangle_\AB =0$ and
$\sum_\si \lan a_{{\bf k},A} |  Q_\si(t) |  a_{{\bf k},A}\ran \,
=\, 1$.}. In the present case we are dealing with a neutral field
and thus the charge operator vanishes identically. Nevertheless
the quantities in Eq.(\ref{one}) are well defined and are the
neutral-field counterparts of the corresponding ones for the case
of charged fields. Thus we look for a physical interpretation of
such oscillating quantities.

Let us consider the momentum operator, defined as the diagonal
space part of the energy-momentum tensor \cite{Itz}: \bea
P^j\equiv\int d^{3}{\bf x}\Te^{0j}(x),\eea with \bea
\Te^{\mu\nu}\equiv\pa^\mu\phi\pa^\nu\phi -
g^{\mu\nu}\lf[\frac{1}{2}(\pa\phi)^2-\frac{1}{2}m^2\phi^2\ri].\eea
For the free fields $\phi_i$ we have:
\bea\label{pmass} {\bf P}_i&=&\int d^{3}{\bf x} \;\pi_{i}(x){\bf
\nabla}\phi_{i}(x) \, =\, \int d^{3}{\bf k} \, \frac{\bf k}{2}\,
\left(a_{{\bf k},i}^{\dag}a_{{\bf k},i} \, -\,
 a_{-{\bf k},i}^{\dag}a_{-{\bf k},i} \ri),\eea with $i=1,2.$

In a similar way we can define the momentum operator for mixed
fields:
\bea\non\label{pflav} {\bf P}_\si(t)&=&\int d^{3}{\bf x}
\;\pi_{\si}(x){\bf \nabla}\phi_{\si}(x) \, =\, \int d^{3}{\bf k}
\, \frac{\bf k}{2}\, \left(a_{{\bf k},\si}^{\dag}(t)a_{{\bf
k},\si}(t) \, -\,
 a_{-{\bf k},\si}^{\dag}(t)a_{-{\bf k},\si}(t) \ri),\\\eea with
 $\si=A,B$.
The two operators are obviously related: $ {\bf P}_\si(t) =
G^{-1}_\te(t)\,{\bf P}_i\, G_\te(t)$. Note that the total momentum
is conserved in time since commutes with the generator of mixing
transformations (at any time):
\bea {\bf P}_A(t) \, + \, {\bf P}_B(t) \, =\,{\bf P}_1 \, + \,
{\bf P}_2 \, \equiv\, {\bf P}\\ \lf[{\bf P}\,, \,G_\te(t)\ri] \,
=\, 0 \quad , \quad   \lf[ {\bf P}\,, \,H\ri] \, =\, 0\,. \eea
Thus in the mixing of neutral fields, the momentum operator plays
an analogous r{\^o}le to that of the charge for charged fields
\cite{Blaspalm}. For charged fields, the total charge operator,
associated with the $U(1)$ invariance of the Lagrangian, is
proportional to the Casimir of the $SU(2)$ group associated to the
generators of the mixing transformations. This is not true anymore
for the case of neutral fields, although the $SU(2)$ structure
persists in this case as well (see
Eqs.(\ref{su2charges1}),(\ref{su2charges2})).

We now consider the expectation values of the momentum operator
for flavor fields on the flavor state $|a_{{\bf k},A}\ran$ with
definite momentum ${\bf k}$. Obviously, this  is an eigenstate of
${\bf P}_A(t)$ at time $t=0$:
\bea {\bf P}_A(0) \,|a_{{\bf k},A}\ran \, =\, {\bf k}\,|a_{{\bf
k},A}\ran\,, \eea
which follows from ${\bf P}_1 \,|a_{{\bf k},1}\ran \, =\, {\bf
k}\,|a_{{\bf k},1}\ran$ by application of $G^{-1}_\te(0)$. At time
$t\neq 0$, the expectation value of the momentum (normalized to
the initial value) gives:
\bea\non {\cal P}_\si^A(t)&\equiv& \frac{\lan a_{{\bf k},A} | {\bf
P}_\si(t) | a_{{\bf k},A}\ran} {\lan a_{{\bf k},A} | {\bf
P}_\si(0) |  a_{{\bf k},A}\ran} \, =\, \lf|\lf[ a_{{\bf
k},\si}(t),  a^{\dag}_{{\bf k},A}(0) \ri]\ri|^2 \; - \;
\lf|\lf[a^{\dag}_{-{\bf k},\si}(t), a^{\dag}_{{\bf k},A}(0)
\ri]\ri|^2,\\ \eea with $\si=A,B$,
which is of the same form as the expression one obtains for the
charged field.

One can explicitly check that the (flavor) vacuum expectation
value of the momentum operator ${\bf P}_\si(t)$ does vanish at all
times:
\bea {}_\AB\lan 0 | {\bf P}_\si(t) | 0\ran_\AB \, =\, 0 \quad,
\qquad \si=A,B\, \eea
which can be understood intuitively by realizing that the flavor
vacuum $| 0\ran_\AB$  does not carry momentum since it is a
condensate of pairs carrying zero total momentum (like the BCS
ground state, for example).

The explicit calculation of the oscillating quantities ${\cal
P}_{{\bf k},\si}^A(t)$ gives:
\bea\non\label{Acharge} {\cal P}_{{\bf k},A}^A(t) &=& 1 -
\sin^{2}( 2 \theta) \lf[ |U_{{\bf k}}|^{2} \; \sin^{2} \lf(
\frac{\omega_{k,2} - \omega_{k,1}}{2} t \ri) -|V_{{\bf k}}|^{2} \;
\sin^{2} \lf(
\frac{\omega_{k,2} + \omega_{k,1}}{2} t \ri) \ri]\\
\\ [4mm] \non\label{Bcharge}
{\cal P}_{{\bf k},B}^A(t)&=& \sin^{2}( 2 \theta) \lf[ |U_{{\bf
k}}|^{2} \; \sin^{2} \lf( \frac{\omega_{k,2} - \omega_{k,1}}{2} t
\ri) -|V_{{\bf k}}|^{2} \; \sin^{2} \lf( \frac{\omega_{k,2} +
\omega_{k,1}}{2} t \ri) \ri].\\ \eea
in complete agreement with the charged field case \cite{BCRV01}.

The Eqs.(\ref{Acharge}), (\ref{Bcharge}) are the flavor
oscillation formulas for the neutral mesons, such as $\eta-\eta'$,
$\phi-\omega$. By definition of the momentum operator, the
Eqs.(\ref{Acharge}), (\ref{Bcharge}) are the relative population
densities of flavor particles in the beam.

We present in Figure 8.5 a plot of momentum oscillations for the
system of $\eta$ and $\eta^\prime$ ($m_\eta = 547 MeV$,
$\Gamma_\eta = 1.18 keV$, $m_{\eta^\prime} = 958 MeV$,
$\Gamma_{\eta^\prime} = 0.2 MeV$ and $\theta \approx -54^{\circ}$
\cite{JM01}) with zero momenta (i.e. $k = 0 GeV$).

\begin{figure}
\vspace{1cm} \centerline{\epsfysize=3.0truein\epsfbox{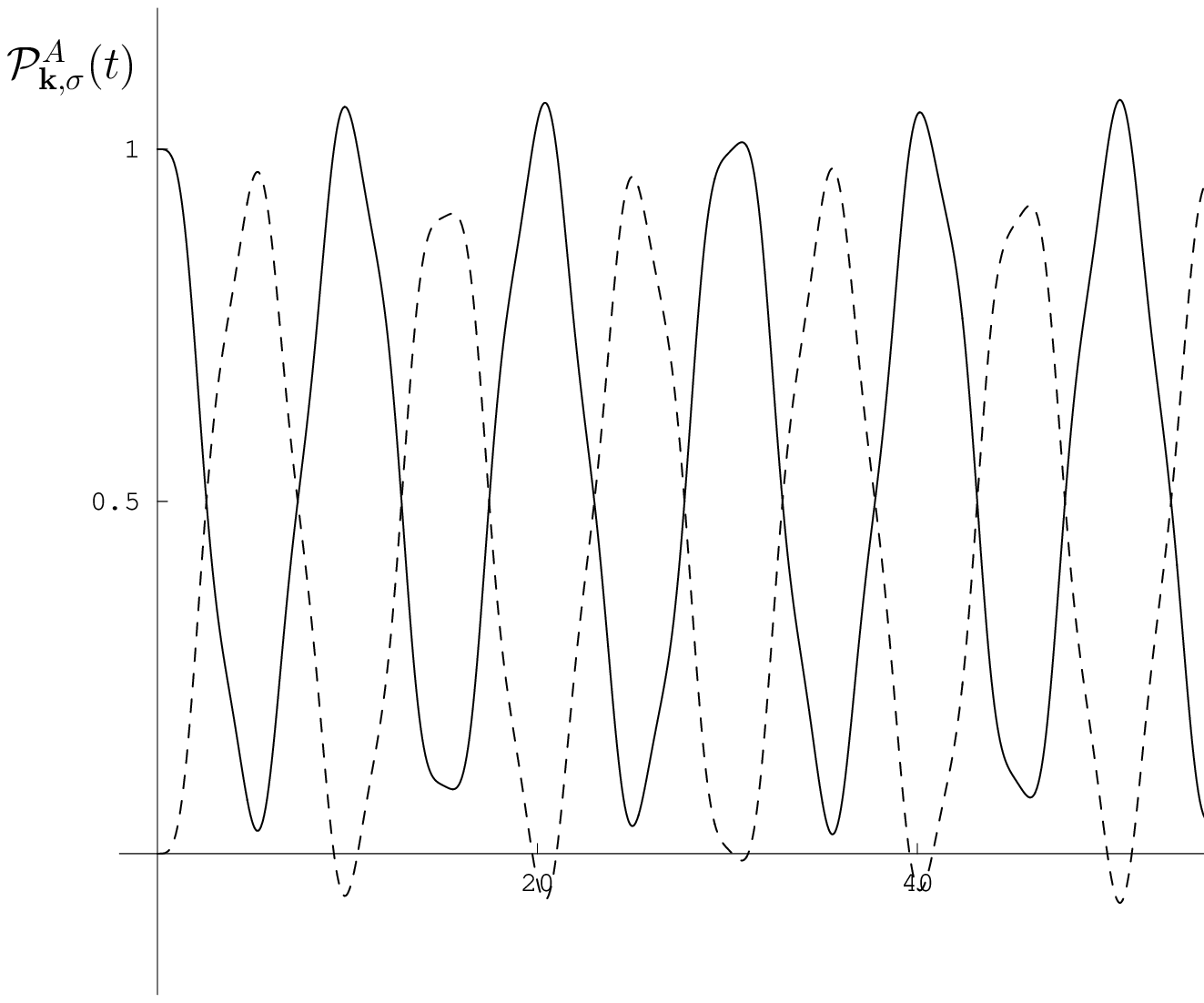}}
\vspace{.2cm} \caption{ Plot of ${\cal P}^{\eta}_{{\bf
k},\eta}(t)$ (solid line) and ${\cal P}^{\eta}_{{\bf k},\eta'}(t)$
(dashed line) in function of time }
 \centerline{\small for an initially pure $\eta$ state and
 for $k =0$, $m_1=549$ MeV, $m_2=958$ MeV and  $\te=-54^{o}$.}
\vspace{.1cm}

\hrule
\end{figure}

%%%%%%%%%%%%%%%%%%%%%%%%%%%%%%%%%%%%%%%%%%%%%%%%%%%%%%%%%%%%%%%%%%
\section{Conclusions}

In this chapter we have considered the quantum field theoretical
formulation of spin-zero (charged and neutral) boson field  mixing
and obtained the exact oscillation formula which does not depend
on arbitrary mass parameters which can be introduced in full
generality in the theory. We have also studied the structure of
the currents and charges for the charged mixed fields. In order to
make our discussion more transparent, we have neglected the
instability of the oscillating particles. This does not affect the
general validity of our result which rests on the intrinsic
features of QFT.

A crucial point in our analysis is the disclosure of the fact that
the space for the mixed field states is unitarily inequivalent to
the state space where the unmixed field operators are defined.
This is a common feature with the QFT structure of the fermion
mixing, which has recently been established
\cite{BHV98,Blasone:1999jb,BV95,fujii1,hannabus}). The vacuum for
the mixed fields turns out to be a generalized $SU(2)$ coherent
state. Of course, in the boson case the condensate structure for
the ``flavor'' vacuum is found to be very much different from the
one in the fermion case. Besides the intrinsic mathematical
interest, our analysis provides interesting phenomenological
insights. It leads to the exact oscillation formula for bosons
which predicts oscillation behaviors susceptible of being
experimentally tested.

In the framework of the QFT analysis of Refs.
\cite{lathuile,BV95}, a study of the meson mixing and oscillations
has been already carried out in Ref.\cite{binger}. However, the
results of Ref.\cite{binger} give observable quantities which are
dependent on arbitrary mass parameters, and this is of course
physically not acceptable \cite{fujii1}. In the present chapter we
have pointed out the origin of such a pathology and have shown how
to obtain results which are independent from arbitrary parameters
\cite{BCRV01}. The oscillation formula obtained  in
Ref.\cite{binger} has to be actually replaced with the exact one
here presented \cite{BCRV01}. Moreover we have studied the case of
the neutral bosons \cite{Blaspalm,CJMV2003}.

 Let us close by observing that although our
QFT analysis discloses features which cannot be ignored in any
further study of the field mixing and oscillations, since they are
intrinsic to the structure of the QFT formalism, nevertheless
there are many aspects of the physics of mixing which are not
fully understood and many features are still obscure
\cite{zralek}, as already observed in the introduction. The mixing
of neutrinos and their oscillations seem to be now experimentally
established and quark mixing and meson mixing are widely accepted
and verified. However, several questions
\cite{grossman,lipkin,srivastava1,lowe,kiers,burkhardt,giunti,Takeuchi:1999as}
are the object of active discussion in the framework of the
quantum mechanics formalism for neutrino oscillations. As a matter
of fact, such a state of affairs has been a strong motivation for
our searching in the structural aspects of QFT a possible hint to
the understanding of particle mixing and oscillations.

\vspace{.5cm}
%%%%%%%%%%%%%%%%%%%%%%%%%%%%%%%%%%%%%%%%%%%%%%%%%%%%%%%%%%%%%%%%%%%

%%%%%%%%%%%%%%%%%%%%%%%%%%%%%%%%%%%%%%%%%%%%

%%%%%%%%%%%%%%%%%%%%%%%%%%%%%%%%%%%%%%%%%%%%%%%%%%%%%%%%%%
%

\chapter{Phenomenology of flavor oscillations with nonperturbative
Quantum Field Theoretic effects}

We consider the quantum field theoretical formulation of meson
mixing and obtain the exact oscillation formula in the presence of
the decay. This formula is different from quantum mechanical
formula by additional high-frequency oscillation terms. We analyze
phenomenological aspects of this nonperturbative effect and find
that the systems where this phenomena could be detected are
$\eta-\eta^\prime$ and $\phi-\omega$ mesons.

%%\vspace{8mm}
%]
%\tableofcontents\newpage
%P.A.C.S.:

%\maketitle

\section{Introduction}

 In the chapters 2, 3, 8 it was found that the
fermions and bosons oscillation formula in quantum field
theoretical formulation is modified by additional high-frequency
terms and simpler quantum mechanical result is reproduced only in
the relativistic limit.

In this chapter, we attempt the phenomenological analysis of this
nonperturbative phenomenon. The mixing of particles and
antiparticles in the meson sector (e.g., $K^{0}-\bar{K}^{0}$,
$B^{0}-\bar{B}^{0}$) requires specific adjustments to the results
obtained in the case of charged fields. Moreover, except
neutrinos, all known mixed systems are subject to decay and thus
the effect of particle life-time should also be taken into
account.

Specifically, in Section 9.2, we analyze the adjustments needed
for the general formulation in order to make applications for the
known systems. We also study the effect of the finite particle
life-time on the field-theoretical oscillation formula. Finally we
estimate the magnitudes of the nonperturbative corrections in
various systems and discuss the systems in which the
field-theoretical effect may be most significant \cite{CJMV2003}.
 The Section 9.3 is devoted to conclusions.

%%%%%%%%%%%%%%%%%%%%%%%%%%%%%%%%%%%%%%%%%%%%%%%%%%%%%%%%%%%%%%%%%%

\section{Phenomenological application
of the nonperturbative oscillation formula}

\subsection{$K^{0}-\bar{K}^{0}$, $B^{0}-\bar{B}^{0}$ and
$D^{0}-\bar{D}^{0}$ mixing}

The typical systems for flavor mixing are $K^{0}-\bar{K}^{0}$,
$B^{0}-\bar{B}^{0}$ and $D^{0}-\bar{D}^{0}$. The
$K^{0}-\bar{K}^{0}$ and $B^{0}-\bar{B}^{0}$ systems provide the
evidence of CP-violation in weak interaction and the
$D^{0}-\bar{D}^{0}$ is also important for the analysis of CKM
mixing matrix. As an illustration of the mixing in these systems,
we consider in particular the $K^{0}-\bar{K}^{0}$ system here.

In $K^{0}-\bar{K}^{0}$ mixing, $K^{0}$ may not be treated as
neutral since $K^{0} \neq \bar{K}^{0}$. Of course, this is not the
mixing of two different charged particles either. Rather, the
particle here is mixed with its antiparticle. In this case it is
important to identify the mixed degrees of freedom properly. Note
that in $K^{0}-\bar{K}^{0}$ mixing there are three distinct modes,
namely the strange eigenstates $K^{0}-\bar{K}^{0}$, the mass
eigenstates $K_L-K_S$ and the CP eigenstates $K_1-K_2$. Each pair
can be written as a linear combination of the other ones, e.g.
\begin{equation}\label{yuri01}
\begin{array}{c}
K_1=\frac{1}{\sqrt{2}}(K^0 + \bar K^0),\quad
K_2=\frac{1}{\sqrt{2}}(K^0 - \bar K^0);

\cr K^0=\frac{e^{i\delta}}{\sqrt{2}}(K_L+K_S), \quad \bar
K^0=\frac{e^{-i\delta}}{\sqrt{2}}(K_L-K_S);

\cr K_1=\frac{1}{\sqrt{1+|\epsilon|^2}}(K_S+\epsilon K_L), \quad
K_2=\frac{1}{\sqrt{1+|\epsilon|^2}}(K_L+\epsilon K_S);
\end{array}
\end{equation}
with $e^{i\delta}$ being a complex phase and $\epsilon=i\delta$
being the imaginary CP-violation parameter. $K^{0}-\bar{K}^{0}$
are produced as strange eigenstates and propagate as the mass
eigenstates $K_L, K_S$.

The mass eigenstates $K_L$ and $K_S$ are defined as the +1 and -1
CPT eigenstates, respectively, so that they can be represented in
terms of self-adjoint scalar fields $\phi_1, \phi_2$ as
\begin{equation}
K_L=\phi_1, \quad K_S = i\phi_2.
\end{equation}
Therefore the mixing in this system is similar to the case of
neutral fields with {\em complex} mixing matrix. Since the complex
mixing matrix in SU(2) can be always transformed into the real one
by suitable redefinition of the field phases which does not affect
the expectation values, the mixing in this case is equivalent to
the mixing of neutral fields.

The oscillating observables may be that of the strange charge (in
the system $K^0$ and $\bar K^0$ taken as flavor A and B,
respectively) with the trivial mixing angle $\theta=\pi/4$ from
Eq.(\ref{yuri01}).

Phenomenologically relevant, however, is the oscillation of
CP-eigenvalue which determines the ratio of experimentally
measured $\pi\pi$ to $\pi\pi\pi$ decay rates. CP-oscillations are
given in terms of $K_1$ and $ K_2$ flavors with small mixing angle
\bea \cos(\theta)=1/\sqrt{1+|\epsilon|^2}.\eea

\subsection{The effect of the particle decay}

To analyze the phenomenological aspects of the quantum field
theory oscillation formulas for fermions and bosons, we should
remember that, excluding the mixing of neutrinos, most mixed
particles are unstable. Hence, we need to consider the effect of
the finite lifetime in the oscillation formulas of
Eqs.(\ref{Acharge}) and (\ref{Bcharge}).

The particle decay is taken in account by inserting, by hand, as
usually done, the factor $e^{-\Gamma t}$ in the annihilation
(creation) operators \cite{CJMV2003}: \bea a_{{\bf k},i}
\rightarrow a_{{\bf k},i}e^{-\frac{\Gamma_{i}}{2}t}, \eea \bea
b_{{\bf -k},i} \rightarrow b_{{\bf -k},i}
e^{-\frac{\Gamma_{i}}{2}t}. \eea

Then, the oscillation formulas can be written as \bea \non {\cal
P}^{A}_{{\bf k},A}(t)&=& \lf|\lf[a_{{\bf k},A}(t), a^{\dag}_{{\bf
k},A}(0) \ri]\ri|^2 \; - \; \lf|\lf[b^\dag_{-{\bf k},A}(t),
a^{\dag}_{{\bf k},A}(0) \ri]\ri|^2
\\[2mm] \label{Achargedecay}
&=& \lf(\cos^{2}\theta e^{-\frac{\Gamma_{1}}{2}t}+\sin^{2}\theta
e^{-\frac{\Gamma_{2}}{2}t} \ri)^{2}
\\[2mm] \non
&-& \sin^{2}(2\theta)e^{-\frac{\Gamma_{1} + \Gamma_{2}}{2}t} \lf[
|U_{{\bf k}}|^{2} \; \sin^{2} \lf( \frac{\omega_{k,2} -
\omega_{k,1}}{2} t \ri) -|V_{{\bf k}}|^{2} \; \sin^{2} \lf(
\frac{\omega_{k,2} + \omega_{k,1}}{2} t \ri) \ri] \, ,
\\[4mm]  \non
{\cal P}^{A}_{{\bf k},B}(t)&=& \lf|\lf[a_{{\bf k},B}(t),
a^{\dag}_{{\bf k},A}(0) \ri]\ri|^2 \; - \; \lf|\lf[b^\dag_{-{\bf
k},B}(t), a^{\dag}_{{\bf k},A}(0) \ri]\ri|^2
\\[2mm] \label{Bchargedecay}
&=& \sin^{2}( 2 \theta) \Big(
\lf[\frac{e^{-\frac{\Gamma_{1}}{2}t}-e^{-\frac{\Gamma_{2}}{2}t}}{2}
\ri]^{2}
\\[2mm] \non &+&
e^{-\frac{\Gamma_{1} + \Gamma_{2}}{2}t}\lf[ |U_{{\bf k}}|^{2} \;
\sin^{2} \lf( \frac{\omega_{k,2} - \omega_{k,1}}{2} t \ri)
-|V_{{\bf k}}|^{2} \; \sin^{2} \lf( \frac{\omega_{k,2} +
\omega_{k,1}}{2} t \ri) \ri] \Big) \,. \eea

We note the difference between these oscillation formulas and the
quantum mechanical Gell-Mann--Pais formulas. Essentially the
quantum field theoretic corrections are proportional to $|V_{\bf
k}|^2$ and appear as the additional high-frequency oscillation
terms.

%%%%%%%%%%%%%%%%%%%%%%%%%%%%%%%%%%%%%%%%%%%%%%%%%%%%%%%%%%%%%%%%%%

\subsection{The nonperturbative effect in the boson sector}

We are now in the position to analyze the known mixing systems and
compute the magnitude of the nonperturbative correction in each
system \cite{CJMV2003}. In the boson sector the typical mixed
systems are $K^{0}-\bar{K}^{0}$, $B^{0}-\bar{B}^{0}$,
$D^{0}-\bar{D}^{0}$, $\eta-\eta'$ and $\phi-\omega$. One may
expect that the nonperturbative high-frequency effects may be
observed adequately in some of these systems since not only the
mass difference for mesons may be as large as $400 MeV$ (e.g.
$\eta-\eta'$) but also in principle the mesons may be produced at
low momenta where the flavor vacuum effect is most prominent. In
practice, however, we note that one may encounter certain
experimental difficulty due to either extremely short lifetime of
mixed mesons or extremely small mass difference in
mass-eigenstates for the systems except $\eta-\eta'$ and
$\phi-\omega$.

\subsubsection{The magnitude of the nonperturbative term}

In all of field-theoretical derivations (See
Eq.(\ref{Achargedecay})-(\ref{Bchargedecay})), the
field-theoretical effect (or the high-frequency oscillation term)
is proportional to $|V_{\bf k}|^2$. In estimating the maximal
magnitude of this term, it is useful to write $|V_{{\bf k}}|^{2}$
in terms of the dimensionless momentum \bea p\equiv\sqrt{\frac{2
|{\bf k}|^2}{m_1^2 +m_2^2}}\eea and the dimensionless parameter
\bea a\equiv \frac{m_2^2 -m_1^2}{m_1^2 +m_2^2}\eea as follows:
\bea |V(p,a)|^2 & =& \frac{p^2 +1}{2\sqrt{(p^2 + 1)^2 - a^2}}
-\frac{1}{2}, \eea
from which we see that $|V_{{\bf k}}|^{2}$ is maximal at $p=0$
($|V_{max}|^2 = \frac{(m_1 -m_2)^2}{4 m_1 m_2}$) and goes to zero
for large momenta (i.e. for $|{\bf k}|^2\gg\frac{m_1^2
+m_2^2}{2}$).

\begin{figure}
\begin{center}
\epsfig{width=300pt, file=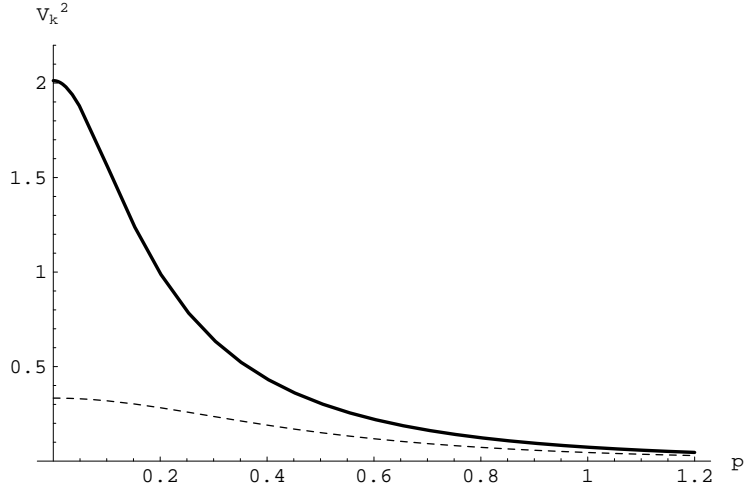} \caption{The bosonic
condensation density $|V(p,a)|^2$ as a function of $p$ for
$a=0.98$ (for instance: $m_{1}=5$, $m_{2}=50$ in dimensionless
units, solid line) and $a=0.8$ (for instance: $m_{1}=5$,
$m_{2}=15$ in dimensionless units, dashed line). }
\vspace{.1cm}

\hrule
\end{center}
\end{figure}

A plot of the condensation density $|V(p,a)|^2$ as a function of
$p$ is presented in Figure 9.1 for two sample values of the
parameter $a$.

The optimal observation scale for field theoretical effect in
meson mixing, therefore, is $k=0$ and the maximum correction is of
the order of $|V|^2\sim \frac{\Delta m^2}{m^2}$. For known
parameters of meson mixing we find
\begin{itemize}
\item   $K^0 - \bar K^0$: $\quad$   $|V|^2\sim (\frac{3.5
10^{-12}MeV} {500MeV})^2\sim 10^{-28}$, \item  $D^0 - \bar D^0$:
$\quad$    $|V|^2\sim (\frac{10^{-12}MeV} {1000MeV})^2\sim
10^{-30}$, \item  $B^0 - \bar B^0$: $\quad$    $|V|^2\sim (\frac{5
10^{-10}MeV} {5000MeV})^2\sim 10^{-26}$, \item  $B^0_s - \bar
B^0_s$: $\quad$    $|V|^2\sim (\frac{2 10^{-8}MeV}
{10^4MeV})^2\sim 10^{-28}$, \item  $\eta-\eta'$: $\quad$
$|V|^2\sim (\frac{400MeV}{700MeV})^2\sim 0.2$, \item
$\omega-\phi$: $\quad$     $|V|^2\sim(\frac{200MeV}{900MeV})^2\sim
0.05$.
\end{itemize}

\noindent From the above discussion we observe hence that the
nontrivial flavor vacuum effect can be maximally seen in the mixed
systems such as $\eta-\eta'$ and $\omega-\phi$, thus one needs to
be careful about taking them into account should these systems
ever be used in some sort of mixing experiments.

\subsection{The nonperturbative effect in fermion sector}

We can employ the similar method in the fermion sector. Since
neutrinos are stable, no additional adjustments are necessary to
the known results \cite{BV95}. We can write the field-theoretical
correction amplitude  $|V_{{\bf k}}|^{2}$ in the fermion case as a
function of the dimensionless momentum \bea p=\frac{|{\bf
k}|}{\sqrt{m_{1} m_{2}}}\eea and of the dimensionless parameter
\bea a=\frac{(\Delta m)^{2}} {m_{1} m_{2}}\;\;\;,\;\;\;0\leq a < +
\infty ~, \eea as follows:
\bea |V_{{\bf k}}|^{2} \equiv |V(p,a)|^2 & =&
\frac{1}{2}\lf(1-\frac{p^2+1} {\sqrt{(p^2 + 1)^2 + a p^2}}\ri)
~.\label{Vpa} \eea where $\Delta m \equiv m_{2}-m_{1}$ (we take
$m_{1}\leq m_{2}$).

From Eq.(\ref{Vpa}), we see that the effect is maximal when $p=1$
($|V(1,a)|^{2}=|V_{max}|^2 \approx \frac{(m_1 -m_2)^2}{16 m_1
m_2}$). $|V_{{\bf k}}|^{2}$ goes asymptotically to 1/2 when $a
\rightarrow \infty$ and goes to zero for large momenta (i.e. for
$|{\bf k}|^2\gg \frac{m_1^2 +m_2^2}{2}$) as $|V|^2\approx
\frac{\Delta m^2}{4 k^2}$.

In Figure 9.2 we show the fermion condensation density $|V_{\bf
k}|^2$ in function of ${|\bf k|}$ and for sample values of the
parameters $m_1$ and $m_2$.

%%%%%%%%%%%%%%%%%%%%%%%%%%%%%%%%%%%%%%%%%%%%%%%%%%%%%%%%%%%%%
\begin{figure}[t]
\setlength{\unitlength}{1mm} \vspace*{80mm} %%
%% dvips
\includegraphics{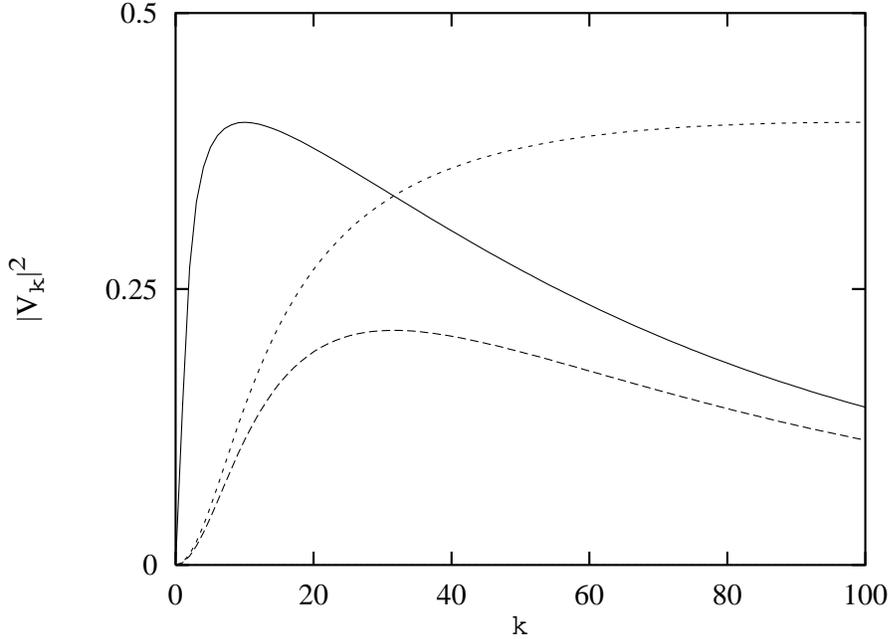}
\caption{The fermion condensation density $|V_{\bf k}|^2$ in
function of  ${|\bf k|}$ and for sample values of the parameters
$m_1$ and $m_2$. Solid line: $m_1=1 \;,\; m_2=100\;$; Long-dashed
line: $m_1=10 \;,\; m_2=100\;$; Short-dashed line: $m_1=10 \;,\;
m_2=1000$.} \vspace{.1cm}

\hrule
\end{figure}
%%%%%%%%%%%%%%%%%%%%%%%%%%%%%%%%%%%%%%%%%%%%%%%%%%%%%%%%%%%%%%%

 For the
currently known neutrino mixing parameters, i.e. $\Delta m \sim
10^{-3}eV$, the best observation scale is of the order of $k
\approx 10^{-3}eV$. However, experimentally observed neutrinos are
always in extremelly relativistic domain and therefore the value
of $|V_{\bf k}|^2$ is as small as  $|V_{\bf k}|^2 \sim 10^{-18}$.
Only for extremely low energies (like those in neutrino
cosmological background) the field-theoretical corrections might
be large and account for few percent. In this connection, we
observe that the non-perturbative field theory effects, in spite
of the small corrections they induce in the oscillation
amplitudes, nevertheless they may contribute in a specific and
crucial way in other physical contexts or phenomena. Indeed, as
shown in chapter 6, the mixing of neutrinos may specifically
contribute to the value of the cosmological constant exactly
because of the non-perturbative effects expressed by the non-zero
value of $|V_{\bf k}|^2$ \cite{cosmolog2003}.

\section{Conclusions}

In this chapter we have considered phenomenological aspects of the
quantum field theoretical formalism of flavor mixing. A crucial
point in this analysis is the disclosure of the fact that the
space for the mixed field states is unitarily inequivalent to the
state space where the unmixed field operators are defined. This is
a common feature with the QFT structure of mixing, which has
recently been established \cite{BHV98,Blasone:1999jb,BV95,JM011}.
%%hannabus,Fujii:1999xa,
The vacuum for the mixed fields turns out to be a generalized
SU(2) coherent state.

We have estimated the magnitude of the field-theoretical effect in
known mixed systems. We found that for most mixed systems both in
meson and neutrino sectors this effect is negligible. Only in
strongly mixed systems such as $\omega-\phi$ or $\eta-\eta'$, or
for very low-energy neutrino effects the corrections may be as
large as 5-20\% \cite{CJMV2003} and thus additional attention may
be needed to see if these systems can be used in oscillation
experiments. The non-perturbative vacuum effect is the most
prominent when the particles are produced at low momentum.

Moreover, we recall that the neutrino mixing may contribute to the
value of the cosmological constant because of the non-perturbative
effects.

\newpage

%%%%%%%%%%%%%%%%%%%%%%%%%%%%%%%%%%%%%%%%%%%%%%%%%%%%%%%%%%%%%%%%%%%%%%%%%%%%%
%%%%%%%%%%%%%%%%%%%%%%%%%%%%%%%%%%%%%%%%%%%%%%%%%%%%%%%%%%%%%%%%%%%%%%%%%%%%%
\appendix
\chapter{Double beta decay}

\medskip

Double beta decay, the rarest known nuclear decay process, can
occur in different modes: \bea\label{A1} &&
2\nu\beta\beta-\text{decay}:\qquad\;\qquad A(Z,N)\rightarrow
A(Z+2,N-2)+2 e^{-} + 2\bar{\nu_{e}} \\ \label{A2}&&
0\nu\beta\beta-\text{decay}:\qquad\;\qquad A(Z,N)\rightarrow
A(Z+2,N-2)+2 e^{-}\\\label{A3} &&
0\nu(2)\chi^{0}\beta\beta-\text{decay}:\qquad A(Z,N)\rightarrow
A(Z+2,N-2)+2 e^{-}+(2)\chi^{0}. \eea

While the two-neutrino mode Eq.(\ref{A1}) is allowed by the
Standard Model of particle physics, the neutrinoless mode
Eq.(\ref{A2}) requires violation of lepton number $\Delta L = 2$.
This mode is possible only if the neutrino is a Majorana particle.

Moreover, the study of double beta decay gives insight on the
coupling of the neutrino to hypothetical light neutral bosons,
named Majorons $\chi^{0}$. Such Majorons could be emitted in the
$0\nu(2)\chi^{0}\beta\beta-\text{decay}$, Eq.(\ref{A3}).

Neutrinoless double beta decay can probe not only a Majorana
neutrino mass, but various new scenarios beyond the Standard
Model, such as violation of Lorentz invariance, R-parity violating
supersymmetric model, R-parity conserving SUSY model, and
leptoquarks.

A sensitive double beta decay experiment is the HEIDELBERG--MOSCOW
experiment in the Gran Sasso Underground Laboratory that search
for double beta decay of \bea ^{76}Ge \rightarrow ^{76}Se + 2
e^{-}+ (2\bar{\nu_{e}}). \eea

\bigskip
\medskip

\chapter{Useful formulas for the generation of the flavor vacuum}

\medskip

Using the algebra Eq.(\ref{su(2)k1}) and the relations
Eq.(\ref{S3Spm}), we have:
$$\begin{array}{ll}
S_{+}^{{\bf k}}S_{-}^{{\bf k}}|0\rangle_{1,2} = S_{-}^{{\bf
k}}S_{+}^{{\bf k}}|0\rangle_{1,2} &(S_{+}^{{\bf
k}})^{2}(S_{-}^{{\bf k}})^{2}|0\rangle_{1,2} =
(S_{-}^{{\bf k}})^{2}(S_{+}^{{\bf k}})^{2}|0\rangle_{1,2} \\ \\
(S_{+}^{{\bf k}})^{2}S_{-}^{{\bf k}}|0\rangle_{1,2} = S_{-}^{{\bf
k}}(S_{+}^{{\bf k}})^{2}
|0\rangle_{1,2} +2S_{+}^{{\bf k}}|0\rangle_{1,2}\\ \\
(S_{-}^{{\bf k}})^{2}S_{+}^{{\bf k}}|0\rangle_{1,2} = S_{+}^{{\bf
k}}(S_{-}^{{\bf k}})^{2} |0\rangle_{1,2} +2S_{-}^{{\bf
k}}|0\rangle_{1,2}
\\ \\
(S_{+}^{{\bf k}})^{3}S_{-}^{{\bf k}}|0\rangle_{1,2} =
6(S_{+}^{{\bf k}})^{2}|0\rangle_{1,2}
& (S_{+}^{{\bf k}})^{4}S_{-}^{{\bf k}}|0\rangle_{1,2} =0\\ \\
(S_{+}^{{\bf k}})^{3}(S_{-}^{{\bf k}})^{2}|0\rangle_{1,2} =
6S_{-}^{{\bf k}}(S_{+}^{{\bf k}})^{2}|0\rangle_{1,2} &(S_{-}^{{\bf
k}})^{3}(S_{+}^{{\bf k}})^{2}|0\rangle_{1,2} =
6S_{+}^{{\bf k}}(S_{-}^{{\bf k}})^{2}|0\rangle_{1,2}\\ \\
(S_{+}^{{\bf k}})^{4}(S_{-}^{{\bf k}})^{2}|0\rangle_{1,2}
=24(S_{+}^{{\bf k}})^{2}|0\rangle_{1,2}
&(S_{+}^{{\bf k}})^{5}(S_{-}^{{\bf k}})^{2}|0\rangle_{1,2} =0\\ \\
S_{+}^{{\bf k}}S_{-}^{{\bf k}}(S_{+}^{{\bf
k}})^{2}|0\rangle_{1,2}= 4(S_{+}^{{\bf k}})^{2}|0\rangle_{1,2}
&S_{-}^{{\bf k}}S_{+}^{{\bf k}}(S_{-}^{{\bf
k}})^{2}|0\rangle_{1,2}=
4(S_{-}^{{\bf k}})^{2}|0\rangle_{1,2}\\ \\
S_{3}^{{\bf k}}S_{-}^{{\bf k}}S_{+}^{{\bf k}}|0\rangle_{1,2}=0
&S_{3}^{{\bf k}}(S_{+}^{{\bf k}})^{2}(S_{-}^{{\bf k}})^{2}|0\rangle_{1,2} = 0\\ \\
(S_{3}^{{\bf k}})^{n}S_{-}^{{\bf k}}|0\rangle_{1,2} = (-1)^{n}
S_{-}^{{\bf k}}|0\rangle_{1,2} &(S_{3}^{{\bf k}})^{n}(S_{-}^{{\bf
k}})^{2}|0\rangle_{1,2} =
(-2)^{n} (S_{-}^{{\bf k}})^{2}|0\rangle_{1,2}\\ \\
(S_{3}^{{\bf k}})^{n}S_{+}^{{\bf k}}|0\rangle_{1,2} = S_{+}^{{\bf
k}}|0\rangle_{1,2} &(S_{3}^{{\bf k}})^{n}(S_{+}^{{\bf
k}})^{2}|0\rangle_{1,2} =
2^{n} (S_{+}^{{\bf k}})^{2}|0\rangle_{1,2}\\ \\
S_{3}^{{\bf k}}S_{-}^{{\bf k}}(S_{+}^{{\bf
k}})^{2}|0\rangle_{1,2}= S_{-}^{{\bf k}}(S_{+}^{{\bf
k}})^{2}|0\rangle_{1,2} &S_{3}^{{\bf k}}S_{+}^{{\bf
k}}(S_{-}^{{\bf k}})^{2}|0\rangle_{1,2}= -S_{+}^{{\bf
k}}(S_{-}^{{\bf k}})^{2}|0\rangle_{1,2}
\end{array}$$
Use of the above relations gives Eq.(\ref{flavvac3}).

%%%%%%%%%%%%%%%%%%%%%%%%%%%%%%%%%%%%%%%%%%%%%%%%%%%%%%%%%%%%%%%%%%%%%%%%%%%%%

\bigskip
\medskip

\chapter{Wave function and $V_{{\bf k}}$}

\medskip

Wave functions:

\bea u_{{\bf k},i}^{r}(t)=u_{{\bf k},i}^{r} e^{-i\omega_{k,i} \;
t}=A_{i} \left(\begin{array}{c}
 \xi^{r} \\ \\
\frac{\bar{\sigma}\cdot\bar{k}}{\omega_{k,i}+m_{i}} \xi^{r}
\end{array}\right)e^{-i\omega_{k,i} \; t}\\
v_{{\bf k},i}^{r}(t)=v_{{\bf k},i}^{r} e^{i\omega_{k,i} \;
t}=A_{i} \left(\begin{array}{c}
\frac{\bar{\sigma}\cdot\bar{k}}{\omega_{k,i}+m_{i}} \xi^{r} \\ \\
\xi^{r}
\end{array}\right)e^{i\omega_{k,i} \; t}\eea
with \bea\non\xi_{1}=\left(\begin{array}{c} 1 \\ 0
\end{array}\right)\;\;,\;\; \xi_{2}=\left(\begin{array}{c} 0 \\ 1
\end{array}\right)\;\;\;,\;\;\; A_{i}\equiv
\left(\frac{\omega_{k,i}+m_{i}}{2\omega_{k,i}}\right)^{\frac{1}{2}}
\;\;\;,\;\;\; i=1,2\;,\;\; r=1,2\;.\\\eea
\bea v_{-{\bf k},1}^{1\dag} u_{{\bf k},2}^{1}= - v_{-{\bf
k},1}^{2\dag} u_{{\bf k},2}^{2}= A_{1}A_{2}
\left(\frac{-k_{3}}{\omega_{k,1}+m_{1}}
+\frac{k_{3}}{\omega_{k,2}+m_{2}} \right)\eea
\bea v_{-{\bf k},1}^{1\dag} u_{{\bf k},2}^{2}= \left(v_{-{\bf
k},1}^{2\dag} u_{{\bf k},2}^{1}\right)^{*} =A_{1}A_{2}
\left(\frac{-k_{1}+ik_{2}}{\omega_{k,1}+m_{1}}
+\frac{k_{1}-ik_{2}}{\omega_{k,2}+m_{2}}\right)\eea

Using the above relations Eq.(\ref{Vk}) is obtained.

Eq.(\ref{s+s-}) follows if one observe that
\bea\non (S_{-}^{{\bf k}})^{2} |0\rangle_{1,2}  &=& 2\Big[
\left(u^{1\dag}_{{\bf k},2}v^{2}_{-{\bf k},1}\right)
\left(u^{2\dag}_{{\bf k},2}v^{1}_{-{\bf k},1}\right)- \\\non
&-&\left(u^{1\dag}_{{\bf k},2}v^{1}_{-{\bf k},1}\right)
\left(u^{2\dag}_{{\bf k},2}v^{2}_{-{\bf k},1}\right) \Big]
\alpha^{1\dag}_{{\bf k},2}\beta^{2\dag}_{-{\bf k},1}
\alpha^{2\dag}_{{\bf k},2}\beta^{1\dag}_{-{\bf k},1}
|0\rangle_{1,2}=
\\ \non &=& 2|V_{{\bf
k}}|^{2}\;e^{2i(\omega_{k,1}+\omega_{k,2} ) t}\;
\alpha^{1\dag}_{{\bf k},2}\beta^{2\dag}_{-{\bf k},1}
\alpha^{2\dag}_{{\bf k},2}\beta^{1\dag}_{-{\bf k},1}
|0\rangle_{1,2}\;.\eea

%%%%%%%%%%%%%%%%%%%%%%%%%%%%%%%%%%%%%%%%%%%%%%%%%%%%%%%%%%%%%%%%%%%%%%%%%%%%%
\bigskip
\medskip
\chapter{The flavor vacuum}

\medskip

For the calculation  of $|0\rangle_{e,\mu}^{{\bf k}}$ it is useful
to choose ${\bf k}=(0,0,|{\bf k}|)$. In this reference frame the
operators $S_{+}^{{\bf k}}$, $S_{-}^{{\bf k}}$, $S_{3}^{{\bf k}}$
are written as follows:
\bea\non S_{+}^{{\bf k}}\equiv\sum_{{\bf k},r}S_{+}^{{\bf k},r}=
\sum_{r} \left( U^{*}_{{\bf k}}\; \alpha^{r\dag}_{{\bf k},1}
\alpha^{r}_{{\bf k},2} - \epsilon^{r}\; V^{*}_{{\bf k}}\;
\beta^{r}_{-{\bf k},1} \alpha^{r}_{{\bf k},2} + \epsilon^{r}\;
V_{{\bf k}}\;\alpha^{r\dag}_{{\bf k},1}\beta^{r\dag}_{-{\bf k},2}
+ U_{{\bf k}}\; \beta^{r}_{-{\bf k},1}\beta^{r\dag}_{-{\bf
k},2}\right)\\\eea \bea\non S_{-}^{{\bf
k}}\equiv\sum_{k,r}S_{-}^{k,r}= \sum_{r} \left( U_{{\bf k}}\;
\alpha^{r\dag}_{{\bf k},2} \alpha^{r}_{{\bf k},1} + \epsilon^{r}\;
V^{*}_{{\bf k}}\; \beta^{r}_{-{\bf k},2} \alpha^{r}_{{\bf k},1} -
\epsilon^{r}\; V_{{\bf k}}\;\alpha^{r\dag}_{{\bf
k},2}\beta^{r\dag}_{-{\bf k},1} + U_{{\bf k}}^{*}\;
\beta^{r}_{-{\bf k},2}\beta^{r\dag}_{-{\bf k},1}\right)\\\eea
\bea\non S_{3}^{{\bf k}}\equiv \sum_{{\bf k},r} S_{3}^{{\bf k},r}
=\frac{1}{2}\sum_{{\bf k},r}\left(\alpha^{r\dag}_{{\bf
k},1}\alpha^{r}_{k,1} -\beta^{r\dag}_{-{\bf k},1}\beta^{r}_{-{\bf
k},1} -\alpha^{r\dag}_{{\bf k},2}\alpha^{r}_{{\bf k},2} +
\beta^{r\dag}_{-{\bf k},2}\beta^{r}_{-{\bf k},2} \right)\;,\\ \eea
where $U_{{\bf k}}$, $V_{{\bf k}}$ have been defined in
Eqs.(\ref{Vk2})-(\ref{Vk}) and $\epsilon^{r}=(-1)^{r}$. It is easy
to show that the $su(2)$ algebra holds for $S_{\pm}^{{\bf k},r}$
and $S_{3}^{{\bf k},r}$, which means that the $su_{{\bf k}}(2)$
algebra given in Eqs.(\ref{su(2)k1}) splits into $r$ disjoint
$su_{{\bf k},r}(2)$ algebras. Using the Gaussian decomposition,
$|0\rangle_{e,\mu}^{{\bf k}}$ can be written as
\bea|0\rangle_{e,\mu}^{{\bf k}} = \prod_{r} exp(-tan\theta \;
S_{+}^{{\bf k},r}) exp(-2 ln \: cos\theta \; S_{3}^{{\bf k},r})
\;exp(tan\theta \; S_{-}^{{\bf k},r})|0\rangle_{1,2} \eea
where $0\leq \theta < \frac{\pi}{2}$.
The final expression for $|0\rangle_{e,\mu}^{{\bf k}}$ in terms of
$S^{{\bf k},r}_{\pm}$ and $S^{{\bf k},r}_{3}$ is then
\bea|0\rangle_{e,\mu}^{{\bf k}}= \prod_{r}\left[ 1 + \sin\theta
\cos\theta \left(S_{-}^{{\bf k},r} - S_{+}^{{\bf k},r}\right)
-\sin^{2}\theta \; S_{+}^{{\bf k},r}S_{-}^{{\bf
k},r}\right]|0\rangle_{1,2}\;,\eea
from which we finally obtain Eq.(\ref{0emu}).

%%%%%%%%%%%%%%%%%%%%%%%%%%%%%%%%%%%%%%%%%%%%%%%%%%%%%%%%%%%%%%%%%%%%%%%%%%%%%%
\chapter{Orthogonality of flavor vacuum states and flavor
states at different times}

\smallskip

The product of two vacuum states at different times $t\neq t'$ (we
put for simplicity $t'=0$) is
\begin{eqnarray}
_\flav\langle0|0(t)\rangle_\flav = \prod_{k} C_{k}^{2}(t)=
e^{2\sum_{k}ln C_{k}(t)}
\end{eqnarray}
with
\begin{eqnarray}\nonumber C_{k}(t) &\equiv&
(1-\sin^{2}\theta\;|V_{k}|^{2})^{2}
+\;2\;\sin^{2}\theta\;\cos^{2}\theta\; |V_{k}|^{2} \;  e^{-i
(\omega_{k,2} + \omega_{k,1}) t}+
 \\
\label{bit} &+&\;\sin^{4}\theta \;|V_{k}|^{2} \;|U_{k}|^{2}
\left(e^{- 2 i \omega_{k,1} t} \; + \;e^{-2 i \omega_{k,2} t}\;
\right) +
 \\ \non
&+& \sin^{4}\theta \; |V_{k}|^{4} \; e^{- 2 i (\omega_{k,2} +
\omega_{k,1}) t}\;.
\end{eqnarray}
In the infinite volume limit we obtain (note that $|C_{k}(t)| \leq
1$ for any value of $k$, $t$, and of the parameters $\theta$, $
m_1$, $m_2$ ):
\begin{equation}\label{lim}
\lim_{V \rightarrow \infty}\;_\flav\langle0|0(t)\rangle_\flav =
\lim_{V \rightarrow \infty}\; exp\left[\frac{2V}{(2\pi)^{3}}\int
d^{3}k \; \left( ln\;|C_{k}(t)|\; + \; i \alpha_{k}(t)
\right)\right]=0
\end{equation}
 with $|C_{k}(t)|^{2}= Re[C_{k}(t)]^{2} + Im[C_{k}(t)]^{2} $ and
 $\alpha_{k}(t)= \tan^{-1}\left(Im[C_{k}(t)]/ Re[C_{k}(t)]
 \right)$.

Thus we have orthogonality of the vacua at different times.

Now we can show the orthogonality of flavor states at different
times.

We define the electron neutrino state at time $t$ with momentum
${\bf k}$ as \bea |\nu_{{\bf k},e}(t)\rangle = \alpha^{r
\dag}_{{\bf k},e}(t) |0(t)\rangle_\flav. \eea The flavor vacuum is
explicitly given by\footnote{To be precise, the mass vacuum is to
be understood as $|0\rangle_\mass= |0\rangle^{{\bf
k}_{1}}_\mass\bigotimes |0\rangle^{{\bf k}_{2}}_\mass\bigotimes
|0\rangle^{{\bf k}_{3}}_\mass..... $} \bea |0(t)\rangle_\flav=
\prod_{\bf p}G^{-1}_{{\bf p},\theta}(t)|0\rangle_\mass,\eea then,
we have \bea \langle \nu_{{\bf k},e}(0)|\nu_{{\bf k},e}(t)\rangle
&=&_\flav\langle0|\alpha^{r }_{{\bf k},e}(0)\alpha^{r \dag}_{{\bf
k},e}(t)|0(t)\rangle_\flav
\\
\non &=& \prod_{\bf p}\prod_{\bf q}\; _\mass\langle 0|G_{{\bf
p},\theta}(0)\alpha^{r }_{{\bf k},e}(0)\alpha^{r \dag}_{{\bf
k},e}(t)G^{-1}_{{\bf q},\theta}(t)|0\rangle_\mass. \eea With ${\bf
p}\neq{\bf q}$ the mixing generators commute, then we put ${\bf
p}={\bf q}$:
 \bea \langle \nu_{{\bf k},e}(0)|\nu_{{\bf
k},e}(t)\rangle = \prod_{\bf p}\;_\mass\langle 0|G_{{\bf
p},\theta}(0)\alpha^{r }_{{\bf k},e}(0)\alpha^{r \dag}_{{\bf
k},e}(t)G^{-1}_{{\bf p},\theta}(t)|0\rangle_\mass. \eea $\alpha^{r
\dag}_{{\bf k},e}$ acts only on vacuum with momentum ${\bf k}$,
then \bea \non \langle \nu_{{\bf k},e}(0)|\nu_{{\bf
k},e}(t)\rangle &\propto&_\flav\langle 0^{\bf k}|\alpha^{r }_{{\bf
k},e}(0)\alpha^{r \dag}_{{\bf k},e}(t)|0^{\bf k}(t)\rangle_\flav
\prod_{\bf p\neq k}\;_\mass\langle 0 |G_{{\bf
p},\theta}(0)G^{-1}_{{\bf p},\theta}(t)|0\rangle_\mass
\\ &=&
 _\flav\langle 0^{\bf k}|\alpha^{r }_{{\bf
k},e}(0)\alpha^{r \dag}_{{\bf k},e}(t)|0^{\bf
k}(t)\rangle_\flav\;_\flav\langle0|0(t)\rangle_\flav.\eea By using
the Eq.(\ref{lim}), in the infinite volume limit we obtain the
orthogonality of flavor states at different times.

\chapter{Useful relations for three flavors fermion
mixing}

\smallskip

We work in the frame $k=(0,0,|k|)$ and for simplicity we omit the
$k$ and the helicity indices.

\bea\non\left\{\begin{array}{l}
G_{23}^{-1} \alpha_{1} G_{23}=\alpha_{1} \\ \\
G_{13}^{-1} \alpha_{1} G_{13}=
    c_{13}\;\alpha_{1}\;+e^{i\delta}\;s_{13}\;
\left( U_{13}^{*}\; \alpha_{3}\;
+\epsilon_{r}\;V_{13}\; \beta^{\dag}_{3}\right) \\ \\
G_{12}^{-1} \alpha_{1} G_{12}=
    c_{12}\;\alpha_{1}\;+\;s_{12}\;
\left( U_{12}^{*}\; \alpha_{2}\;+\epsilon_{r}\;V_{12}\;
\beta^{\dag}_{2}\right)
\end{array} \right.\eea

\smallskip

%%%%%%%%%%%%%%%%%%%%%%%%%%%%%%% 2
\bea\non\left\{\begin{array}{l} G_{23}^{-1} \alpha_{2} G_{23}=
    c_{23}\;\alpha_{2}\;+\;s_{23}\;
\left( U_{23}^{*}\; \alpha_{3}\;
+\epsilon_{r}\;V_{23}\; \beta^{\dag}_{3}\right) \\ \\
G_{13}^{-1} \alpha_{2} G_{13}=\alpha_{2} \\ \\
G_{12}^{-1} \alpha_{2} G_{12}=
    c_{12}\;\alpha_{2}\;-\;s_{12}\;
\left( U_{12}\; \alpha_{1}\;-\epsilon_{r}\;V_{12}\;
\beta^{\dag}_{1}\right)
\end{array}\right.\eea

\smallskip

%%%%%%%%%%%%%%%%%%%%%%%%%%% 3
\bea\non\left\{\begin{array}{l} G_{23}^{-1} \alpha_{3} G_{23}=
    c_{23}\;\alpha_{3}\;-\;s_{23}\;
\left( U_{12}\; \alpha_{2}\;
-\epsilon_{r}\;V_{12}\; \beta^{\dag}_{2}\right) \\ \\
G_{13}^{-1} \alpha_{3} G_{13}=
    c_{13}\;\alpha_{3}\;-e^{-i\delta}\;s_{13}\;
\left( U_{13}\; \alpha_{1}\;-\epsilon_{r}\;V_{13}\;
\beta^{\dag}_{1}\right)
\\ \\
G_{12}^{-1} \alpha_{3} G_{12}=\alpha_{3}
\end{array}\right.\eea

\smallskip

%%%%%%%%%%%%%%%%%%%%%%%%%%%%%%% 4
\bea\non\left\{\begin{array}{l}
G_{23}^{-1} \beta^{\dag}_{1} G_{23}=\beta^{\dag}_{1} \\ \\
G_{13}^{-1} \beta^{\dag}_{1} G_{13}=
    c_{13}\;\beta^{\dag}_{1}\;+e^{i\delta}\;s_{13}\;
\left( U_{13}\; \beta^{\dag}_{3}\;-\epsilon_{r}\;V_{13}^{*}\;
\alpha_{3}\right) \\ \\
G_{12}^{-1} \beta^{\dag}_{1} G_{12}=
    c_{12}\;\beta^{\dag}_{1}\;+\;s_{12}\;
\left( U_{12}\; \beta^{\dag}_{2}\;-\epsilon_{r}\;V_{12}^{*}\;
\alpha_{2}\right)
\end{array}\right.\eea

\smallskip

%%%%%%%%%%%%%%%%%%%%%%%%%%%%%%%
\bea\non\left\{\begin{array}{l} G_{23}^{-1} \beta^{\dag}_{2}
G_{23}=
    c_{23}\;\beta^{\dag}_{2}\;+\;s_{23}\;
\left( U_{23}\; \beta^{\dag}_{3}\;
-\epsilon_{r}\;V_{23}^{*}\; \alpha_{3}\right)  \\ \\
G_{13}^{-1} \beta^{\dag}_{2} G_{13}=\beta^{\dag}_{2} \\ \\
G_{12}^{-1} \beta^{\dag}_{2} G_{12}=
    c_{12}\;\beta^{\dag}_{2}\;-\;s_{12}\;
\left( U_{12}^{*}\; \beta^{\dag}_{1}\; +\epsilon_{r}\;V_{12}^{*}\;
\alpha_{1}\right)
\end{array}\right.\eea

\smallskip

%%%%%%%%%%%%%%%%%%%%%%%%%%%%%%% 6
\bea\non\left\{\begin{array}{l} G_{23}^{-1} \beta^{\dag}_{3}
G_{23}=
    c_{23}\;\beta^{\dag}_{3}\;-\;s_{23}\;
\left( U_{23}^{*}\; \beta^{\dag}_{2}\;
+\epsilon_{r}\;V_{23}^{*}\; \alpha_{2}\right) \\ \\
G_{13}^{-1} \beta^{\dag}_{3} G_{13}=
    c_{13}\;\beta^{\dag}_{3}\;-e^{-i\delta}\;s_{13}\;
\left( U_{13}^{*}\; \beta^{\dag}_{1}\;
+\epsilon_{r}\;V_{13}^{*}\; \alpha_{1}\right) \\ \\
G_{12}^{-1} \beta^{\dag}_{3} G_{12}=\beta^{\dag}_{3}
\end{array}\right.\eea

$$ $$
with
\bea
V^{k}_{ij}=|V^{k}_{ij}|\;e^{i(\omega_{k,j}+\omega_{k,i})t}\;\;\;\;,\;\;\;\;
U^{k}_{ij}=|U^{k}_{ij}|\;e^{i(\omega_{k,j}-\omega_{k,i})t}\eea
\bea
|U^{k}_{ij}|=\left(\frac{\omega_{k,i}+m_{i}}{2\omega_{k,i}}\right)
^{\frac{1}{2}}
\left(\frac{\omega_{k,j}+m_{j}}{2\omega_{k,j}}\right)^{\frac{1}{2}}
\left(1+\frac{k^{2}}{(\omega_{k,i}+m_{i})(\omega_{k,j}+m_{j})}\right)\eea
\bea
|V^{k}_{ij}|=\left(\frac{\omega_{k,i}+m_{i}}{2\omega_{k,i}}\right)
^{\frac{1}{2}}
\left(\frac{\omega_{k,j}+m_{j}}{2\omega_{k,j}}\right)^{\frac{1}{2}}
\left(\frac{k}{(\omega_{k,j}+m_{j})}-\frac{k}{(\omega_{k,i}+m_{i})}\right)\eea
where $i,j=1,2,3$ and $j>i$, and

\bea |U^{k}_{ij}|^{2}+|V^{k}_{ij}|^{2}=1 \;\;\;\;\;,\;\;\;\;
i=1,2,3 \;\; j>i \eea
$$\left(V^{k}_{23}V^{k*}_{13}+U^{k*}_{23}U^{k}_{13}\right)
= U^{k}_{12}\;\;\;\;\;,\;\;\;\;\;
\left(V^{k}_{23}U^{k*}_{13}-U^{k*}_{23}V^{k}_{13}\right) =-
V^{k}_{12}$$
$$\left(U^{k}_{12}U^{k}_{23}-V^{k*}_{12}V^{k}_{23}\right)
= U^{k}_{13}\;\;\;\;\;,\;\;\;\;\;
\left(U^{k}_{23}V^{k}_{12}+U^{k*}_{12}V^{k}_{23}\right) =
V^{k}_{13}$$
$$\left(V^{k*}_{12}V^{k}_{13}+U^{k*}_{12}U^{k}_{13}\right)
= U^{k}_{23}\;\;\;\;\;,\;\;\;\;\;
\left(V^{k}_{12}U^{k}_{13}-U^{k}_{12}V^{k}_{13}\right) =-
V^{k}_{23}\;.$$

\chapter{Anti-neutrino oscillation formulas}

\smallskip

If we consider an initial electron anti-neutrino state defined as
$|\overline{\nu}_e\ran \equiv \bt_{{\bf k},e}^{r\dag}(0)
|0\ran_{f}$, we obtain the anti-neutrino oscillation formulas as
\bea  {\cal Q}^{\bar e}_{{\bf k},e}(t)
 \,= - {\cal Q}^{ e}_{{\bf k},e}(t)\,,
\eea

\bea\non {\cal Q}^{\bar e}_{{\bf k},\mu}(t) &=&2 J_{\CP}
 \Big[|U_{12}^{\bf k}|^2\, \sin(2\De_{12}^{\bf k}t)
- |V_{12}^{\bf k}|^2\, \sin(2\Om_{12}^{\bf k} t) + (|U_{12}^{\bf
k}|^2 - |V_{13}^{\bf k}|^2 ) \sin(2\De_{23}^{\bf k}t)
\\ \non
&+& (|V_{12}^{\bf k}|^2 - |V_{13}^{\bf k}|^2 ) \sin(2\Om_{23}^{\bf
k}t)
  - |U_{13}^{\bf k}|^2\,
\sin(2\De_{13}^{\bf k}t)+ |V_{13}^{\bf k}|^2\, \sin(2\Om_{13}^{\bf
k}t)\Big]
\\ \non
&-&\, \cos^{2}\te_{13} \sin\te_{13}
\Big[\cos\de\sin(2\te_{12})\sin(2\te_{23}) + 4
\cos^2\te_{12}\sin\te_{13}\sin^2\te_{23}\Big]\\\non &\times &
\Big[|U_{13}^{\bf k}|^2\sin^{2} \lf(\De_{13}^{\bf k} t \ri) +
|V_{13}^{\bf k}|^2\ \sin^{2} \lf( \Om_{13}^{\bf k} t \ri)\Big]
\\ \non
& +& \cos^{2}\te_{13}\sin\te_{13}
 \Big[\cos\de\sin(2\te_{12})\sin(2\te_{23}) -
4 \sin^2\te_{12}\sin\te_{13}\sin^2\te_{23}\Big]\\\non &\times &
\Big[|U_{23}^{\bf k}|^2\ \sin^{2} \lf( \De_{23}^{\bf k} t \ri)
 + |V_{23}^{\bf k}|^2\
\sin^{2} \lf( \Om_{23}^{\bf k} t \ri)\Big]
\\ \non
& -&\cos^{2}\te_{13} \sin(2\te_{12}) \Big[ (\cos^2\te_{23} -
\sin^2\te_{23}\sin^2\te_{13})\sin(2\te_{12})\\\non
&+&\cos\de\cos(2\te_{12})\sin\te_{13}\sin(2\te_{23})\Big]
 \Big[|U_{12}^{\bf k}|^2\ \sin^{2} \lf(\De_{12}^{\bf k} t
\ri) + |V_{12}^{\bf k}|^2\ \sin^{2} \lf( \Om_{12}^{\bf k} t
\ri)\Big]\, ,\\ \eea

\bea \non {\cal Q}^{\bar e}_{{\bf k},\tau}(t) &=& - 2 J_{\CP}
 \Big[|U_{12}^{\bf k}|^2\ \sin(2\De_{12}^{\bf k}t)
- |V_{12}^{\bf k}|^2\, \sin(2\Om_{12}^{\bf k} t) + (|U_{12}^{\bf
k}|^2\, - |V_{13}^{\bf k}|^2 ) \sin(2\De_{23}^{\bf k}t)
\\ \non
&+& (|V_{12}^{\bf k}|^2\, - |V_{13}^{\bf k}|^2 )
\sin(2\Om_{23}^{\bf k}t)
 \, - |U_{13}^{\bf k}|^2\,
\sin(2\De_{13}^{\bf k}t)+ |V_{13}^{\bf k}|^2\, \sin(2\Om_{13}^{\bf
k}t)\Big]
\\ \non
&+& \cos^{2}\te_{13} \sin\te_{13}
\Big[\cos\de\sin(2\te_{12})\sin(2\te_{23}) -4
\cos^2\te_{12}\sin\te_{13}\cos^2\te_{23} \Big]\\\non &\times&
\Big[|U_{13}^{\bf k}|^2\, \sin^{2} \lf( \De_{13}^{\bf k} t \ri) +
|V_{13}^{\bf k}|^2\, \sin^{2} \lf( \Om_{13}^{\bf k}  t \ri)\Big]
\\ \non
& -&\cos^{2}\te_{13}\sin\te_{13}
 \Big[\cos\de\sin(2\te_{12})\sin(2\te_{23}) +
  4 \sin^2\te_{12}\sin\te_{13}\cos^2\te_{23}\Big]\\\non &\times &
  \Big[|U_{23}^{\bf k}|^2 \,
\sin^{2} \lf(\De_{23}^{\bf k} t \ri) +
 |V_{23}^{\bf k}|^2\,
\sin^{2} \lf(\Om_{23}^{\bf k}  t \ri)\Big]
\\ \non
&-& \cos^{2}\te_{13} \sin(2\te_{12}) \Big[ (\sin^2\te_{23} -
\sin^2\te_{13}\cos^2\te_{23})\sin(2\te_{12})
\\\non &-&\cos\de\cos(2\te_{12})\sin\te_{13}\sin(2\te_{23})\Big]
\Big[|U_{12}^{\bf k}|^2\, \sin^{2} \lf( \De_{12}^{\bf k} t \ri) +
|V_{12}^{\bf k}|^2\, \sin^{2} \lf(\Om_{12}^{\bf k} t \ri)\Big]\,
.\\ \eea

\chapter{Useful formulas for the generation of the
mixing matrix }

\smallskip

In deriving the ${\cal U}_i$ mixing matrices of Sections 3.2 and
3.3, we use the following relationships
\bea [\nu_{1}^{\al}(x),L_{12}]=\nu_{2}^{\al}(x)e^{-i\de_{12}}\;,
\;\;\; [\nu_{1}^{\al}(x),L_{23}]=0\;, \;\;\;
[\nu_{1}^{\al}(x),L_{13}]=\nu_{3}^{\al}(x)e^{-i\de_{13}}\;, \;
\eea \bea
[\nu_{2}^{\al}(x),L_{12}]=-\nu_{1}^{\al}(x)e^{i\de_{12}}\;,\;\;\;
[\nu_{2}^{\al}(x),L_{23}]=\nu_{3}^{\al}(x)e^{-i\de_{23}}\;, \;\;\;
[\nu_{2}^{\al}(x),L_{13}]=0\;, \; \eea\bea
[\nu_{3}^{\al}(x),L_{12}]=0\;, \;
[\nu_{3}^{\al}(x),L_{23}]=-\nu_{2}^{\al}(x)e^{i\de_{23}}\;, \;\;\;
[\nu_{3}^{\al}(x),L_{13}]=- \nu_{1}^{\al}(x)e^{i\de_{13}}\;,
\;\;\; \eea and \bea\ &&G_{23}^{-1}(t)\nu_{1}^{\al}(x)G_{23}(t)=
\nu_{1}^{\al}(x)\;,\;
\\
&&G_{13}^{-1}(t)\nu_{1}^{\al}(x)G_{13}(t)=
\nu_{1}^{\al}(x)c_{13}+\nu_{3}^{\al}(x)e^{-i\de_{13}}s_{13}\;,\;
\\
&&G_{12}^{-1}(t)\nu_{1}^{\al}(x)G_{12}(t)=
\nu_{1}^{\al}(x)c_{12}+\nu_{2}^{\al}(x)e^{-i\de_{12}}s_{12}\;,\;
\eea

\bea &&G_{23}^{-1}(t)\nu_{2}^{\al}(x)G_{23}(t)=
\nu_{2}^{\al}(x)c_{23}+\nu_{3}^{\al}(x)e^{-i\de_{23}}s_{23}\;,\;
\\
&&G_{13}^{-1}(t)\nu_{2}^{\al}(x)G_{13}(t)= \nu_{2}^{\al}(x)\;,\;
\\
&&G_{12}^{-1}(t)\nu_{2}^{\al}(x)G_{12}(t)=
\nu_{2}^{\al}(x)c_{12}-\nu_{1}^{\al}(x)e^{i\de_{12}}s_{12}\;,\;
\eea

\bea &&G_{23}^{-1}(t)\nu_{3}^{\al}(x)G_{23}(t)=
\nu_{3}^{\al}(x)c_{23}-\nu_{2}^{\al}(x)e^{i\de_{23}}s_{23}\;,\;
\\
&&G_{13}^{-1}(t)\nu_{3}^{\al}(x)G_{13}(t)=
\nu_{3}^{\al}(x)c_{13}-\nu_{1}^{\al}(x)e^{i\de_{13}}s_{13}\;,\;
\\
&&G_{12}^{-1}(t)\nu_{3}^{\al}(x)G_{12}(t)= \nu_{3}\; \eea

\chapter{Arbitrary mass parameterization and physical
quantities}

\smallskip

In Ref. \cite{fujii1,fujii2} it was noticed that  expanding the
flavor fields in the same basis as the (free) fields with definite
masses (cf. Eq.(\ref{exnue123})) is actually a special choice, and
that a more general possibility exists. In other words, in the
expansion Eq.(\ref{exnue123}) one could  use eigenfunctions with
arbitrary masses $\mu_\sigma$, and therefore not necessarily the
same as the masses which appear in the Lagrangian.  On this basis,
the authors of Ref.\cite{fujii1,fujii2} have generalized the
Eq.(\ref{exnue123}) by writing the flavor fields as
\bea\label{exnuf2} \nu_{\sigma}(x)     &=& \sum_{r} \int
\frac{d^3{\bf k}}{(2\pi)^{\frac{2}{3}}} \left[ u^{r}_{{\bf
k},\sigma} {\widetilde \alpha}^{r}_{{\bf k},\sigma}(t) +
v^{r}_{-{\bf k},\sigma} {\widetilde \beta}^{r\dag}_{-{\bf
k},\sigma}(t) \right] e^{i {\bf k}\cdot{\bf x}} , \eea
where $u_{\sigma}$ and $v_{\sigma}$ are the helicity
eigenfunctions with mass $\mu_\sigma$. We denote by a tilde the
generalized flavor operators introduced in
Ref.\cite{fujii1,fujii2} in order to distinguish them from the
ones in Eq.(\ref{exnue123}).  The expansion Eq.(\ref{exnuf2}) is
more general than the one in Eq.(\ref{exnue123}) since the latter
corresponds to the particular choice $\mu_e\equiv m_1$, $\mu_\mu
\equiv m_2$, $\mu_\tau\equiv m_3$. Of course, the flavor fields in
Eq.(\ref{exnuf2}) and Eq.(\ref{exnue123}) are the same fields. The
relation, given in Ref.\cite{fujii1,fujii2}, between the general
flavor operators and the flavor operators Eq.(\ref{BVoper}) is
\bea\label{FHYBVa} &&\left(\begin{array}{c}
{\widetilde \alpha}^{r}_{{\bf k},\sigma}(t)\\
{\widetilde \beta}^{r\dag}_{{-\bf k},\sigma}(t)
\end{array}\right)
\;=\; J^{-1}_{\mu_{\sigma}}(t)  \left(\begin{array}{c}
\alpha^{r}_{{\bf k},\sigma}(t)\\ \beta^{r\dag}_{{-\bf
k},\sigma}(t)
\end{array}\right)J_{\mu_{\sigma}}(t) ~~,
\\ [2mm]\label{FHYBVb}
&&J_{\mu_{\sigma}}(t)\,=\, \prod_{{\bf k}, r}\, \exp\left\{ i
\mathop{\sum_{(\sigma,j)}} \xi_{\sigma,j}^{\bf k}\left[
\alpha^{r\dag}_{{\bf k},\sigma}(t)\beta^{r\dag}_{{-\bf
k},\sigma}(t) + \beta^{r}_{{-\bf k},\sigma}(t)\alpha^{r}_{{\bf
k},\sigma}(t) \right]\right\}\,, \eea
with $(\sigma,j)=(e,1) , (\mu,2), (\tau,3)$, $\xi_{\sigma,j}^{\bf
k}\equiv (\chi^{\bf k}_\sigma - \chi^{\bf k}_j)/2$ and
$\cot\chi^{\bf k}_\sigma = |{\bf k}|/\mu_\sigma$, $\cot\chi_j^{\bf
k} = |{\bf k}|/m_j$. For $\mu_\si\equiv m_j$, one has
$J_{\mu_{\sigma}}(t)=1$.

As already noticed in Ref.\cite{Blasone:1999jb}, the flavor charge
operators are the Casimir operators for the Bogoliubov
transformation (\ref{FHYBVa}), i.e. they are free from arbitrary
mass parameters : ${\wti Q}_\si(t) =Q_\si(t) $. This is obvious
also from the fact that they can be expressed in terms of flavor
fields (see Ref.\cite{comment}).

Physical quantities should not carry any dependence on the
$\mu_\si$: in the two--flavor case, it has been shown
\cite{Blasone:1999jb} that the expectation values of the flavor
charges on the neutrino states are free from the arbitrariness.
For three generations, the question is more subtle due to the
presence of the CP violating phase. Indeed, in Ref.\cite{fujii2}
it has been found that the corresponding generalized quantities
depend on the arbitrary mass parameters.

In order to understand better the nature of such a dependence, we
consider the identity:
\bea\lab{gen1}\non
 \langle {\ti \psi}|{\ti Q}_\si(t)| {\ti
\psi}\ran  &=& \langle \psi| J(0) \, Q_\si(t)\, J^{-1}(0)|
\psi\ran \, =\, \\&=&\langle \psi|  Q_\si(t) | \psi\ran + \langle
\psi| \lf[J(0) , Q_\si(t)\ri]\, J^{-1}(0)|  \psi\ran\,. \eea
valid on any vector $|\psi\ran$ of the flavor Hilbert space (at
$t=0$). From the explicit expression for $J(0)$ we see that the
commutator $\lf[J(0) , Q_\si(t)\ri]$ vanishes for $\mu_\rho=m_j$ ,
$(\rho,j)=(e,1) , (\mu,2), (\tau,3)$.

It is thus tempting to define the (effective) physical flavor
charges as:
\bea {\wti {Q}}^{phys}_{\si}(t)\, \equiv \,{Q}_{\si}(t) -
J^{-1}(0)\lf[J(0) , Q_\si(t)\ri]\,= \, J^{-1}(0) \, {Q}_{\si}(t)\,
J(0),
 \eea
such that for example:
\bea \langle {\ti \nu_\rho}|{\wti {Q}}^{phys}_{\si}(t) \,| {\ti
\nu_\rho}\ran \, =\,\langle \nu_\rho| Q_\si(t) |  \nu_\rho\ran.
 \eea

It is clear that the operator ${\wti {Q}}^{phys}_{\si}(t)$ does
depend on the arbitrary mass parameters and this dependence is
such to compensate the one arising from the flavor states. The
choice of physical quantities (flavor observables) as those not
depending on the arbitrary mass parameters is here adopted,
although different possibilities are explored by other authors,
see Refs.\cite{fujii2,JM01,JM011}.

\chapter{Temporal component of spinorial derivative}

\smallskip

For the sake of completeness, we derive the 0th component
spinorial derivative used to get the results (\ref{sttress energy
tensor esplicito}). We start from the Cartan Equations (\ref{eq
forma di connessione}) writing the tetrads one-forms in the FRW
metric (\ref{metrica di friedmann}) as in (\ref{tetradi di
Friedmann}).

The exterior derivative of $e^{a}$ is
\begin{eqnarray}\label{differenziale esterno tetradi}
d e^{0}&=& dt\wedge dt=0\\ d e^{1}&=& \frac{\dot{a}}{\rho}\;
dt\wedge d x^{1}= \frac{\dot{a}}{a}\; e^{0} \wedge e^{1}\\ d
e^{2}&=& \dot{a}r\; dt\wedge d \theta +a\; dr\wedge d \theta =
\frac{\dot{a}}{a}\; e^{0} \wedge e^{2}+ \frac{\rho}{a r}\; e^{1}
\wedge e^{2}\\\nonumber d e^{3}&=& \dot{a}r\sin(\theta)\; dt\wedge
d \phi +a\sin(\theta) \; dr\wedge d \phi+ a r \cos(\theta)\;
d\theta\wedge d \phi=\\ && = \frac{\dot{a}}{a}\; e^{0} \wedge
e^{3}+ \frac{\rho}{a r}\; e^{1} \wedge e^{3}+\frac{\tan(\theta)}{a
r}\;e^{2} \wedge e^{3}.
\end{eqnarray}
Since the connection forms are antisymmetric in the tetradic
(latin) indexes Eq. (\ref{eq forma di connessione}) gives
\begin{eqnarray}\label{espressione delle unoforme di connessione}
\omega^{0}_{0}&=&0\\ \omega^{0}_{i}&=&-\frac{\dot{a}}{a} \;
e^{i}={\cal H}\;e^{i}\qquad i=1,2,3\\
\omega^{1}_{j}&=&\frac{\rho}{a r}\;e^{j}\qquad \qquad \qquad
j=2,3\\ \omega^{2}_{3}&=&\frac{\tan(\theta)}{a r}\;e^{3}\\
\omega^{i}_{i}&=&0.
\end{eqnarray}
Using the definition of spinorial derivative and spinorial
connection given in (\ref{der spin}), we have
\begin{equation}\label{derivata covariante spinoriale}
 D_{\mu}=\partial_{\mu}+\frac{1}{4}[\gamma_{i},\gamma_{0}]\;\omega^{i 0}_{\mu}+
 \frac{1}{4}[\gamma_{1},\gamma_{j}]\;\omega^{1 j}_{\mu}+
 \frac{1}{4}[\gamma_{2},\gamma_{3}]\;\omega^{2 3}_{\mu},
\end{equation}
where we have used the antisymmetry of the commutators. The 0th
component of Eq.(\ref{derivata covariante spinoriale}) gives
 Eq.(\ref{derivata covariante spinoriale componente
zero}).

\chapter{Orthogonality between mass and flavor vacua in boson
mixing}

\smallskip

We calculate here  $_{1,2}\langle 0| 0(\te , t)\rangle _\AB\ $. In
the following we work at finite volume (discrete ${\bf k}$) and
suppress the time dependence of the operators when $t=0$. Let us
first observe that \bea | 0(\te , t)\ran_\AB= e^{i H t}|0(\te
,0)\ran_\AB,\eea with \bea H=\sum\limits_{i=1}^2 \sum\limits_{\bf
k}\, \om_{ k,i}\left( a_{{\bf k},i}^{\dagger }a_{{\bf
k},i}+b_{-{\bf k},i}^{\dagger }b_{- {\bf k},i}\right).\eea

Thus we have \bea_{1,2}\langle 0| 0(\te , t)\rangle _\AB\,= \,
_{1,2}\langle 0| 0(\te , 0)\rangle _\AB.\eea

We then define \bea \prod\limits_{\bf k}f_0^{\bf k}(\te) \equiv
\prod \limits_{\bf k} \, _{1,2}\lan 0|  G_{{\bf
k},\te}^{-1}(0)|0\ran_{1,2}\eea and observe that
\bea\non \frac{d}{d\te} f_0^{\bf k}(\te) &=& |V_{\bf k}| \,
_{1,2}\lan 0| (b_{{-\bf k},1} a_{{\bf k},2} +b_{{-\bf k},2}
a_{{\bf k},1} )G_{{\bf k},\te}^{-1} |0\ran_{1,2}
\\
&=& - |V_{\bf k}| \, _{1,2}\lan 0|G_{{\bf k},\te}^{-1} ( a_{{\bf
k},2}^\dag b_{{-\bf  k},1}^\dag  + a_{{\bf k},1}^\dag b_{{-\bf
k},2}^\dag) |0\ran_{1,2} ~, \eea
where, we recall,  $|V_{\bf k}| \equiv  V_{\bf k}(0)$  in our
notation of Section II. We now consider the identity
\bea \non (b_{{-\bf k},1} a_{{\bf k},2} +b_{{-\bf k},2} a_{{\bf
k},1})G_{{\bf k},\te}^{-1} &=& G_{{\bf k},\te}^{-1} G_{{\bf
k},-\te}^{-1} (b_{{-\bf k},1} a_{{\bf k},2} +b_{{-\bf k},2}
a_{{\bf k},1}) G_{{\bf k},-\te}
\\ \non
&=& G_{{\bf k},\te}^{-1}[ b_{{-\bf k},A}(-\te) a_{{\bf k},B}(-\te)
+ b_{{-\bf k},B}(-\te) a_{{\bf k},A}(-\te)] ~. \eea
Then the equation follows
\bea \non \frac{d}{d\te} f_0^{\bf k}(\te) &=& -2\, |V_{\bf k}|^2
\cos\te \sin\te  f_0^{\bf k}(\te) +\sin^{2}\te \,|V_{\bf k}|^3 \,
_{1,2}\lan 0|G_{{\bf k},\te}^{-1} ( a_{{\bf k},2}^\dag b_{{-\bf
k},1}^\dag + a_{{\bf k},1}^\dag b_{{-\bf k},2}^\dag) |0\ran_{1,2}
\\
&=&  -2 \,|V_{\bf k}|^{2} \cos\te \sin\te f_0^{\bf k}(\te ) -
\sin^{2}\te |V_{\bf k}|^{2} \frac{d}{d\te} f_0^{\bf k}(\te) \eea
and
\bea \frac{d}{d\te} f_0^{\bf k}(\te) &=& -\frac{2 |V_{\bf k}|^{2}
\cos\te \sin\te }{1+\sin^{2}\te |V_{\bf k}|^{2}}  f_0^{\bf k}(\te)
~, \eea
which is solved by
\bea\label{A74} f_0^{\bf k}(\te) &=& \frac{1 }{1+\sin^{2}\te
|V_{\bf k}|^{2}} ~, \eea
with the initial condition $f_0^{\bf k}(0)=1$.

We observe that we can  operate in a similar fashion  directly
with $f_0^{\bf k}(\te , t) \equiv \, _{1,2}\lan 0|  G_{{\bf
k},\te}^{-1}(t)|0\ran_{1,2}$. We then find that $f_0^{\bf k}(\te ,
t)$ is again given by Eq.(\ref{A74}) and thus it is actually
time-independent. We also note that by a similar procedure it can
be proved that
 $\lim\limits_{V\rar\infty}\, {}_\AB\langle
0(\te,t)|0(\te',t)\ran_\AB \rar 0$ for $\te'\neq\te$.

\vspace{.5cm}
%%%%%%%%%%%%%%%%%%%%%%%%%%%%%%%%%%%%%%%%%%%%%%%%%%%%%%%%%%%%%%%%%%%
%

\end{document}